\documentclass[a4paper,14pt]{extreport}

\usepackage[14pt]{extsizes} 
\usepackage{cmap} 
\usepackage[T2A]{fontenc}
\usepackage[utf8]{inputenc}

\usepackage{latexsym}
\usepackage{amsmath}
\usepackage{amssymb}
\usepackage{relsize}

\usepackage{geometry}
\usepackage{tensor}
\usepackage{physics} 
\usepackage{csquotes} 

\usepackage{epsfig}
\usepackage{graphicx}       	
\graphicspath{{./img/}{./img/feynman/}{./img/numerical/}{./img/bmeson/}} 
\usepackage[font=small,labelfont=bf]{caption}
\usepackage{subcaption}
\usepackage[toc,page, title, titletoc]{appendix}

\usepackage{mathtools}
\usepackage{hyperref}
\usepackage{xcolor}
\hypersetup{
    colorlinks,
    linkcolor={blue!60!black},
    citecolor={blue!60!black},
    urlcolor={blue!80!black}
}

\usepackage{tikz-cd}
\usepackage{indentfirst} 

\makeatletter
\newcommand{\customlabel}[2]{%
   \protected@write \@auxout {}{\string \newlabel {#1}{{#2}{\thepage}{#2}{#1}{}} }%
   \hypertarget{#1}{}
}
\makeatother

\usepackage{fancyhdr}
\fancypagestyle{plain}{ 
    \fancyhf{}
    \fancyhead[C]{\thepage}
    }

\usepackage{titlesec}

\titleformat{\chapter}[display]
    {\filcenter}
    {\MakeUppercase{\chaptertitlename} \thechapter}
    {8pt}
    {\bfseries \Large}{}
 
\titleformat{\section}
    {\large\bfseries}
    {\thesection}
    {1em}{}
\titleformat{\subsection}
    {\normalsize\bfseries}
    {\thesubsection}
    {1em}{}

\usepackage{tocloft}

\makeatletter
	\newcommand{\l@likechapter}[2]{{\bfseries\@dottedtocline{0}{0pt}{0pt}{#1}{#2}}}
\makeatother

\usepackage{calc}
\newcommand{\setupname}[1]{%
  \addtocontents{toc}{%
    \unexpanded{\unexpanded{%
      \renewcommand{\cftchappresnum}{#1 }%
      \setlength\cftchapnumwidth{\widthof{\bfseries #1 }}%
      \addtolength\cftchapnumwidth{\fixedchapnumwidth}%
    }}%
  }%
}
\AtBeginDocument{\edef\fixedchapnumwidth{\the\cftchapnumwidth}}

\usepackage[figure,table]{totalcount}

\usepackage{totcount}
\newtotcounter{citnum} 
\def\oldbibitem{} \let\oldbibitem=\bibitem
\def\bibitem{\stepcounter{citnum}\oldbibitem}

\def\beqn{\begin{eqnarray}}
\def\eeqn{\end{eqnarray}}

\def\beq{\begin{equation}}
\def\eeq{\end{equation}}
\def\ba{\beq\new\begin{array}{c}}
\def\ea{\end{array}\eeq}

\def\Tr{{\rm Tr}}
\newcommand{\gsim}{\lower.7ex\hbox{$
\;\stackrel{\textstyle>}{\sim}\;$}}
\newcommand{\lsim}{\lower.7ex\hbox{$
\;\stackrel{\textstyle<}{\sim}\;$}}

\newcommand{\ntwo}{${\mathcal N}=2$ }

\newcommand{\ntwot}{${\mathcal N}= \left(2,2\right) $ }
\newcommand{\ntwoo}{${\mathcal N}= \left(0,2\right) $ }
\newcommand{\none}{${\mathcal N}=1$ }

\newcommand{\vp}{\varphi}
\newcommand{\pt}{\partial}

\newcommand{\qt}{\tilde q}
\newcommand{\bren}{{\beta_\text{ren}}}
\newcommand{\wcpt}{$\mathbb{WCP}(2,2)\;$}
\newcommand{\wcp}{$\mathbb{WCP}(N,\tilde N)\;$}
\newcommand{\cpone}{$\mathbb{CP}(1)\;$}
\newcommand{\tN}{\widetilde{N}}

\numberwithin{equation}{section}

\newcommand{\p}{\partial}
\newcommand{\wt}{\widetilde}
\newcommand{\ov}{\overline}
\newcommand{\mc}[1]{\mathcal{#1}}
\newcommand{\md}{\mathcal{D}}

\newcommand{\lgr}{\left\lgroup}
\newcommand{\rgr}{\right\rgroup}

\newcommand{\ue}{{\rm U}(1)}
\def\slashed#1{\setbox0=\hbox{$#1$}             
   \dimen0=\wd0                                 
   \setbox1=\hbox{/} \dimen1=\wd1               
   \ifdim\dimen0>\dimen1                        
      \rlap{\hbox to \dimen0{\hfil/\hfil}}      
      #1                                        
   \else                                        
      \rlap{\hbox to \dimen1{\hfil$#1$\hfil}}   
      /                                         
   \fi}                                        %

\newcommand{\sun}{{\rm SU(}N{\rm )}}

\newcommand{\mUp}{m_{\rm U(1)}^{+}}
\newcommand{\mUm}{m_{\rm U(1)}^{-}}
\newcommand{\mNp}{m_\text{SU($N$)}^{+}}
\newcommand{\mNm}{m_\text{SU($N$)}^{-}}
\newcommand{\AU}{\mc{A}^{\rm U(1)}}
\newcommand{\AN}{\mc{A}^\text{SU($N$)}}
\newcommand{\aU}{a^{\rm U(1)}}
\newcommand{\aN}{a^\text{SU($N$)}}
\newcommand{\baU}{\ov{a}{}^{\rm U(1)}}
\newcommand{\baN}{\ov{a}{}^\text{SU($N$)}}
\newcommand{\lU}{\lambda^{\rm U(1)}}
\newcommand{\lN}{\lambda^\text{SU($N$)}}

\newcommand{\nbar}{\ov{n}}

\newcommand{\CP}{$\mathbb{CP}(N-1)$~}

\begin{document}

\sloppy 


%
%

\begin{titlepage}

\begin{center}
SAINT PETERSBURG STATE UNIVERSITY

\vspace{20pt}

PETERSBURG NUCLEAR PHYSICS INSTITUTE \\
NAMED BY B.P. KONSTANTINOV OF NATIONAL RESEARCH CENTRE \\
KURCHATOV INSTITUTE
\end{center}

\vspace{10pt}

\begin{flushright}
Manuscript copyright
\end{flushright}

\vspace{30pt}

\begin{center}
Ievlev Evgenii Albertovich

\vspace{50pt}

\textbf{ \Large
Dynamics of non-Abelian strings 
}

\vspace{10pt}

\textbf{ \Large
in supersymmetric gauge theories
}

\vspace{30pt}

Specialisation 01.04.02 --- Theoretical physics

\vspace{30pt}

Dissertation is submitted for the degree \\
of Candidate of Physical and Mathematical Sciences

\vspace{30pt}

\end{center}

\vspace{40pt}

\begin{flushright}
Thesis supervisor: \\
Alexei Viktorovich Yung \\
Doctor in Physical and Mathematical Sciences
\end{flushright}

\vspace*{\fill}

\begin{center}
Saint Petersburg \\
2020
\end{center}

\end{titlepage}

\clearpage

\setcounter{page}{2}
\linespread{1.3}
\selectfont

\chapter*{Abstract}

This thesis is devoted to studying strong coupling phenomena (and confinement in particular) in supersymmetric gauge theories.
The central object of investigation is the non-Abelian string that is responsible for the "instead-of-confinement" phase for monopoles in 4D
${\mathcal N} = 2$ supersymmetric QCD with the U($N$) gauge group and $N_f$ flavors of quark hypermultiplets, $N \leqslant N_f \leqslant 2 N$.

Here it is shown that the non-Abelian strings and confined monopoles survive when we transition to the ${\mathcal N} = 1$ supersymmetric QCD.
To this end we consider a mass term $\mu$ for the adjoint matter, and in the limit of large $\mu$ the bulk theory flows to ${\mathcal N} = 1$.
We consider this transition both from the bulk point of view and from the world sheet theory, which is the two-dimensional $\mathbb{CP}(N-1)$ model. Survival of monopoles in the ${\mathcal N} = 1$ supersymmetric QCD is important for the "instead-of-confinement" phase, and also for the Seiberg-Witten picture of confinement.

Apart from that, we also consider non-Abelian vortex strings in 4D $\mathcal{N} = 2$ supersymmetric QCD with U$(N=2)$ gauge group and $N_f=4$ flavors of quark hypermultiplets. It has been recently shown that these vortices behave as critical superstrings. In particular, the lowest  string state appears to be a massless BPS "baryon." Here we show the occurrence of this stringy baryon using a purely field-theoretic method. Moreover, we explicitly demonstrate the "instead-of-confinement" phase, when the screened quarks and gauge bosons of weak coupling are replaced by the confined monopole-antimonopole pairs of strong coupling.

\vspace{30pt}

English version. Full text (Russian and English) is available at the official web page
\url{https://go.spbu.ru/20a2711}.

\clearpage

\tableofcontents
\clearpage

\setupname{Chapter}

%
%

\chapter*{Introduction} 
\addcontentsline{toc}{likechapter}{Introduction}

Understanding confinement phenomenon is one of the major unsolved problems in modern theoretical physics.
This phenomenon is a feature of strongly interacting particles, and the underlying mechanism is still undetermined.

Quantum chromodynamics (QCD) is a well-established theory of strong interactions. 
Confinement of quarks and gluons, or color confinement, is a low energy effect, but at low energies QCD is in the strong coupling regime. This circumstance is a serious obstruction to a detailed theoretical study of the confinement phenomenon from the point of view of the QCD itself. 
However, one of the promising approaches that can nevertheless help us to understand strong coupling phenomena is consideration of supersymmetric cousins of QCD.

%
%


In the seminal works of Seiberg and Witten \cite{SW1,SW2} it was shown that in theories with \ntwo supersymmetry it is possible to observe monopole condensation.
This is a realization of the so-called dual Mei{\ss}ner effect suggested by 't Hooft and Mandelstam \cite{tHooft:1981bkw,Mandelstam:1974pi}.
The effect is that when magnetic charges condense, the electric field between two probes is compressed in a thin tube, which results in a linear potential between the electric charges.
However, confinement in this model is inherently Abelian.

{\em Non-Abelian flux tubes} (strings) were discovered in \ntwo supersymmetric quantum chromodynamics (SQCD) with the gauge group U$(N)$ and $N_f=N$ flavors of quark hypermultiplets \cite{HT1,ABEKY,SYmon,HT2} (see also \cite{Trev,Jrev,SYrev,Trev2} for a review).
When the theory under consideration is in the Higgs phase with respect to the scalar quarks (i.e. in the so-called quark vacuum), non-Abelian strings are formed.
They lead to the confinement of monopoles at weak coupling, and to the so-called \textquote{instead-of-confinement} phase at strong coupling, see \cite{SYdualrev} for a review.
Thus, it is a non-Abelian generalization of the Seiberg-Witten mechanism \cite{SW1,SW2}.

Apart from usual translational zero modes that are characteristic of Abrikosov-Nielsen-Olesen (ANO) vortices \cite{ANO}, the non-Abelian strings possess also orientational zero modes. 
Dynamically these new modes can be described by an \ntwot supersymmetric sigma model with \CP target space.
Of course, the coordinate space of this model is the two-dimensional string world sheet \cite{HT1,ABEKY,SYmon,HT2}. 

Since the bulk SQCD is in the Higgs phase for scalar quarks monopoles are confined by non-Abelian strings.
However, 
the monopoles cannot be attached to the string endpoints. In fact, in the U$(N)$ theories confined  
 monopoles 
are  junctions of two distinct elementary non-Abelian strings. From the point of view of 
\CP model living on the string world sheet confined monopoles are seen as kinks interpolating between
different vacua of \CP model \cite{SYmon,HT2,Tong} (see \cite{SYrev} 
for a review).

{\em The aim} of this work is to generalize these constructions to the theories with less supersymmetry, and also
to deepen the understanding of the non-Abelian string in the \ntwo theory. 
We start with the first of these goals.

The \ntwo supersymmetric quantum chromodynamics is a nice theoretical laboratory to study non-perturbative non-Abelian dynamics. However, since we wish to learn more about the \textquote{real world}, we are interested in studying more realistic theories. \none
supersymmetric QCD is one of the most promising examples. Much in the same way as the real world QCD it has no adjoint scalars and no Abelianization of the theory can occurs due to their condensation.

A lot of work has been done to generalize the construction of non-Abelian strings to QCD-like theories with less supersymmetry, in particular to \none SQCD \cite{SYnone,Edalati,SY02,YIevlevN=1} see \cite{SYrev} for a review.
The author of this thesis also had a hand in these developments, see
\cite{YIevlevN=1,Ievlev:2018rxo,Ievlev:2018xub,Gorsky:2019shz}.
One promising approach is to deform \ntwo SQCD by the mass $\mu$ of the adjoint matter (one then obtains the so-called $\mu$-deformed SQCD) and study what happen
to non-Abelian strings upon this deformation. This deformation breaks \ntwo supersymmetry. 
In the limit
$\mu\to\infty$ 
the adjoint matter decouples and the theory flows to \none QCD.

%
%


We started this journey in Chapter~\ref{sec:none} from the simplest case when the number of flavors of quark hypermultiplets is the same as the number of colors, $N_f = N$.
The $\mu$-deformed \ntwo SQCD equipped with the Fayet-Iliopoulos (FI) $D$-term was already considered in the literature \cite{SYnone,Edalati,SY02,Tongd,SYhetN,BSYhet}. In this case, the solitonic vortex string saturates the Bogomol'nyi–Prasad–Sommerfield (BPS) bound, which simplifies the analysis.
However, in the large $\mu$ limit this theory does not flow to the \none SQCD.

Here we take a different route
and consider the $\mu$-deformation of \ntwo QCD without a FI term in a quark vacuum.
It is more \textquote{realistic} theory, since there is no FI term in the \none supersymmetric QCD. 
And indeed, in the large $\mu$ limit this deformed theory flows to the \none SQCD.
%
%
The squark condensate here is triggered by $\sqrt{\mu m}$, were $m$ is a quark mass. This makes the non-Abelian strings to loose their BPS saturation property. This makes it a lot harder to investigate such solitons, but still it can be done.

The question of the crucial physical importance is whether monopoles survive the limit
of large $\mu$ when the the bulk theory flows to \none QCD.
From a quasiclassical point of view, the very existence of 't Hooft-Polyakov monopoles relies on the presence of
adjoint scalars which develop vacuum expectation values (VEV). 
These adjoint scalar VEVs make possible such solitonic solutions of classical equations of motion.
The adjoint fields are also essential in the Seiberg-Witten picture, where the adjoint field VEV leads to formation of monopoles, which in turn condense and are responsible for confinement.
At large $\mu$ adjoint fields become heavy and decouple in our 
bulk theory, and their VEVs go to zero. So, quasiclassically we do not expect monopoles to survive.

In the large $\mu$ limit we managed to derive the effective theory on the string world sheet. Translational sector again trivial, but something happens with the orientational modes of the string. 
Turns out, that while the bosonic sector of this theory is still given by the \CP model, the fermionic sector completely decouples. This happens because the string superorientational fermionic zero modes acquire mass and become lifted. This ensures that the
world sheet theory is in the Coulomb/confinement phase, at least at large $N$, see \cite{W79}.
Moreover, quark mass differences induce a potential in the effective theory, which effectively destroys the monopoles. Therefore, in order for monopoles to survive, the bulk quarks must have equal masses.

These results show that 
non-Abelian strings and confined monopoles of the $\mu$-deformed \ntwo SQCD can survive the large $\mu$ limit when the bulk theory flows to the \none SQCD, which is an important and somewhat unexpected result. 
It serves as an evidence for a physically important conclusion made previously
(see e.g. \cite{SYdualrev} for a review) that the \textquote{instead-of-confinement} phase survives the
large $\mu$ limit in the quark vacuum of deformed SQCD.

Next, we move on and consider the case $N_f > N$ in Chapter~\ref{sec:semiloc}. Here, the non-Abelian string acquires new size moduli and turns into the so-called semilocal string.
We study the fate of the semilocal string in the $\mu$-deformed SQCD. As could be anticipated, we find that the solitonic vortex under consideration is again no longer BPS saturated. The world sheet theory is no longer supersymmetric.

Somewhat surprising is the fact, that the \textquote{semilocality} of the string is also lost. The string size moduli acquire a potential and become heavy, which makes the string to shrink and become \textquote{local}. In the large $\mu$ limit the world sheet theory becomes exactly the same as in the $N_f = N$ case.
And again, the presence of monopoles connected to non-Abelian string supports the \textquote{instead-of-confinement} picture.

%
%


{\em Next logical step} on this road is a closeup consideration of the world sheet effective theory.
This is done in Chapter~\ref{sec:large_N} by means of the $1/N$ expansion.
Large $N$ approximation was first used by Witten to solve both non-supersymmetric 
and \ntwot supersymmetric two-dimensional \CP models \cite{W79}. 

Here we use the large $N$ approximation to study  a  phase structure of the world sheet theory on the non-Abelian string in $\mu$-deformed SQCD with respect to the deformation parameter $\mu$ and quark
mass differences $\Delta m$. We find a rich phase structure which  includes two strong coupling phases and two Higgs phases. 

The \ntwot theory has a family of degenerate vacua, which correspond to different non-Abelian strings in the bulk.
It turns out that if we start from small $\mu$ and go on to increase the deformation parameter, the world sheet theory inevitably goes through one or more phase transitions. 

At large $\Delta m$ the former degenerate vacua split and become quasivacua, which eventually disappear when the deformation parameter $\mu$ is sufficiently large. 
On the contrary, when we set $\Delta m$ to zero, the splitting quasivacua do not disappear. Even in the large $\mu$ limit the theory still has $N$ quasivacua corresponding to non-Abelian strings with different tensions. The kinks interpolating between these vacua survive. 

This allows us to conclude that the confined monopoles survive the $\mu$-deformation if the quark masses are equal to each other.
Thus we confirm the results obtained from the bulk SQCD. This also serves as a consistency check of our approach.

%
%


{\em The second goal} pursued in this work is a better understanding of non-Abelian strings in the \ntwo case \cite{Ievlev:2020qch}.

Consider the \ntwo SQCD with the U$(N=2)$ gauge group, $N_f=4$ flavors of quarks and a Fayet-Iliopoulos $D$-term \cite{FI}.  
It was discovered earlier in \cite{SYcstring} that the non-Abelian semilocal string in this theory is very special.  
In the bulk, the gauge coupling renormalization exactly cancels, and the $\beta$-function is zero.
The world sheet theory turns out to be both superconformal and critical.

From the analysis of the non-Abelian vortex, a \textquote{thin string} hypothesis was put forward \cite{SYcstring}.
Basically it states that in the strong coupling limit, the vortex string transverse size goes to zero, and the string can be treated as the critical superstring. This allows one to apply the advanced machinery of string theory to calculating e.g. the spectrum of states in this theory. The vortex string at hand was identified as the string theory of Type IIA \cite{KSYconifold}.

Hadrons of the \ntwo SQCD are pictured as closed string states\footnote{There are no open strings in our theory since the non-Abelian string cannot end on a monopole. Instead, the monopole always is a junction of two vortices. This is a fortunate circumstance; otherwise, we would have \none supersymmetry in four dimensions instead of \ntwo.}.
In particular, in \cite{KSYconifold,KSYcstring} a massless hypermultiplet was found, which was identified with a baryon of the four dimensional \ntwo SQCD. It was called the $b$-baryon.

In view of these stringy results, we would like to test and explain them from the field theory point of view. In the present work we do just that, see Chapter~\ref{sec:b_meson}.
In order to do that we exploit the so-called 2D-4D correspondence, i.e. the coincidence of BPS spectra in four-dimensional (4D) \ntwo SQCD and in the string world-sheet theory \cite{SYmon,HT2,Dorey}. This allows us to essentially study only the two dimensional effective theory and then translate the results into the four dimensional language.

We explore the BPS protected sector of the world-sheet \wcpt model, starting from the weak coupling regime where we can compare with quasiclassical results. After we've established firm ground there, we progress into the strong coupling regime. We confirm that the theory enters the so-called \textquote{instead-of-confinement} phase found earlier, see \cite{SYdualrev} for a review.
This phase  is qualitatively similar  to
the conventional QCD confinement: the quarks and gauge bosons screened at
weak coupling, at strong coupling evolve into monopole-antimonopole pairs
confined by non-Abelian strings. They form mesons and baryons.

At very strong coupling a new short BPS massless hypermultiplet arises, which turns out to be the $b$-baryon found earlier from the string theory picture.   
In this way we demonstrate that the massless \textquote{baryon} state which had been previously observed using string theory
arguments \cite{KSYconifold} is seen in the field-theoretical approach too. We believe this is the first example of this type.

These results also serve as another confirmation of the \textquote{thin string hypothesis} mentioned above. Treatment of the solitonic vortex as a critical superstring appears to be consistent. 
Thus, on this strong coupling voyage, we seem to have pretty good paddles to travel.

%
%



%
%

\section*{The statements and results put forward for defense}

The statements and results are:
\begin{enumerate}
\item 
It is shown that
non-Abelian strings and confined monopoles of the $\mu$-deformed \ntwo SQCD can survive the large $\mu$ limit when the bulk theory flows to the \none SQCD. 
%
They survive if the SQCD quarks have equal masses. 

\item 
It is shown that the low energy effective theory on the world sheet of the non-Abelian string in the \none SQCD with the U($N$) gauge group and $N_f = N$ quark hypermultiplets is the non-supersymmetric sigma model with the \CP target space in the orientational sector. The translational sector is trivial and decoupled. 

\item 
It is shown that
the semilocal string of the $\mu$-deformed \ntwo SQCD with $N < N_f < 2N$ degenerates in the large $\mu$ limit when the bulk theory flows to the \none SQCD. Namely, the size moduli develop a potential and decouple, while the semilocal string becomes a local one. The world sheet theory in this limit is the same as in the $N_f = N$ case.

\item The low energy effective theory on the world sheet of the non-Abelian string in the theory interpolating from \ntwo to \none SQCD  is solved to the leading order in the large $N$ approximation. The solution of this world sheet model confirms the results derived from the bulk theory, namely, that the 
non-Abelian strings and confined monopoles survive in the limit of equal quark masses. 
Moreover, the phase diagram of the world sheet model is obtained. 

\item The massless $b$-baryon hypermultiplet of the \ntwo SQCD with the U(2) gauge group and $N_f = 4$ flavors of quark hypermultiplets, derived previously using the string theory approach, is found here by pure field theory methods. This is yet another confirmation of the \textquote{thin string} hypothesis for the non-Abelian string in this theory. 

\item The \textquote{instead-of-confinement} mechanism is demonstrated explicitly in the \ntwo SQCD with the U(2) gauge group and $N_f = 4$ flavors of quark hypermultiplets. It is shown that, after a wall-crossing, the screened quarks and gauge bosons of weak coupling are replaced by the confined monopole-antimonopole pairs of strong coupling.
\end{enumerate}

%
%

\section*{Thesis structure}

The thesis consists of Introduction, five Chapters, Conclusion, seven Appendices and a list of references. 
The thesis contains 
\pageref{last_one_ever}
pages, \totalfigures\ figures. The list of references includes \total{citnum} items.
\begin{itemize}
\item In {\bf Introduction} we describe the general idea of this thesis. Also, the main statements to defend are formulated, and the approbation of this research is discussed.

\item In {\bf Chapter~\ref{sec:ntwo}} we review the necessary background on non-Abelian string in supersymmetric gauge theories.

\item In {\bf Chapter~\ref{sec:none}} we consider the $\mu$-deformed \ntwo SQCD with the number of colors equal to the number of flavors, $N_f = N$. We focus on the large $\mu$ limit, when the theory flows to the \none SQCD. We investigate the fate of the non-Abelian strings and confined monopoles in this limit.

\item In {\bf Chapter~\ref{sec:semiloc}} we generalize the construction of Chapter~\ref{sec:none} and move on to the $N_f > N$ case. We investigate what happens to the semilocal non-Abelian strings in the \none limit.

\item {\bf Chapter~\ref{sec:large_N}} presents the study of the low energy effective theory on the non-Abelian string of SQCD that interpolates between \ntwo and \none supersymmetry.
We focus on the $N_f = N$ case. Using the large $N$ expansion, we solve the world sheet theory and obtain a phase diagram of the model.

\item In {\bf Chapter~\ref{sec:b_meson}} we move in another direction and consider again the \ntwo bulk theory, that is \ntwo SQCD with the U(2) gauge group and $N_f = 4$ quark hypermultiplets. In this case, the world sheet theory is superconformal. We find there a massless state corresponding to the $b$-baryon found previously by using string theory methods. We also study the \textquote{instead-of-confinement} mechanism in action.

\item {\bf Conclusion} presents the main results of this work and outlines possible future directions.

\item Quite extensive {\bf Appendices} contain useful information and additional results. They explain and complement some of the material presented in the core of this thesis, while at the same time not obstructing reading of the main text.

\end{itemize}

%
%

\section*{Personal contribution of the author}

All of the main findings submitted for defense were obtained personally by the applicant or in work	of joint authorship.

%
%

\section*{Approbation of the research}

The findings of the investigation were reported and discussed at the following conferences:
\begin{enumerate}
\item 2015 09-12 November, SPbSU: International Student Conference "Science and Progress"
\item 2016 29 February - 05 March, Roschino, Russia: 50th PNPI Winter School
\item 2016 17-21 October, SPbSU: International Student Conference "Science and Progress"
\item 2018 27 May - 2 June, Valday, Russia: XXth International Seminar on High Energy Physics "Quarks-2018"
\item 2018 14-23 June, Erice, Italy: 56th Course of the International School of Subnuclear Physics "From gravitational waves to QED, QFD and QCD"
\item 2018 27-31 August, SPbSU: VI International Conference "Models in Quantum Field Theory" 
\item 2019 2-7 March, Roschino, Leningrad Oblast, Russia: 53th PNPI Winter School
\item 2019 21-30 June, Erice, Italy: 57th Course of the International School of Subnuclear Physics on "In Search for the Unexpected"
\item 2020 10-15 March, Roschino, Leningrad Oblast, Russia: 54th PNPI Winter School
\item 2020 13-24 July, online: "QFT and Geometry Summer School" 
\item 2020 24-28 August, online: "Hamilton School on Mathematical Physics"
\item 2020 9-13 November, online: "The XXIV International Scientific Conference of Young Scientists and Specialists"
\item 2020 16-20 November, online: International Conference "YITP workshop Strings and Fields"
\end{enumerate}
In addition, the results were reported and discussed at the meetings of the Theory Division of the NRC Kurchatov Institute --- PNPI 
and at the meetings of the 
High Energy and Elementary Particles Physics Department
Department of Saint Petersburg State University.

The results obtained within this study were published in 5 articles (and are included in the RSCI, Web of Science and Scopus databases):
\begin{enumerate}
\item E. Ievlev, A. Yung,
{\em Non-Abelian strings in N=1 supersymmetric QCD,}
Phys. Rev. D \textbf{95}, 125004 (2017)
\item E. Ievlev, A. Yung,
{\em  What Becomes of Semilocal non-Abelian strings in N=1 supersymmetric QCD,}
Phys. Rev. D \textbf{98}, 094033 (2018)
\item E. Ievlev, A. Yung,
{\em  Non-Abelian strings in N=1 supersymmetric QCD (Conference Paper),}
EPJ Web of Conferences \textbf{191}, 06003 (2018)
\item A. Gorsky, E. Ievlev, A. Yung,
{\em  Dynamics of non-Abelian strings in the theory interpolating from N=2 to N=1 supersymmetric QCD,}
Phys. Rev. D \textbf{101}, 014013 (2020)
\item E. Ievlev, M. Shifman, A. Yung,
{\em  String Baryon in Four-Dimensional N=2 Supersymmetric QCD from the 2D-4D Correspondence,}
Phys. Rev. D \textbf{102}, 054026 (2020)
\end{enumerate}

%
%

\section*{Acknowledgments}

The author would like to express his deepest gratitude to his advisor Dr. Alexei Viktorovich Yung for his patience, friendly attitude and support.
Over the years he guided me towards understanding modern theoretical physics and shared countless ideas.
The applicant would also like to thank his article co-authors, Mikhail Arkadyevich Shifman and Aleksandr Sergeevich Gorsky, without whom this work would also have been hardly possible.

This work was supported by the Foundation for the Advancement of Theoretical Physics and Mathematics \textquote{BASIS} according to the research project No. 19-1-5-106-1 \textquote{PhD Student}, 
and by the Russian Foundation for Basic Research (RFBR) according to the research projects No. 18-32-00015 (for young scientists) and No. 18-02-00048. 
The author wishes that RFBR would continue to support small research groups and ambitious young scientists.

The applicant appreciates the hospitality of Saint Petersburg State University and NRC Kurchatov Institute --- Petersburg Nuclear Physics Institute, where this work was carried out.
The applicant also thanks the members of the PNPI Theory Division and SPbSU High Energy and Elementary Particles Physics Department, who participated in discussions of the results presented here.

%
%

\chapter{Review of non-Abelian strings in \ntwo supersymmetric QCD} \label{sec:ntwo}

This Chapter complements the Introduction.
It presents a more detailed background on the subject of the non-Abelian strings.

\section{Four-dimensional \boldmath{${\mathcal N}=2\;$} SQCD}
\label{SQCD}

Non-Abelian vortex strings were first found in 4D
\ntwo SQCD with the gauge group U$(N)$ and $N_f \ge N$ quark flavors  
\cite{HT1,ABEKY,SYmon,HT2}, see  \cite{Trev,Jrev,SYrev,Trev2} for   review.
In particular, the matter sector of  the U$(N)$ theory contains
 $N_f$ quark hypermultiplets  each consisting
of   the complex scalar fields
$q^{kA}$ and $\widetilde{q}_{Ak}$ (squarks) and
their  fermion superpartners -- all in the fundamental representation of 
the SU$(N)$ gauge group.
Here $k=1,..., N$ is the color index
while $A$ is the flavor index, $A=1,..., N_f$. We also introduce quark masses $m_A$.
In addition, we introduce the Fayet–Iliopoulos parameter $\xi$ corresponding to the $D$-term in the U(1) factor of the gauge group.
It does not break \ntwo supersymmetry.

At weak coupling, $g^2\ll 1$ (here $g^2$ is the SU$(N)$ gauge coupling), this theory is in the Higgs regime in which squarks develop vacuum 
expectation values.
The squark VEV's  are  
\beqn
\langle q^{kA}\rangle &=& \sqrt{\xi}\,
\left(
\begin{array}{cccccc}
1 & \ldots & 0 & 0 & \ldots & 0\\
\ldots & \ldots & \ldots  & \ldots & \ldots & \ldots\\
0 & \ldots & 1 & 0 & \ldots & 0\\
\end{array}
\right), \qquad  \langle\bar{\widetilde{q}}^{kA}\rangle= 0,
\nonumber\\[4mm]
k&=&1,..., N\,,\qquad A=1,...,N_f\, ,
\label{Nf_qvev}
\eeqn
where the squark fields are presented as matrices in the color ($k$) and flavor ($A$) indices (the small Latin letters mark the lines in
this matrix while capital letters mark the rows).

These VEVs break 
the U$(N)$ gauge group. As a result, all gauge bosons are
Higgsed. The Higgsed gauge bosons combine with the screened quarks to form 
long \ntwo multiplets, with the mass 
\beq
m_G \sim g\sqrt{\xi}\,.
\label{ximG}
\eeq

In addition to  the U$(N)$ gauge symmetry, the squark condensate (\ref{Nf_qvev}) 
breaks also the flavor SU$(N_f)$ symmetry. If the quark masses vanish,
a diagonal global SU$(N)$ combining the gauge SU$(N)$ and an
SU$(N)$ subgroup of the flavor SU$(N_f)$
group survives, however.  This is a well known phenomenon of color-flavor locking. 

Thus, the unbroken global symmetry of our 4D SQCD is  
\beq
 {\rm SU}(N)_{C+F}\times {\rm SU}(\tN)\times {\rm U}(1)_B\,.
\label{globgroup_d=4}
\eeq
Above,  
$$\tN=N_f-N\,.$$ 
This U(1) in (\ref{globgroup_d=4}) is associated with $\tN$ quarks with the flavor indices $A= N+1, \, N+2, ..., N_F$, see \cite{SYrev} for more details. 
More exactly,  our U(1)$_B$
is  an unbroken (by the squark VEVs) combination of two U(1) symmetries:  the first is a subgroup of the flavor 
SU$(N_f)$ and the second is the global U(1) subgroup of U$(N)$ gauge symmetry.

 The unbroken global U(1)$_B$ factor in Eq. (\ref{globgroup_d=4})  is identified with a \textquote{baryonic} symmetry. Note that 
what is usually identified as the baryonic U(1) charge is a {\em part} of  our 4D SQCD  {\em gauge} group.

The 4D theory has a Higgs branch ${\cal H}$ formed by massless quarks which are in  the bifundamental representation
of the global group \eqref{globgroup_d=4} and carry baryonic charge, see \cite{KSYconifold} for more details.
The dimension of this branch is 
\beq
{\rm dim}\,{\cal H}= 4N \tN.
\label{dimH}
\eeq
This perturbative Higgs branch is an exact property of the theory and can be continued all the way to strong coupling.
 
\begin{figure}
\centerline{\includegraphics[width=10cm]{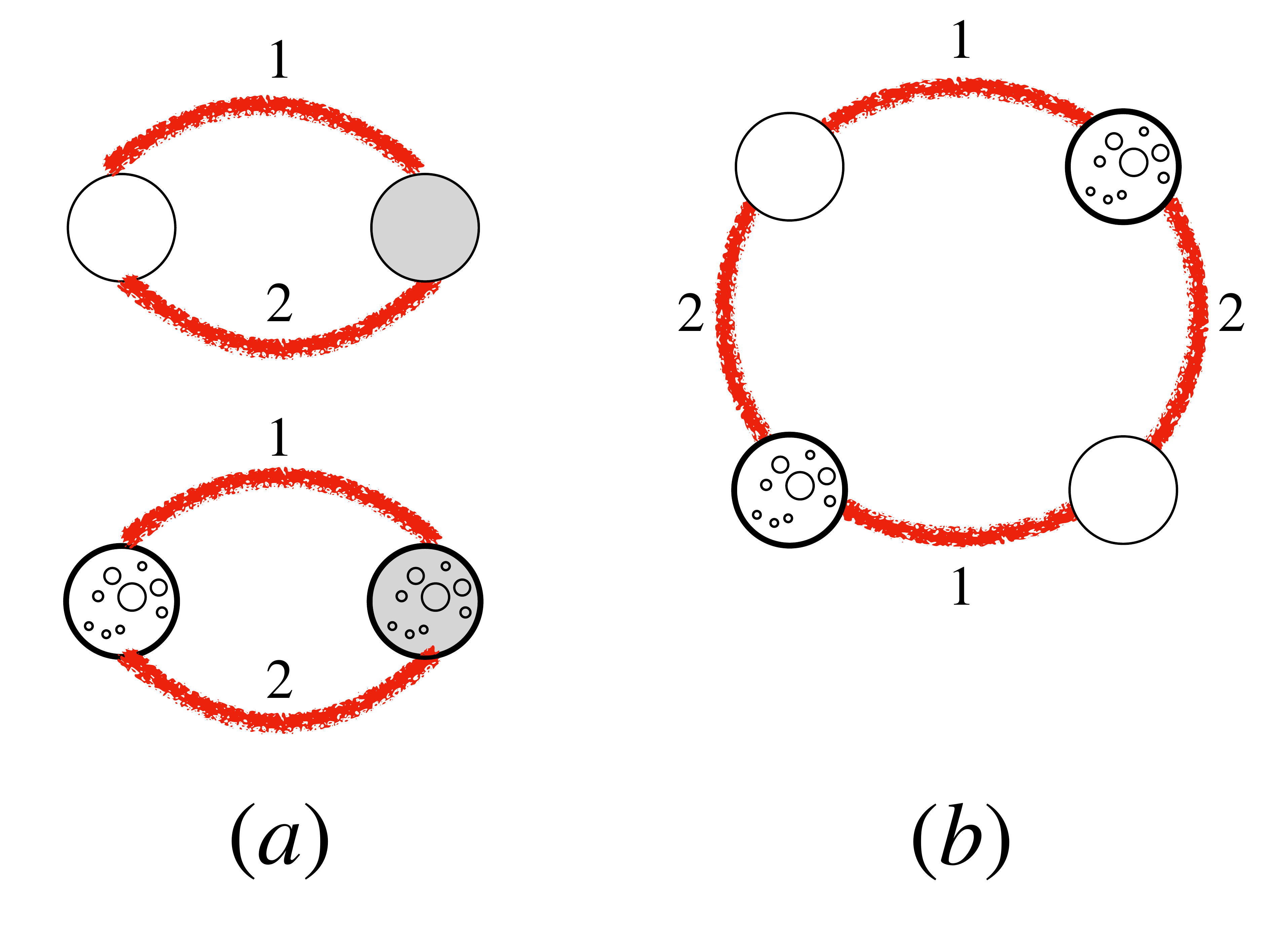}}
\caption{ Examples of the monopole \textquote{necklaces}: (a) mesonic; (b) baryonic. 1,2 refer to two types of strings corresponding to two vacua on the 
string world sheet. The shaded circles are antimonopoles.
The two types of kinks are the $n^P$-kinks and $\rho^K$-kinks.}
\label{monmb} 
\end{figure}

As was already noted, we consider \ntwo SQCD  in the Higgs phase:  $N$ squarks  condense. 
Therefore, the non-Abelian 
vortex strings at hand confine monopoles. In the \ntwo 4D bulk theory the above strings are 1/2 BPS-saturated; hence,  their
tension  is determined  exactly by the FI parameter,
\beq
T=2\pi \xi\,.
\label{ten_N=2}
\eeq
However, 
the monopoles cannot be attached to the string endpoints because in U$(N)$ theories strings are topologically stable. In fact, in the U$(N)$ theories confined  
 monopoles 
are  junctions of two distinct elementary non-Abelian strings \cite{SYmon,HT2,T} (see \cite{SYrev} 
for a review). As a result,
in  4D \ntwo SQCD we have 
monopole-antimonopole mesons in which monopole and antimonopole are connected by two confining strings, see Fig.~\ref{monmb}a.
 In addition, in the U$(N)$  gauge theory we can have baryons  appearing as  a closed 
\textquote{necklace} configurations of $N\times$(integer) monopoles \cite{SYrev}. For the U(2) gauge group the 
important example of a baryon consists of four monopoles as shown in Fig.~\ref{monmb}b.

Both stringy monopole-antimonopole mesons and monopole baryons with spins $J\sim 1$ have masses determined 
by the string tension,  $\sim \sqrt{\xi}$ and are heavier at weak coupling $g^2\ll 1$ than perturbative states with masses
$m_G\sim g\sqrt{\xi}$. 
Thus they can decay into perturbative states \footnote{Their quantum numbers with respect to the global group 
\eqref{globgroup_d=4} allow these decays, see \cite{SYrev}.} and in fact at weak coupling we do not 
expect them to appear as stable  states.

Only in the   strong coupling domain $g^2\sim 1$  we can expect that (at least some of) stringy mesons and baryons
shown in Fig.~\ref{monmb} become stable. 

\section{World-sheet sigma model}
\label{sec:wcp}

The presence of the color-flavor locked group SU$(N)_{C+F}$ is the reason for the formation of the
non-Abelian vortex strings \cite{HT1,ABEKY,SYmon,HT2}.
The most important feature of these vortices is the presence of the  orientational  zero modes.
As was already mentioned, in \ntwo SQCD these strings are 1/2 BPS saturated. 

Let us briefly review the model emerging on the world sheet
of the non-Abelian  string \cite{SYrev}.

The translational moduli fields  are described by the Nambu–Goto action\,\footnote{In the supersymmetrized form.} and  decouple from all other moduli. Below we focus on
 internal moduli.

If $N_f=N$  the dynamics of the orientational zero modes of the non-Abelian vortex, which become 
orientational moduli fields 
 on the world sheet, are described by two-dimensional (2D)
\ntwot supersymmetric ${\mathbb{CP}}(N-1)$ model.

If one adds additional quark flavors, non-Abelian vortices become semilocal --
they acquire size moduli \cite{AchVas}.  
For the non-Abelian semilocal vortex in U($N$) \ntwo SQCD with $N_f$ flavors,  in 
addition to  the complex orientational moduli  $n^P$ (here $P=1,..,N$), we must add the size moduli   
$\rho^K$ (where $K=N+1,..,N_f$), see \cite{HT1,HT2,AchVas,SYsem,Jsem,SVY}. The size moduli are also 
complex.  

The effective theory on the string world sheet is a two-dimensional \ntwot weighted \CP sigma model, which we denote
$\mathbb{WCP}(N,\tilde{N})\;$ \footnote{Both the orientational and the size moduli
have logarithmically divergent norms, see e.g.  \cite{SYsem}. After an appropriate infrared 
regularization, logarithmically divergent norms  can be absorbed into the definition of 
relevant two-dimensional fields  \cite{SYsem}.
In fact, the world-sheet theory on the semilocal non-Abelian string is 
not exactly the \wcp  model \cite{SVY}, there are minor differences. The actual theory is called the $zn$ model. Nevertheless it has the same infrared physics as the model (\ref{wcpNtN}) \cite{KSVY}, see also \cite{CSSTY}.} %
\cite{SYcstring,KSYconifold,KSYcstring}%
.
This model describes internal dynamics of the non-Abelian semilocal string. 
For details see e.g. the review \cite{SYrev}.

The $\mathbb{WCP}(N,\tilde{N})\;$ sigma model 
can be  defined  as a low energy limit of the  U(1) gauge theory \cite{W93}.  The bosonic part of the action reads
 \footnote{Equation 
(\ref{wcpNtN}) and similar expressions below are given in Euclidean notation.}
\begin{equation}
\begin{aligned}
	&S = \int d^2 x \left\{
	\left|\nabla_{\alpha} n^{P}\right|^2 
	+\left|\tilde{\nabla}_{\alpha} \rho^K\right|^2
	+\frac1{4e^2}F^2_{\alpha\beta} + \frac1{e^2}\,
	\left|\pt_{\alpha}\sigma\right|^2
	\right.
	\\[3mm]
	&+\left.
	2\left|\sigma+\frac{m_P}{\sqrt{2}}\right|^2 \left|n^{P}\right|^2 
	+ 2\left|\sigma+\frac{m_{K}}{\sqrt{2}}\right|^2\left|\rho^K\right|^2
	+ \frac{e^2}{2} \left(|n^{P}|^2-|\rho^K|^2 - r \right)^2
	\right\},
	\\[4mm]
	&
	P=1,..,N\,,\qquad K=N+1,..,N_f\,.
\end{aligned}
\label{wcpNtN}
\end{equation}
Here, $m_A$ ($A=1,..,N_f$) are the so-called twisted masses (they come from 4D quark masses),
while $r$ is the inverse coupling constant (2D FI term). Note that $r$ is the real part of the complexified coupling constant 
introduced in Eq. (\ref{beta_complexified}), $$r = {\rm Re}\,\beta\,.$$

The fields $n^{P}$ and $\rho^K$ have
charges  $+1$ and $-1$ with respect to the auxiliary U(1) gauge field, and the corresponding  covariant derivatives in (\ref{wcpNtN}) are defined as 
\begin{equation}
	\nabla_{\alpha}=\p_{\alpha}-iA_{\alpha}\,,
	\qquad 
	\tilde{\nabla}_{\alpha}=\p_{\alpha}+iA_{\alpha}\,,	
\end{equation}
respectively. The complex scalar field $\sigma$ is a superpartner of the U(1) gauge field $A_{\alpha}$.

The number of real bosonic degrees of freedom in the model \eqref{wcpNtN} is $2N_f-1-1=2(N_f-1)$.  Here $2N_f$ is the number of real degrees of 
freedom of $n^P$ and $\rho^K$ fields and we subtracted one real constraint imposed by the last  term in \eqref{wcpNtN} in the limit $e^2\to \infty$  and one  gauge phase eaten by the Higgs mechanism.

Apart from the U(1) gauge symmetry, the sigma model (\ref{wcpNtN}) in the massless limit has a global symmetry group
\begin{equation}
	 {\rm SU}(N)\times {\rm SU}(\tN)\times {\rm U}(1)_B \,,
\label{globgroup}
\end{equation}
i.e. exactly the same as the unbroken global group in the 4D theory \eqref{globgroup_d=4}. 
The fields $n$ and $\rho$ 
transform in the following representations:
\begin{equation}
	n:\quad \left(\textbf{N},\,\textbf{1},\, 0\right), \qquad \rho:\quad \left(\textbf{1},\, \widetilde{\textbf{N}},\, 1\right)\,.
\label{repsnrho}
\end{equation}
%
Here  the global \textquote{baryonic}  U(1)$_B$ symmetry is a classically unbroken (at $\beta >0$) combination of the 
global U(1) group which
rotates $n$ and $\rho$ fields with the same phases plus  U(1) gauge symmetry which  rotates them with the opposite phases, see
\cite{KSYconifold} for details.
Non-zero twisted masses $m_A$ break each of the SU factors in \eqref{globgroup} down to a product of U(1)'s.

The 2D coupling constant $r$ can be naturally complexified if we
include the $\theta$ term in the action,
\begin{equation}
	\beta = r + i \, \frac{\theta_{2d}}{2 \pi} \,,
\label{beta_complexified}	
\end{equation}
where $\theta_{2d}$ is the two-dimensional $\theta$ angle.

\section{2D-4D correspondence}
\label{2-4}

Previous studies of the vortex strings  supported in four-dimensional  ${\cal N} = 2$ super-QCD at weak coupling showed that the non-Abelian vortices confine monopoles. The elementary monopoles are junctions of two distinct elementary non-Abelian strings \cite{SYmon,HT2}. In the 4D bulk theory 
we have monopole-antimonopole mesons in which monopole and antimonopole are connected by two confining strings 
(see Fig.~\ref{monmb}). For the U($N$) gauge group we can have also \textquote{baryons} consisting of $N\times$number of  monopoles rather than of the monopole-antimonopole pair. 

The monopoles acquire quantum numbers with respect to the global group \eqref{globgroup_d=4}
of the 4D SQCD, see \cite{SYrev} for a review. Indeed, in the world-sheet model on the vortex string, confined monopole are seen as kinks interpolating between two different vacua \cite{SYmon,HT2}. These kinks are described at strong coupling by  $n^P$ and $\rho^K$ fields \cite{W79,SYtorkink} (for $\mathbb{WCP}(N,\tilde{N})\;$ model $P=1,..,N$, $K=N+1,..,N_f$).
These two types of kinks correspond to two types of monopoles -- both have the same magnetic charge but different global charges. This is seen from the fact that
the global symmetry in the world-sheet theory on the string is exactly the same as given in Eq. (\ref{globgroup_d=4})
and the U(1) charges of the $n^P$ and $\rho^K$ fields are 0 and 1, respectively.
 One of them   is a fundamental field in the first SU group and the other  in the second, see \eqref{repsnrho}.
This refers to confined 4D monopoles too.

As was mentioned above confined monopoles of 4D SQCD are junctions of two different elementary non-Abelian strings. In  the  world-sheet theory they are seen as kinks interpolating between different vacua of
$\mathbb{WCP}(N,\tN)$ model. This ensures 2D-4D correspondence: the coincidence between the BPS spectrum of 
monopoles in 4D SQCD at a particular singular point on the Coulomb branch (which becomes  the quark vacuum \eqref{Nf_qvev} once we introduce non-zero $\xi$)  and the spectrum of kinks in 2D $\mathbb{WCP}(N,\tN)$ model. The masses of (dyonic) monopoles in  4D SQCD are 
given by the exact Seiberg-Witten solution \cite{SW2}, while the kink spectrum in $\mathbb{WCP}(N,\tN)$ model can be 
derived from exact twisted effective superpotential  \cite{Dorey,W93,AdDVecSal,ChVa,HaHo,DoHoTo}. This effective superpotential is written in terms of the twisted chiral superfield which has the complex scalar  field $\sigma$ 
(see \eqref{wcpNtN})
as its lowest component \cite{W93}, see Sec.~\ref{sec:kink_mass} where we introduce this superpotential and study
the kink spectrum for the $\mathbb{WCP}(N,\tilde{N})\;$ model.

This coincidence was observed in \cite{Dorey,DoHoTo} and   explained later 
in \cite{SYmon,HT2} using the picture of confined bulk monopoles which are seen as kinks in the world 
sheet theory. A crucial point is that both the monopoles and the kinks are BPS-saturated states\,\footnote{Confined
 monopoles, being junctions of two distinct 1/2-BPS strings, are 1/4-BPS states in 4D SQCD 
\cite{SYmon}.},
and their masses cannot depend on the non-holomorphic parameter $\xi$ \cite{SYmon,HT2}. This means that,
although the confined monopoles look physically very different from unconfined monopoles on the Coulomb branch
of 4D SQCD (in a particular singular point that becomes the isolated vacuum at nonzero $\xi$),
their masses are the same. Moreover, these masses coincide with the masses of kinks in the world-sheet 
theory.

Note that  VEVs of $\sigma$ given by the exact twisted superpotential  coincide
with the double roots of the Seiberg-Witten curve \cite{SW2} in the quark vacuum of
4D SQCD \cite{Dorey,DoHoTo}. This is the key technical reason that leads to the
coincidence of the  2D and 4D BPS spectra.

%
%

\chapter{Non-Abelian strings in \none supersymmetric QCD} \label{sec:none}

In much the same way as the real world QCD,
 \none supersymmetric QCD does not have  adjoint scalars.  Therefore it is believed to have an essentially non-Abelian dynamics. On the other hand, due to supersymmetry it is more tractable then non-supersymmetric QCD. One may hope that, starting
from \ntwo QCD and decoupling the adjoint scalars, one can arrive at a non-Abelian
regime. In particular, it was shown that the non-Abelian ''instead-of-confinement'' phase
survives in the limit where the adjoint matter (present in  \ntwo QCD)  decouples, see review \cite{SYdualrev} and references therein.

In this Chapter we make this step  and study what happens to the non-Abelian confining strings upon decoupling of the adjoint matter. Namely, we consider a deformation of  \ntwo supersymmetric QCD with the U$(N)$ gauge group and $N_f=N$ quark flavors by a mass term $\mu$ of the adjoint matter. The $\mu$-deformation breaks the \ntwo supersymmetry and in the limit of large $\mu$ the theory flows to \none supersymmetric QCD.

%
%

\section{Outline}

In addition to the
translational zero modes typical for ANO strings, non-Abelian strings have  orientational moduli associated with rotations of their fluxes inside the non-Abelian SU$(N)$ group. The dynamics of the orientational moduli in \ntwo QCD is described by the two dimensional $\mathbb{CP}(N-1)$ model living on the world sheet of the non-Abelian string. In this Chapter we study the solution for the  non-Abelian string  and derive an effective theory on the string world sheet in the
limit of large $\mu$.

Similar problem was  addressed in \cite{SYnone,Edalati,SY02,Tongd,SYhetN,BSYhet} where the $\mu$-deformation was considered in \ntwo supersymmetric QCD with the U$(N)$ gauge group and $N_f=N$ flavors of massless  quarks supplemented by the Fayet-Iliopoulos $D$-term.  In the limit of large $\mu$ this theory flows to
a theory which differs from \none QCD by the presence of the  FI term. In particular,
in this theory a scalar quark (squark) condensation is triggered by the FI $D$-term.

It was shown in the aforementioned papers that bosonic profile functions of the non-Abelian string stay intact
upon the $\mu$-deformation, while the fermionic zero modes are changed as compared to the ones in the \ntwo limit. The string remains BPS saturated and the world sheet theory becomes the heterotic $\mathbb{CP}(N-1)$ model with \ntwoo supersymmetry \cite{Edalati,SY02,Tongd,BSYhet}. In this model, the supertranslational fermionic moduli interact with the superorientational ones. Large $N$ solution of the world sheet
model shows that \ntwoo supersymmetry  is spontaneously broken \cite{SYhetN}. The model has $N$ vacua corresponding to $N$ different non-Abelian strings and the discrete  $Z_{2N}$ symmetry is
spontaneously broken.

In this Chapter we consider the $\mu$-deformation of \ntwo QCD without a FI term in a quark vacuum. Squark condensate is determined by $\sqrt{\mu m}$, were $m$ is a quark mass. 
Note that while in the presence of the FI $D$-term it was not really possible to introduce quark masses, in the present setup quarks are massive.
Moreover, as was mentioned above, this theory is more \textquote{realistic} also because there is no FI term.
In the large $\mu$ limit the theory flows to  \none QCD in the quark vacuum. Non-Abelian strings cease to be BPS
saturated and both bosonic and fermionic profile of the string are modified.

We study solutions for the  non-Abelian string profile functions in the large $\mu$
limit and derive the effective theory on the string world sheet. The bosonic sector of this theory is still given by the $\mathbb{CP}(N-1)$ model. The $\mathbb{CP}(N-1)$ model is asymptotically free, and it
is determined by its scale $\Lambda_{CP}$ (position of the infra-red pole of the coupling constant). 
At small $\mu$ $\Lambda_{CP} = \Lambda_{{\cal N}=2}$, where $\Lambda_{{\cal N}=2}$ is the scale of
four dimensional \ntwo QCD, see, for example, review \cite{SYrev}.
We show that in the in the large $\mu$
limit $\Lambda_{CP}$ is exponentially small.
We also derive a potential in  two dimensional world sheet theory induced by quark mass differences.

 Next we study  the fermionic sector of the world sheet theory. Upon the $\mu$-deformation the
 fermionic superorientational zero modes are all lifted. This leaves us with the pure bosonic $\mathbb{CP}(N-1)$ model on the string world sheet in the limit when the bulk theory becomes \none QCD. This ensures that the
 world sheet theory is in the Coulomb/confinement phase, at least at large $N$, see \cite{W79}.

We also address a question of what happen to the confined 't Hooft-Polyakov monopoles 
present in the \ntwo limit,
when we go to the large $\mu$. Studying the world sheet potential we show that 
confined monopoles seen in the world sheet theory
as kinks \cite{SYmon,HT2} become unstable at large $\mu$ if quark masses are not equal.  However, if 
 quarks have equal masses the confined monopoles survive in the limit of \none QCD.

\section{$\mu$-deformed \ntwo supersymmetric QCD}

In this section we review our four dimensional bulk theory, see review \cite{SYrev} for more details. The bulk theory  is $\mu$-deformed \ntwo supersymmetric QCD  with the gauge group U$(N)=$SU$(N)\times$U(1).
The field content of the theory is as follows.
The \ntwo vector multiplet consists of the U(1) gauge field $A_\mu$ and SU($N$) gauge field $A^a_\mu$, complex scalar fields $a^{U(1)}$ and $a^a$ in the adjoint representation, and their  fermion superpartners ($\lambda^{1}_{\alpha}$, $\lambda^{2}_{\alpha}$) and ($\lambda^{1a}_{\alpha}$, $\lambda^{2a}_{\alpha}$). The adjoint index $a$ runs from 1 to $N^2 - 1$, while the spinorial index $\alpha = 1,2$.
The adjoint scalars and fermions $\lambda^{2}$ can be combined into the \none adjoint matter chiral multiplets $ \AU $ and $ \AN = \mc{A}^a T^a $, where $T^a$ are generators of the SU($N$) gauge group normalized as
$\Tr \left( T^a T^b \right) ~=~ (1/2) \, \delta^{ab}~$.

The matter sector consists of $N_f = N$ flavors of quark hypermultiplets in the fundamental representation and scalar components (squarks) $q^{kA}$ and $\wt{q}_{Ak}$, while the fermions are represented by $\psi^{kA}$ and $\wt{\psi}_{Ak}$. Here $A = 1,..,N$ is a flavor index and $k=1,..,N$ is a color index.

The  superpotential of \ntwo supersymmetric QCD reads
\begin{equation}
 \mc{W}_{\mc{N}=2} ~~=~~  \sqrt{2}\, \Bigl\{
	 \frac12 \wt{q}{}_A \AU q^A ~+~
	 \wt{q}{}_A \mc{A}^a T^a q^A \Bigr\}  ~+~
	 m_A\, \wt{q}{}_A q^A \ ,
\label{supN2}
\end{equation}
where we use the same notations for quark multiplets $ q^A $ and $ \wt{q}{}_A $ and their scalar components, while $m_A$ are quark masses.

The $\mu$-deformation is the mass term for the adjoint matter
\begin{equation}
 \mc{W}_{\mc{N}=1} ~~=~~ \sqrt{\frac{N}{2}}\,\frac{\mu_1}{2} \left(\mc{A}^{\rm U(1)}\right)^2  ~~+~~
	 \frac{\mu_2}{2} \left( \mc{A}^a \right)^2 ~,
\label{none_superpotential_general}
\end{equation}
which breaks \ntwo supersymmetry down to \none.

In the special case when
\begin{equation}
\mu ~~\equiv~~\mu_2 ~~=~~ \mu_1 \sqrt{\frac{2}{N}}  ,
\label{mu_condition}
\end{equation}
superpotential  \eqref{none_superpotential_general} becomes a single trace operator
\begin{equation}
\mc{W}_{\mc{N}=1} = \mu \Tr (\Phi^2) \,
\end{equation}
where we defined a scalar adjoint matrix as
\beq
\Phi = \frac12 a^{U(1)} + T^a\, a^a.
\label{Phidef}
\eeq

We will consider bulk QCD in the limit of large $\mu_1$ and $\mu_2$, when the
adjoint matter decouples and the theory becomes \none QCD.
 Integrating  out the adjoint matter in a sum of superpotentials (\ref{supN2}) and (\ref{none_superpotential_general}) we get a quark superpotential of our $\mu$-deformed
 bulk theory
\begin{equation}
 \mc{W} (q, \tilde{q}) = - \frac{1}{2\mu_2} \left[ (\tilde{q}_A q^B)(\tilde{q}_B q^A) - \frac{\alpha}{N} (\tilde{q}_A q^A)^2 \right] + m_A (\tilde{q}_A q^A) \,,
\label{superpotential-adjoints_integrated}
\end{equation}
where
\begin{equation}
\alpha= 1- \sqrt{\frac{N}{2}} \frac{\mu_2}{\mu_1} \,.
\label{alpha}
\end{equation}
In the case of single trace deformation (\ref{mu_condition}) $\alpha =0$.

The bosonic action of the theory is %
 \footnote{From here further on in this Chapter we use a  Euclidean notation, that is $F_{\mu\nu}^2 = 2F_{0i}^2 + F_{ij}^2$, $\, (\partial_\mu a)^2 = (\partial_0 a)^2 +(\partial_i a)^2$, etc.  Furthermore, the sigma-matrices are defined as $\sigma^{\alpha\dot{\alpha}}=(1,-i\vec{\tau})$,
	$\bar{\sigma}_{\dot{\alpha}\alpha}=(1,i\vec{\tau})$. Lowering and raising
	of spinor indices are performed by  virtue of an anti-symmetric tensor
	defined as $\varepsilon_{12}=\varepsilon_{\dot{1}\dot{2}}=1$,
	$\varepsilon^{12}=\varepsilon^{\dot{1}\dot{2}}=-1$.
	The same raising and lowering convention applies to the flavor SU($N$)
	indices $f$, $g$, etc. }
\begin{align}
\label{theory}
S_{\rm bos} ~~=~~ & \int d^4 x
\lgr
\frac{1}{2g_2^2}\Tr \left(F_{\mu\nu}^\text{SU($N$)}\right)^2  ~+~
\frac{1}{4g_1^2} \left(F_{\mu\nu}^{\rm U(1)}\right)^2 ~+~
\right.
\\[2mm]
\notag
&
\phantom{int d^4 x \lgr\right.}
\left.
\left| \nabla_\mu q^A \right|^2 ~+~ \left|\nabla_\mu \ov{\wt{q}}{}^A \right|^2
~+~ V(q^A, \wt{q}_A)
\rgr .
\end{align}
Here $ \nabla_\mu $ is a covariant derivative
\beq
\nabla_\mu  ~~=~~ \p_\mu ~-~ \frac{i}{2}\,A^{\rm U(1)}_\mu ~-~ i\, A_\mu^a T^a~,
\label{covder}
\eeq
while the scalar potential $V(q^A, \wt{q}_A)$ is a sum of the $D$-term and $F$-term potentials,
\beq
V(q^A, \wt{q}_A)=V_D (q^A, \wt{q}_A) + V_F (q^A, \wt{q}_A).
\label{scpot}
\eeq
The $D$-term potential reads
\begin{equation}
	V_D ~~=~~ \frac{g_2^2}{2}\left(\bar{q}_A T^a q^A - \qt_A T^a \bar{\qt}^A\right)^2 ~+~
				\frac{g_1^2}{8} \left(|q^A|^2 - |\qt^A|^2\right)^2,
\label{potential:3+1:D:original}
\end{equation}
while the $F$ term potential is determined by superpotential (\ref{superpotential-adjoints_integrated}). It has the form
\begin{equation}
	\begin{aligned}
		V_F ~~=~~ \frac{1}{|\mu_2|^2}\,
		&\Bigg\{ (\bar{q}_A q^B)\, \Big[ (\bar{q}_C \bar{\qt}^A) - \frac{\bar{\alpha}}{N} \delta^A_C (\bar{q}_F \bar{\qt}^F) - \bar{\mu_2}\bar{m}_A\delta_C^A \Big]	\\
		&\times \Big[ (\qt_B q^C) - \frac{\alpha}{N} \delta_B^C (\qt_F q^F) - \mu_2 m_B\delta_B^C \Big]\\
		&+ (\qt_A \bar{\qt}^B)\, \Big[ (\bar{q}_B \bar{\qt}^C) - \frac{\bar{\alpha}}{N} \delta_B^C (\bar{q}_F \bar{\qt}^F) - \bar{\mu_2}\bar{m}_B\delta_B^C \Big]	\\
		&\times \Big[ (\qt_C q^A) - \frac{\alpha}{N} \delta_C^A (\qt_F q^F) - \mu_2 m_A\delta_C^A \Big]
		\Bigg\} \,.
	\end{aligned}
\label{potential:3+1:F:original}
\end{equation}

In this work we will consider the vacuum (zero of the potential (\ref{scpot})) where the maximal possible number of quark flavors equal to $N$ condense (the so called $r=N$ vacuum, where $r$ is the number of condensed squark flavors at weak coupling, see \cite{SYdualrev} for  a review). In this vacuum squark VEVs are given by
\beq
\langle q^{kA} \rangle ~~=~~ \langle \ov{\wt{q}}{}^{kA} \rangle = \frac1{\sqrt{2}}
\begin{pmatrix}
\sqrt{\xi_1}  &   0  &  ... \\
... &  ... &  ... \\
... &   0  &  \sqrt{\xi_N}
\end{pmatrix}  , \\
\label{qVEVdifferent}
\eeq
where we wrote down the squark field as an $N\times N$ matrix in color and flavor indices, and the parameters $\xi_A$ are defined as
\begin{equation}
	\xi_A ~~=~~ 2\left( \sqrt{\frac{2}{N}} \mu_1 \widehat{m} ~+~ \mu_2 (m_A - \widehat{m}) \right) ,
	\label{xi-general}
\end{equation}
while
\begin{equation}
	\widehat{m} ~~=~~ \frac{1}{N} \sum_{A=1}^{N} m_A .
\label{averagemass}
\end{equation}

For single trace deformation (\ref{mu_condition}) expressions for the parameters $\xi_A$ simplify:
\beq
\xi_A= 2 \,\mu_2 m_A
\label{xisingletrace}
\eeq

In this Chapter we will mostly consider the non-Abelian limit when all quark masses are equal,
\begin{equation}
m_1=m_2=...=m_N \equiv m ,
\label{equalmasses}
\end{equation}
so that the parameters $\xi_A$ degenerate, $\xi_A\equiv\xi$, and the squark VEVs become
\beq
\langle q^{kA} \rangle ~~=~~ \langle \ov{\wt{q}}{}^{kA} \rangle =
\sqrt{\frac{\xi}{2}}
\begin{pmatrix}
1  &   0  &  ... \\
... &  ... &  ... \\
... &   0  &  1
\end{pmatrix}
\label{qVEV}
\eeq

Note that if we take the limit $\mu\to\infty$ (keeping the quark masses fixed) the parameters
$\xi\sim \mu m$ also go to infinity, and our quark vacuum becomes a run-away vacuum (all the $r$ vacua
with the non-zero $r$ become run-away vacua). In this case \none QCD is a theory with only $N$ vacua
which originate from $N$ monopole vacua ($r=0$ vacua) of \ntwo QCD.

Here we define \none QCD in a different way. By taking the limit of large $\mu$ we make the quark masses
small so that the product $\mu m$ (and the quark VEVs) are fixed,
\beq
\mu \to \infty, \qquad m\to 0, \qquad \mu\,m = {\rm fixed}.
\label{mutoinfty}
\eeq
 This way we keep track of all the $r$ vacua present in \ntwo QCD. In this Chapter we will study non-Abelian strings particularly in the
$r=N$ quark vacuum (\ref{qVEV}) assuming the limit of large $\mu$ when the bulk theory flows to the
generalized \none QCD defined above.

In order to keep our bulk theory at weak coupling we assume that the squark VEVs are large as compared with the scale $\Lambda_{{\cal N}=1}$ of the SU$(N)$ sector of \none QCD. Namely, we assume that
\beq
\sqrt{\mu m} \gg \Lambda_{{\cal N}=1}.
\label{weakcoupl}
\eeq

Squark VEVs (\ref{qVEV}) result in  a spontaneous
breaking of both gauge and flavor SU($N$)'s.
The diagonal global SU($N$) survives, however,
\beq
{\rm U}(N)_{\rm gauge}\times {\rm SU}(N)_{\rm flavor}
\to {\rm SU}(N)_{C+F}\,,
\label{c+f}
\eeq
cf. \eqref{globgroup_d=4}.
A color-flavor locking takes place in the vacuum. This fact  leads to an emergence of non-Abelian strings, see \cite{SYrev} for a review.

Let us briefly summarize a perturbative spectrum of our bulk theory in the large $\mu$ limit,
cf. \cite{SYrev}. Consider for simplicity the case of equal quark masses. The U$(N)$ gauge group is completely Higgsed and  the masses of the gauge bosons are
\begin{equation}
	m_G^{SU(N)} ~~=~~ g_2|\sqrt{\xi}| ~
\label{massGSUN}
\end{equation}
for the SU$(N)$ gauge bosons and
\begin{equation}
	m_G^{U(1)} ~~=~~ g_1 \sqrt{\frac{N}{2}} |\sqrt{\xi}|~
\label{massGU1}
\end{equation}
for the U(1) one. Below we also assume that the gauge boson masses are of the same order,
\beq
m_G^{U(1)} \sim m_G^{SU(N)} \equiv m_G
\label{massG}
\eeq

Extracting a quark mass matrix from potentials (\ref{potential:3+1:D:original}), (\ref{potential:3+1:F:original}) we find that out of $4N^2$ real degrees of freedom of the $q^{kA}$ and $\ov{\wt{q}}{}^{kA}$ squarks $N^2$ phases are eaten by the Higgs mechanism, $(N^2-1)$ real squarks have mass (\ref{massGSUN}), while one real squark has mass (\ref{massGU1}). These squarks
are scalar superpartners of the SU$(N)$ and U$(1)$ gauge bosons in massive vector \none supermultiplets, respectively.

Other  $2N^2$ squarks become much lighter in the large $\mu$ limit. The masses of $2(N^2-1)$ of
them forming the adjoint representation of the global color-flavor SU$_{C+F}(N)$ (\ref{c+f}) are given by
\beq
m_L^{SU(N)} =\left|\frac{\xi}{\mu_2}\right|,
\label{massLSUN}
\eeq
while  two real SU$_{C+F}(N)$ color-flavor singlets have mass
\beq
m_L^{U(1)} =\sqrt{\frac{N}{2}}\,\left|\frac{\xi}{\mu_1}\right|,
\label{massLU1}
\eeq

If $\mu_2$ and $\mu_1$ are of the same order (more exactly, we assume  below that  $\alpha = {\rm const}$, see \eqref{alpha}), then
\beq
 m_L^{U(1)} \sim m_L^{SU(N)} \equiv m_L   \sim m \ll m_G.
\label{masshierarchy}
\eeq
Below we will heavily use this mass hierarchy of the perturbative spectrum.

In particular, in the limit (\ref{mutoinfty}) $m_L\to 0$, and $2N^2$ squarks become massless.
This reflects the presence of the Higgs branch which develops  in this limit. The presence of
massless scalars developing VEVs makes the string solution ill-defined \cite{PennRubak,Y99}, see also next section.
Below we use the $\mu$-deformed \ntwo QCD at large $\mu$ as an infra-red (IR) regularization of \none
QCD. At large but finite $\mu$ the Higgs branch present in \none QCD is lifted and the IR
divergences are regularized, cf. \cite{EvlY}.

%
%

\section{Non-Abelian strings}
\label{sec:bosonic_solution}

In this section we derive a vortex solution, assuming the equal quark mass limit (\ref{equalmasses}). First we review a general {\it ansatz} for the non-Abelian string and
present equations for string profile functions. Then we  solve these equations
assuming the mass hierarchy \eqref{masshierarchy} in the large $\mu$ limit.

\subsection{Equations of motion}

 We consider a static string stretched along the $x_3$ axis so that the corresponding profile functions depend only on coordinates in the $(x_1,x_2)$ plane. Closely following the strategy developed for \ntwo supersymmetic QCD (see review \cite{SYrev}) we first assume that
only those squark fields which develop VEVs have non-trivial profile functions in a string solution. Therefore we set
\begin{equation}
q^{kA}=\bar{\tilde{q}}^{kA}=\frac1{\sqrt{2}}\vp^{kA}  .
\label{q-ansatz}
\end{equation}
and look for the string solutions using the following {\em ansatz} \cite{ABEKY,SYmon,SYrev}:
\begin{equation}
	\begin{aligned}
		\varphi & ~~=~~
		\phi_2 ~+~ n\nbar\, \bigl( \phi_1 ~-~ \phi_2 \bigr)
		\\[2mm]
		&
		~~=~~
		\frac{1}{N}\bigl( \phi_1 ~+~ (N-1)\phi_2 \bigr)
		~+~ \bigl( \phi_1 ~-~ \phi_2 \bigr)
		\lgr n\nbar ~-~ 1/N \rgr \,,
		\\[2mm]
		A_i^\text{SU($N$)} & ~~=~~ \varepsilon_{ij}\, \frac{x^j}{r^2}\, f_{W}(r)
		\lgr n\nbar ~-~ 1/N \rgr\,,
		\\[2mm]
		A_i^{\rm U(1)} & ~~=~~ \frac{2}{N}\varepsilon_{ij}\, \frac{x^j}{r^2}\, f(r)~,
	\end{aligned}
\label{string-solution}
\end{equation}
where the index $i$ runs over 1, 2. The profile functions $ \phi_1(r) $ and $ \phi_2(r) $ determine the profiles of the squarks
in the plane orthogonal to the string at rest, while $ f(r) $ and $ f_{W}(r) $ are the profiles of the gauge
fields. The profile functions  depend on the distance $r$ from a given point $x^i$
to the center of the string $x^i_0$ in the $(x_1,x_2)$ plane.

Here we have also introduced the orientational complex  vector $n^l$, $l=1,...,N$, subject to the condition
\begin{equation}
\ov{n}{}_l \cdot n^l ~~=~~ 1 \, .
\end{equation}
 Vector $n^l$ parametrizes the orientational modes of the non-Abelian vortex string. It arises due to a possibility to rotate a given particular string solution with respect to the unbroken color-flavor global group SU$(N)_{C+F}$, see (\ref{c+f}).

Boundary conditions for the gauge and scalar profile functions are		
\begin{align}
\label{boundary}
\phi_1(0) & ~~=~~  0\text,                   & \phi_2(0) & ~~\neq~~ 0\text,  &
\phi_1(\infty) & ~~=~~ \sqrt{\xi} \text,     & \phi_2(\infty) & ~~=~~ \sqrt{\xi}\text, \\
\notag
f_{W}(0) & ~~=~~ 1\text,                   & f(0) & ~~=~~ 1\text,   &
f_{W}(\infty) & ~~=~~ 0 \text,            &  f(\infty) & ~~=~~ 0\text.
\end{align}

  Substituting {\it ansatz} (\ref{string-solution}) into action \eqref{theory} we get an energy functional (tension of the string):
\begin{multline}
	T ~~=~~ 2\pi \int rdr \Bigg( \frac{2}{g_1^2 N^2} \frac{f'^2}{r^2} ~+~ \frac{N-1}{N}\frac{1}{g_2^2} \frac{f_W'^2}{r^2} ~+~ \phi_1'^2 ~+~ (N-1)\phi_2'^2 \\
	~+~ \frac{1}{N^2} \frac{\left[f + (N-1)f_W\right]^2}{r^2}\phi_1^2 ~+~ \frac{N-1}{N^2} \frac{\left[f - f_W\right]^2}{r^2}\phi_2^2
	~+~ V(\phi_1, \phi_2) ,
	\Bigg)
\label{tenfunct}
\end{multline}
where the potential $V(\phi_1, \phi_2)$ is
\begin{multline}
	V(\phi_1, \phi_2) ~~=~~ \frac{1}{4|\mu_2|^2} \Bigg(
	\phi_1^2 \left[\phi_1^2 ~+~ \frac{\alpha}{N}(\phi_1^2 + (N-1)\phi_2^2) - 2\mu_2 m  \right]^2 \\
	~+~ (N-1)\phi_2^2 \left[\phi_2^2 ~+~ \frac{\alpha}{N}(\phi_1^2 + (N-1)\phi_2^2) - 2\mu_2 m  \right]^2
	\Bigg)
\label{potphi}
\end{multline}
and we assume that $\mu_2 m$ is real \footnote{If it is in fact a complex quantity, we should modify relation
\eqref{q-ansatz} inserting there the phase of $\mu_2 m$.}.

String tension functional \eqref{tenfunct} gives  equations for the profile functions. We get
\begin{align}
\label{profileeqs}
	f'' ~-~ \frac{f'}{r} ~-~ \frac{g_1^2}{2}(f + (N-1)f_W)\phi_1^2 ~-~ (N-1) \frac{g_1^2}{2} (f - f_W) \phi_2^2 &~~=~~ 0\\
\notag
	f_W'' ~-~ \frac{f_W'}{r} ~-~ \frac{g_2^2}{N}(f + (N-1)f_W)\phi_1^2 ~+~ \frac{g_2^2}{N} (f - f_W) \phi_2^2 &~~=~~ 0 \\
\notag
	\phi_1'' ~+~ \frac{\phi_1'}{r} ~-~ \frac{1}{N^2}\frac{(f + (N-1)f_W)^2}{r^2}\phi_1 ~-~ \frac{1}{2} \frac{\p V}{\p \phi_1} &~~=~~ 0 \\
\notag
	\phi_2'' ~+~ \frac{\phi_2'}{r} ~-~ \frac{1}{N^2}\frac{(f - f_W)^2}{r^2}\phi_2 ~-~ \frac{1}{2(N-1)} \frac{\p V}{\p \phi_2} &~~=~~ 0
\end{align}
These equations are of the second order rather than the first order. This is because  our string is not  BPS saturated. Note, that for a BPS string the masses of the scalars forming the string are equal to masses of the gauge bosons \eqref{massGSUN} and \eqref{massGU1}, see \cite{SYrev}.
For our $\mu$-deformed theory this is not the case. Masses of singlet and adjoint scalars in the scalar matrix $\varphi^{kA}$ in \eqref{q-ansatz} are given by \eqref{massLSUN} and \eqref{massLU1}, and in the large $\mu$ limit they are much smaller than the masses of gauge bosons.
In particular, as we mentioned already, in the limit \eqref{mutoinfty} $m_L\to 0$, and our $\mu$-deformed theory develops a Higgs branch.

\subsection{String profile functions}
\label{sec:profile}

It is quite often that supersymmetric gauge theories have Higgs branches. These
are flat directions of the scalar potential on which charged scalar fields can
develop  VEVs breaking the gauge symmetry. In
many instances this breaking provides topological reasons behind formation
of  vortex strings. A dynamical side of the problem of the vortex string
formation in theories with Higgs branches was addressed in \cite{PennRubak,Y99,EvlY}. A
priori it is not clear at all whether or not stable string solutions exist in
this class of theories. The fact is that a theory with a Higgs branch
represents a limiting case of type I superconductor with vanishing Higgs
mass. In particular, it was shown in \cite{PennRubak} that infinitely long strings cannot be
formed in this case due to infrared divergences.

Later this problem was studied in \cite{Y99,EvlY}. It was shown that vortices on Higgs branches become logarithmically ''thick'' due to the presence of massless scalars in the bulk. Still,
they are well defined if IR divergences are regularized. One way of regularization is to
consider a vortex string of the finite length $L$ \cite{Y99}. This setup is typical for the confinement problem. It was shown in \cite{Y99} that confining potential between heavy trial charges becomes nonlinear,
\beq
V(L) \sim \frac{L}{\log{L}} ,
\eeq
in theories with Higgs branches.

Another way of IR regularization is to lift the Higgs branch so that scalar fields forming the string have small but non-zero masses $m_L$, cf. \cite{EvlY}. We use this approach here, see
Eqs. \eqref{massLSUN} and \eqref{massLU1}, assuming that $\mu$ is large but finite.

To the leading order in $\log{m_G/m_L}$ the vortex solution has the following structure in the
$(x_1,x_2)$ plane \cite{Y99}. The gauge fields are localized inside the core region of the radius
$R_g$ and almost zero outside this region\footnote{We will determine $R_g$ shortly.}. In contrast, scalar profiles are almost constant inside the core. In particular, the $\phi_1$ profile function associated with winding of the vortex is almost zero
inside the core (see \eqref{boundary}),
\begin{equation}
\begin{aligned}
\phi_1 &~~\approx~~ 0 \\
\phi_2 &~~\approx~~ (1-c)\sqrt{\xi},
\end{aligned}
\label{string:boson:large_mu:dm=0:small_r:phi}
\end{equation}
where $c$ is a constant to be determined.

Then the two first equations for gauge profile functions in \eqref{profileeqs} have solutions
\begin{equation}
	f ~~=~~ f_W ~~\approx~~ 1 ~-~ \frac{r^2}{R_g^2}
\label{string:boson:large_mu:dm=0:small_r:f}
\end{equation}
inside the core.

Outside the core in a logarithmically wide region
\beq
1/m_G \lesssim r \lesssim 1/m_L
\label{logregion}
\eeq
gauge fields are almost zero and two last equations in \eqref{profileeqs} reduce to the equations for free massless scalars. Their solutions have a logarithmic form
\begin{equation}
	\begin{aligned}
		\phi_1 &~~\approx~~ \sqrt{\xi} \left(1 ~-~ \frac{\ln \displaystyle\frac{1}{r m_L}}{\ln \displaystyle\frac{1}{m_L R_g}} \right) \,, \\
		\phi_2 &~~\approx~~ \sqrt{\xi} \left(1 ~-~ c\cdot\frac{\ln \displaystyle\frac{1}{r m_L}}{\ln \displaystyle\frac{1}{m_L R_g}} \right) \,,
	\end{aligned}
\end{equation}
where the normalization is fixed by matching with the behavior inside the core \eqref{string:boson:large_mu:dm=0:small_r:phi} and with the boundary conditions at infinity
\eqref{boundary}.

In the region of very large $r$, $r\gg 1/m_L$, the scalar fields exponentially approach  their
VEVs ($\sim \exp{-m_Lr}$), see \eqref{boundary}).
This region gives a negligible contribution to the string tension and a particular form of the scalar potential \eqref{potphi} is not important.

Upon a substitution of the above  solution into tension functional \eqref{tenfunct} one arrives at
\begin{equation}
	T ~~\approx~~ \frac{{\rm const}}{ R^2_g}\left(\frac2{g_1^2 N^2} + \frac{N-1}{g_2^2N}\right)+ \frac{2\pi|\xi|}{\ln \displaystyle\frac{1}{R_g m_L}} \left[ 1 ~+~ (N-1)\,c^2\right],
\label{T}
\end{equation}
where the first term comes from the gauge fields inside the core while the second term is produced by
the logarithmic integral over the region \eqref{logregion} coming from the kinetic terms of scalars.

Minimization of this expression with respect to the constant $c$ yields
\begin{equation}
	c ~=~ 0 ,
\end{equation}
so that the profile function $\phi_2$ does not depend on $r$ and is given by its VEV $\sqrt{\xi}$.

Minimizing \eqref{T} with respect to $R_g$ we find
\beq
R_g \sim \frac{\rm{const}}{m_G} \ln{\frac{m_G}{m_L}}.
\label{Rg}
\eeq

The solutions for string profile functions in the intermediate region \eqref{logregion} becomes
\begin{equation}
	\begin{aligned}
		\phi_1 &~~\approx~~ \sqrt{\xi} \left(1 ~-~ \frac{\ln \displaystyle\frac{1}{r m_L}}{\ln \displaystyle\frac{m_G}{m_L}} \right) \,, \\
		\phi_2 &~~\approx~~ \sqrt{\xi} \,, \\
		f &~~\approx~~ f_W ~~\approx~~ 0 \, ,
	\end{aligned}
\label{string:boson:large_mu:dm=0:intermediate}
\end{equation}
while  the final result for the tension of a non-Abelian string takes the form
\beq
T = \frac{2\pi|\xi|}{\ln \displaystyle\frac{m_G}{m_L}} +\cdots,
\label{ten}
\eeq
where corrections are suppressed by powers of large logarithm $\log{m_G/m_L}$.
The leading term here comes from quark kinetic energy ($(\phi_1')^2$) integrated over
intermediate region \eqref{logregion}, see \eqref{tenfunct}.
Note, that the logarithmic suppression of the string tension is not specific for non-Abelian strings. Similar expression was found for the ANO string on a Higgs branch \cite{Y99,EvlY}.

\subsection{Non-equal quark masses}

In this section we relax condition \eqref{equalmasses} and consider a string solution
assuming that quark mass differences are small,
\beq
\Delta m_{AB}=m_A-m_B \ll \widehat{m},
\label{smallmassdiff}
\eeq
where $\widehat{m}$ is the average quark mass, \eqref{averagemass}.

Non-equal quark masses break color-flavor symmetry \eqref{c+f} down to U$(1)^N$, so the orientational modes of the non-Abelian string are no longer zero modes. They become quasizero
modes in the approximation of small quark mass differences \eqref{smallmassdiff}, cf. \cite{SYrev}. In fact, in Sect. \ref{sec:worldsheetpot} we will derive a shallow world sheet potential
with $N$   extreme points associated with $Z_N$ strings.

Now we generalize the {\it ansatz} for the string solution \eqref{string-solution} as follows. First we
set an orientational vector
\beq
n^l = \delta^{lA_0}, \qquad A_0=1,...,N
\label{nA0}
\eeq
separating the $A_0$-th $Z_N$ string (the string associated with the winding of $A_0$ squark flavor, see \cite{SYrev}).

We expect that, much in the same way as for the equal quark masses case, the main contribution
to the string tension comes from logarithmically wide intermediate region \eqref{logregion},
while the string core does not contribute to the leading order. Then taking into account \eqref{smallmassdiff} we can neglect mass differences of
different gauge bosons setting
\beq
m_G \approx g_2\sqrt{|\widehat{\xi}|},
\label{averagemG}
\eeq
where
\begin{equation}
	\widehat{\xi} ~~=~~ \frac{1}{N} \sum_{A=1}^{N} \xi_A .
\label{averagexi}
\end{equation}

In this approximation we can use the same {\it ansatz} for the gauge fields as for the case of equal quark masses, see two last equations in \eqref{string-solution} with $n^l$ from \eqref{nA0}. Gauge fields are still parametrized by only two gauge profile functions $f(r)$ and $f_W$, which are non-zero inside the string core determined by $m_G$ \eqref{averagemG}.

The {\it ansatz} for the squark fields in \eqref{string-solution} is generalized as follows
\begin{equation}
	\begin{aligned}
		\vp &~~=~~
		\begin{pmatrix}
		\phi_1(r)	&	0		&	\ldots	&	0			&	\ldots	&	0	 \\
			0		&\phi_2(r)	&	\ldots	&	0			&	\ldots	&	0	 \\
		\ldots		&\ldots		&	\ldots	&	\ldots		&	\ldots	&\ldots	 \\
			0		&	0		&	\ldots	&	\phi_{A_0}(r)	&	\ldots	&	0	 \\
		\ldots		&\ldots		&	\ldots	&	\ldots		&	\ldots	&\ldots	 \\
			0		&	0		&	\ldots	&	0			&	\ldots	&\phi_N(r)	 \\
		\end{pmatrix}	
	\end{aligned},
\label{ansatz-dm_neq_0-n=1}
\end{equation}
where we introduce the profile functions $\phi_1,...,\phi_N$ for non-winding flavors $A\neq A_0$
while the profile function $\phi_{A_0}$ is associated with $A_0$-th winding flavor.

Boundary conditions for gauge profile functions are the same as in \eqref{boundary} while for
quarks we require
\begin{align}
	\phi_{A_0}(0) &~=~ 0\\
	\phi_A(\infty) &~=~ \sqrt{\xi_A}, \qquad A=1,...,N ,
\end{align}
where $\xi_A$ are given by \eqref{xi-general}.

Equations for the profile functions now read
\begin{align}
\notag
f'' ~-~ \frac{f'}{r} ~-~ \frac{g_1^2}{2}(f + (N-1)f_W)\phi_{A_0}^2 ~-~  \frac{g_1^2}{2} (f - f_W) \sum_{A \neq A_0}  \phi_A^2 &~~=~~ 0\\
\notag
f_W'' ~-~ \frac{f_W'}{r} ~-~ \frac{g_2^2}{N}(f + (N-1)f_W)\phi_{A_0}^2 \phantom{\frac{g_2^2}{N(N-1)}} & \\
 ~+~ \frac{g_2^2}{N(N-1)} (f - f_W)\sum_{A \neq A_0}  \phi_A^2 &~~=~~ 0 \\
\notag
\phi_{A_0}'' ~+~ \frac{\phi_{A_0}'}{r} ~-~ \frac{1}{N^2}\frac{(f + (N-1)f_W)^2}{r^2}\phi_{A_0} ~-~ \frac{1}{2} \frac{\p V}{\p \phi_{A_0}} &~~=~~ 0 \\
\notag
\text{For $A \neq A_0$,   } \quad
\phi_A'' ~+~ \frac{\phi_A'}{r} ~-~ \frac{1}{N^2}\frac{(f - f_W)^2}{r^2}\phi_A ~-~ \frac{1}{2} \frac{\p V}{\p \phi_A} &~~=~~ 0 \ .
\end{align}

Solving these equations in much the same way as we did in the previous subsection we get
\beq
\phi_{A} ~~\approx~~ \sqrt{\xi_A}, \qquad A\neq A_0.
\label{phiA}
\eeq
Moreover,  gauge profile functions are determined by \eqref{string:boson:large_mu:dm=0:small_r:f}
inside the core, while they are zero outside it. Here the size of the core is still given by \eqref{Rg}.

In much the same way as for the case of equal quark masses the profile function $\phi_1$ is
almost zero inside the core and is given by
\beq
\phi_{A_0} ~~\approx~~ \sqrt{\xi_{A_0}} \left(1 ~-~ \frac{\ln \displaystyle\frac{1}{r m_L}}{\ln \displaystyle\frac{m_G}{ m_L}} \right) \,
\label{phi1massdiff}
\eeq
in the region \eqref{logregion} of intermediate $r$.

The results for the string tensions of  $Z_N$ strings have the form
\beq
T_{A_0} ~~=~~ \frac{2\pi |\xi_{A_0}|}{\ln \displaystyle\frac{m_G}{m_L}}+\cdots, \qquad A_0=1,...,N.
\label{tenA0}
\eeq
We see that now string tensions of $N$ $Z_N$ strings are split.

%
%

\section{World sheet effective theory}
 \label{sec:effective_bosonic_action}

A non-Abelian string has both translational  and orientational zero modes. If we allow a slow dependence of the associated moduli on the world sheet coordinates $z=x_3$ and $t$ they become fields in the effective two dimensional low energy theory on the  string world sheet \cite{ABEKY,SYmon}, see \cite{SYrev} for a review.
Namely, we will have translational moduli $x_{0}^i(t,z)$ (position of the string in the $(x_1,x_2)$ plane, $i=1,2$) and orientational moduli $n^l(t,z)$, $l=1,...,N$. Translational sector is free and decouples therefore we will focus on the orientational sector.

In this section we will derive the bosonic part of the effective world sheet theory on the string.

\subsection{$\mathbb{CP}(N-1)$ model on the string world sheet}

First we consider the limit of equal quark masses \eqref{equalmasses}. In this limit color-flavor symmetry \eqref{c+f} is unbroken and the orientational moduli $n^l$ describe the zero modes of the non-Abelian string. Namely, consider a particular $Z_N$ solution \eqref{string-solution} with $n^l=\delta^{lA_0}$, $A_0=1,...,N$. It  breaks the SU$(N)_{C+F}$ group down to SU$(N-1)\times$U(1). Therefore the SU$(N)_{C+F}$
rotation of the $Z_N$ string solution generates the whole family of solutions (non-Abelian string) parametrized by the vector $n^l$ from the moduli space
\beq
\frac{\text{SU}(N)_{C+F}}{\text{SU}(N-1)\times U(1)} = \mathbb{CP}(N-1).
\label{modspace}
\eeq

Since for equal quark masses the SU$(N)_{C+F}$ group is unbroken there is no world sheet potential for orientational moduli $n^l$. To derive the kinetic term we closely follow the general procedure developed in \cite{ABEKY,SYmon} (see \cite{SYrev} for a review) for the \ntwo case.
We substitute  solution \eqref{string-solution} into four-dimensional action \eqref{theory} assuming slow $t$ and $z$ dependence of the orientational moduli $n^l$.

Once the moduli $n^l$ cease to be constant, the gauge field components $A^{{\rm SU}(N)}_0$ and $A^{{\rm SU}(N)}_3$ also become non-vanishing. We use the {\it ansatz} \cite{ABEKY,SYmon,GSY05} 
\begin{equation}
	A^{{\rm SU}(N)}_{k}=-i\,  \big[ \pt_{k} n \,\cdot \bar{n} -n\,\cdot
	\pt_{k} \bar{n} -2n\,
	\cdot \bar{n}(\bar{n}\,\pt_{k} n)
	\big] \,\rho (r)\, , \quad k=0, 3\,,
	\label{An}
\end{equation}
for these components, where we assume a contraction of the color indices inside the parentheses
in the third term.  We also introduced a new profile function $\rho (r)$ which will be determined through a minimization procedure. 

Substituting \eqref{string-solution} and \eqref{An} into \eqref{theory} we get $\mathbb{CP}(N-1)$ model
\begin{equation}
	S^{(1+1)}= 2 \beta\,   \int d t\, dz \,  \left\{(\pt_{k}\, \bar{n}\,
	\pt_{k}\, n) + (\bar{n}\,\pt_{k}\, n)^2\right\}\,,
	\label{cp}
\end{equation}
with the coupling constant $\beta$  given by
\begin{equation}
	\beta=\frac{2\pi}{g^2_2}\, I \,,
	\label{betaI}
\end{equation}
where $I$ is a normalization integral determined by the string profile functions 
integrated over $(x_1,x_2)$ plane.
\begin{equation}
	\begin{aligned}
		I & = &	\int_0^{\infty}	rdr\left\{\left(\frac{d}{dr}\rho (r)\right)^2 + \frac{1}{r^2}\, f_{W}^2\,\left(1-\rho \right)^2
		\right.\\[4mm]
		& + &
		\left.  g_2^2\left[\frac{\rho^2}{2}\left(\phi_1^2
		+\phi_2^2\right)+
		\left(1-\rho \right)\left(\phi_2-\phi_1\right)^2\right]\right\}\, .
	\end{aligned}
\label{I}
\end{equation}
The above functional determines an equation of motion for $\rho$,
\begin{equation}
	-\frac{d^2}{dr^2}\, \rho -\frac1r\, \frac{d}{dr}\, \rho
	-\frac{1}{r^2}\, f_{W}^2 \left(1-\rho\right)
	+
	\frac{g^2_2}{2}\left(\phi_1^2+\phi_2^2\right)
	\rho
	-\frac{g_2^2}{2}\left(\phi_1-\phi_2\right)^2=0\, ,
	\label{rhoeq-none}
\end{equation}
while  boundary conditions for $\rho$ read 
\begin{equation}
	\rho (\infty)=0 , \ \ \rho (0)=1 ,
\end{equation}
see \cite{SYrev} for the details.

The formulas above are valid for a non-Abelian string in  both \ntwo QCD and  $\mu$-deformed
QCD. For our large $\mu$  limit we use the string profile functions found in Sec.~\ref{sec:profile}. To find the solution for $\rho$ we first note that inside the core
$\rho\approx 1$. In the intermediate region \eqref{logregion} 
$f_W \approx 0$. The terms of equation \eqref{rhoeq-none} which involve derivatives of $\rho$ are negligible compared to the others (we will check that afterwards), and so an approximate solution can be easily found:
\begin{equation}
	\rho ~~\approx~~ \frac{\left(\phi_1-\phi_2\right)^2}{\left(\phi_1^2+\phi_2^2\right)}
		 ~~\approx~~ \frac{\left( \frac{\ln \frac{1}{r m_L}}{\ln \frac{1}{R_g m_L}} \right)^2}%
		 {2 ~-~ 2\frac{\ln \frac{1}{r m_L}}{\ln \frac{1}{R_g m_L}} ~+~ \left( \frac{\ln \frac{1}{r m_L}}{\ln \frac{1}{R_g m_L}} \right)^2},
\end{equation}
where we used solutions \eqref{string:boson:large_mu:dm=0:intermediate} for quark profile functions.

It is easy to check that $\rho' \sim m_L\rho$ and $\rho'' \sim m_L^2\rho$, so it was indeed consistent to drop the derivatives out of the equation \eqref{rhoeq-none}.

Our next step is to substitute this solution into the \eqref{I} and calculate $I$. As we will see, only the region $r \lesssim 1/m_L$ gives a significant contribution to this integral. We have:
\begin{multline}
	I ~~\approx~~ g_2^2 \int r dr \left( \frac{1}{2} \frac{\left(\phi_1-\phi_2\right)^4}{\left(\phi_1^2+\phi_2^2\right)} ~+~
	 \frac{2 \phi_1 \phi_2 \left(\phi_1-\phi_2\right)^2}{\left(\phi_1^2+\phi_2^2\right)} \right) \\
	 ~~=~~ \frac{g_2^2}{2}  \int r dr \frac{\left(\phi_1^2-\phi_2^2\right)^2}{\left(\phi_1^2+\phi_2^2\right)}
\end{multline}
Calculation yields
\begin{equation}
	I ~~\approx~~ c \,\frac{m_G^2}{m_L^2}\,\frac{1}{\ln^2 \frac{m_G^2}{m_L^2}} ~~\sim~~  \frac{g^2|\mu|}{|m|}\,\frac{1}{\ln^2 \frac{g^2|\mu|}{|m|}},
	\label{Iresult}
\end{equation}
where we used Eqs. \eqref{massGSUN} and  \eqref{massLSUN} while the constant $c$ is associated with the ambiguity of the upper limit ($\sim 1/m_L$) of the integral above.
As for the region of large $r$, $r \gg 1/m_L$, the function $\rho$ falls off exponentially, and a contribution from this region is therefore negligible.

Substituting \eqref{Iresult} into \eqref{betaI}  we get the final result for the coupling $\beta$ of the world sheet $\mathbb{CP}(N-1)$ model  \eqref{cp}
\begin{equation}
	\beta ~~\approx ~~ c\,\frac{2\pi}{g^2_2}  \,\frac{m_G^2}{m_L^2}\,\frac{1}{\ln^2 \frac{m_G}{m_L}}
	\sim \frac{|\mu|}{|m|}\,\frac{1}{\ln^2 \frac{g^2|\mu|}{|m|}}.
	\label{beta}
\end{equation}

 $\mathbb{CP}(N-1)$ model \eqref{cp} is a low  energy effective theory on the string world sheet. It describes
 the  dynamics of massless orientational moduli at energies much below the inverse thickness of the string proportional to $m_L$. If we go to higher energies we have to take into account higher derivative 
 corrections to \eqref{cp}. 

Relation \eqref{beta} is derived at the classical level. In quantum theory 
the coupling constant $\beta$ runs. Relation \eqref{beta} defines the $\mathbb{CP}(N-1)$ model coupling at
 a scale of the ultra-violet (UV) cutoff of the world sheet theory equal to $m_G$.
 In fact,  $\mathbb{CP}(N-1)$ model is an asymptotically free theory. Its coupling at the UV scale 
$m_G\sim \sqrt{\xi}$
 at one loop is given by 
\beq
4\pi \beta(\sqrt{\xi})= N\ln{\frac{\sqrt{\xi}}{\Lambda_{CP}}},
\eeq
where  $\Lambda_{CP}$ is the scale of the $\mathbb{CP}(N-1)$ model. This gives for the scale $\Lambda_{CP}$
\beq
\Lambda_{CP} \approx \sqrt{\xi} \exp{- {\rm const}\,\frac{|\mu|}{|m|}\frac{1}{\ln^2 \frac{g^2|\mu|}{|m|}}}.
\label{LambdaCP}
\eeq

We see that the scale of $\mathbb{CP}(N-1)$ model $\Lambda_{CP}$ is exponentially small, so the world sheet theory
is weаkly coupled in a wide region of energies $\gg \Lambda_{CP}$. This should be  contrasted to non-Abelian string in \ntwo QCD where world sheet theory has a scale $\Lambda_{CP}$ equal to scale 
$\Lambda_{{\cal N}=2}$ of
the bulk QCD \cite{SYrev}.

\subsection{World-sheet potential  at large $\mu$}
\label{sec:worldsheetpot}

In this subsection we relax the condition of equal quark masses \eqref{equalmasses} and consider the effect
of quark mass differences to the leading order in $\Delta m_{AB}$, see \eqref{smallmassdiff}.
Non-equal quark masses break color-flavor symmetry \eqref{c+f} down to U$(1)^N$ so 
as we already mentioned above the orientational modes of the non-Abelian string are no longer zero modes. They become quasizero
modes in the approximation of small quark mass differences \eqref{smallmassdiff}. We still can introduce the orientational quasimoduli $n^l$, $l=1,...,N$ and consider a shallow  potential in the $\mathbb{CP}(N-1)$ world sheet theory \eqref{cp} generated by the mass differences.
 We neglect effects of small mass differences in the kinetic term assuming that it is still given by 
 Eq. \eqref{cp}.
 
Our general strategy is to take string solution \eqref{string-solution} with the unperturbed string profile functions of Sec.~\ref{sec:profile} and substitute it into potential \eqref{potential:3+1:F:original}
 taking into account explicit $m_A$ dependence of this potential to the leading order in $\Delta m_{AB}$.
 After a rather involved  calculation we arrive at the potential of the world sheet theory
 \begin{equation}
 	\delta V_{1+1} ~~=~~ \chi   \sum_{A=1}^{N} \frac{\Re \left[(\xi_A - \hat{\xi}) \bar{\hat{\xi}}\right]}{|\hat{\xi}|} |n^A|^2 ,
 	\label{V1+1-general}
 \end{equation}
 where $\delta V_{1+1}$  is the potential up to a constant, $\xi_P$ are given by \eqref{xi-general}, while the factor $\chi$ is determined  by the
string profile functions integrated over $(x_1,x_2)$ plane,
\begin{equation}
	\chi ~~=~~ \frac{\pi }{|\mu_2|^2} \int\limits_{0}^{\infty} rdr\, (\phi_2^2 - \phi_1^2)\,
	\left[\phi_1^2 - \frac{\alpha}{N}(\phi_1^2 - \phi_2^2)\right].
	\label{chi}
\end{equation}

Now we use our solutions for $\phi_1$ and $\phi_2$ (see Sect. \ref{sec:profile}) and integrate here over the region $r\lsim 1/m_L$. We also  assume that
$\mu_1$ and $\mu_2$ scale in such a way that the parameter $\alpha$ in \eqref{alpha} is fixed. More explicitly,	 we assume that
\begin{equation}
\mu ~~\equiv~~\mu_2 ~~=~~ {\rm const} \cdot \mu_1 \sqrt{\frac{2}{N}}  ,
\label{mus}
\end{equation} 
This gives for $\chi$
\begin{equation}
	\chi ~~\approx~~ {\rm const} \cdot \frac{2\pi}{\ln \frac{m_G }{m_L}} .
\end{equation}
 Moreover, the region of integration $r\gg 1/m_L$ in \eqref{chi} does not contribute to the leading order. The unknown constant
 above appears due to an ambiguity of the upper limit of the integral over $r$, $r\sim 1/m_L$.

Substituting this into \eqref{V1+1-general} we get
\begin{equation}
	\delta V_{1+1} ~~\approx~~ {\rm const} \cdot 2\pi   \sum_{A=1}^{N}
			\frac{ \Re \left[(\xi_A - \hat{\xi}) \bar{\hat{\xi}}\right]}{|\hat{\xi}|
			\ln \frac{m_G }{m_L}} |n^A|^2 .
			\label{deltaV1+1}
\end{equation} 

Now let us  fix the unknown constant in the equation above comparing it with expressions \eqref{tenA0} for the string tensions of $Z_N$ strings. $Z_N$ strings are extreme points of the world sheet potential $V_{1+1}$, therefore a 
value of $V_{1+1}$ at the extreme point $n^l = \delta^{lA_0}$ corresponding to $A_0$-th $Z_N$ 
string should be equal
to string tension \eqref{tenA0}.
This gives ${\rm const} =1$ in \eqref{deltaV1+1} and leads us to the final expression for the potential in the world sheet $\mathbb{CP}(N-1)$ theory\footnote{Although \eqref{V1+1} seems to be logical, it cannot be correct for arbitrary mass differences $\Delta m_{AB}$. This can be seen from e.g. considering the U($N=2$) model with masses $m_1 = - m_2$. The resolution of this issue is still unknown, but note that at large $\Delta m_{AB}$ it makes less sense to even introduce the moduli $n^A$.}
\begin{equation}
		V_{1+1} ~~\approx~~ \frac{4\pi}{\ln \frac{ m_G}{m_L}}  \,  \abs{\sqrt{\frac{2}{N}} \mu_1 \hat{m} ~+~ \mu_2 \left( \sum\limits_{A=1}^{N} m_A |n^A|^2 - \hat{m} \right) }.
					\label{V1+1}
\end{equation} 
This potential integrated over world sheet coordinates $t$ and $z$ should be added to the kinetic term 
in \eqref{cp}.
Note, that this potential is a generalization of our result in \eqref{deltaV1+1} since it includes all terms in the expansion in powers of $(m_A-\hat{m})/\hat{m}$. For the single trace $\mu$-deformation \eqref{mu_condition}
the world sheet potential takes a particularly simple form
\begin{equation}
		V_{1+1} ~~\approx~~ \frac{4\pi}{\ln \frac{ m_G}{m_L}} \,   |\mu_2|\left| \sum\limits_{A=1}^{N} m_A |n^A|^2  \right| .
					\label{V1+1singletr}
\end{equation} 

The potential \eqref{V1+1} has only one minimum and one maximum at generic $\Delta m_{AB}$. Other $(N-2)$ extreme points are saddle points. All these extreme points  are located at
\beq
n^A = \delta^{AA_0}, \qquad A_0=1,...,N,
\label{saddleloc}
\eeq
and associated with the $Z_N$ strings. 
A value of the potential at a given extreme point coincides with the tension of the $A_0$-th $Z_N$ string,
\beq
V_{1+1}(n^A = \delta^{AA_0}) = T_{A_0}, \qquad A_0=1,...,N.
\label{minima}
\eeq
Absolute minimum (the unique vacuum)  of \eqref{V1+1} corresponds to the $Z_N$ string associated with winding of a squark with the smallest mass.

Note that our derivation of Eq. \eqref{deltaV1+1} reproduced the logarithmic suppression typical
for string tensions in extreme type I superconductors (with small Higgs mass $m_L$), see \eqref{tenA0}
and \cite{Y99}.

Potential \eqref{V1+1} is similar to the potential in the world sheet theory on the non-Abelian string
derived in \cite{SYfstr} for $\mu$-deformed \ntwo QCD in the limit of  small $\mu$. In this case the world sheet
theory is heterotic $\mathbb{CP}(N-1)$ model with \ntwoo supersymmetry. For small $\mu$ the world sheet potential
obtained in \cite{SYfstr} can be written in the form
\beq
	V_{1+1}^{\mu\to 0} ~~=~~ 4\pi     \abs{\sqrt{\frac{2}{N}} \mu_1 \hat{m} ~+~ \mu_2 \left( \sum\limits_{A=1}^{N} m_A |n^A|^2 - \hat{m} \right) }.
					\label{V1+1SYfstr}
\eeq
It differs from the one in \eqref{V1+1} by the absence of the logarithmic suppression. This has a natural explanation. At small $\mu$ in much the same way as in our case the saddle points of the world sheet potential correspond to the $Z_N$ strings and relation \eqref{minima} is still valid.
On the other hand in the limit of small $\mu$ the $Z_N$ strings are BPS saturated and their tensions are given by
$T_{A_0}^{\mu\to 0} = 2\pi|\xi_{A_0}|$, see \cite{SYfstr}. This explains the absence of the logarithmic suppression in the potential \eqref{V1+1SYfstr}.

\subsection{Mass spectrum on the string}

Let us assume that $m_1$ is the smallest quark mass. Then the vacuum of the world sheet potential \eqref{V1+1} is located at 
\beq
n^A = \delta^{A 1}
\label{minimumloc}
\eeq
and the minimum value $V_{1+1}^{\rm min} = T_{A=1}$. Let us calculate a perturbative mass spectrum of 
the world sheet theory in this vacuum. Expanding 
\begin{equation}
	|n^{1}|^2 ~~=~~ 1 ~-~ \sum_{A\neq 1} |n^A|^2
\end{equation}
and extracting the quadratic in fluctuations $n^A$ terms from potential \eqref{V1+1}, 
we get masses of world sheet excitations $n^A$, $A\neq 1$
\begin{equation}
	m_{A \neq 1}^2 ~~=~~ \frac{\pi}{\beta \ln \frac{ m_G}{m_L}} 
	\frac{ \Re \left[(\xi_A - \xi_{1}) \bar{\xi_{1}}\right]}{|\xi_{1}|} .
	\label{masses2Dgen}
\end{equation}
Note the factor $1/(2\beta)$ here that comes from the kinetic term in \eqref{cp}. Substituting here the coupling $\beta$ from \eqref{beta} we see that the masses of the perturbative world sheet excitations behave as
\beq
m_{A \neq 1}^2 \sim m (m_A-m_1)\,\ln \frac{ m_G}{m_L} \,.
\label{masses2D}
\eeq
 
The coupling constant of $\mathbb{CP}(N-1)$ model grows at low energies and gets frozen at the scale of 
the masses calculated above. If these masses are much larger than $\Lambda_{CP}$ \eqref{LambdaCP} then the world sheet theory is at weak coupling. Since $\Lambda_{CP}$ is exponentially small we see that world sheet theory is 
at weak coupling even at rather small mass differences $\Delta m_{AB}$. However in the equal quark mass limit
\eqref{equalmasses} when $\Delta m_{AB} =0$ the world sheet $\mathbb{CP}(N-1)$ model becomes strongly coupled.

Our result \eqref{V1+1} for the world sheet potential on the non-Abelian string in $\mu$-deformed theory can be  compared with the world sheet potential for the non-Abelian string in \ntwo supersymmetric QCD with FI $D$-term
generated by quark mass differences, see \cite{SYrev} for a review. In the \ntwo case all the $Z_N$ strings are degenerate, with tensions given by the FI parameter. The world sheet potential in this case has $N$  minima
located at \eqref{saddleloc} separated by shallow barriers quadratic in $\Delta m_{AB}$. The world sheet
theory has \ntwot supersymmetry and the presence of $N$ vacua is ensured by the Witten index for $\mathbb{CP}(N-1)$ 
supersymmetric model. There are kinks interpolating between these vacua which are interpreted as confined monopoles of bulk QCD \cite{SYmon,HT2}, see Sec.~\ref{physics} and \cite{SYrev} for a review.

In the limit of large $\mu$ potential \eqref{V1+1} dominates over the quadratic in $\Delta m_{AB}$
potential, and one can neglect the latter one. We see that most of the vacua present in the \ntwo case are lifted and
the world sheet theory has a single vacuum at non-zero  $\Delta m_{AB}$. Moreover, the lifted vacua are 
saddle points rather than  local minima and therefore classically they are unstable. This means that there are no kinks 
in the world sheet theory. 

Thus we come to the  conclusion that confined monopoles present in \ntwo QCD with FI term do not survive
large $\mu$ limit when $\mu$-deformed theory flows to \none QCD, provided that $\Delta m_{AB}$ are non-zero.
Only when $\Delta m_{AB}=0$, the potential \eqref{V1+1} vanishes (and the world sheet theory enters into the strong
coupling), and we can consider kinks/confined monopoles. We will discuss this case below in Sec.~\ref{physics}.

%
%

\section{Fermion zero modes
\label{sec:superorient-zero}}

In this section we consider the fermion zero modes of the non-Abelian string. First we briefly review  
the limit of 
small $\mu$, see \cite{SYrev} for a more detailed review. In this limit deformation superpotential 
\eqref{none_superpotential_general} reduces to the FI $F$-term and does not break \ntwo supersymmetry
\cite{HSZ,VY}. In the \ntwo limit both superorientational and supertranslational fermion zero modes 
of the non-Abelian string can be obtained by a supersymmetry transformation of the bosonic string solution
\cite{SYmon,SY02,BSYhet}. Next we gradually increase $\mu$ and study perturbations of 
superorientational zero modes at small $\mu$. We   show that all the superorientational fermion zero 
modes are lifted by the $\mu$-deformation. As a result fermionic moduli which become fermion fields in the 
two dimensional low energy $\mathbb{CP}(N-1)$ model on the string acquire masses. Eventually they disappear from
the world sheet theory in the large $\mu$ limit. Finally we  comment on supertranslational
fermion zero modes which in much the same way as in \ntwo theory can be obtained by supersymmetry 
transformations from the bosonic string solution.

\subsection{Superorientational modes in \ntwo limit}

The fermionic part of the \ntwo QCD defined by superpotentials \eqref{supN2} and 
\eqref{none_superpotential_general} (before integrating out adjoint fields)  is as follows:
\begin{align}
\notag
&\mc{L}_{\rm 4d}  ~~=~~ 
\\
\notag
&
\frac{2i}{g_2^2} \Tr\, \ov{\lN_f \slashed{\md}} \lambda^{f\text{SU($N$)}}
~+~ \frac{i}{g_1^2} \ov{\lU_f \slashed{\p}} \lambda^{f{\rm U(1)}}
~+~ \Tr\, i\, \ov{\psi \slashed{\nabla}} \psi
~+~ \Tr\, i\, \wt{\psi} \slashed{\nabla} \ov{\wt{\psi}}
\\[3mm]
\notag
&
~+~
i\sqrt{2}\, \Tr \lgr \ov{q}{}_f \lambda^{f{\rm U(1)}}\psi
~+~ \wt{\psi} \lU_f q^f
~+~ \ov{\psi \lU_f} q^f
~+~ \ov{q^f \lU_f \wt{\psi}}
\rgr
\\[3mm]
\label{fermact}
&
~+~
i\sqrt{2}\, \Tr \lgr \ov{q}{}_f \lambda^{f\text{SU($N$)}} \psi
~+~ \wt{\psi} \lN_f q^f
~+~ \ov{\psi \lN_f} q^f
~+~ \ov{q^f \lN_f \wt{\psi}}
\rgr
\end{align}
\begin{align}
\notag
&
~+~
i\sqrt{2}\, \Tr\; \wt{\psi} \left( \frac12 \aU ~+~ \frac{m_A}{\sqrt{2}} ~+~ \aN \right) \psi 
\\[3mm]
\notag
&
~+~
i\sqrt{2}\, \Tr\; \ov{\psi} \left( \frac12\baU ~+~ \frac{m_A}{\sqrt{2}} ~+~ \baN \right) \ov{\wt{\psi}}
\\[3mm]
\notag
&
~-~
2 \sqrt{\frac{N}{2}} \mu_1 \lgr \left( \lambda^{2\,{\rm U(1)}} \right)^2
~+~ \left( \ov{\lambda}{}^{\rm U(1)}_2 \right)^2 \rgr
~-~
\mu_2 \Tr \lgr \left( \lambda^{2\,\text{SU($N$)}} \right)^2
~+~ \left( \ov{\lambda}{}^\text{SU($N$)}_2 \right)^2 \rgr\,,
\end{align}
where derivatives acting on fermion fields are defined by the $\sigma$-matrices, for example
  $\bar{\slashed{\nabla}}= \nabla_{\mu}\overline{\sigma}{}^\mu_{\dot{\alpha}\alpha}$, and  
a color-flavor matrix notation is used for the quark fermions $\psi_{\alpha}^{kA}$, $\tilde{\psi}^{\alpha}_{Ak}$. 
Index $f$ 
is SU(2)$_R$ index of the \ntwo theory,  $ q^f ~=~ (q, \ov{\wt{q}}) $, 
$\lambda_{\alpha}^f=(\lambda_{\alpha}^1,\lambda_{\alpha}^2)$. Note the 
$\mu$-deformation mass terms for the $f = 2$ gauginos in \eqref{fermact}. In the \ntwo limit these terms 
vanish.

A string solution in the \ntwo limit at small $\mu$ is 1/2 BPS, which means that half of the supercharges 
of the \ntwo theory act trivially on solution \eqref{string-solution}, provided the orientational vector
$n^l$ is a constant vector. Namely, the four supercharges (out of eight supercharges $Q^{\alpha f}$)
 that satisfy the constraints 
\beq
Q^{21} ~=~ Q^{22}, \qquad\qquad   Q^{11} ~=~ - Q^{12} \ .
\label{trivQ}
\eeq
act trivially on the BPS string in the \ntwo theory with the FI $F$-term \cite{SYmon,SYrev,VY}.

The other four supercharges  generate 
four supertranslational modes which are superpartners of the two translational modes.

However once the orientational vector
$n^l$ acquires a slow $t$ and $z$ dependence, the supercharges selected by \eqref{trivQ} 
become supersymmetry generators acting in the \ntwot supersymmetric $\mathbb{CP}(N-1)$ model on the 
string world sheet \cite{SYmon}. This allows one to obtain the orientational fermionic zero modes
from a bosonic solution using supersymmetry transformations selected by \eqref{trivQ} \cite{SYmon,SYrev}.
The result is 
\begin{align}
\notag
\ov{\psi}{}_{\dot{2}} & ~~=~~  \phantom{-\, }
\frac{\phi_1^2 ~-~ \phi_2^2}{\phi_2}\cdot n \bar{\xi}_L~,
\\[2mm]
\notag
\ov{\wt{\psi}}{}_{\dot{1}} & ~~=~~   -\,
\frac{\phi_1^2 ~-~ \phi_2^2}{\phi_2}\cdot \xi_R\bar{n} ~,
\\[2mm]
\label{N2_sorient}
\lambda^{11\ \text{SU($N$)}}  & ~~=~~ \phantom{-\, } i \frac{\phi_1}{\phi_2}\, f_W \, \frac{\displaystyle x^1 - i\, x^2}{\displaystyle r^2}\cdot
n \bar{\xi}_L
\\[2mm]	
\notag
\lambda^{22\ \text{SU($N$)}}  & ~~=~~ -\,  i \frac{\phi_1}{\phi_2}\, f_W \, \frac{\displaystyle x^1 + i\, x^2}{\displaystyle r^2}\cdot
\xi_R\bar{n}
\\[2mm]
\notag
\lambda^{12\ \text{SU($N$)}}  & ~~=~~ \lambda^{11\ \text{SU($N$)}}
\, ,\qquad    \lambda^{21\ \text{SU($N$)}}   ~~=~~ -\, \lambda^{22\ \text{SU($N$)}}\,,
\end{align}	
where we suppress color and flavor indices while superscripts of adjoint fermions mean $\lambda^{\alpha f}$.

Note that the bosonic profile functions of the string $\phi_{1,2}(r)$, $f(r)$ and $f_W$  in this 
section are the profile 
functions of the BPS string in the \ntwo limit of small $\mu$ rather than the string profile 
functions of 
Sec.~\ref{sec:profile}, which corresponds to the large $\mu$ limit. Former are solutions of first order 
equations rather than second order equations \eqref{profileeqs}. They satisfy boundary conditions 
\eqref{boundary} and were found numerically in \cite{ABEKY}, see \cite{SYrev} for a review.

The Grassmann variables $ \xi_{R,L}^l$, $l=1,...,N$ in \eqref{N2_sorient} are proportional to  the parameters of 
the supersymmetry transformations $\epsilon^{\alpha f}$ selected by \eqref{trivQ}, namely
\beq
 \xi_L^l \sim \epsilon^{21}+\epsilon^{22}, \qquad \xi_R^l \sim \epsilon^{12}-\epsilon^{11}.
\label{xi's}
\eeq
These parameters become fermion fields (superpartners of $n^l$) in the effective world sheet $\mathbb{CP}(N-1)$
 model once we allow their slow 
dependence on the world sheet coordinates $t$ and $z$ \cite{SYmon,SYrev}.
They are subject to the conditions
\begin{equation}
 \label{constr}
    \nbar_l\, \xi^l_{L,R}   ~~=~~ 0~,
\end{equation}
which are a supersymmetric generalization of the \CP condition $ |n|^2 = 1 $.

\subsection{Small $\mu$ expansion for fermion orientational zero modes}

As we switch on the mass terms for the $f=2$ gauginos (see the last line in \eqref{fermact}), 
the theory becomes \none supersymmetric and half of the supercharges $Q^{\alpha f=2}$ are lost. 
There are no SUSY 
transformations which act trivially on the string with constant $n^l$ (they were used to 
generate superorientational modes in \ntwo limit), and the string is no longer BPS. Therefore, to 
calculate zero modes one has to solve the Dirac equations. 

Note that for the case of the massless $\mu$-deformed theory with FI $D$-term considered in \cite{BSYhet}
the supercharges that act trivially on the string with constant $n^l$ in the \ntwo limit
 are $Q^{12}$ and $Q^{21}$ instead of the linear combinations selected by \eqref{trivQ}. 
 Therefore as we switch on the $\mu$-deformation only one (two real) of the above supercharges 
is lost, namely $Q^{12}$. The other one (two real), $Q^{21}$, still 
acts trivially and ensures that the string is 
still BPS-saturated. In our case all four supercharges of \none theory $Q^{\alpha 1}$ act non-trivially
on the string. This is the reason why the string ceases to be a BPS one as we switch on $\mu$.

Dirac equations which follow from action \eqref{fermact} read
\begin{align}
\notag
&\phantom{-i}
\frac{i}{g_1^2}\, \left( \ov{\slashed{\p}}\lambda^{f {\rm U(1)}} \right)
+  i \sqrt{2}\, \Tr\lgr \ov{\psi} q^f  + \ov{q}{}^f \ov{\wt{\psi}} \rgr
~-~ 4\, \delta_2^{\ f}\sqrt{\frac{N}{2}} \mu_1\, \ov{\lambda}^{U(1)}_2  = 0\,,
\\[2mm]
\notag
&\phantom{-i}
\frac{i}{g_2^2}\, \left( \ov{\slashed{\md}}\lambda^{f \text{SU($N$)}}\right)^a
+ i \sqrt{2}\, \Tr\lgr \ov{\psi}T^a q^f  +  \ov{q}{}^f T^a \ov{\wt{\psi}} \rgr
- \delta_2^{\ f} \mu_2\, \ov{\lambda}{}_2^{a \text{SU($N$)}}  = 0 \,,\\[2mm]
\notag
&
-i\, \ov{\psi} \overleftarrow{\ov{\slashed{\nabla}}}
+ i \sqrt{2} \lgr \ov{q}{}_f \left\{ \lambda^{f {\rm U(1)}} + \lambda^{f \text{SU($N$)}} \right\}
+ \wt{\psi} \left\{ \frac12\baU + \frac{m_A}{\sqrt{2}} + \baN \right\} \rgr    = 0\,, \\[2mm]
\label{dirac}
&\phantom{-i}
i\, \slashed{\nabla} \ov{\wt{\psi}}
+ i \sqrt{2} \lgr \left\{ \lambda_f^{\rm U(1)} + \lambda_f^\text{SU($N$)} \right\} q^f
+ \left\{ \frac12\baU + \frac{m_A}{\sqrt{2}} + \baN \right\}\psi \rgr  = 0\,, \\[2mm]
\notag
&\phantom{-i}
i \ov{\slashed{\nabla}}\psi
+ i \sqrt{2} \lgr \left\{ \ov{\lambda}{}^{\rm U(1)}_f + \ov{\lambda}{}^\text{SU($N$)}_f \right\} q^f
+ \left\{\frac12 \aU + \frac{m_A}{\sqrt{2}} + \aN \right\} \ov{\wt{\psi}} \rgr   = 0 \,,\\[2mm]
\notag
&
-i\, \wt{\psi} \overleftarrow{\slashed{\nabla}}
+ i \sqrt{2} \lgr \ov{q}{}^f \left\{ \ov{\lambda}{}_f^{\rm U(1)} + \ov{\lambda}{}_f^\text{SU($N$)} \right\}
+ \ov{\psi} \left\{ \frac12\aU + \frac{m_A}{\sqrt{2}} + \aN \right\} \rgr  = 0\,.
\end{align}

To simplify the problem we will use below the following strategy. We consider the region of small $\mu$
and look for solutions of the Dirac equations above  perturbatively in $\mu$.
Of course,  zero modes \eqref{N2_sorient}  satisfy  Dirac equations \eqref{dirac} \cite{SY02,BSYhet}
at $\mu=0$. 
We take these modes as a zero order solutions and solve for perturbations proportional to $\mu$.

Similar method was used in \cite{SY02,BSYhet} for a massless $\mu$-deformed theory with a FI $D$-term.
In that case it was  shown that the orientational fermion zero modes survive the $\mu$-deformation,
however, their profile functions become deformed. Below we will show that in our case of $\mu$-deformed 
theory  with massive quarks without the FI $D$-term the answer is different: orientational fermion 
zero modes do not survive the $\mu$-deformation.

In analogy with the method of \cite{BSYhet} we will use an {\it ansatz} for 
the superorientational modes:
\begin{equation}
 \begin{aligned}
  \lambda^{1f\ \text{SU($N$)}} & ~~=~~ 2\, \frac{x^1 - ix^2}{r}\, \lambda^{1f}_+(r) \; n\ov{\xi}_L
  ~~+~~  2\, \lambda^{1f}_-(r)\; \xi_L\ov{n}\,,
  \\[2mm]
  \lambda^{2f\ \text{SU($N$)}} & ~~=~~ 2\, \frac{x^1 + ix^2}{r}\, \lambda^{2f}_+(r) \; \xi_R\ov{n}
  ~~+~~  2\, \lambda^{2f}_-(r)\; n\ov{\xi}{}_R\,,
 \end{aligned}
 \label{small_mu-supertransl-ansatz-lambda}
\end{equation}
\begin{equation}
\begin{aligned}
  \ov{\wt{\psi}}{}_{\dot{1}} & ~~=~~ 2\, \ov{\wt{\psi}}_{\dot{1}+}(r)\;  \xi_R \ov{n}
  ~~+~~  2\, \frac{x^1 - i x^2}{r}\, \ov{\wt{\psi}}_{\dot{1}-}(r)\; n\ov{\xi}{}_R~.
  \\[2mm]
  \ov{\wt{\psi}}{}_{\dot{2}} & ~~=~~ 2\, \ov{\wt{\psi}}_{\dot{2}+}(r)\; n\ov{\xi}_L
  ~~+~~  2\, \frac{x^1 + i x^2}{r}\, \ov{\wt{\psi}}_{\dot{2}-}(r)\; \xi_L\ov{n}~.
  \\[2mm]
  \ov{\psi}{}_{\dot{1}} & ~~=~~ 2\, \ov{\psi}_{\dot{1}+}(r)\;  \xi_R \ov{n}
  ~~+~~  2\, \frac{x^1 - i x^2}{r}\, \ov{\psi}_{\dot{1}-}(r)\; n\ov{\xi}{}_R~.
  \\[2mm]
  \ov{\psi}{}_{\dot{2}} & ~~=~~ 2\, \ov{\psi}_{\dot{2}+}(r)\; n\ov{\xi}_L
  ~~+~~  2\, \frac{x^1 + i x^2}{r}\, \ov{\psi}_{\dot{2}-}(r)\; \xi_L\ov{n}~.
\end{aligned}
\label{small_mu-supertransl-ansatz-psi}
\end{equation}
Here $ \lambda_+(r) $ and $ \psi_+(r) $ represent "undeformed" profile functions  present 
in the \ntwo case, while $ \lambda_-(r) $ and $ \psi_-(r) $ are the "perturbations" due to $\mu$-deformation.
 Of course this terminology makes sense only in the small $\mu$ limit, then "-" -components will 
be of order $\mu$. More generally, the ''+'' profile functions are expanded in even powers of $\mu$,
while ''-'' components are expanded in odd powers of $\mu$.

Let us consider the equations for the perturbative "-" -components (solutions for the "+" -components 
are given   by \eqref{N2_sorient} up to the $O(\mu^2)$ terms). Half of them are very similar to those 
solved in \cite{BSYhet}. If we denote
\[
\lambda^{22}_- - \lambda^{21}_- \equiv \lambda_-\, ,
\]
then two of these equations for $\lambda_{-} $ and $\ov{\wt{\psi}}_{\dot{1}-}$ take the form
\beqn
&&
\p_r \ov{\wt{\psi}}_{\dot{1}-}(r) ~+~ \frac{1}{r}\ov{\wt{\psi}}_{\dot{1}-}(r) ~-~ 
\frac{1}{Nr}\left(f+f_W(N-1)\right) \, \ov{\wt{\psi}}_{\dot{1}-}(r) ~+~ i\phi_2\,\lambda_-~~=~~ 0
\nonumber \\
&&
-\,  \p_r\lambda_- ~-~ \frac{f_W}{r}\lambda_-
~+~ i\,g_2^2 \phi_2\, \ov{\wt{\psi}}_{\dot{1}-}(r) ~-~ \mu_2 g_2^2\, \frac{i}{2} \frac{f_W}{r} \frac{\phi_1}{\phi_2} ~~=~~ 0
\label{easy}
\eeqn
These equations can be solved in much the same way as in \cite{BSYhet}. The solutions are
\begin{align}
\notag
\ov{\wt{\psi}}_{\dot{1}-} & ~~=~~ -\, \mu_2 g_2^2 \,\frac{r}{8\phi_1} \left( \phi_1^2 ~-~ \phi_2^2 \right)  
~+~ O(\mu^3)\,,
\\[2mm]
\label{firstorder-ovwt1}
\lambda_- & ~~=~~ \lambda^{22}_- - \lambda^{21}_- ~~=~~ -\, \mu_2 g_2^2 \,\frac{i}{4} \lgr (f_W - 1) \frac{\phi_2}{\phi_1} ~+~ \frac{\phi_1}{\phi_2} \rgr ~+~ O(\mu^3)~.
\end{align}
Another pair of the profile functions $\ov{\psi}_{\dot{2}-}$ and $(\lambda^{12}_- ~+~ \lambda^{11}_-)$
satisfies the same equations \eqref{easy}.  Hence the solution reads
\begin{align}
\notag
\ov{\psi}_{\dot{2}-} & ~~=~~ \ov{\wt{\psi}}_{\dot{1}-}\, , 
\\[2mm]
\label{firstorder-ov2}
\lambda^{12}_- ~+~ \lambda^{11}_- &~~=~~ \lambda_- .
 \end{align}

Let us study behavior of these solutions in the limits $r\to\infty$ and $r\to 0$.
The bosonic  profile functions fall off exponentially at infinity
\begin{equation}
 f_W(r) \sim \exp{-m_G r}, \qquad \phi_{1,2}-\sqrt{\xi}\sim \exp{-m_G r},
\label{infbehb}
\end{equation}
while their behavior at $r\to 0$ is as follows:
\beq
 f_W(r)-1 \sim r^2, \qquad \phi_{1}\sim r, \qquad \phi_2 \sim {\rm const},
\label{zerobehb}
\eeq
see \eqref{boundary}.

From this behavior we see that fermion zero modes \eqref{firstorder-ovwt1}, \eqref{firstorder-ov2}
are normalizable. They fall off exponentially at $r\to\infty$ and are regular at $r\to 0$.

Now consider solutions  for the other components. They turn out to be more complicated. Denoting
\[
\lambda^{22}_- + \lambda^{21}_- \equiv \lambda_{(1)}\, ,
\]
one gets:
\begin{equation}
\begin{aligned}
\p_r \ov{\psi}_{\dot{1}-}(r) ~+~ \frac{1}{r}\ov{\psi}_{\dot{1}-}(r) ~+~ \frac{1}{Nr} (f-f_W) \, 
\ov{\psi}_{\dot{1}-}(r) ~+~ i\phi_1\,\lambda_{(1)} &~~=~~ 0\,,\\[2mm]
-\,  \p_r\lambda_{(1)} ~-~ \frac{f_W}{r}\lambda_{(1)}
~+~ i\,g_2^2 \phi_1\, \ov{\psi}_{\dot{1}-}(r) ~-~ \mu_2 g_2^2\, 
\frac{i}{2} \frac{f_W}{r} \frac{\phi_1}{\phi_2} &~~=~~ 0.
\end{aligned}
\label{fermeqs_ovpsi1_lplus_small-1}
\end{equation}
 So far solutions for our equations were given by certain algebraic combinations of the bosonic profile functions. 
However, for the functions $\ov{\psi}_{\dot{1}-}$ and $\lambda_{(1)}$ it is not the case. 
The above equations are solved in Appendix~\ref{app:dirac}. The solutions are given by
Eqs.~ \eqref{fermsol_ovpsi1_lplus_small} and \eqref{fermsol_lamplus_lplus_small}.

Two remaining modes   $\ov{\wt{\psi}}_{\dot{2}-}$ and $(\lambda^{12}_- ~-~ \lambda^{11}_-)$
satisfy the same equations \eqref{fermeqs_ovpsi1_lplus_small-1}.  Therefore  these modes are 
given by the same expressions,


\beq
\ov{\wt{\psi}}_{\dot{2}-}(r)=\ov{\psi}_{\dot{1}-}(r), \qquad 
\lambda^{12}_{-} ~-~ \lambda^{11}_{-}= \lambda_{(1)}(r).
\label{remainingferm}
\eeq

Solutions \eqref{fermsol_ovpsi1_lplus_small} and \eqref{fermsol_lamplus_lplus_small} 
fall off exponentially at infinity, however, the behavior of the field $\lambda$ 
in  \eqref{fermsol_lamplus_lplus_small}  is singular at $r\to 0$, namely
it is proportional to $1/r$.
   This means that these modes are non-renormalizable. Our perturbative approach does not work: 
the corrections
to \eqref{N2_sorient} proportional to $\mu$ turn out to be non-normalizable. We will show in the next 
subsection that
the resolution of this puzzle is that the fermion orientational modes get lifted by the $\mu$-deformation.

 \subsection{Lifted fermion orientational modes}

Let us consider instead of Dirac equations \eqref{dirac}  equations with a non-zero eigenvalue for 
quark fermions, namely
\begin{multline}
  -i\, \ov{\psi} \overleftarrow{\ov{\slashed{\nabla}}}
 ~+~ i \sqrt{2} \Bigg( \ov{q}{}_f \left\{ \lambda^{f {\rm U(1)}} + \lambda^{f \text{SU($N$)}} \right\} \\
 ~+~ \wt{\psi} \left\{ \frac12 \baU ~+~ \frac{m_A}{\sqrt{2}} ~+~ \baN \right\} \Bigg)   
 ~~=~~ - m_{or } \, \wt{\psi} \,,
 \label{dirac_modifyed-ovpsi}
\end{multline}
\begin{multline}
  i\, \slashed{\nabla} \ov{\wt{\psi}}
 ~+~ i \sqrt{2} \Bigg( \left\{ \lambda_f^{\rm U(1)} ~+~ \lambda_f^\text{SU($N$)} \right\} q^f \\
 ~+~ \left\{ \frac12\baU ~+~ \frac{m_A}{\sqrt{2}} ~+~ \baN \right\}\psi \Bigg)  ~~=~~ - m_{or } \,\psi \, .
 \label{app:dirac_modifyed-ovwtpsi}
\end{multline}
with the mass $m_{or}$  to be determined from the condition of normalizability 
of superorientational modes.

Proceeding exactly as it was done in the previous subsection, instead of 
Eqs.~\eqref{fermeqs_ovpsi1_lplus_small-1}
we arrive  at
\begin{equation}
\begin{aligned}
\p_r \ov{\psi}_{\dot{1}-}(r) ~+~ \frac{1}{r}\ov{\psi}_{\dot{1}-}(r) ~+~ \frac{1}{Nr} (f-f_N) \, \ov{\psi}_{\dot{1}-}(r) ~+~ i\phi_1\,\lambda_{(1)} &~~=~~ m_{or }\,\frac{\phi_1^2 ~-~ \phi_2^2}{2\phi_2}\,,\\[2mm]
-\,  \p_r\lambda_{(1)} ~-~ \frac{f_N}{r}\lambda_{(1)}
~+~ i\,g_2^2 \phi_1\, \ov{\psi}_{\dot{1}-}(r) ~-~ \mu_2 g_2^2\, \frac{i}{2} \frac{f_N}{r} \frac{\phi_1}{\phi_2} &~~=~~ 0.
\end{aligned}
\label{fermeqs_ovpsi1_lplus_small-2}
\end{equation}
We consider these equations in the  Appendix~\ref{app:dirac}. The solutions are given by  \eqref{solution_orient_ov_psi_1-}
and \eqref{solution_orient_lambda(1)}.
The condition of regularity of these solutions at $r\to 0$ gives the eigenvalue 
\begin{equation}
m_{or } ~~=~~ -\frac{\displaystyle{\mu_2 g_2^2}\,\int\limits_0^\infty {\mathrm d}y \frac{f_N^2(y)\phi_1^2(y)}{y\phi_2^2(y)}}{ 1 ~-~ 2\displaystyle\int\limits_0^\infty {\mathrm d}y \frac{f_N^2(y)\phi_1^2(y)}{y\phi_2^2(y)}}.
\label{solution_orient_alpha}
\end{equation}
Solutions for $\ov{\wt{\psi}}_{\dot{2}-}$ and the combination $(\lambda^{12}_{-} ~-~ \lambda^{11}_{-})$ 
satisfy the same equations \eqref{fermeqs_ovpsi1_lplus_small-2} and are related to  solutions
\eqref{solution_orient_ov_psi_1-} and \eqref{solution_orient_lambda(1)} via
 \eqref{remainingferm}.

\subsection{Effective action in the orientational sector \label{superorient-eff_action}}

Now to see the effect of lifting of the orientational fermion zero modes 
let us derive  a fermionic part of the two-dimensional effective action on the string 
world sheet  with the $O(\mu)$ accuracy. In order to do so, we assume a slow $t$ and $z$ dependence
of the fermionic moduli $\xi^l_{L,R}$, 
substitute our {\it ansatz} \eqref{small_mu-supertransl-ansatz-lambda},  
\eqref{small_mu-supertransl-ansatz-psi}  into the four dimensional fermion action \eqref{fermact}
 and integrate over $x_1,\ x_2$. Kinetic terms for bulk fermions (containing derivatives
 $\p_0$ and $\p_3$) produce corresponding kinetic terms for two-dimensional fermions, 
and mass terms are generated because fermionic modes are now lifted. 
The result for the quadratic terms in the two dimensional fermionic action is
\begin{equation}
\mc{S}_{\rm 2d} ~~=~~ \int dt dz \,\left\{ \frac{4\pi}{g_2^2} (\bar{\xi}_L i\p_R \xi_L ~+~  \bar{\xi}_R i\p_L \xi_R) ~+~ m_{or }\gamma\, (\bar{\xi}_R \xi_L ~+~  \bar{\xi}_L \xi_R) + \cdots \right\},
\label{effective_action-superorient}
\end{equation}
where dots stand for higher order terms in fields and
\begin{equation}
\gamma ~~=~~ -4 \,  \int dx_1 dx_2\, \ov{\wt{\psi}}_{\dot{1}+} \, \ov{\psi}_{\dot{2}+} ~~=~~ 4\,  \int dx_1 dx_2\, |\ov{\psi}_{\dot{2}+}|^2 \, ,
\end{equation}
while
\[	
\p_R ~~=~~ \p_0 ~+~ i\, \p_3 ~, \qquad\qquad  \p_L ~~=~~ \p_0 ~-~ i\, \p_3~.
\]
We see that all two-dimensional fermionic fields $\xi^l_{L,R}$ become massive with mass
$m_{or}$ proportional to $\mu$. We expect that in the limit of large $\mu$ these fermions decouple
from bosonic $\mathbb{CP}(N-1)$ model \eqref{cp}.

\subsection{Supertranslational zero modes}

As we already mentioned, supertranslational modes  
can be obtained via a supersymmetry transformation from bosonic string solution even in the 
$\mu$-deformed theory. String solution ceases to be a BPS one, and all of the four remaining supercharges 
$Q^{\alpha 1}$ of the \none theory act non-trivially on the string solution. Much in the same way as the 
bosonic translational modes, the supertranslational ones decouple from orientational $\mathbb{CP}(N-1)$
model and are described by free fermions on the string world sheet. This can be anticipated
on general grounds.  To see this note that the orientational
fermion fields $\xi^l_{L,R}$ become heavy at large $\mu$ and without them  
we cannot construct interaction terms of $n^l$ and  supertranslational moduli $\zeta_{L,R}$
compatible with symmetries of the theory (if we do not consider higher derivative corrections). 

For the sake of completeness
we construct explicitly supertranslational zero modes in the large $\mu$ limit,
acting by \none supersymmetry
transformations on the string solution of Sec.~\ref{sec:bosonic_solution}.
The \none supersymmetry
transformations have the form
\begin{eqnarray}
\delta\bar{\tilde\psi}_{\dot{\alpha}}^{kA}
&=&
i\sqrt2\
\bar\nabla\hspace{-0.65em}/_{\dot{\alpha}\alpha}q^{kA}\epsilon^{\alpha } ~+~ \sqrt{2}\, \bar{\epsilon}^{\alpha }\bar{\tilde F}^{kA}  \  \ ,
\nonumber\\[4mm]
\delta\bar\psi_{\dot{\alpha}Ak}
&=&
i\sqrt2\
\bar\nabla\hspace{-0.65em}/_{\dot{\alpha}\alpha}\bar{q}_{\,Ak}\epsilon^{\alpha } ~+~ \sqrt{2}\, \bar{\epsilon}^{\alpha }\bar{F}_{kA}  \ ,
\label{transf_n=1}
\end{eqnarray}
where the $F$-terms
are given by derivatives of the superpotential \eqref{superpotential-adjoints_integrated}, 
\begin{equation}
\begin{aligned}
\bar{F}_{Ak} ~~=~~ -\frac{\p \mc{W}}{\p q^{kA}} &~=~ \frac{i}{\mu_2} \left( \tilde{q}_{Ck} (\tilde{q}_A q^C) - \frac{\alpha}{N} (\tilde{q}_C q^C)\tilde{q}_{Ak} \right) + m \tilde{q}_{Ak} \\
&~=~ \frac{i}{\mu_2} \left( \vp_{Ck} (\vp_A \vp^C) - \frac{\alpha}{N} (\vp_C \vp^C)\vp_{Ak} \right) + m \vp_{Ak} \,,
\end{aligned}
\end{equation}
\begin{equation}
\begin{aligned}
\bar{\tilde F}^{kA} ~~=~~ -\frac{\p \mc{W}}{\p \tilde{q}_{Ak}} &~=~ \frac{i}{\mu_2} \left( q^{kC} (\tilde{q}_C q^A) - \frac{\alpha}{N} (\tilde{q}_C q^C)q^{kA} \right) + m q^{kA} \\
&~=~ \frac{i}{\mu_2} \left( \vp^{kC} (\vp_C \vp^A) - \frac{\alpha}{N} (\vp_C \vp^C)\vp^{kA} \right) + m \vp^{kA}  \,,
\end{aligned}
\end{equation} 
where we also used \eqref{q-ansatz}.

Consider first the region of intermediate $r$,
 in the 
range $1/m_G \lesssim r \lesssim 1/m_L$. 
As we will see, the fermion zero modes  behave as $1/r$. This will give us leading logarithmic 
contributions to the kinetic terms for fermions of two dimensional
effective theory on the string world sheet. 

To calculate the fermionic modes one should substitute bosonic solutions
 \eqref{string:boson:large_mu:dm=0:intermediate} into the 
transformations \eqref{transf_n=1}.  In \eqref{transf_n=1}, the first terms in the first and second 
  lines give $1/r$ contributions, whereas the $F$-terms adds  constant and logarithmic terms, which 
does not produce leading logarithmic terms in the effective action. 
 We neglect these last terms, and get non-zero fermionic 
profiles 
%
%
%
\begin{equation}
\begin{aligned}
\ov{\psi}{}_{\dot{1}} & ~~\approx~~   (n\nbar)\ \frac{\displaystyle x^1 - i\, x^2}{\displaystyle r}  \frac{1}{r} \ \frac{\sqrt{\xi}}{\ln \frac{g_2\sqrt{\xi}}{m_L}} \, \zeta_R
~, \\[2mm]
\ov{\psi}{}_{\dot{2}} & ~~\approx~~  (n\nbar)\ \frac{\displaystyle x^1 + i\, x^2}{\displaystyle r}  \frac{1}{r} \ \frac{\sqrt{\xi}}{\ln \frac{g_2\sqrt{\xi}}{m_L}} \, \zeta_L
~, \\[2mm]
\ov{\wt{\psi}}{}_{\dot{1}} & ~~\approx~~ (n\nbar)\ \frac{\displaystyle x^1 - i\, x^2}{\displaystyle r}  \frac{1}{r} \ \frac{\sqrt{\xi}}{\ln \frac{g_2\sqrt{\xi}}{m_L}}  \, \zeta_R
~, \\[2mm]
\ov{\wt{\psi}}{}_{\dot{2}} & ~~\approx~~ (n\nbar)\ \frac{\displaystyle x^1 + i\, x^2}{\displaystyle r}  \frac{1}{r} \ \frac{\sqrt{\xi}}{\ln \frac{g_2\sqrt{\xi}}{m_L}}  \, \zeta_L .
\end{aligned}
\label{supertranslational_N=1_psi_solutions}
\end{equation}
One can see that these modes are indeed proportional to $1/r$.
Here $\zeta_{L,R}$ are the Grassmann parameters generated by supersymmetry transformations,
$\zeta_L = \frac{1}{\sqrt{2}}\epsilon^{1}, \  \zeta_R = -\frac{1}{\sqrt{2}} \epsilon^{2}$.
These parameters become fermionic fields in the two dimensional effective 
theory on the string world sheet.

The region of small $r$, $r\ll 1/m_G$ does not contribute because 
quark fields vanish in this limit.

To find  the effective world sheet action, one should substitute  solutions \eqref{supertranslational_N=1_psi_solutions} into four-dimensional fermionic action \eqref{fermact}. For the kinetic term, we obtain:
\begin{equation}
\mc{L}_{\rm 2d} ~~=~~ 2\pi\xi \, I_\xi (\bar{\zeta}_L i\p_R \zeta_L ~+~  \bar{\zeta}_R i\p_L \zeta_R) \,,
\label{action:supertrans_2d}
\end{equation}
where the normalization constant is
\begin{equation}
I_\xi ~~=~~ \frac{2N}{\ln{\frac{m_W}{m_L}}} \,.
\end{equation}
As we already mentioned,  deriving this effective action we integrated over the transversal coordinates in the 
range $1/m_G \lesssim r \lesssim 1/m_L$. The integral over $r$ is logarithmically enhanced.
 The contributions of other regions do not have  logarithmic enhancement and can be 
neglected.

We see that \eqref{action:supertrans_2d} is the action for free fermions which decouples from
the orientational sector given by $\mathbb{CP}(N-1)$ model \eqref{cp}.

%
%

\section{Physics of the world sheet theory and confined monopoles
\label{physics}}

As we have seen above the fermionic fields $\xi^l$ of the effective world sheet theory 
 become heavy in the large $\mu$ limit and decouple. Moreover, the translational sector
is free and does not interact with the orientational sector. Thus, our effective world sheet theory on the 
non-Abelian string is given by  bosonic 
CP($N$-1) model \eqref{cp} without fermions  in the large $\mu$ limit. If quark masses are 
small but not equal, the orientational moduli $n^l$ are lifted by shallow  potential \eqref{V1+1}.

As we already mentioned, our four dimensional bulk theory is in the Higgs phase where squarks
develop condensate \eqref{qVEV}. Therefore 't Hooft-Polyakov monopoles present in the  theory
in the \ntwo limit of small $\mu$ are confined by non-Abelian strings. In fact in U$(N)$ gauge theories
confined monopoles are junctions of two distinct  strings \cite{SYmon,HT2,T}. In the 
effective world sheet theory on the non-Abelian string they are seen as kinks interpolating between 
different vacua of $\mathbb{CP}(N-1)$ model, see \cite{SYrev} for a review.

The question of the crucial physical importance is whether monopoles survive the limit
of large $\mu$ when the the bulk theory flows to \none QCD. Quasiclassically we do not expect this to happen.
From a quasiclassical point of view, the very existence of 't Hooft-Polyakov monopoles relies on the presence of
adjoint scalars which develop VEVs. At large $\mu$ adjoint fields become heavy and decouple in our 
bulk theory, so quasiclassically we do not expect monopoles to survive.

We will see now that in quantum theory the story becomes more interesting. Confined monopoles 
are represented by kinks of $\mathbb{CP}(N-1)$ model on the non-Abelian string. Therefore to address the above
problem we have to study kinks in the world sheet theory. Certain results in this direction 
were already obtained. As we mentioned before, in the framework of massless $\mu$-deformed 
\ntwo QCD with FI $D$-term it was shown that the effective theory on the string world sheet is 
heterotic \ntwoo supersymmetric $\mathbb{CP}(N-1)$ model \cite{Edalati,SY02,Tongd,BSYhet}. 
This model has $N$ degenerative vacua and kinks interpolating between them. This means that
kinks/confined monopoles do survive the large $\mu$ limit in the above mentioned theory.

In this Chapter we study a more ''realistic'' version of $\mu$-deformed theory without the FI $D$-term.
This theory flows to \none QCD in the large $\mu$-limit. As we have shown  the 
world sheet theory on the non-Abelian string reduces to non-supersymmetric $\mathbb{CP}(N-1)$ model without
fermions in the large $\mu$ limit. 
If quark mass differences  are non-zero, a potential  \eqref{V1+1} is generated.
It does not have multiple local minima, therefore kinks (confined monopoles of the bulk theory) 
become unstable and disappear.

Consider the case when quark masses are equal. Then $\mathbb{CP}(N-1)$ model is at strong coupling.
This model was solved by Witten \cite{W79} in the large $N$ approximation. It was shown that 
kinks in this model are in a confinement phase. In terms more suitable for application to monopole physics 
of the bulk theory this can be understood as follows, see also \cite{SYrev} for a more detail review.

\begin{figure}
\centering
\includegraphics[width=0.6\linewidth]{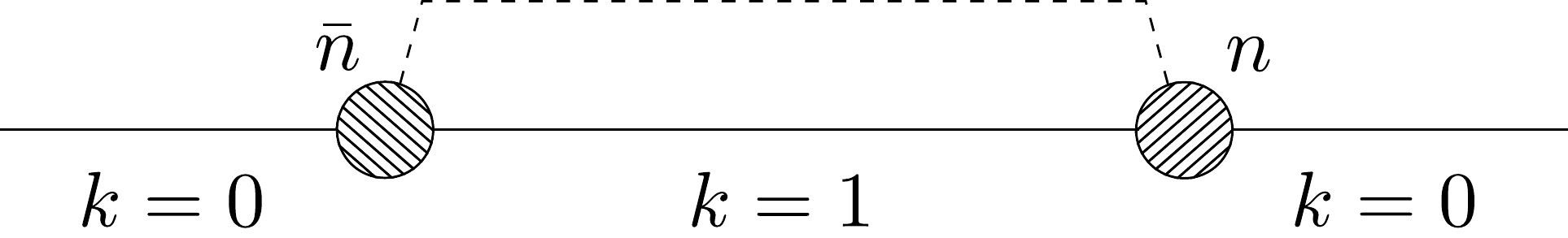}
\caption{Configuration of a string with kink and anti-kink on it. $k=0$ and $k=1$ is the notation for 
the true vacuum and the first quasi-vacuum respectively.}
\label{diag3} 
\end{figure}

The vacuum structure of the $\mathbb{CP}(N-1)$ model was studied in  \cite{W2}. It was shown that the genuine vacuum 
is unique. There are, however, of order $N$ quasi-vacua, which become stable in the 
limit $N\rightarrow\infty\,,$ since an energy split between the neighboring quasi-vacua 
is $O(1/N)$. Thus, one can imagine a kink interpolating between the true 
vacuum  and the first quasi-vacuum and the anti-kink returning to the true vacuum as 
in Fig.(\ref{diag3}). Linear confining potential between 
kink and anti-kink is associated with excited quasi-vacuum.

This two dimensional confinement of kinks was interpreted in terms of strings and monopoles of the bulk
theory in \cite{GSY05}. The fine structure of vacua in $\mathbb{CP}(N-1)$ model on the non-Abelian string means that
$N$ elementary strings are split by quantum effects and  have slightly different tensions. The difference 
between the tensions of ''neighboring'' strings is proportional to $\Lambda_{CP}^2$, see \eqref{LambdaCP}.
Therefore monopoles, in addition to the four dimensional confinement (which ensures that
they are attached to the string), acquire two-dimensional confinement along the string. 
Monopole and antimonopole connected by a string with larger tension form a mesonic bound state.

Fig.~\ref{diag3} represents a monopole-antimonopole pair interpolating between strings 0 and 1.
Energy of the excited part of the string (labeled as $1$) is proportional to the distance $R$ 
between the kink and anti-kink as 
\beq
V(R) \sim \Lambda_{CP}^2\,R.
\eeq
 When it exceeds the 
mass of  two monopoles which is of order of $\Lambda_{CP}$ then the second monopole-antimonopole pair 
emerges breaking the excited part of the string. This gives an estimate for the typical 
length of the excited part of the string, $R\sim N/\Lambda_{CP}$. Since this length grows in the large 
$N$ limit, kinks are metastable with an exponentially small decay rate $\exp{-N}$.

The results of this Chapter are published in the papers \cite{YIevlevN=1,Ievlev:2018rxo}.

%
%

\chapter{\none supersymmetric QCD: investigating the semilocal string} \label{sec:semiloc}

In the previous Chapter we studied $\mu$-deformed \ntwo supersymmetric QCD with the gauge group U($N$) and $N_f=N$ flavors.
To the leading order in small $\mu$ the mass term for the adjoint matter 
reduces to Fayet-Iliopoulos $F$-term which does not break \ntwo supersymmetry \cite{VY,Hanany:1997hr}.
In the quark vacuum squark condensate is determined by $\sqrt{\mu m}$, where $m$ is a quark mass. 
In this setup non-Abelian strings were first found \cite{HT1,ABEKY,SYmon,HT2} and 
their dynamics was well studied, see \cite{SYrev} for a review. 
 In addition to the
translational zero modes typical for Abelian ANO vortex strings  \cite{ANO}, 
non-Abelian strings have  orientational moduli associated with rotations of their fluxes inside the non-Abelian SU$(N)$ group. The dynamics of the orientational moduli in \ntwo QCD is described by the two dimensional 
$\mathbb{CP}(N-1)$ model living on the world sheet of the non-Abelian string. 

It turns out that at large $\mu$
the non-Abelian string ceases to be BPS, and world sheet supersymmetry is completely lost. Fermionic sector of the low energy world sheet  theory decouples at large $\mu$, while the bosonic sector is given by two dimensional  $\mathbb{CP}(N-1)$ model. It was also shown in Chapter~\ref{sec:none} that in the case of equal quark masses confined monopoles seen in the world sheet theory as kinks \cite{SYmon,HT2} survive $\mu$ deformation and present in the limit of \none 
SQCD. The  potential in  two dimensional world sheet theory induced by quark mass differences was also found. 

Non-Abelian strings in \ntwo
SQCD with \textquote{extra} quark flavors  ($N_f > N$) were well studied in the literature. In this setting the  string 
 develops size moduli and becomes semilocal. In particular, in the Abelian case these strings interpolate between ANO local strings and sigma-model lumps \cite{AchVas,Vachaspati:1991dz,Hindmarsh:1991jq,Hindmarsh:1992yy,Preskill:1992bf}. World-sheet theory on the semilocal non-Abelian string  was first 
considered from a D-brane prospective \cite{HT1,HT2}, and later from a field theory side \cite{Shifman:2006kd,Eto:2006uw,Eto:2007yv,ShYVinci}. In particularly, in \cite{ShYVinci}  
it was found that the world sheet theory is the so-called  ${\mathcal N}= (2,2)$ supersymmetric $zn$ model.

In this Chapter we continue studies of non-Abelian strings in SQCD with additional quark flavors, $N_f>N$
and consider $\mu$ deformed theory. In particular, we study what becomes of semilocal non-Abelian strings
as we increase $\mu$ and take the large $\mu$ limit where the theory flows to \none SQCD. 
First we found that much in the same way as for $N_f=N$ case of Chapter~\ref{sec:none} the string is no longer  BPS and the world sheet supersymmetry is lost. 

Moreover, 
 as we switch on the deformation parameter $\mu$ the string itself ceases to be semilocal. Considering the world sheet theory at small $\mu$ we show that string  size moduli develop a potential which forces them to shrink. Eventually in the large $\mu$ limit size moduli decouple and the effective theory on the string
reduces to $\mathbb{CP}(N-1)$ model. 

We also briefly discuss the  physics of confined monopoles.

%
%

\section{Theoretical setup \label{sec:basics}}

\subsection{Bulk theory}

In this section we briefly describe our initial theory in the bulk. The basic model is four-dimensional \ntwo supersymmetric QCD with the gauge group SU$(N)\times$U(1). 
The field content of the theory is as follows. The matter consists of $N_f = N + \tilde{N}$ flavors of quark hypermultiplets in the fundamental representation, scalar components being $q^{kA}$ and $\wt{q}_{Ak}$.  Here, $A = 1,..,N_f$ is the flavor index and $k=1,..,N$ is the color index.
The vector multiplets consist of U(1) gauge field $A_\mu$ and SU($N$) gauge field $A^a_\mu$, complex scalar fields $a$ and $a^a$ in the adjoint representation of the color group, and their Weyl fermion superpartners. Index $a$ runs from 1 to $N^2 - 1$, and the spinorial index $\alpha = 1,2$. 

Superpotential of the \ntwo SQCD is 
\begin{equation}
 \mc{W}_{\mc{N}=2} ~~=~~  \sqrt{2}\, \left\{ 
	 \frac{1}{2}\wt{q}{}_A \AU q^A ~+~
	 \wt{q}{}_A \mc{A}^a T^a q^A \right\}  ~+~
	 m_A\, \wt{q}{}_A q^A \ ,
\label{ntwo_superpotential_general}
\end{equation}
which includes  adjoint matter chiral \none multiplets $ \AU $ and $ \AN = \mc{A}^a T^a $, and the quark chiral \none multiplets $ q^A $ and $ \wt{q}{}_A $ (here we use the same notation for the quark superfields and their scalar components). The $\mu$ deformation considered in this Chapter is given by the superpotential \eqref{none_superpotential_general},
\begin{equation*}
 \mc{W}_{\mc{N}=1} ~~=~~ \sqrt{\frac{N}{2}}\,\frac{\mu_1}{2} \left(\mc{A}^{\rm U(1)}\right)^2  ~~+~~
	 \frac{\mu_2}{2} \left( \mc{A}^a \right)^2 ~.
\label{semiloc:none_superpotential_general}
\end{equation*}
We assume the deformation parameters to be of the same order, $\mu_1 \sim \mu_2 \sim \mu$. When we increase $\mu\to\infty$, \ntwo supersymmetry becomes broken, and the theory flows to \none SQCD. 
Instead in the limit of small $\mu$ this superpotential does not break the \ntwo supersymmetry and reduces to a Fayet--Iliopoulos $F$-term \cite{VY,Hanany:1997hr}. 

In order to control  the theory and stay at weak coupling as we take this limit, we require the product $\sqrt{\mu m}$ to stay fixed and well above $\Lambda_{{\cal N}=1}$, which is the scale of the \sun~sector of \none QCD .

The bosonic part of the action is basically the same as \eqref{theory}:
%
\begin{align}
\label{semiloc-theory}
S_{\rm bos} ~~=~~ & \int d^4 x 
\lgr
\frac{1}{2g_2^2}\Tr \left(F_{\mu\nu}^\text{SU($N$)}\right)^2  ~+~
\frac{1}{4 g_1^2} \left(F_{\mu\nu}^{\rm U(1)}\right)^2 ~+~ 
\right. 
\\[2mm]
\notag
&
\phantom{int d^4 x \lgr\right.}
\frac{2}{g_2^2}\Tr \left|\nabla_\mu \aN \right|^2   ~+~
\frac{1}{g_1^2} \left|\p_\mu \aU \right|^2
~+~
\left| \nabla_\mu q^A \right|^2 ~+~ \left|\nabla_\mu \ov{\wt{q}}{}^A \right|^2 
~+~
\\[2mm]
\notag
&
\phantom{int d^4 x \lgr\right.}
\left.
V(q^A, \wt{q}{}_A, \aN, \aU)
\rgr .
\end{align}
Here $ \nabla_\mu $ is the covariant derivative in the corresponding representation:
\begin{align*}
\nabla_\mu^{\rm adj} & ~~=~~ \p_\mu  ~-~ i\, [ A_\mu^a T^a, \;\cdot\;]~, \\[3mm]
\nabla_\mu^{\rm fund} & ~~=~~ \p_\mu ~-~ \frac{i}{2} \,A^{\rm U(1)}_\mu ~-~ i\, A_\mu^a T^a~,
\end{align*}
with the SU($N$) generators normalized as
$\Tr \left( T^a T^b \right) ~=~ (1/2) \, \delta^{ab}~$.
%
Superpotentials \eqref{ntwo_superpotential_general}, \eqref{semiloc:none_superpotential_general} contribute to the scalar potential $V$ which is given by the sum of $F$ and $D$ terms, 
\begin{align}
\notag
& V(q^A, \wt{q}{}_A, \aN, \aU) ~~=~~ 
\\[3mm]
\notag
&\qquad\quad ~~=~~
\frac{g_2^2}{2} \left( \frac{1}{g_2^2}\,f^{abc}\ov{a}{}^b a^c 
~+~ \ov{q}{}_A\, T^a q^A ~-~ \wt{q}{}_A\, T^a \ov{\wt{q}}{}^A \right)^2 
\\[3mm]
\label{V}
&\qquad\quad ~~+~~
\frac{g_1^2}{8}\, (\ov{q}{}_A q^A ~-~ \wt{q}{}_A \ov{\wt{q}}{}^A)^2
\\[3mm]
\notag
&\qquad\quad ~~+~~
2g_2^2\, \Bigl| \wt{q}{}_A\,T^a q^A ~+~ 
\frac{1}{\sqrt{2}}\, \frac{\p\mc{W}_{\mc{N}=1}}{\p a^a} \Bigr|^2
~+~
\frac{g_1^2}{2}\, \Bigl| \wt{q}{}_A q^A ~+~ 
\sqrt{2}\, \frac{\p\mc{W}_{\mc{N}=1}}{\p\aU} \Bigr|^2
\\[3mm]
\notag
&\qquad\quad ~~+~~
2 \sum_{A=1}^{N_f} \Biggl\{  
\left| \left( \frac{1}{2}\,\aU ~+~ \frac{m_A}{\sqrt{2}} ~+~ a^a T^a \right) q^A \right|^2  ~+~
\\[3mm]
\notag
&\phantom{\qquad\quad ~~+~~ 2 \sum_{A=1}^{N_f} \Biggl\{  }
\left| \left( \frac{1}{2}\,\aU ~+~ \frac{m_A}{\sqrt{2}} ~+~ a^a T^a \right) \ov{\wt{q}}{}^A \right|^2  
\Biggr\}~,
\end{align}
where summation is implied over the repeated flavor indices $A$ (and over omitted color indices, too). 

Consider the case when we have one \textquote{extra} flavor, $N_f = N + 1$. Scalar potential \eqref{V} has a set of supersymmetric vacua, but in this Chapter we concentrate on a particular  vacuum where the maximal number of squarks equal to $N$ condense. Up to a gauge transformation, the squark vacuum expectation values are given by 
%
%
\begin{eqnarray}
\langle q^{kA} \rangle ~=~ \langle \ov{\wt{q}}^{kA} \rangle& =& \frac{1}{\sqrt{2}}
\left(
\begin{array}{cccc|c}
  \sqrt{\xi_1} 	& 0  		& 0 				&0 								& 0 		\\
   0 			& \ddots   	& \vdots 			& \vdots 						& \vdots	\\  
  \vdots        &  \dots 	& \sqrt{\xi_{N-1}} 	&0 								& 0 		\\
 0 				& \dots 	& 0  				&  \sqrt{\xi_N}    				& 0 
\end{array}
\right)\,,
\label{semi-qVEV}
\end{eqnarray}
where  we write quark fields as a rectangular matrices $N\times N_f$ and $\xi_P$ are defined as \eqref{xi-general},
\begin{equation}
	\xi_P ~~=~~ 2\left( \sqrt{\frac{2}{N}} \mu_1 \widehat{m} ~+~ \mu_2 (m_P - \widehat{m}) \right) ,
\end{equation}
\begin{equation}
	\widehat{m} ~~=~~ \frac{1}{N} \sum_{A=1}^{N} m_P .
\end{equation}

If we define a scalar adjoint matrix as \eqref{Phidef},
\begin{equation*}
	\Phi =  \frac{1}{2}\,a + T^a\, a^a \,,
\end{equation*}
then the adjoint fields VEVs are given by
\beq
\langle \Phi\rangle = - \frac1{\sqrt{2}}
\left(
\begin{array}{ccc}
	m_1 & \ldots & 0 \\
	\ldots & \ldots & \ldots\\
	0 & \ldots & m_N\\
\end{array}
\right)\,.	
\label{avev}
\eeq
The  vacuum field (\ref{semi-qVEV}) results in  the spontaneous
breaking of both gauge U$(N)$ and flavor SU($N$). However, in the equal mass limit $m_A \equiv m$, $A=1,...,N_f$ all
parameters $\xi$ become equal, $\xi_P\equiv  \xi$, $P=1,...,N$ and 
a diagonal global ${\rm SU}(N)_{C+F}$ survives, or, more exactly (cf. \eqref{c+f}, \eqref{globgroup_d=4}):
\begin{equation*}
	{\rm U}(N)_{\rm gauge}\times {\rm SU}(N)_{\rm flavor}
	\to {\rm SU}(N)_{C+F} \times {\rm SU}(\wt{N})_{F} \times {\rm U}(1) \,.
\label{semiloc:c+f}
\end{equation*}
Thus, a color-flavor locking takes place in the vacuum. The presence of the  color-flavor ${\rm SU}(N)_{C+F}$ global symmetry is the reason for the formation of non-Abelian strings, see \cite{SYrev} for a review.

In the special case when
\begin{equation*}
\mu_2 ~~=~~ \mu_1 \sqrt{2/N} ~~\equiv~~ \mu \ ,
\end{equation*}
superpotential \eqref{semiloc:none_superpotential_general} is simplified and becomes a single-trace operator
\begin{equation}
\mc{W}_{\mc{N}=1} ~=~ \mu \Tr (\Phi^2) \,.
\label{eq:none_superpotential_singletrace}
\end{equation}



\subsection{Mass spectrum  \label{sec:mass_spectrum}}

In this section we review the  mass spectrum of our bulk SQCD taking all quark masses equal, cf. 
\cite{SYrev,BSYhet,VY}. Due to squark condensation, the gauge bosons acquire masses%
\footnote{Here we assume for simplicity that  $\xi, \, \mu_1,\, \mu_2 $  are real}
\begin{equation}
\begin{aligned}
	m_{\ue} ~~=~~ g_1 \sqrt{\frac{N}{2}} \xi \,, \\
	m_{\sun} ~~=~~ g_2\sqrt{\xi} \,.	
\end{aligned}
\label{eq:gauge_mass}
\end{equation}

Scalar states masses are to be read off from the potential \eqref{V}. Expanding and diagonalizing the mass matrix one can find  $N^2 - 1$ real scalars with the masses $m_{\sun}$ and one scalar with the mass $m_{\ue}$. These are \none superpartners of SU$(N)$ and U(1)  gauge bosons. Other $N^2$ components are eaten by the Higgs mechanism. Another $2\times 2N^2 $ real scalars (  adjoint scalars $a^a$, $a$ and the half of squarks) become scalar components of  the following \none chiral multiplets: one  with mass
\begin{equation}
	\mUp ~~=~~ g_1 \sqrt{\frac{N}{2}\xi\lambda_1^+}~,
\end{equation}
and another one  with mass
\begin{equation}
	\mUm ~~=~~ g_1 \sqrt{\frac{N}{2}\xi\lambda_1^-}~.
\end{equation}
The remaining $ 2(N^2 - 1)$ chiral multiplets  have masses
\begin{equation}
	\mNp ~~=~~ g_2 \sqrt{\xi\lambda_2^+} ~,
\end{equation}
\begin{equation}
	\mNm ~~=~~ g_2 \sqrt{\xi\lambda_2^-} ~.
\end{equation}
Here  $ \lambda_i^\pm $ are roots of the quadratic equation \cite{SYrev,VY}
\begin{equation}
	\lambda_i^2  -  \lambda_i(2 + \omega^2_i)  +   1  =  0
\end{equation}
with
\begin{equation}
	\omega_1  =  \frac{g_1\mu_1}{\sqrt{\xi}}\,,\qquad
	\omega_2  =  \frac{g_2\mu_2}{\sqrt{\xi}}\,.
	\label{omega}
\end{equation}
Once $N_f>N$ apart from the above  massive scalars, we also have $4 N (N_f - N)$ scalars which come from the extra squark flavors $q^K$ and $\wt{q}_K$, $K=(N+1),...,N_f$. In the equal mass limit  these extra scalars are massless, and the theory enjoys a Higgs branch
\begin{equation}
{\cal H} = T^*\textrm{Gr}^{\mathbb{C}}(N_f, N) 
\label{HiggsbranchGr:3}
\end{equation}
of real dimension
\begin{equation}
	{\rm dim}{\cal H}= 4 N (N_f-N) \,,
	\label{dimH:3}
\end{equation}
cf. \eqref{dimH}.

In the large $\mu$ limit, states with masses $\mUp$ and $\mNp$ become heavy with masses $\sim g^2\mu$ and decouple. They correspond to the adjoint matter multiplets. Instead states  with masses $\mUm$ and $\mNm$ become light with masses $\sim \xi/\mu$. Scalar components of these multiplets are Higgs scalars. They develop VEVs
\eqref{semi-qVEV}. 
In the opposite limit of small $\mu$ their masses are given by
\begin{equation}
\begin{aligned}
	\mUm &~~=~~ g_1 \sqrt{\frac{N}{2}\xi} \left(1 - \frac{g_1\mu_1}{2 \sqrt{\xi}} + \cdots\right)~,	\\
	\mNm &~~=~~ g_2 \sqrt{\xi} \left(1 - \frac{g_2\mu_2}{2 \sqrt{\xi}} + \cdots \right) ~.
\label{eq:m_scalar-small_mu}
\end{aligned}
\end{equation}
As we already mentioned  \ntwo supersymmetry is not broken in our theory to the leading order at small $\mu$
\cite{VY,Hanany:1997hr}. The leading order corresponds to sending parameters $\omega$ in \eqref{omega} to zero while keeping
FI parameter $\xi \sim \mu m$ fixed.
One can see that in the \ntwo limit  Higgs scalars are degenerate  with the gauge fields \footnote{They belong to the same long vector \ntwo supermultiplet \cite{VY}},  but become lighter as we switch on the $\mu$-deformation. 

The ratio of squares of Higgs and gauge boson masses $\beta$ is an important 
parameter\footnote{To avoid confusion we clarify, that in this Chapter the 2D inverse coupling is denoted as $\gamma$, whereas the letter $\beta$ is reserved for the Higgs-gauge mass ratio.} in the theory of superconductivity. Type I superconductors correspond to $\beta< 1$, while type II
superconductors correspond to $\beta > 1$. BPS strings arise on the border at $\beta =1$. We see that in 
our theory both parameters $\beta$,
\beq
\beta_{U(1)} = \left(\frac{m^{-}_{U(1)}}{m_{U(1)}}\right)^2, 
\qquad \beta_{SU(N)} = \left(\frac{m^{-}_{SU(N)}}{m_{SU(N)}}\right)^2 ,
\label{betas}
\eeq
are less than unity, and thus our theory is in the  type I  superconducting phase at non-zero $\mu$.
This will turn out to be important later.

%
%

\section{Semilocal non-Abelian vortices  \label{sec:string-small_mu}}

In this section we study a vortex string solution in the equal quark mass limit. 
First we review  previous results \cite{ShYVinci} for the BPS semilocal non-Abelian vortex string and then consider  a small $\mu$-deformation. We   derive the world-sheet effective theory for the string moduli fields
in this case. For simplicity we consider the theory with one extra quark flavor, $N_f = N + 1$.

\subsection{BPS semilocal non-Abelian string}

We start by reviewing the semilocal non-Abelian string in the \ntwo limit \cite{ShYVinci}.
Once number of flavors exceed number of colors vortices  have no longer the conventional 
 exponentially small tails of the
profile functions. The presence of the Higgs branch and associated massless fields in the bulk makes them semilocal, see detail review of the Abelian case in \cite{AchVas}.
The semilocal  strings have a power fall-off at large distances from the string axis. For example, the  semilocal Abelian BPS string
interpolates between ANO string \cite{ANO} and two-dimensional  O(3) sigma-model instanton
uplifted to four dimensions (also known as the lump). For one extra flavor the semilocal string possesses
 two additional zero modes parametrized by the complex modulus $\rho$. The string's  transverse size 
is associated with $|\rho|$.
In the limit  $|\rho|\to 0$ in the Abelian case we recover the ANO string while
at $|\rho| \gg 1/m_{U(1)}$ it becomes a lump.

Consider an infinite static string stretched along the $x_3$ axis.
We can start with basically the same ansatz as \eqref{q-ansatz}, \eqref{string-solution}:
\begin{equation}
q^{kA}=\bar{\tilde{q}}^{kA}=\frac1{\sqrt{2}}\vp^{kA}  ,
\label{semi-q-ansatz}
\end{equation} 
\begin{equation}
	\vp ~~=~~ \bigg(\phi_{2}(r) + n\nbar (\phi_{1}(r)-\phi_{2}(r)) \,|\, n\,\phi_{3}(r) e^{- i \alpha}\bigg)
	\label{ansatz:vp:singular}
\end{equation}
for quarks, while the gauge fields are given by
\begin{equation}
	\begin{aligned}
		A_i^\text{SU($N$)} & ~~=~~ \varepsilon_{ij}\, \frac{x^j}{r^2}\, f_G(r)
		\lgr n\nbar ~-~ 1/N \rgr\,,
		\\[2mm]
		A_i^{\rm U(1)} & ~~=~~ \frac{2}{N}\varepsilon_{ij}\, \frac{x^j}{r^2}\, f(r)~.
	\end{aligned}
\label{semi-string-solution}
\end{equation}
Index $i$ runs $i=1, 2$, all other components are zero; $\alpha$, $r$ are polar angle and radius in the 
$(x_1,x_2)$ plane respectively. The complex parameters $n^l, l=1,..,N$ obey the \CP constrain $\ov{n}n = 1$. They parametrize the orientational zero modes of the non-Abelian string which appear due to the presence of the 
color-flavor group \eqref{semiloc:c+f}, see \cite{SYrev} for a review.

The string profile functions entering \eqref{ansatz:vp:singular} and \eqref{semi-string-solution} satisfy first order BPS equations. For  the case 
\begin{equation}
	\frac{g_1^2}{2} = \frac{g_2^2}{N} \equiv \frac{g^2}{N} 
\label{single_gauge}
\end{equation}
 the solution is particularly simple \cite{ShYVinci}. It is  is parametrized by a complex size modulus $\rho$:
\begin{equation}
\begin{aligned}
	\phi_1 ~~\approx~~&	\sqrt\xi\frac{r}{\sqrt{r^{2}+|\rho|^{2}}}									\,,		\\
	\phi_2 ~~\approx~~&	\sqrt\xi																	\,,		\\
	\phi_3 ~~=~~& \frac{\rho}{r}\phi_1 ~~\approx~~ \sqrt\xi\frac{\rho}{\sqrt{r^{2}+|\rho|^{2}}}		\,,		\\
	f ~~=~~ f_G	   ~~\approx~~&	\frac{|\rho^{2}|}{r^{2}+|\rho|^{2}} 										\,.
\end{aligned}
\label{eq:solution-semiloc-bps}
\end{equation}
This solution is valid in the limit $|\rho| \gg 1/(g_2\sqrt{\xi}|\rho|)$, i.e. when the scalar fields approach the vacuum manifold (Higgs branch). Tension of the BPS string is given by
\begin{equation}
	T_{BPS} = 2\pi\xi.
\label{eq:tension-semilocal}
\end{equation}

To obtain the low energy effective two dimensional theory living on the string world sheet, one should assume $n^P$ and  $\rho$ to be slowly varying functions of the transversal coordinates $t, z$, and substitute the solution \eqref{eq:solution-semiloc-bps} into the action \eqref{semiloc-theory}. This procedure yields  the effective action
\begin{equation}
	S^{2d}_{SUSY} ~=~ \int d^2 x \left\{ 2\pi\xi \,	|\pt_k(\rho n_P)|^2	\,\,\ln{\frac{L}{|\rho|}}\,
		~+~ \frac{4\pi}{g^2} \Big[|\partial_{k}n_P |^2+(\nbar_P \pt_k n_P)^{2} \Big]
		\right\} ,
\label{eq:ws-bps}
\end{equation}
where the integration is carried over the coordinates $x_0, x_3$,
see the detailed derivation in \cite{ShYVinci}.
Here $k = 0, 3$, and $L$ is an infra-red  cutoff introduced for the regularization of the logarithmic divergences
of orientational and size zero modes of the string. More exactly we introduce the  string of a large but finite length $L$.  This also regularize  the spread of string profile functions in the transverse plane 
\cite{Shifman:2006kd}.
The IR divergences arise due to the slow (power) fall-off  of the string profile functions associated with the presence of the Higgs branch \cite{Shifman:2006kd,ShYVinci}.

\subsection{Deformed world-sheet theory}

When we take into account higher order $\mu$-corrections, supersymmetry in the bulk reduces to \none, and 
as we already explained our theory becomes that of the type I superconductor, cf. \cite{VY}. The string  is no
longer BPS saturated. To mimic this we consider a simplified version of our theory with the bosonic action given by
\begin{multline}
	S_{\rm 0} =\int d^4x \Bigg\{ \frac1{4g_2^2}\left(F^{a}_{\mu\nu}\right)^{2}
		+ \frac1{4g_1^2}\left(F_{\mu\nu}\right)^{2}
		+   |\nabla_\mu \vp^A|^2														\\
		+	\lambda_N	\left(\bar{\vp}_A T^a \vp^A\right)^2
		+   \lambda_1 \left( | \vp^A|^2 - N\xi \right)^2
		\Bigg\}
	\, .
\end{multline}
This model depends on two parameters -- ratios of the squires of \ue and \sun Higgs and  gauge  boson masses given by
\begin{equation}
\begin{aligned}
	\beta_{U(1)} &= \frac{8\lambda_1}{g_1^2}\,, \\
	\beta_{SU(N)} &= \frac{2\lambda_N}{g_2^2}\,,
\end{aligned}
\label{eq:beta-qed}
\end{equation}
which we identify with $\beta$-parameters \eqref{betas} of our original theory .
The model above is a non-Abelian generalization  the one considered in \cite{GorskyShY_skyrmion}, where the  scalar QED was studied see also \cite{AchVas}.  

In \ntwo supersymmetric QCD parameters $\beta$ are exactly equal to one. In this case the  the Bogomol'nyi representation produces first order equations for the string profile functions. World sheet theory in this case is given by \eqref{eq:ws-bps}.

As we switch on $\mu$-corrections parameters $\beta$ are no longer equal to unity. Let us write the 
Bogomol'nyi representation for  the tension of the string
\begin{multline}
	T_\beta ~=~ \int d^2 x_{\perp}  \Bigg\{
			  \left[\frac1{\sqrt{2}g_2}F_{12}^{a} +	\frac{g_2}{\sqrt{2}}    \left(\bar{\vp}_A T^a \vp^A\right)\right]^2
			+ \left[\frac1{\sqrt{2}g_1}F_{12} +  \frac{g_1 }{2\sqrt{2}}	\left(|\vp^A|^2-N\xi \right)\right]^2			\\
			+  \left|\nabla_1 \,\vp^A +  i\nabla_2\, \vp^A\right|^2
			+ \frac{N}{2}\xi\,  F^{*}_3 																					\\
			+ \frac{g_2^2}{2} (\beta_{SU(N)} - 1) \left(\bar{\vp}_A T^a \vp^A\right)^2 
			+ \frac{g_1^2}{8} (\beta_{U(1)} - 1) \left( | \vp^A|^2 - N\xi \right)^2
			\Bigg\} \,,
			\label{bogomolny}
\end{multline}
where $\vec{x}_{\perp}$ represents the coordinates in the transverse plane.
Two extra terms written in the last line above appear. The Bogomol'nyi bound is no longer valid. But if the values $\beta_{U(1)}$ and $\beta_{SU(N)}$ only slightly differ from unity, then we can use the first order  equations to rewrite expressions in these extra terms as follows
\beq 
g_2^2 \left(\bar{\vp}_A T^a \vp^A\right) = - F_{12}^a, 
\qquad \frac{g_1^2}{2}\left( | \vp^A|^2 - N\xi \right) =- F_{12}.
\label{foe}
\eeq
In the case \eqref{single_gauge} we can use \eqref{eq:solution-semiloc-bps} to calculate the effective action.
Substituting \eqref{ansatz:vp:singular}, \eqref{semi-string-solution}, \eqref{eq:solution-semiloc-bps} into 
\eqref{foe}, \eqref{bogomolny}  one arrives at the  deformed world-sheet theory,
\begin{multline}
 S^{2d}_{\beta} ~=~ \int d^2 x\left\{ 2\pi\xi \,	|\pt_k(\rho n_P)|^2	\,\,\ln{\frac{L}{|\rho|}}\, 
		+	\frac{4\pi}{g^2} \Big[|\partial_{k}n_P |^2+(\nbar_P \pt_k n_P)^{2} \Big]
		\right.
		\\
		\left.
		+ \frac{\beta - 1}{g^2} \, \frac{4\pi}{3|\rho|^2}\, +\cdots \, , 
	\right\} ,
\label{eq:ws-small_mu0}
\end{multline}
%
where now $\beta \equiv \beta_{U(1)} = \beta_{SU(N)}$ and the dots represent corrections in powers of $1/g^2\xi |\rho|^2$.

We see that for non-BPS string $\rho$ is no longer a modulus. It develops a potential proportional to
 the deviation of $\beta$ from unity. In particular, for type I superconductor ($\beta <1$) the size
$\rho$ tends to shrink, while for type II superconductor ($\beta >1$) the size
$\rho$ tends to expand making the vortex unstable, cf. \cite{AchVas}.

In our case, the value of $\beta$ is less then unity and is given
  by \eqref{eq:m_scalar-small_mu} at small $\mu$, namely
\begin{equation}
	\beta ~~=~~ 1 - \frac{g \mu}{\sqrt{\xi}} + \cdots\,.
\end{equation}
This gives the effective world sheet action on the string
\beqn
	&& S^{2d}_{\beta} ~=~ \int d^2 x\left\{ 2\pi\xi\,	|\pt_k(\rho n_P)|^2	\,\,\ln{\frac{L}{|\rho|}}\, 
		+	\frac{4\pi}{g^2} \Big[|\partial_{k}n_P |^2+(\nbar_P \pt_k n_P)^{2} \Big]
	\right.
	\nonumber\\
	&&
	\left.
	- \,4\pi \frac{\mu}{3 g\sqrt{\xi}} \, \frac{1}{|\rho|^2}\, + \cdots\, 
	\right\} .
\label{eq:ws-small_mu}
\eeqn

We see that the size of the semilocal string tends to shrink and at large $\mu$ we expect that 
the long-range tails of the string are not developed. The string becomes a  local non-Abelian string with only orientational moduli $n^l$, whose world sheet dynamics is described by $\mathbb{CP}(N-1)$ model.

In fact we can argue on general grounds that as we turn on $\mu$ and make it large  the semilocal string  become unstable. The semilocal string solution
\eqref{eq:solution-semiloc-bps} is ''made'' of massless fields associated with the Higgs branch of the theory.
As we already mentioned say, in the Abelian case this solution  correspond to the instanton of the two dimensional
O(3) sigma model uplifted to four dimensions. The instanton is  essentially  a BPS solution and therefore it is natural to expect that it becomes unstable once we increase $\mu$ breaking the world sheet  supersymmetry.

In particular, as we see from Bogomol'ny representation \eqref{bogomolny} extra terms arising at $\beta <1$
reduce the tension of the string. This is forbidden for BPS lump (uplifted instanton) since its tension
is exactly determined by the central charge and given by $2\pi \xi$, see \eqref{eq:tension-semilocal}.
As we increase $\mu$ the string is no longer BPS, $\rho$ develops instability and shrinks leading at large $\mu$
to much lower tension, see \eqref{eq:T_local} below.

%
%

\section{Summary of  results \label{sec:results}}

In this Chapter we studied what happens to the non-Abelian semilocal string in \ntwo supersymmetric QCD as we switch on the $\mu$ deformation and go to the large $\mu$ limit. We showed that the size modulus $\rho$ develops a potential and eventually decouples as the theory flows to the \none SQCD at large $\mu$. 
Note that the Higgs brunch is still there, just the string is no longer ''made'' of massless fields,
so the long-range ''tails'' of the string disappear.

Thus,  the semilocal string degenerates into the local one. Non-Abelian local strings  in the large $\mu$ limit
of \none SQCD  were studied in Chapter~\ref{sec:none}, and now we see that those results can be directly applied to our case $N_f >N$ as well.
Below we briefly summarize these results.

In the large $\mu$ limit, the string tension is logarithmically suppressed, 
\begin{equation}
	T_{local} = \frac{4\pi|\xi|}{\ln \displaystyle\frac{g^2|\mu|}{|m|}}  \,.
\label{eq:T_local}
\end{equation}
This should be contrasted with the BPS formula \eqref{eq:tension-semilocal} valid to 
the leading order at small $\mu$.

As usual the world sheet  theory contains  translational moduli but they decouple from the orientational sector.  The orientational sector is described by  $\mathbb{CP}(N-1)$ model with the action \eqref{cp},
\begin{equation*}
	S^{(1+1)}=    \int d t\, dz \,  
			\Big\{ 
				\gamma\, \left[ (\pt_{k}\, \bar{n}\,	\pt_{k}\, n) + (\bar{n}\,\pt_{k}\, n)^2 \right]
				~+~  V_{1+1}
			\Big\}\,.
\end{equation*}
Note, that orientational fermionic zero modes are all lifted (see Chapter~\ref{sec:none}) and do not enter 
the low energy world sheet theory. The above world sheet theory is purely bosonic.

Here two dimensional inverse coupling constant $\gamma$ is large\footnote{Only in this Chapter the 2D coupling is denoted as $\gamma$. In the Chapters~\ref{sec:none} and \ref{sec:large_N} it's $2\beta$, while in Chapters~\ref{sec:ntwo} and \ref{sec:b_meson} it's $\beta$.}, given by \eqref{beta}:
\begin{equation*}
	\gamma  \sim \frac{|\mu|}{|m|}\,\frac{1}{\ln^2 \frac{g^2|\mu|}{|m|}}. 
\end{equation*}
At the quantum level $\mathbb{CP}(N-1)$ model is asymptotically free, so the coupling $\gamma$ runs and at the energy  $E$ is given by
\beq
2\pi\gamma (E)= N \log{\left(\frac{E}{\Lambda_{CP}}\right)},
\eeq
where the scale of 
the  world sheet theory is given by \eqref{LambdaCP},
\begin{equation*}
	\Lambda_{CP} \approx \sqrt{\xi} \exp{\left(- {\rm const}\,\frac{|\mu|}{|m|}\frac{1}{\ln^2 \frac{g^2|\mu|}{|m|}}\right)}.
\end{equation*}
We see that the scale $\Lambda_{CP}$ of $\mathbb{CP}(N-1)$ model above  is exponentially small, so the world sheet theory
is weakly coupled in a wide region of energies $\gg \Lambda_{CP}$. This should be  contrasted to non-Abelian string in \ntwo QCD where world sheet theory has a scale $\Lambda_{CP}$ equal to scale 
$\Lambda_{{\cal N}=2}$ of the bulk SQCD \cite{SYrev}.

In the case when the quark masses entering the Lagrangian \eqref{semiloc-theory} are non-identical, a potential for $n^P$ is generated. In the simplest case when all quark masses are positive, this potential is given by \eqref{V1+1},
\begin{equation}
	 V_{1+1} ~~\approx~~ \frac{8\pi |\mu|}{\ln \frac{g^2|\mu|}{|m|}}     \sum_{P=1}^{N} 
			 m_P |n^P|^2 \,.
\end{equation} 
The potential \eqref{V1+1} has only one minimum and one maximum at generic $\Delta m_{AB}$. Other $(N-2)$ extreme points are saddle points. For equal quark masses this potential reduces to the constant equal to the tension of the string \eqref{eq:T_local}.

Since our four-dimensional theory is in the Higgs phase for squarks, 't Hooft-Polyakov monopoles present in the  theory
in the \ntwo limit of small $\mu$ are confined by non-Abelian strings and serve as junctions of two distinct strings \cite{SYmon,HT2,T}. In the 
effective world sheet theory on the non-Abelian string they are seen as kinks interpolating between 
different vacua of $\mathbb{CP}(N-1)$ model, see \cite{SYrev} for a review.

In the large $\mu$ limit adjoint fields decouple. Therefore we could expect quasiclassically that the confined monopoles disappear in this limit. This indeed happen for non-equal quark masses. If quark mass differences  are non-zero, a potential  \eqref{V1+1} is generated.
It does not have multiple local minima, therefore kinks (confined monopoles of the bulk theory) 
become unstable and disappear.

However, in the equal quark mass case the potential \eqref{V1+1} is absent and the bosonic $\mathbb{CP}(N-1)$ model supports kinks. Thus, in this case confined monopoles do survive the large $\mu$ limit, as follows from Chapter~\ref{sec:none}. The monopoles are represented by kinks in the effective $\mathbb{CP}(N-1)$ model on the non-Abelian string, see \cite{SYrev} for a detail review.

The results of this Chapter are published in the papers \cite{Ievlev:2018rxo,Ievlev:2018xub}.

%
%

\chapter{Large $N$ solution of the worldsheet theory} \label{sec:large_N}

In this Chapter we present a large $N$ solution of the world sheet theory for the non-Abelian string in the 
$\mu$-deformed SQCD, which was derived in Chapter \ref{sec:none}. Large $N$ approximation was first used by Witten to solve both non-supersymmetric 
and \ntwot supersymmetric two-dimensional \CP models \cite{W79}. In particular, large-$N$ Witten's solution
shows that an auxiliary U(1) gauge field $A_{\mu}$ introduced to formulate \CP model becomes physical.
The \ntwot supersymmetric \CP model
has $N$ degenerate vacua as dictated by its Witten index. The order parameter which distinguishes between these vacua is the vacuum expectation value of the scalar superpartner $\sigma$ of the gauge field $A_{\mu}$ \cite{W79}. 

In the non-supersymmetric
\CP model these vacua are split with splittings proportional to $1/N$ and become quasivacua. The theory has a single true vacuum \footnote{We assume below that the $\theta$-angle is zero.}. The order parameter which distinguish between these quasivacua is the value of the constant  field strength of the gauge field $A_{\mu}$ which  is massless in the non-supersymmetric case \cite{W79}, see also \cite{GSY05} and review \cite{SYrev}.

In this Chapter we use the large $N$ approximation to study  a  phase structure of the world sheet theory on the non-Abelian string in $\mu$-deformed SQCD with respect to the deformation parameter $\mu$ and quark
mass differences $\Delta m$. We find a rich phase structure which  includes two strong coupling phases and two Higgs phases. 

Strong coupling phases appear at small $\Delta m$. The first strong coupling phase appears at small $\mu$.
It is qualitatively similar to \ntwot supersymmetric phase at $\mu=0$. Although $N$ vacua are split
and become quasivacua the order parameter is still the VEV of the field $\sigma$. In the second 
strong coupling phase at large $\mu$ quasivacua are distinguished by the value of the constant electric field.
This phase is qualitatively similar to the non-supersymmetric \CP model.

At large $\Delta m$ we find two weakly coupled Higgs phases. At small $\mu$ $N$ vacua present in \ntwot case
split and become quasivacua. Still we have kinks interpolating between them.
As we increase $\mu$ above certain critical value, these lifted quasivacua disappear one by one, so we have a cascade of phase transitions. In the end we are left with a single vacuum and no kinks at all.

From the point of view of the bulk SQCD we interpret this as follows.
At large $\Delta m$ and small $\mu$ we have monopoles confined by non-Abelian strings while as we increase $\mu$
 monopoles disappear.

%
%

\section{Review of \CP sigma models \label{sec:revCP}}

In this section we review basic \CP models that are of interest to us.
First, we will briefly review the non-supersymmetric and the \ntwot supersymmetric models, which were considered before, see for example \cite{W79,W93,Gorsky:2005ac,BSY3}. After that, we will introduce the model that we will be working with, namely the $\mu$-deformed \CP model which is an effective  theory living on the world sheet
of non-Abelian string in $\mu$-deformed SQCD considered in Chapter~\ref{sec:none}. 

\subsection{Non-supersymmetric model}
\label{sec:no-susy}

Throughout this Chapter we will be working with the gauge formulation \cite{W79} of the \CP models. In this formalism, the model is formulated via  $N$ complex scalar fields $n^i$, $i = 1, \ldots, N$ interacting with auxiliary U(1) gauge field $A_{\mu}$. The Lagrangian is written as
\begin{equation}
\mathcal L =
	\left|\nabla_\mu n^i\right|^2  
	+ i\,D\left(\bar{n}_i n^i -2\beta_0 \right)+ \sum_i\left|\sqrt 2\sigma-m_i\right|^2\, |n^i|^2,
\label{cpn_lagr_simplest}	
\end{equation}
where $\nabla_\mu = \partial_\mu -i\,  A_\mu$. Fields $\sigma$ and $D$ come without kinetic energy and are also auxiliary. They can be eliminated via their equations of motion. In particular integrating out $D$ imposes 
the constraint
\begin{equation}
	\nbar_i \, n^i = 2\beta_0, 
\label{cp_constraint_n}	
\end{equation}
which together with gauge invariance reduces the number of real degrees of freedom of the $n^i$ field down to
$2(N-1)$.

This is the non-supersymmetric  version of the \CP model, and it arises as a world sheet theory on the non-Abelian string in a non-supersymmetric QCD-like theory, see \cite{GSY05} and review \cite{SYrev}. The mass parameters $m_i$ are equal to quark masses in the four-dimensional  theory. 

Throughout this Chapter we will consider the masses placed uniformly on a circle,
\begin{equation}
	m_k = m - \Delta m \, \exp(\frac{2 \pi i \, k}{N}), \quad k=0,\, \ldots,\, N-1 \,.
\label{masses_ZN}	
\end{equation}
Here $m \in \mathbb{R}$ is the average mass, and $\Delta m > 0$ is effectively the mass scale of the  model.
Note that by  a shift of $\sigma$ one can always add a  constant to all 
 $m_i$. In particular one can get rid of the average mass $m$.

The bare coupling constant $\beta_0$ in quantum theory becomes a running coupling $\beta$. It is asymptotically free and defines the scale $\Lambda_{CP}$ via 
\begin{equation}
	\Lambda_{CP}^2 = M_\text{uv}^2 \, \exp(- \frac{8 \pi \beta_0}{N}),
\label{2d_4d_coupling_nosusy}	
\end{equation}
where $M_\text{uv}$ is the ultra-violet (UV) cutoff.

Let us review phases of this theory. It is known 
that in the case of vanishing masses $\Delta m = 0$ this non-supersymmetric \CP model is at strong coupling with vanishing VEV $\langle n^i \rangle = 0$. It can be solved by means of the $1 / N$ expansion \cite{W79}. It turns out that at the quantum level spontaneous breaking of the global
SU(N) (flavor) symmetry present at the classical level disappears. There are no massless Goldstone bosons in the physical
spectrum. The $n^i$ fields acquire mass of the order of $\Lambda_{CP}$. 

Moreover,  composite degree of freedom -- the would-be auxiliary photon $A_\mu$ acquires a kinetic term at the one-loop level and becomes dynamical. The presence of massless photon ensures long range
forces in the non-supersymmetric \CP model. The Coulomb potential is linear
in two dimensions, namely 
\begin{equation}
	V(r) \sim \frac{\Lambda_{CP}^2}{N} \, r \,,
\end{equation}
where $r$ is the separation between the charges.
This leads to the Coulomb/confinement phase \cite{W79}. Electric charges
are confined. The lightest electric charges are the $n^i$ quanta
which become kinks at strong coupling \cite{W79}. Confinement of kinks
means that they are not present in the physical spectrum of the theory in isolation.
They form bound states, kink-antikink \textquote{mesons}. 

\begin{figure}[h]
	\centering
	\includegraphics[width=0.5\linewidth]{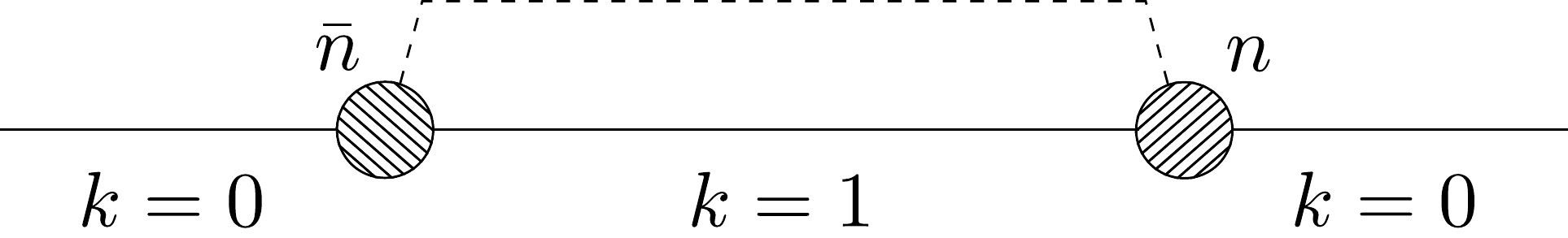}
	\caption{
	Linear confinement of the $n$-$\bar{n}$ pair.
	The solid straight line represents the ground state ($k=0$ vacuum).
	The dashed line shows
	the vacuum energy density in the first quasivacuum.}
\label{fig:conf}
\end{figure}

Masses of kinks are of order of $\Lambda_{CP}$ while the confining potential is weak, proportional to $1/N$.
Therefore kink and antikink in the ''meson'' are well separated forming a quasivacuum inside the ''meson''. Thus,
 beside the single ground state, there is a family of quasivacua with energy splittings of order $\sim \Lambda_{CP}^2 / N$. The order parameter which distinguish different quasivacua is the value of the constant
electric field or topological density
\beq
Q=\frac{i}{2\pi}\,\varepsilon_{\mu\nu}\,\pt^{\mu}A^{\nu} =\frac1{8\pi\beta} \,\varepsilon_{\mu\nu}
\,\pt^{\mu}\bar{n}_i\pt^{\nu}n^i
\label{topdensity}
\eeq
The picture of
confinement of $n$'s is shown on Fig.~\ref{fig:conf}.


The kinks interpolate between the adjacent vacua. They are confined
monopoles of the bulk theory. Since the excited string tensions are larger than the
tension of the lightest one, these monopoles, besides four-dimensional confinement,
are confined also in the two-dimensional sense: a monopole is necessarily attached to
an antimonopole on the string to form a meson-like configuration \cite{GSY05,MMY}
 
On the other hand, at large mass scales $\Delta m \gg \Lambda_{CP}$ the coupling constant is small, frozen at the scale $\Delta m$, and semiclassical calculations are applicable. The field $n^i$ develops a non-zero VEV, and there is no massless photon and no long-range interactions. That is why this phase is usually called \textquote{Higgs phase} as opposed to the Coulomb/confinement strong coupling phase. More exactly \CP model
in this phase gives a low energy description of a Higgs phase below the scale of the photon mass. Essentially
this weakly coupling  Higgs phase is similar to the ''classical phase'' described by the classical Lagrangian 
\eqref{cpn_lagr_simplest}.

It was shown that at intermediate mass scales $\Delta m \sim \Lambda_{CP}$ there is a phase transition between the Higgs and Coulomb phases, see \cite{GSY05,Gorsky:2005ac,Ferrari,Ferrari2}.

\subsection{\ntwot model}

Supersymmetric generalization of the above model \cite{W79,W93} has additional fermionic field $\xi^i$, $i = 1, \ldots, N$, which are superpartners of the $n_i$ fields. 
The Euclidean version of the full \ntwot Lagrangian is 
\begin{equation}
\begin{aligned}
\mathcal L =&
	\frac{1}{e_0^2}\left(\frac{1}{4} F_{\mu\nu}^2 +\left|\pt_\mu\sigma\right|^2 + \frac{1}{2}D^2
	+\bar\lambda \, i\bar{\sigma}^\mu\pt_\mu\,\lambda
	\right) + i\,D\left(\bar{n}_i n^i -2\beta_0
	\right)
	\\
	&+
	\left|\nabla_\mu n^i\right|^2+ \bar{\xi}_i\, i\bar{\sigma}^\mu\nabla_\mu\,\xi^i
	+ 2\sum_i\left|\sigma-\frac{m_i}{\sqrt 2}\right|^2\, |n^i|^2
	\\
	&+
	i\sqrt{2}\,\sum_i \left( \sigma -\frac{m_i}{\sqrt 2}\right)\bar\xi_{Ri}\, \xi^i_L 
	- i\sqrt{2}\,\bar{n}_i \left(\lambda_R\xi^i_L - \lambda_L\xi^i_R \right)
	\\
	&+
	i\sqrt{2}\,\sum_i \left( \bar\sigma -\frac{\bar{m}_i}{\sqrt 2}\right)\bar\xi_{Li}\, \xi^i_R 
	- i\sqrt{2}\,{n}^i \left(\bar\lambda_L\bar\xi_{Ri} - \bar\lambda_R\bar\xi_{Li} \right),
\end{aligned}	
\label{lagrangian_N=2}
\end{equation}
where  $m_i$ are twisted masses and  the limit $e_0^2\to\infty$ is implied. Moreover, $\bar\sigma^\mu = \{1,\,i\sigma_3\}$ . Fermions $\xi_L, \xi_R$ are respectively left and right components of the $\xi$ field. Here again one can add a uniform constant to all the $m_i$ by shifting the $\sigma$ field.

The gauge field $A_\mu$, complex scalar superpartner $\sigma$, real scalar $D$ 
and a two-component complex fermion  $\lambda$ form a vector  auxiliary supermultiplet. 
In particular, integrating over $D$ and fermion $\lambda$ give
the  constraints 
\begin{equation}
	\nbar_i \, n^i = 2\beta_0,
\end{equation}
\begin{equation}
	\bar{n}_i\,\xi^i =0\,,\qquad \bar\xi_i\,n^i = 0\,
\end{equation}
in the limit $e_0\to\infty$.

This model was derived as a world sheet theory on the non-Abelian string in \ntwo SQCD. The $n_i$ fields parametrize  the orientational moduli of the non-Abelian string \cite{HT1,ABEKY,SYmon,HT2} 
The mass parameters $m_i$ are in fact masses of the bulk quark fields. The bare coupling constant $\beta_0$ is related to the bulk gauge coupling constant $g^2$ normalized at the scale of the bulk gauge boson mass $m_G \sim g\sqrt{\mu m}$ via (see e.g.~\cite{SYrev})
\begin{equation}
	2\beta_0 = \frac{4\pi}{g^2(m_G)} = \frac{N}{2\pi}\ln{\frac{m_G}{\Lambda_{CP}}}\,,
\label{2d_4d_coupling_susy}	
\end{equation}
In order to keep the bulk theory at weak coupling we assume that $m_G \gg \Lambda_{CP}$. 

Witten solved this model in the large $N$ approximation in the zero mass case  \cite{W79}.
Large-$N$ solution of this model at non-zero masses shows two different regimes at weak and  strong coupling \cite{Bolokhov:2010hv}. At small mass scales $\Delta m < \Lambda_{CP}$ the theory is in the strong coupling phase with zero VEV $\langle n^i \rangle = 0$ and with a dynamical photon (Witten's phase). However the photon  now is massive due to the presence of the chiral anomaly. There are no long-range forces and no confinement of kinks. 

In both strong and weak coupling regimes the 
 theory has $N$ degenerate vacuum states as dictated by its Witten index. They are  labeled by the VEV of 
$\sigma$ \cite{Bolokhov:2010hv}. At $\Delta m <\Lambda_{CP}$ we have 
\begin{equation}
	\sqrt{2}\sigma ~=~ \exp\left( \frac{2\pi\,i\, k}{N}
	\right)\times \Lambda_{CP}
	 \qquad k=0, ..., N-1
	\label{22sigmaapp}
\end{equation}
This result can be understood as follows.
The chiral anomaly breaks U(1) $R$-symmetry present at zero masses down to $Z_{2N}$ which is then broken spontaneously down to $Z_2$ by VEV of the $\sigma$ field (which has $R$ charge equal to two). 
In particular, from the large-$N$ solution it follows that VEV of  $\sqrt{2}|\sigma| = \Lambda_{CP}$.
Then 
$Z_{2N}$
symmetry ensures presence of $N$ vacua as shown in  Eq.~\eqref{22sigmaapp}. 

At large masses located on a circle
(see \eqref{masses_ZN}) the $Z_{2N}$ symmetry is still unbroken. This  leads to to the similar structure of the $\sigma$  VEVs at $\Delta m > \Lambda_{CP}$, namely
\begin{equation}
	\sqrt{2}\sigma ~=~ \exp\left( \frac{2\pi\,i\, k}{N}
	\right)\times \Delta m,
	 \qquad k=0, ..., N-1
	\label{sigmavevLargem}
\end{equation}

The above  formulas show a phase transition at $\Delta m = \Lambda_{CP}$. 
As follows from the large-$N$ solution the model above this point  is in the Higgs phase with a nonzero VEV for say, zero component of $n$, $\langle n^0 \rangle \neq 0$.  In both phases there is no confinement, in contrast  to the non-supersymmetric case.

In fact the above phase transition is a consequence of the large-$N$ approximation. At finite $N$ the transition 
between two regimes is smooth. This follows from the exact effective superpotential known for \ntwot \CP
model \cite{W93}.

\subsection{$\mu$-Deformed \CP model \label{sec:intro_deformed}}

Now let us pass on to the case of interest, namely, the $\mu$-deformed \CP model. 
This model appears as a world sheet theory on a non-Abelian string in \ntwo SQCD deformed by the adjoint field mass $\mu$. It was derived in 
Chapter \ref{sec:none}
in two cases, for small and large values of the deformation $\mu$.
Here and throughout this Chapter we will take the mass parameters to lie on the circle \eqref{masses_ZN}, and we also assume that the deformation parameter is real and positive, $\mu > 0$. 
%

The first effect derived  in Chapter~\ref{sec:none} is  that  $n_i$ fields entering the \ntwot \CP model \eqref{lagrangian_N=2} develop an additional potential  upon $\mu$ deformation which depends on mass differences. This potential in the small $\mu$ limit was first found in \cite{SYfstr}.
The second effect is that superorientational modes of the non-Abelian string are lifted. In other words the
two-dimensional fermions $\xi^i$ (fermionic superpartners of $n^{i}$)  were massless in the supersymmetric version of the model at $\mu=0$. However, at  small $\mu$ they  acquire a mass $\lambda(\mu) \sim \mu$. At large deformations they become heavy and decouple.

In order to capture these features, we write the following Lagrangian for the deformed \CP model:
\begin{equation}
\begin{aligned}
\mathcal L &= 
\left|\nabla_\mu n^i\right|^2+ \bar{\xi}_i\, i\bar{\sigma}^\mu\nabla_\mu\,\xi^i 
+ i\,D\left(\bar{n}_i n^i -2\beta \right)
\\[3mm]
&+ \sum_i\left|\sqrt 2\sigma-m_i\right|^2\, |n^i|^2
+ \upsilon (\mu) \sum_i \Re\Delta m_{i0} |n^i|^2
\\[3mm]
&+
i\,\sum_i \left( \sqrt{2}\sigma -m_i - \lambda (\mu) \right)\bar\xi_{Ri}\, \xi^i_L 
- i\sqrt{2}\,\bar{n}_i \left(\lambda_R\xi^i_L - \lambda_L\xi^i_R \right)
\\
&+
i\,\sum_i \left( \sqrt{2}\bar\sigma - \bar{m}_i - \ov{\lambda (\mu)}  \right)\bar\xi_{Li}\, \xi^i_R 
- i\sqrt{2}\,{n}^i \left(\bar\lambda_L\bar\xi_{Ri} - \bar\lambda_R\bar\xi_{Li} \right),
\end{aligned}
\label{lagrangian_init}
\end{equation}
where $\Delta m_{i0} = m_i - m_0$,  $m_i$ are  quark masses $i=1,...N$, and $m_0$ is the mass with the smallest real part.

The coefficient functions $\upsilon (\mu)$ and $\lambda (\mu)$ were derived in Chapter~\ref{sec:none} at the classical level for small and large values of $\mu$:
\begin{equation}
\upsilon (\mu) = 
	\begin{cases}
		\frac{4\pi\mu}{2\beta} 											\,, \quad &\mu \to 0 ,\\
		\frac{1}{2\beta} \frac{8\pi \mu}{\ln \frac{g^2\mu}{m}} 	\,, \quad &\mu \to \infty
	\end{cases}
\label{upsilon_old}	
\end{equation}
\begin{equation}
\lambda (\mu) =
	\begin{cases}
		\lambda_0 \frac{\mu}{2\beta} \,, \quad &\mu \to 0 ,\\
		\text{const  } g\sqrt{\mu m} \sim m_G   \,, \quad &\mu \to \infty
	\end{cases}
\label{lambda_old}	
\end{equation}
%
%
Here $g^2$ is the four-dimensional  bulk coupling constant.
The numerical value for $\lambda_0$ is  $\lambda_0 \approx 3.7$ (this can be computed numerically using the formula \eqref{solution_orient_alpha}). Note that although we can get rid 
of the explicit dependence on the average quark mass $m$ in \eqref{lagrangian_init} by shift of $\sigma$ the above formulas show that  it enters indirectly through definitions of parameters of $\mu$-deformed \CP model \eqref{lagrangian_init} in terms of parameters of the bulk SQCD.

This model interpolates between the supersymmetric and the non-supersymmetric models briefly described above. In the limit $\mu \to 0$ supersymmetry is restored to \ntwot, and we obtain \eqref{lagrangian_N=2}. At large deformations the fermions can be integrated out, and the theory flows to the bosonic model \eqref{cpn_lagr_simplest}. 

Our main tool of investigating this model in the quantum level will be the $1 / N$ expansion. In order to have a smooth large $N$ limit, our parameters should scale as
\begin{equation}
\begin{aligned}
	g^2   \sim 1/N , \quad
	\beta \sim N , \quad
	\mu   \sim N , \\
	m \sim 1, \quad
	\upsilon(\mu) \sim 1 , \quad
	\lambda(\mu)  \sim 1
\end{aligned}
\end{equation}

Below in this Chapter we will use  three independent  physical parameters to describe  our four-dimensional bulk  model. The first one is the bulk gauge boson mass 
\beq
m_G^2=2 g^2 \mu m, 
\label{Wmass}
\eeq
which plays a role of the physical UV cutoff in the world sheet \CP model on the non-Abelian string, see \cite{SYrev}. The second one is 
the quark mass differences $( m_i-m_j)$ and the third parameter is the physical mass of the adjoint matter
\beq
m_\text{adj} = g^2\mu = \frac{\mu}{\frac{N}{8\pi^2}\, \ln\frac{m_G}{\Lambda_\text{4d}}} \equiv \wt{\mu} \,.
\label{tildem}
\eeq
which will be  our actual deformation parameter. All three parameters scales as $N^0$ in the large $N$ limit.
Here $\Lambda_{4d}$ is the scale of \ntwo bulk SQCD.

Thus, in fact, the average quark mass $m$ is not an independent parameter. It can be written  as 
\begin{equation}
	m  = \frac{m_G^2}{2\wt{\mu}} \,.
\end{equation}

At the scale of the gauge boson mass \eqref{Wmass} the world sheet coupling constant  for small $\mu$  is given by \cite{ABEKY,SYmon}, cf. \eqref{2d_4d_coupling_susy}
\begin{equation}
	2\beta = \frac{4\pi}{g^2}  = \frac{N}{2\pi}\, \ln\, \frac{m_G}{\Lambda_{4d}} . 
\end{equation}

For large $\mu$ the world sheet coupling normalized at the scale $m_G$ becomes
\begin{equation}
	2\beta = \text{const  } \frac{\mu}{m}\,\frac{1}{\ln^2 \frac{g^2\mu}{m}}.  
\end{equation}
Expressed in terms of the invariant parameters it reads
\begin{equation}
	2\beta = \text{const  }\frac{ N}{\pi} \, \frac{\wt{\mu}^2}{m_G^2}  \frac{\ln \frac{m_G}{\Lambda_\text{4d}^{{\mathcal N}=1}}}{\ln^2 \frac{2 \wt{\mu}}{m_G}},
	\label{beta_classical_largemu}
\end{equation}
where we take into account that at large $\wt{\mu}$ our bulk theory flows to  \none SQCD  with the scale
$(\Lambda_\text{4d}^{{\mathcal N}=1})^{2N}=\wt{\mu}^N\Lambda_\text{4d}^N$.
 
In terms of the independent parameters the coefficient functions $\upsilon$ and $\lambda$ become
\begin{equation}
\upsilon (\wt{\mu}) = 
	\begin{cases}
		\wt{\mu} 											\,, \quad &\wt{\mu} \to 0 ,\\
		\frac{m_G^2}{\wt{\mu}} \, \ln \frac{2 \wt{\mu}}{m_G} 		 	\,, \quad &\mu \to \infty
	\end{cases}
\label{upsilon}	
\end{equation}
\begin{equation}
\lambda (\wt{\mu}) =
	\begin{cases}
		\wt{\lambda}_0 \wt{\mu} \,, \quad &\wt{\mu} \to 0 ,\\
		m_G   \,, \quad &\wt{\mu} \to \infty
	\end{cases}
\label{lambda}	
\end{equation}
where $\wt{\lambda}_0 = \lambda_0 / 4\pi \approx 0.3$.

As we already mentioned the value of the bulk gauge boson mass $m_G$ plays a role of the UV cutoff of our world sheet theory.
Below $m_G$ our model is asymptotically free (cf. \eqref{2d_4d_coupling_nosusy}) with 
\beq
2\beta(E) = \frac{N}{2\pi}\, \ln\, \frac{E}{\Lambda_{2d}} 
\label{beta_run}
\eeq
at the scale $E$.
This fixes the scale $\Lambda_{2d}$ in terms of the parameters of the bulk theory. At small $\wt{\mu}$ we have 
\beq
\Lambda_{2d}(\wt{\mu}\to 0) = \Lambda_{4d}, 
\label{Lambda2D=4D}
\eeq
while at large $\wt{\mu}$ 
\begin{equation}
	\Lambda_{2d} = \Lambda_\text{4d}^{{\mathcal N}=1} 
	\exp(- \text{const}\,\frac{ \wt{\mu}^2}{m_G^2} \cdot \frac{1}{\ln \frac{2 \wt{\mu}}{m_G}})
\label{Lam_2d}	
\end{equation}


 Note that at  $\wt{\mu} \to \infty$ the scale \eqref{Lam_2d} of our model becomes exponentially
small and the model enters the strong coupling regime only at extremely small energies. We will see below that phase transitions with respect to $\wt{\mu}$ appear at rather small values of $\wt{\mu}$ where the scale $\Lambda_{2d}$ is close to its supersymmetric value $\Lambda_{4d}$. Since the fermion decoupling occurs at very large $\wt{\mu} \gg m_G$, we can use small $\wt{\mu}$ approximation formulas \eqref{upsilon} and \eqref{lambda} while investigating the phase transition.

In the following sections we are going to investigate different phases and vacuum structure of the world sheet  theory. There are two parameters that we can vary -- the SUSY breaking parameter $\wt{\mu}$ and the mass scale $\Delta m$. As we already mentioned  our model \eqref{lagrangian_init} exhibits a rich phase structure  in the $(\Delta m, \ \wt{\mu})$ plane.

%
%

\section{One loop effective potential} \label{sec:Veff}

In this section we proceed with solving the theory \eqref{lagrangian_init} via the $1 / N$ expansion. As we already mentioned the \ntwot model as well as the non supersymmetric \CP model (without mass parameters) were solved by Witten \cite{W79}. This method was also generalized for the case of heterotic \ntwoo model \cite{SYhetN} and for the twisted mass case \cite{Gorsky:2005ac,Bolokhov:2010hv}. Our derivation will closely follow these papers.

\subsection{Derivation of the effective potential}

We want to start by deriving the one-loop effective potential. Our action \eqref{lagrangian_init} is well suited for that since it is quadratic with respect to the dynamical fields $n_i$ and $\xi_i$. However, we do not to integrate over all of them a priori due to  the  following reason.

As was stated in the previous section, our model \eqref{lagrangian_init} is, in a sense, an intermediate case between the \ntwot and the non-supersymmetric \CP models, which were studied before. Therefore we can use the insights derived from these models in order to better understand our case. First of all, we expect that our theory has at least two phases, the strong and weak coupling. The order parameter distinguishing between these two phases is the expectation value of the $n_i$ fields. At weak coupling (so-called Higgs phase \cite{Gorsky:2005ac}) one of the $n_i$ develops a VEV, $\langle n_{i_0} \rangle = 2\beta$. In the strong coupling regime (so-called Coulomb phase), VEVs of all the $n_i$ field vanish.

So, we will use the following strategy. We integrate over  $N-1$ fields $n^{i}$ with $i \ne 0$ (and over the corresponding fermions $\xi_i$). The resulting effective action is a functional of $n^0\equiv n$, $D$ and $\sigma$. To find the vacuum configuration, we will minimize the effective action with respect to $n$, $D$ and $\sigma$.

Note  that this functional also depends on $A_\mu$ and the fermions $\xi_{L,R}^0$, $\lambda_{L,R}$, but  the Lorenz invariance imply that these fields have zero VEVs. We also choose to allow  $n^0$ field to have non-zero VEV because the associated mass $m_0$ has  the minimal real part (see \eqref{masses_ZN} ) and as we will see later  $\langle n^0 \rangle \neq 0$ corresponds to the true vacuum in the Higgs phase rather then a quasivacuum.

Integrating out the $n^i$ and $\xi^i$ fields, we arrive at the following determinants:
\begin{equation}
	 \frac{
	\prod_{i=1}^{N-1} {\rm det}\, \left(-\pt_{k}^2 
	   + \bigl| \sqrt{2}\sigma - m_i - \lambda (\mu) \bigr|^2\right)}{
	\prod_{i=1}^{N-1}{\rm det}\, \left(-\pt_{k}^2 +iD + \upsilon (\mu)\Delta m_{i0} 
	   + \bigl| \sqrt{2}\sigma - m_i \bigr|^2\right)},
\label{det}	
\end{equation}
which gives for the effective potential:
\begin{equation}
\begin{aligned}
	V_\text{eff} &= \int d^2 x\, (iD + |\sqrt{2}\sigma - m_0|^2) |n|^2 - 2\beta \int d^2 x \, iD \\
					&+ \sum_{i=1}^{N-1} \Tr \ln \left(-\pt_{k}^2 +iD + \upsilon (\mu)\Delta m_{i0}  + \bigl| \sqrt{2}\sigma - m_i \bigr|^2\right) \\
					&- \sum_{i=1}^{N-1} \Tr \ln \left(-\pt_{k}^2 + \bigl| \sqrt{2}\sigma - m_i - \lambda (\mu) \bigr|^2\right)
\end{aligned}	
\end{equation}

The next step is to calculate the traces entering this expression. At $\wt{\mu} \to 0$, the supersymmetry is restored, and this expression is well defined. 
However at a non-vanishing deformation, this expression diverges quadratically, and a regularization needs to be performed. 
Below we proceed with the Pauli-Villars regularization (a similar procedure was carried out in \cite{NOVIKOV1984103}). We introduce regulator fields with masses $b_a$, $f_a$, $a=1, 2$, and write the regularized effective potential as
\begin{equation}
\begin{aligned}
	V_\text{eff} &= \int d^2 x\, (iD + |\sqrt{2}\sigma - m_0|^2) |n|^2 - 2\beta\, \int d^2 x \, iD \\
					&+ \sum_{i=1}^{N-1} \Tr \ln \left(-\pt_{k}^2 +iD + \upsilon (\mu)\Delta m_{i0}  + \bigl| \sqrt{2}\sigma - m_i \bigr|^2\right)	\\
					&+ \sum_{a=1}^{2} \sum_{i=1}^{N-1} B_a \Tr \ln \left(-\pt_{k}^2 + b_a^2 \right) \\
					&- \sum_{i=1}^{N-1} \Tr \ln \left(-\pt_{k}^2 + \bigl| \sqrt{2}\sigma - m_i - \lambda (\mu) \bigr|^2\right)	\\
					&- \sum_{a=1}^{2} \sum_{i=1}^{N-1} F_a \Tr \ln \left(-\pt_{k}^2 + f_a^2\right)
\end{aligned}	
\end{equation}
where the coefficients satisfy
\begin{equation}
	\sum_{a=0}^{2} B_a = -1, \quad \sum_{a=0}^{2} B_a b_a^2 = - m_\text{bos}^2
\end{equation}
These equations imply
\begin{equation}
	B_1 = \frac{b_2^2 - m_\text{bos}^2}{b_1^2 - b_2^2}, \quad B_2 = - \frac{b_1^2 - m_\text{bos}^2}{b_1^2 - b_2^2}
\end{equation}
The regulator masses play the role of the UV cutoff. Similar relations hold for the fermionic regulator coefficients.

Moreover, we need to properly normalize our traces by subtracting the contributions in the trivial background, namely 
$\Tr \ln (-\pt_{k}^2)$ from the bosonic and  the fermionic traces. After this procedure we arrive at
\begin{equation}
\begin{aligned}
	V_\text{eff} &= \int d^2 x\, (iD + |\sqrt{2}\sigma - m_0|^2) |n|^2 - 2\beta\, \int d^2 x \, iD \\
					&- \frac{1}{4\pi} \sum_{i=1}^{N-1} \Bigg[ 
						\left( +iD + \upsilon (\mu)\Delta m_{i0}  + \bigl| \sqrt{2}\sigma - m_i \bigr|^2\right) \\
							&\phantom{AAAA} \times \ln \left( +iD + \upsilon (\mu)\Delta m_{i0}  + \bigl| \sqrt{2}\sigma - m_i \bigr|^2\right) \\
					&- \left( +iD + \upsilon (\mu)\Delta m_{i0}  + \bigl| \sqrt{2}\sigma - m_i \bigr|^2\right)
							\frac{b_1^2 \ln b_1^2 - b_2^2 \ln b_2^2}{b_1^2 - b_2^2}
						\Bigg]	\\
					&+ \frac{1}{4\pi} \sum_{i=1}^{N-1} \Bigg[
						\bigl| \sqrt{2}\sigma - m_i - \lambda (\mu) \bigr|^2 \ln \bigl| \sqrt{2}\sigma - m_i - \lambda (\mu) \bigr|^2 \\
					&- \bigl| \sqrt{2}\sigma - m_i - \lambda (\mu) \bigr|^2 \frac{f_1^2 \ln f_1^1 - f_2^2 \ln f_2^2}{f_1^2 - f_2^2}
						\Bigg]
\end{aligned}	
\end{equation}

This is a quite complex expression.
In order to simplify it, let us take the limit \cite{NOVIKOV1984103}
\begin{equation}
	b_1^2 = x  M_\text{uv}^2 , \ b_2^2 = M_\text{uv}^2, \quad f_1^2 = x M_\text{uv}^2 , \ f_2^2 = M_\text{uv}^2, \quad x \to 1,
\end{equation}
where $M_\text{uv}$ is the UV cutoff.
%
%
Moreover, recall from the section \ref{sec:intro_deformed} that the bare coupling constant can be parametrized as
\begin{equation}
	2\beta (M_\text{uv}) ~~=~~\frac{N}{4\pi}\, \ln\, {\frac{M_\text{uv}^2}{\Lambda^2}}\,,
\end{equation}
Here, $\Lambda \equiv \Lambda_{2d}$ is the scale of our model. 
Then the density of the effective potential becomes
\begin{equation}
\begin{aligned}
	\mathcal V_\text{eff} &=  (iD + |\sqrt{2}\sigma - m_0|^2) |n|^2 \\
					&+ \frac{1}{4\pi} \sum_{i=1}^{N-1}  iD
							\left[ 1 - \ln\frac{iD + \upsilon (\mu) \Re\Delta m_{i0}  + \bigl| \sqrt{2}\sigma - m_i \bigr|^2}{\Lambda^2} \right]	\\
					&+ \frac{1}{4\pi} \sum_{i=1}^{N-1}  
						\left(\upsilon (\mu) \Re\Delta m_{i0}  + \bigl| \sqrt{2}\sigma - m_i \bigr|^2\right) \\
						& \phantom{AAAA}\times
						\Bigg[ 1 - 
							\ln \frac{ iD + \upsilon (\mu) \Re\Delta m_{i0}  + \bigl| \sqrt{2}\sigma - m_i \bigr|^2}{M_\text{uv}^2} 
						\Bigg]\\
					&- \frac{1}{4\pi} \sum_{i=1}^{N-1} 
						\bigl| \sqrt{2}\sigma - m_i - \lambda (\mu) \bigr|^2 
						\Bigg[ 1 - 
							\ln \frac{\bigl| \sqrt{2}\sigma - m_i - \lambda (\mu) \bigr|^2}{M_\text{uv}^2}
						\Bigg]
\end{aligned}	
\label{v_eff}
\end{equation}

Note that our regularized effective potential depends on the UV cutoff scale $M_\text{uv}$. We cannot make a subtraction to get rid of it in the model at hand for the following reason. First, when we consider our
$\wt{\mu}$-deformed \CP model \eqref{lagrangian_init} as an effective world sheet theory  on the non-Abelian string the 
the UV cutoff has a clear physical meaning, namely
\beq
M_\text{uv} =m_G,
\label{M_UV}
\eeq
where $m_G$ is the mass of the bulk gauge boson. Moreover, the fermion mass $\lambda(\mu)$ in \eqref{v_eff} interpolates
from zero at $\wt{\mu}=0$ to $m_G=M_\text{uv}$ at $\wt{\mu}\to\infty$, see \eqref{lambda}. Thus $M_\text{uv}$ is in fact a physical parameter in our model and there is no need to get rid of it.

The renormalized coupling constant is 
\begin{equation}
	2\beta_\text{ren} = \frac{1}{4\pi}\sum_{i=1}^{N-1}\ln\frac{iD + \upsilon (\mu) \Re\Delta m_{i0}  + \bigl| \sqrt{2}\sigma - m_i \bigr|^2}{\Lambda^2}
\end{equation}

\subsection{Vacuum equations}

To find the vacuum configuration we minimize the effective potential \eqref{v_eff}. Varying with respect to $D$ we arrive at
\begin{equation}
	|n|^2 = 2\beta_\text{ren} = \frac{1}{4\pi}\sum_{i=1}^{N-1} \ln\frac{iD + \upsilon (\mu) \Re\Delta m_{i0}  + \bigl| \sqrt{2}\sigma - m_i \bigr|^2}{\Lambda^2}
\label{master1}
\end{equation}
Variation with respect to $\nbar$ yields the second equation:
\begin{equation}
	(iD + |\sqrt{2}\sigma - m_0|^2) n = 0
\label{master2}
\end{equation}
Finally, the third equation is obtained by minimizing over the $\sigma$ field,
\begin{equation}
\begin{aligned}
	- (\sqrt{2}\sigma - m_0) |n|^2 
			&+ \frac{1}{4\pi}\sum_{i=1}^{N-1} \left(\sqrt{2}\sigma - m_i\right) 
					\ln \frac{ iD + \upsilon (\mu) \Re\Delta m_{i0}  + \bigl| \sqrt{2}\sigma - m_i \bigr|^2}{m_G^2} \\
			&=  \frac{1}{4\pi}\sum_{i=1}^{N-1}\left(\sqrt{2}\sigma - m_i - \lambda (\mu)\right) 
					\ln \frac{\bigl| \sqrt{2}\sigma - m_i - \lambda (\mu) \bigr|^2}{m_G^2},
\end{aligned}
\label{master3}	
\end{equation}
where here and below we replaced $M_\text{uv}$ by the physical mass $m_G$.

These three equations comprise our {\em master set}. In addition, the vacuum configurations must satisfy the constraint
\begin{equation}
	\beta_\text{ren} \geqslant 0,
\label{beta_positive_condition}	
\end{equation}
which comes from $2\beta_\text{ren} = |n|^2 \geqslant 0$.

From \eqref{master1} and \eqref{master2} it immediately follows that either
\begin{equation}
	n = \beta_\text{ren} =  0
\label{strong_phase_condition}	
\end{equation}
or
\begin{equation}
	iD + |\sqrt{2}\sigma - m_0|^2 = 0 \, .
\label{higgs_phase_condition}	
\end{equation}
The first option corresponds to the strong coupling regime where the VEV of $n$ and the renormalized coupling constant vanish. The second option is realized in the Higgs regime, where the $n$ field develops a VEV. In the following sections we will study each of these regimes in detail.

%
%

\section{Strong coupling regime \label{sec:strong}}

In this section will begin investigation of our model in the strong coupling regime, which is defined by the condition \eqref{strong_phase_condition}. 
This phase occurs when the mass scale of the model $\Delta m \lesssim \Lambda$, see e.g. \cite{Gorsky:2005ac,Bolokhov:2010hv}. 
To start off we will first investigate a simple case $\Delta m = 0$. 
Behavior of our model is different at different values of the deformation parameter: at intermediate  $\wt{\mu}$ we will see a phase transition, while in the limit of large fermion mass $\lambda \to m_G$ we will confirm that the model \eqref{lagrangian_init} flows to non-supersymmetric \CP model \eqref{cpn_lagr_simplest} as expected.
Next, we will generalize our results to the case of distinct masses $m_i$.

\begin{figure}
    \centering
    \begin{subfigure}[t]{0.5\textwidth}
        \centering
        \includegraphics[width=\textwidth]{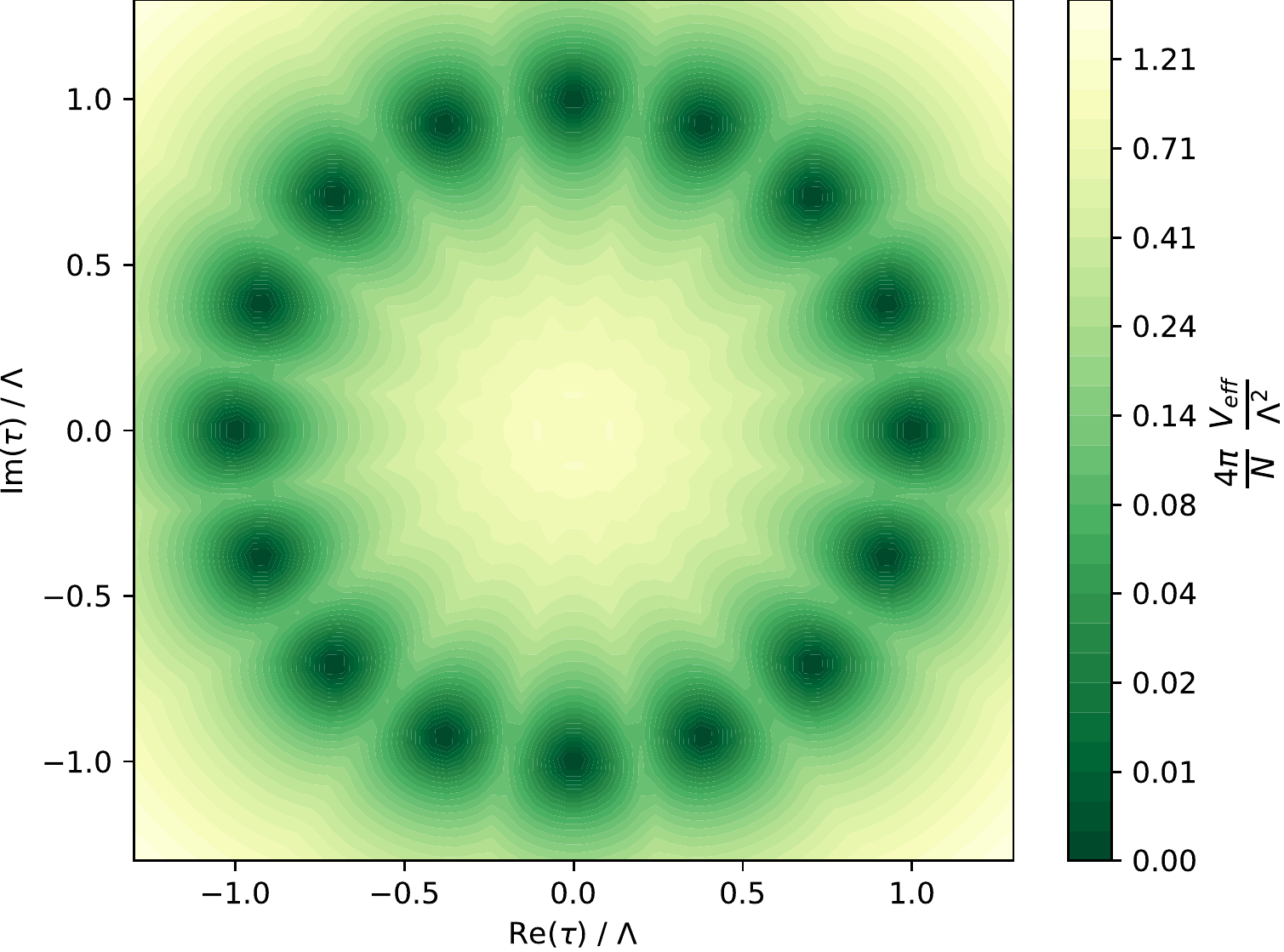}
        \caption{$\lambda = 0$, supersymmetric case, degenerate vacua}
        \label{fig:v_complex_tau_degenerate}
    \end{subfigure}%
    ~ 
    \begin{subfigure}[t]{0.5\textwidth}
        \centering
        \includegraphics[width=\textwidth]{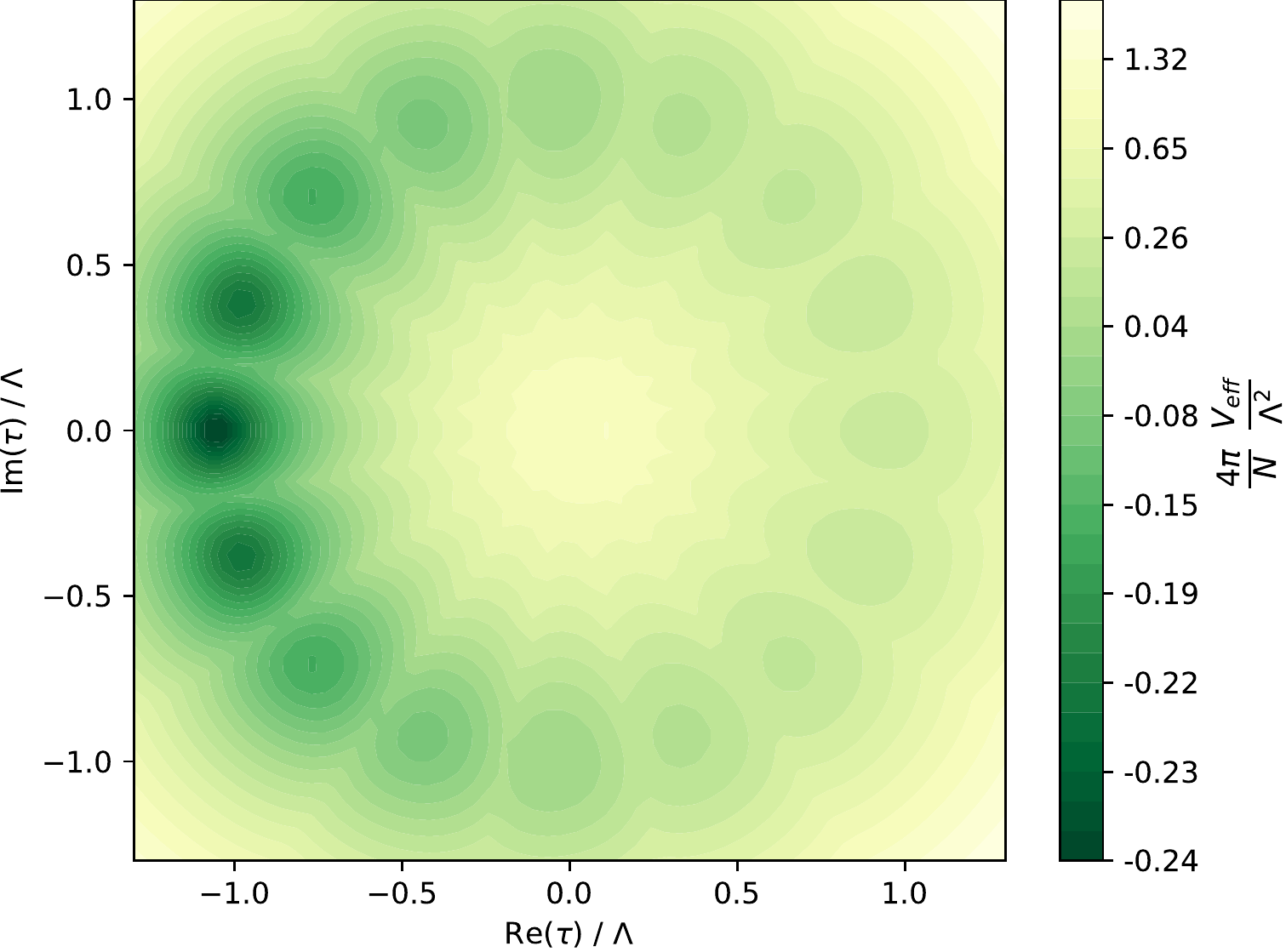}
        \caption{$\lambda > 0$, broken supersymmetry, lifted quasivacua}
        \label{fig:v_complex_tau_lifted}
    \end{subfigure}%
\caption{Effective potential \eqref{V_eff_strong_dm=0} on the complex $\tau = \sqrt{2}\sigma - m_0$ plane, with $D$ integrated out.}
\label{fig:v_complex_tau}
\end{figure}

\subsection{Equal mass case, small deformations}

We start by investigating the simplest case of equal mass parameters,
\begin{equation}
	m_0 = m_1 = \ldots = m_{N-1} \equiv m
\label{equal_masses}	
\end{equation}
Under this assumption  the potential proportional to  $\upsilon (\mu)$ is zero and the only deformation  we are left with is the fermion mass $\lambda$. For now we will not write its dependence on $\wt{\mu}$ explicitly. 


To simplify the equations, let us denote
\begin{equation}
	\tau = \sqrt{2}\sigma - m_0
\end{equation}
%
Then  the effective potential becomes
\begin{equation}
\begin{aligned}
	{\mathcal  V}_\text{eff} 
			&= \frac{N}{4\pi} iD\left[ 1 - \ln\frac{iD + \bigl| \tau \bigr|^2}{\Lambda^2} \right] 
			+ \frac{N}{4\pi} \bigl| \tau \bigr|^2 \left[ 1 -  \ln\frac{iD  + \bigl| \tau \bigr|^2}{m_G^2} \right] \\
			&-  \frac{N}{4\pi} \bigl| \tau - \lambda (\mu) \bigr|^2
				\left[ 1 - \ln\frac{\bigl| \tau - \lambda (\mu) \bigr|^2}{m_G^2} \right] + \Delta V(\arg\tau),
			\end{aligned}
\label{V_eff_strong_dm=0}	
\end{equation}
where $\tau = |\tau|\,e^{i \arg\tau}$. Here we added  a new term $\Delta V(\arg\tau)$ absent in \eqref{v_eff}. 
It takes into account the chiral anomaly and appears already in \ntwot \CP model at $\wt{\mu}=0$. As was shown by Witten
\cite{W79} the photon become massive due to the chiral anomaly with mass equal $2\Lambda$ . The complex scalar 
$\sigma$ is a superpartner of the photon and also acquires mass $2\Lambda$. In particular, its argument $\arg\tau$
becomes massive.

 This effect is taken into account  by the additional potential $\Delta V(\arg\tau)$ in \eqref{V_eff_strong_dm=0}. It is constructed as follows.  At small $\wt{\mu}$ VEVs of $\tau$ are approximately given by their
supersymmetric values,
\begin{equation}
	\tau^\text{SUSY}_k = - \Lambda \exp \left( \frac{2\pi\,i\, k}{N}	\right), \quad k=0, ..., N-1,
\label{vac_tau_susy_dm=0}	
\end{equation}
cf. \eqref{22sigmaapp}. 
We divide $2\pi$ into $N$ patches centered at vacuum values, $\arg\tau^\text{SUSY}_k=2\pi k/N +\pi$,
$k=0,...,(N-1)$ and define the potential $\Delta V(\arg\tau)$ to be quadratic in each patch. Namely, we have
\beq
\Delta V(\arg\tau) = \frac{N}{4\pi} \frac{m_{\arg\tau}^2}{2} (\arg\tau - \arg\tau^\text{SUSY}_k)^2,
\qquad \frac{2\pi (k-\frac12)}{N} < \arg\tau -\pi < \frac{2\pi (k+\frac12)}{N},
\label{DeltaV}
\eeq
where $m_{\arg\tau}$ is the mass of $\arg\tau$. We present its calculation below, in particular showing corrections (see eq. \eqref{m_arg_sigma_leading})
to the Witten's result \cite{W79}
\beq
m_{\arg\tau}^\text{SUSY} =2\Lambda.
\label{wittenmass}
\eeq

Without the additional potential $\Delta V(\arg\tau) $ $N$ discrete vacua \eqref{vac_tau_susy_dm=0} disappear immediately
as we switch on $\wt{\mu}$ due to the lifting of quasivacua. We show below that with $\Delta V(\arg\tau) $ taken into account quasivacua are 
still present at small $\wt{\mu}$ and disappear only at certain finite critical $\wt{\mu}_\text{crit}$ which we identify as a phase transition point. Note that possible higher corrections to the quadratic potential \eqref{DeltaV} are suppressed in 
the large $N$ limit because the width of each patch is small, proportional to $1/N$.

\subsubsection{Vacuum energies}

As we turn on the deformation parameter $\wt{\mu}$ the  mass of $\xi^i$ fermion $\lambda(\wt{\mu})$ is no longer zero.
This breaks explicitly both  chiral symmetry and two-dimensional supersymmetry. As a result the $Z_N$ symmetry is broken and VEVs of  $\sigma$ are no longer located at a circle. 
Moreover, 
at  $\wt{\mu} = 0$  our model has $N$ degenerate vacua given by \eqref{vac_tau_susy_dm=0}.
When we switch on $\wt{\mu}$,  the corresponding vacuum energies split, and all vacua except the one at $k=0$  become quasivacua. The only true vacuum is the one at $k=0$, see Fig.~\ref{fig:v_complex_tau}. As we discussed in Sec.~\ref{sec:no-susy} this leads to the confinement of kinks.

It turns out that there are two mechanisms responsible for the vacuum energy splitting. One is due to the effective potential \eqref{V_eff_strong_dm=0} and dominates at small $\wt{\mu}$.  The other one is typical for the non-supersymmetric
\CP model, see Sec.~\ref{sec:no-susy}. It is  due to the constant electric field of  kinks interpolating between neighboring quasivacua
and dominates at large $\wt{\mu}$. 
We will now study the former mechanism, while the latter one will be considered in the next subsection.

Energy splittings in the small $\wt{\mu}$ limit  can be derived using the small $\lambda (\mu)$ expansion of the effective potential \eqref{V_eff_strong_dm=0}:
\begin{equation}
	{\mathcal  V}_\text{eff} = {\mathcal  V}_\text{SUSY}  + \delta {\mathcal  V},
\end{equation}
where ${\mathcal  V}_\text{SUSY}$ is the supersymmetric effective potential corresponding to $\lambda = 0$, while
\begin{equation}
	\delta {\mathcal  V} \approx \frac{N}{4\pi} \cdot 2  \Re\tau \cdot \lambda  \, \ln\frac{m_G^2}{\bigl| \tau \bigr|^2}
\label{dV_small-lam}	
\end{equation}
is the $O(\lambda)$ deformation. We can immediately infer lifted vacuum energies by plugging unperturbed VEVs \eqref{vac_tau_susy_dm=0}	into \eqref{dV_small-lam}. As we already mentioned the  ground state (true vacuum) is located at 
\beq
\tau_0 = - \Lambda = \Lambda e^{i \pi}, 
\label{yruevacuum}
\eeq
while  the first quasivacuum is at 
\begin{equation}
	\tau_1 = - \Lambda \exp \left(\frac{2\pi\,i}{N} \right) \approx - \Lambda - \Lambda \frac{2\pi\,i}{N} + \Lambda \frac{2\pi^2}{N^2}
\label{first_quasivac_strong}	
\end{equation}
Plugging this into \eqref{dV_small-lam} we get for the vacuum splitting\footnotemark
\begin{equation}
	E_1 - E_0 = \frac{2 \pi}{N} \lambda\Lambda  \ln \frac{m_G}{\Lambda}
\label{splitting_1-0}	
\end{equation}
\footnotetext{Formula \eqref{splitting_1-0} has a correction coming from the energy-momentum trace anomaly, but this correction is of the next order in the small parameter $\frac{\lambda}{\Lambda}  \ln \frac{M_\text{uv}}{\Lambda}$.}
This signifies that kinks interpolating between these vacua are now confined, as opposed to the supersymmetric case.

\subsubsection{Corrections to the VEVs}

\begin{figure}[h]
    \centering
    \begin{subfigure}[t]{0.5\textwidth}
        \centering
        \includegraphics[width=\textwidth]{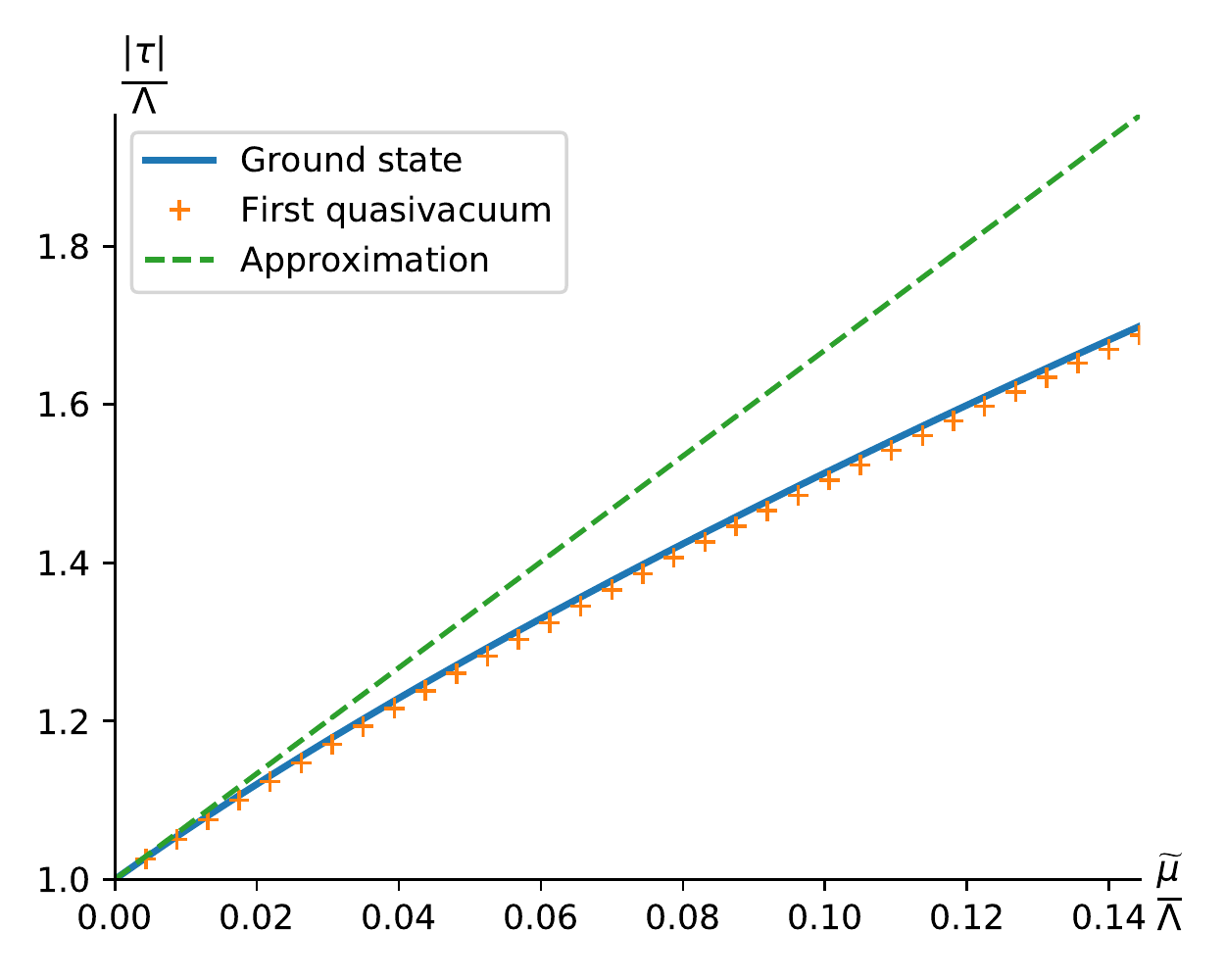}
        \caption{$|\tau_\text{ground}|$ and $|\tau_1|$}
        \label{fig:dm=0_smalllam_tau_abs}
    \end{subfigure}%
    ~ 
    \begin{subfigure}[t]{0.5\textwidth}
        \centering
        \includegraphics[width=\textwidth]{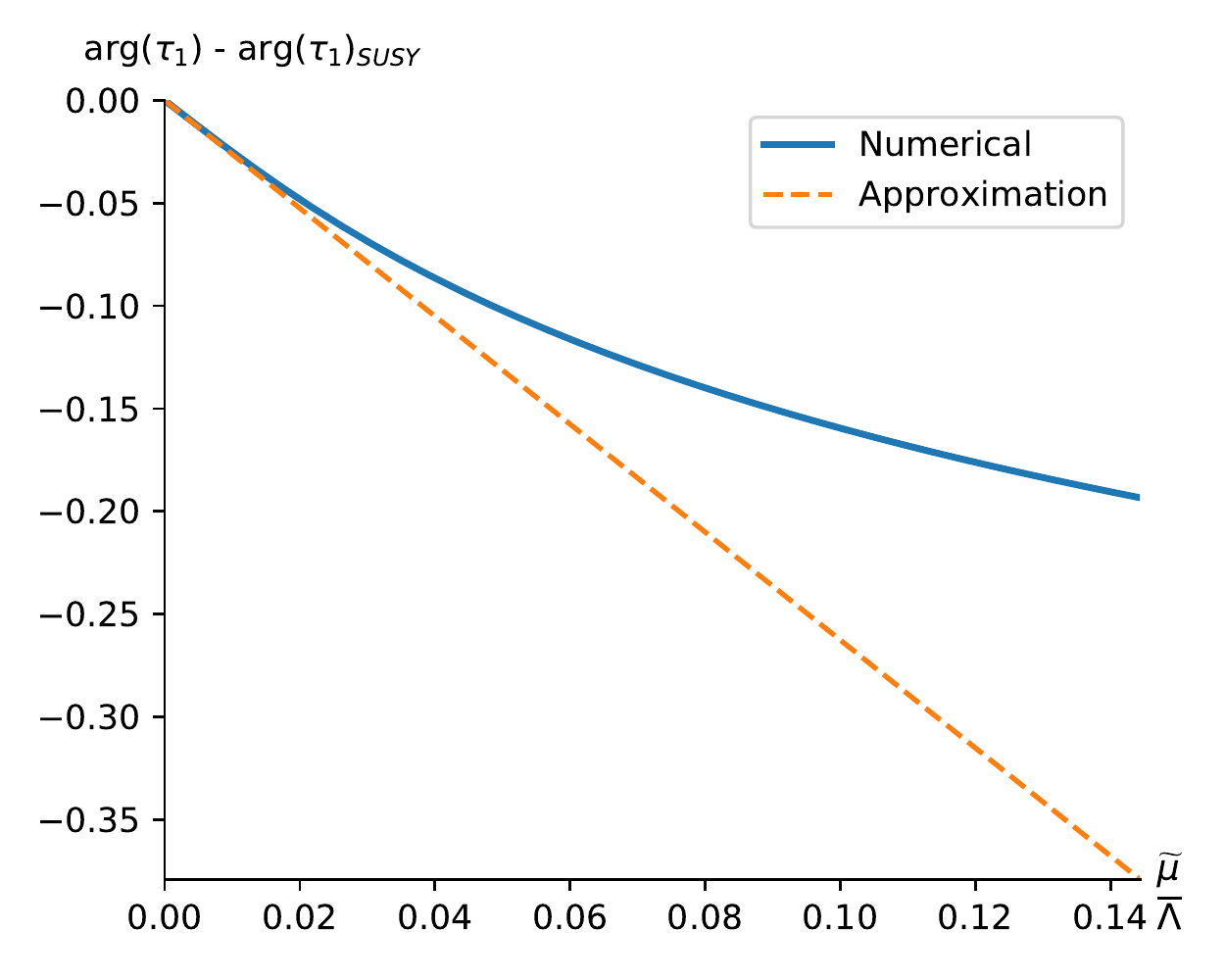}
        \caption{Correction to $\arg\tau_1$}
        \label{fig:dm=0_smalllam_tau_arg}
    \end{subfigure}%
\caption{
	Numerical results for the minima $\tau_\text{ground}$ and $\tau_1$ obtained by directly minimizing \eqref{V_eff_strong_dm=0}. On the figure \subref{fig:dm=0_smalllam_tau_abs}, the green dashed line shows the approximate formula \eqref{ground_state_correction}, the solid blue line is the numerical values of $|\tau_\text{ground}|$, while $|\tau_1|$ is shown by red \textquote{+}. 
	Figure \subref{fig:dm=0_smalllam_tau_arg} shows the approximate correction to $\arg\tau_1$ (\eqref{first_quasivacuum_correction}, the last term) and the numerical results for this quantity. 
}
\label{fig:dm=0_smalllam_tau}
\end{figure}

Now let us derive corrections to the unperturbed VEVs \eqref{vac_tau_susy_dm=0}.
Minimizing the potential \eqref{V_eff_strong_dm=0} we get:
\begin{equation}
	2\beta_\text{ren} = \ln\frac{ iD + \bigl| \tau \bigr|^2}{\Lambda^2} = 0 \Rightarrow  iD + \bigl| \tau \bigr|^2 = \Lambda^2
\label{master1_dm=0}	
\end{equation}
\begin{equation}
	|\tau| \ln\frac{\bigl| \tau - \lambda (\wt{\mu}) \bigr|^2}{\Lambda^2} + \cos(\arg\tau) \lambda (\wt{\mu}) \ln\frac{m_G^2}{\Lambda^2} = 0
\label{stronf_dm=0_taueq}
\end{equation}
\begin{equation}
	-  \sin(\arg\tau) \, \lambda |\tau| \, \ln\frac{m_G^2}{\Lambda^2} + \frac{m_{\arg\tau}^2}{2} \, (\arg\tau - \arg\tau^\text{SUSY}_k) = 0
\end{equation}
The approximate solution  in the limit of small $\wt{\mu}$ is given by
\begin{equation}
	|\tau| \approx \Lambda - \cos(\arg\tau^\text{SUSY}_k) \frac{1}{2} \lambda \ln\frac{m_G^2}{\Lambda^2}
\end{equation}
\begin{equation}
	\arg\tau \approx \arg\tau^\text{SUSY}_k  + \sin(\arg\tau^\text{SUSY}_k) \, \frac{2\lambda\Lambda}{m_{\arg\tau}^2}  \, \ln\frac{m_G^2}{\Lambda^2} 
\end{equation}
In particular, for the $\tau_0 = - \Lambda$ we get the corrected value
\begin{equation}
	\tau_\text{ground} \approx - \Lambda - \frac{1}{2} \lambda \ln\frac{m_G^2}{\Lambda^2}
\label{ground_state_correction}	
\end{equation}
while for the first quasivacuum \eqref{first_quasivac_strong}
\begin{equation}
\begin{aligned}
	|\tau_1| &\approx |\tau_\text{ground}| \approx  \Lambda + \frac{1}{2} \lambda \ln\frac{m_G^2}{\Lambda^2} \\
	\arg\tau_1 &\approx \underbrace{ \left(\pi + \frac{2\pi}{N} \right) }_\text{unperturbed} 
		- \frac{2\pi}{N} \frac{\lambda}{2 \Lambda}  \, \ln\frac{m_G^2}{\Lambda^2} 
\end{aligned}	
\label{first_quasivacuum_correction}
\end{equation}
where we used \eqref{wittenmass} for the non-perturbed  mass of $\sigma$. These results agree with numerical calculations, see Fig.~\ref{fig:dm=0_smalllam_tau}.

Note that when 
\begin{equation}
	\frac{\lambda}{\Lambda}  \, \ln\frac{m_G}{\Lambda} = 1
\label{tau_0=tau_1}	
\end{equation}
we have in our approximation $\arg\tau_1 = \tau_\text{ground} = \pi$, and the quasivacuum at $\tau_1$ effectively disappears. This signifies that around the point \eqref{tau_0=tau_1}	a phase transition might take place. This will turn out to be true, see Sec.~\ref{sec:strong1_strong2} below.

The quasivacuum with the highest energy is located at
\begin{equation}
	\tau_\text{high} \approx  \Lambda - \frac{1}{2} \lambda \ln\frac{m_G^2}{\Lambda^2}
\end{equation}
Further analysis of the equation \eqref{stronf_dm=0_taueq} shows that this solution disappears at
\begin{equation}
	\lambda = \frac{2\Lambda}{e \ln \frac{m_G^2}{\Lambda^2}}
\label{strong1_strong2_trans_potential}	
\end{equation}
which is consistent with \eqref{strong1_strong2_trans_arg-tau}. This suggests that around the critical value of the deformation 
\beq
\lambda_\text{crit} \sim \frac{\Lambda}{ \ln \frac{m_G^2}{\Lambda^2}}
\label{strong1_strong2_trans_arg-tau}
\eeq
all quasivacua have decayed (cf. \eqref{tau_0=tau_1}).

\subsection{Effective action \label{sec:effact}}

As we already mentioned there are two mechanisms of the energy splitting of quasivacua at non-zero $\wt{\mu}$. 
Both lead to the confinement of kinks. The first one
is due to $\wt{\mu}$-corrections present in the effective potential  \eqref{V_eff_strong_dm=0} These corrections  lift 
$\sigma$-quasivacua and lead to the  splitting  described by Eq.~\eqref{splitting_1-0}. The second mechanism is due to the 
constant electric field of kinks interpolating between quasivacua. The photon $A_\mu$ becomes dynamical on the quantum level
\cite{W79}. 
We will see below that, as we turn on the deformation parameter $\wt{\mu}$, the photon acquires a massless component. A linear Coulomb potential is  generated, but the vacuum energy splitting due to the electrical field is much smaller then the one in 
\eqref{splitting_1-0}.
At sufficiently large $\wt{\mu}$ all $N-1$  $\sigma$-quasivacua decay, and the splitting is saturated by the electric field only.
We identify this change of the regime and associated discontinuity in (the derivative of) $(E_1-E_0)$  as a phase transition.

\subsubsection{Derivation of the effective action \label{sec:effact_deriv}}

Consider now the effective action of our $\wt{\mu}$-deformed \CP model \eqref{lagrangian_init} obtained by integrating out 
$n^i$ and $\xi^i$ fields in the large-$N$ approximation. Relaxing the condition that $\sigma$ and $D$ are constant fields
assumed in Sec.~\ref{sec:Veff}  we consider the one loop effective action as a functional of  fields of the vector supermultiplet.

Considering the vicinity of the true vacuum where $\Im\langle\sigma\rangle = 0$  we write down the bosonic part of the action
in the form (Minkowski formulation\footnotemark)
\footnotetext{In this subsection we will use the Minkowski formulation with $g^{\mu\nu} = \text{diag} \{+,-\}$, and for the Levi-Civita symbol $\varepsilon_{01} = - \varepsilon^{01} = +1$. See Appendix~\ref{sec:useful_2d} for the notation and the relationship between the Euclidean and Minkowski formulations.}
%
\begin{multline}
	S_{\rm eff}=
	 \int d^2 x \Bigg\{
	- \frac1{4e_{\gamma}^2}F^2_{\mu\nu} + \frac1{e_{\Im\sigma}^2}	|\pt_{\mu}\Im\sigma|^2 + \frac1{e_{\Re\sigma}^2}	|\pt_{\mu}\Re\sigma|^2 \\
	- V(\sigma) 
	- \sqrt{2} \, b_{\gamma,\Im\sigma} \, \Im\sigma \,F^{*}
	  \Bigg\},
	\label{effaction}
\end{multline}
where 
$F^{*}$ is the  dual gauge field strength,
\begin{equation}
	F^{*}= - \frac12\varepsilon_{\mu\nu}F^{\mu\nu}\,.
\end{equation}
This effective action was first presented  for \ntwot and \ntwoo supersymmetric \CP models in \cite{SYhetN}.
Here we generalize it for the $\wt{\mu}$-deformed \CP model \eqref{lagrangian_init}. The potential 
 $V(\sigma)$ here can be obtained  from \eqref{v_eff} by eliminating $D$ by virtue of its equation of motion.

Coefficients in front of $A_{\mu}$ and $\sigma$ kinetic terms are  finite after renormalization reflecting
 Witten's observation  that these fields become physical \cite{W79}.
The last term in \eqref{effaction} is $A_\mu - \sigma$ induced by the chiral anomaly. Because of this mixing, the would-be massless photon and the phase of $\sigma$ acquire a mass \eqref{wittenmass} already in unperturbed theory at $\wt{\mu}=0$.
This term is also present when we switch on the deformation.

\begin{figure}[h!]
    \centering
    \begin{subfigure}[t]{0.33\textwidth}
        \centering
        \includegraphics[width=\textwidth]{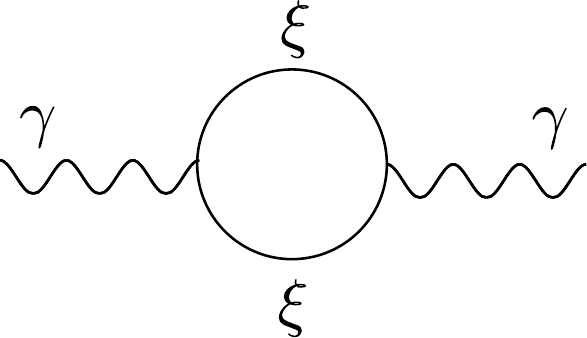}
        \caption{Photon wave function renormalization}
        \label{fig:loops:photon}
    \end{subfigure}%
    \hspace{20pt} %
    \begin{subfigure}[t]{0.33\textwidth}
        \centering
        \includegraphics[width=\textwidth]{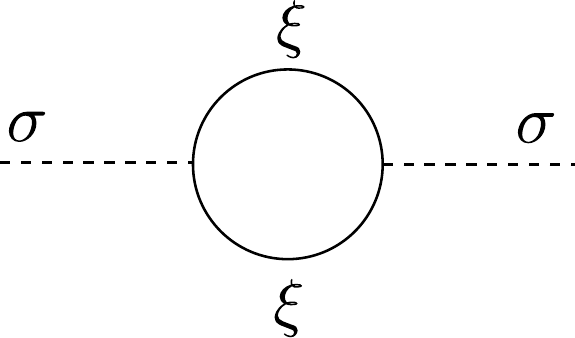}
        \caption{Scalar wave function renormalization}
        \label{fig:loops:scalar}
    \end{subfigure}
    
    \vspace{20pt}
    
    \begin{subfigure}[t]{0.33\textwidth}
        \centering
        \includegraphics[width=\textwidth]{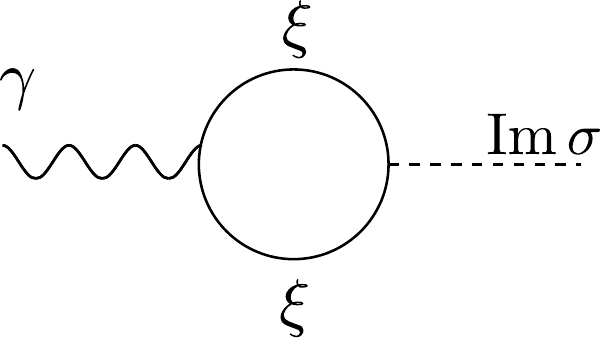}
        \caption{Photon-scalar mixing}
        \label{fig:loops:mixing}
    \end{subfigure}
\caption{Contributions to the effective action}
\label{fig:loops}
\end{figure}

Coefficients in this effective action come from loops. We take the low-energy limit when the external momenta are small. There are several contributions.
Photon wave function renormalization comes from the diagram on Fig.~\ref{fig:loops:photon} and a similar graph with a bosonic loop. 
Wave function renormalizations for $\Re\sigma$ and $\Im\sigma$ come from the diagram on Fig.~\ref{fig:loops:scalar} and also similar graph with a bosonic loop. Finally, the mixing term is given by the diagram on Fig.~\ref{fig:loops:mixing}. For the mass distribution \eqref{masses_ZN} and the vacuum with $\Im\langle\sigma\rangle = 0$, the normalization factors are:
%
%
\begin{equation}
\begin{aligned}
	\frac{1}{e^2_{\Re\sigma}} &= \frac{1}{4\pi}\,\sum_{k=0}^{N-1} 
		\left[ \frac{1}{3} \frac{M_{\xi_k}^2 + 2 \left( \Im m_k \right)^2 }{M_{\xi_k}^4} 
			+ \frac{2}{3} \frac{\left( \sqrt{2}\langle\sigma\rangle - \Re m_k \right)^2}{m_{n_k}^4} \right]
	\,, \\
	\frac{1}{e^2_{\Im\sigma}} &= \frac{1}{4\pi}\,\sum_{k=0}^{N-1} 
			\left[ \frac{1}{3} \frac{3 M_{\xi_k}^2 - 2 \left( \Im  m_k \right)^2 }{M_{\xi_k}^4} 
				+ \frac{2}{3} \frac{\left( \Im m_k \right)^2}{m_{n_k}^4} \right]
	\,, \\
	\frac{1}{e^2_{\gamma}} &= \frac{1}{4\pi}\,\sum_{k=0}^{N-1} \left[ \frac{1}{3} \frac{1}{m^2_{n_k}} + \frac{2}{3} \frac{1}{M^2_{\xi_k}} \right]
	\,, \\
	b_{\gamma,\Im\sigma} &= \frac{1}{2\pi}\,\sum_{k=0}^{N-1} \frac{ \sqrt{2}\langle\sigma\rangle - m_k - \lambda (\wt{\mu}) }{M_{\xi_k}^2}
	\,.
\end{aligned}
\label{eff_normalizations}
\end{equation}
Here, $M^2_{\xi_k}$ and $m^2_{n_k}$ are the masses of the $\xi_k$ and $n_k$ fields respectively:
\begin{equation}
\begin{aligned}
	M^2_{\xi_k} &= | \sqrt{2}\langle\sigma\rangle - m_k - \lambda (\wt{\mu}) \bigr|^2
	\\
	m^2_{n_k} &= i\langle D \rangle + \upsilon (\wt{\mu})\Delta m_k  + \bigl| \sqrt{2}\langle\sigma\rangle - m_k \bigr|^2
\end{aligned}
\end{equation}
We present details of this calculation in Appendix~\ref{sec:loops}.

Next we  diagonalize the photon-$\sigma$ mass matrix  in \eqref{effaction}, see  the next subsection. As we already mentioned this diagonalization
shows that the photon acquires a massless component as soon as we switch on $\wt{\mu}$. This component is responsible for the 
presence of the constant electric field in quasivacua. This constant electric field gives rise to a second mechanism of quasivacua splitting, see \eqref{Evac_electical-1}. This effect is small at small $\wt{\mu}$ but becomes dominant  at larger $\wt{\mu}$ above the phase transition
point. This result can also be derived in a different way which we consider in the next subsection.

\subsubsection{Photon mass}

Here we  diagonalize the photon-$\sigma$ mass matrix  in \eqref{effaction} to find the photon mass\footnotemark.
\footnotetext{Note that the mass generating mechanism here is not the Higgs mechanism. There is no massless Goldstone particle.
Nevertheless, the theory is gauge invariant. See Appendix~\ref{app:2d_photon} for the explanation.}
In order to do that, let us write down bare propagators for $\Im\sigma$ and $A_\mu$ that follow immediately from \eqref{effaction} (in the Minkowski notation):
\begin{equation}
\begin{aligned}
	G^0_\gamma &= -i\, e^2_\gamma \, \frac{g^{\mu\nu} - \frac{k^\mu k^\nu}{k^2}}{k^2}
	\\
	G^0_{\Im\sigma} &= -\frac{i}{2}\, e^2_{\Im\sigma} \frac{1}{k^2 - \delta m_{\Im \sigma}^2}
\end{aligned}
\end{equation}
where we use the Landay gauge, while $\delta m_{\Im \sigma}^2$ is the contribution to the the mass of the $\Im\sigma$ field coming from the potential $V(\sigma)$ in \eqref{effaction}. In the vicinity of the true ground state \eqref{ground_state_correction} we have 
\begin{equation}
	\delta m_{\Im \sigma}^2 \approx 4 \lambda \Lambda \ln \frac{m_G}{\Lambda} \,.
\label{delta_Im_m_sigma}
\end{equation}
 At large $\wt{\mu}$, $\delta m_{\Im \sigma}^2 \sim  \lambda^2 \ln m_G / \Lambda$, see Sec.~\ref{sec:strong_dm=0_largelam}.

\begin{figure}[h]
	\centering
	\includegraphics[width=0.6\linewidth]{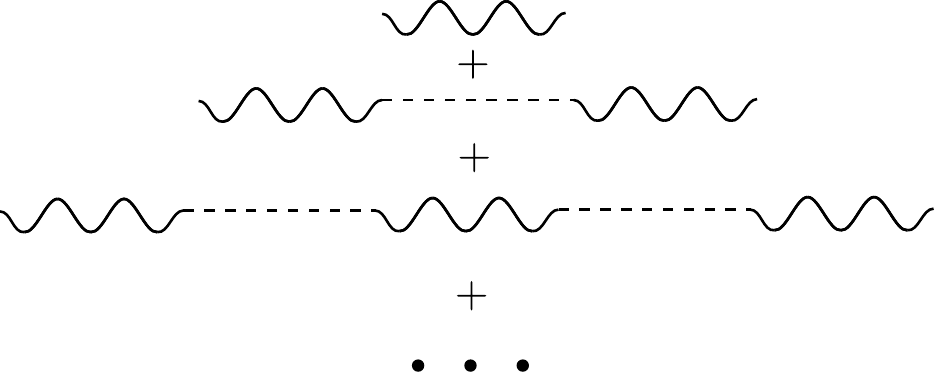}
	\caption{Contributions to the photon propagator}
\label{fig:iter_gamma} 
\end{figure}

Consider the photon propagator. Iterating the scalar $\Im\sigma$ insertions shown in Fig.~\ref{fig:iter_gamma}, we obtain the  full photon propagator, 
\begin{equation}
\begin{aligned}
	\widehat{G}_\gamma 
		&= G^0_\gamma \frac{1}{1 - \frac{e^2_\gamma\, e^2_{\Im\sigma}\, b^2_{\gamma,\Im\sigma}}{k^2 - \delta m^2_{\Im\sigma}}}  \\
		&= -i\, e^2_\gamma \,  \left(g^{\mu\nu} - \frac{k^\mu k^\nu}{k^2}\right) \frac{k^2 - \delta m^2_{\Im\sigma}}{k^2\left(k^2 - \delta m^2_{\Im\sigma} - e^2_\gamma\, e^2_{\Im\sigma}\, b^2_{\gamma,\Im\sigma}\right)}  \\
		&= -i\, e^2_\gamma \,  \left(g^{\mu\nu} - \frac{k^\mu k^\nu}{k^2}\right) 
			\left(A\,\frac{1}{k^2} + (1 - A)\,\frac{1}{k^2 - \delta m^2_{\Im\sigma} - e^2_\gamma\, e^2_{\Im\sigma}\, b^2_{\gamma,\Im\sigma}}\right)
\end{aligned}
\label{photon_propagator_full}
\end{equation}
where the coefficient
\begin{equation}
	A = \frac{\delta m^2_{\Im\sigma}}{\delta m^2_{\Im\sigma} + e^2_\gamma\, e^2_{\Im\sigma}\, b^2_{\gamma,\Im\sigma}}
\end{equation}
increases from 0 to 1 as $\wt{\mu}$ runs from zero to infinity. 
What we see here is that at  non-zero $\wt{\mu}$, the photon acquires a  massless component. In the SUSY case (zero $\wt{\mu}$) the coefficient $A$ vanishes, and we have only the massive component. Note that the number of physical states do not change since
the massless photon has no physical degrees of freedom in two dimensions.
At large $\wt{\mu}$ the  massive component  becomes heavy and decouples ($A\to 1$). We are left with the massless photon much in the same way as in non-supersymmetric \CP model.

If we do a similar calculation for the $\Im\sigma$ propagator, we will get simply
\begin{equation}
\begin{aligned}
	\widehat{G}_{\Im\sigma} 
		&= G^0_{\Im\sigma} \frac{1}{1 - \frac{e^2_\gamma\, e^2_{\Im\sigma}\, b^2_{\gamma,\Im\sigma}}{k^2 - \delta m^2_{\Im\sigma}}}  \\
		&= -i\, e^2_{\Im\sigma} \, 	\frac{1}{k^2 - \delta m^2_{\Im\sigma} - e^2_\gamma\, e^2_{\Im\sigma}\, b^2_{\gamma,\Im\sigma}}
\end{aligned}
\label{Im_sigma_propagator_full}
\end{equation}
%
Just like in \cite{W79}, we see that the would-be massless phase of the $\sigma$ field acquires a mass
\begin{equation}
	m_{\text{arg}\, \tau}^2 =  \delta m^2_{\Im\sigma} + e^2_\gamma\, e^2_{\Im\sigma}\, b^2_{\gamma,\Im\sigma} \,.
\label{m_arg_sigma}
\end{equation}
This effect is taken into account by the additional term \eqref{DeltaV} in the effective potential \eqref{V_eff_strong_dm=0}.
At $\wt{\mu}=0$ $\delta m^2_{\Im\sigma}=0$ and the mass of the phase of $\sigma$  reduces to \eqref{wittenmass}.
Consider the leading correction at small $\lambda$.
For the ground state \eqref{ground_state_correction} at $\Delta m = 0$ we have
\begin{equation*}
	\frac{1}{e^2_{\Im\sigma}} \approx \frac{N}{4\pi\Lambda^2} \left(1 - 2\frac{\lambda}{\Lambda} \ln \frac{m_G}{\Lambda} \right) \,,
	\quad
	\frac{1}{e^2_{\gamma}} \approx \frac{N}{4\pi\Lambda^2} \left(1 - \frac{4}{3}\frac{\lambda}{\Lambda} \ln \frac{m_G}{\Lambda} \right) \,,
\end{equation*}
\begin{equation*}
	b_{\gamma,\Im\sigma} \approx - \frac{N}{2\pi\Lambda} \left(1 - \frac{\lambda}{\Lambda} \ln \frac{m_G}{\Lambda} \right) \,,
\end{equation*}
and, therefore,
\begin{equation}
	m_{\text{arg}\, \tau}^2 \approx 4 \Lambda^2 \left(  1 + \frac{7}{3} \frac{\lambda}{\Lambda} \ln \frac{m_G}{\Lambda} \right) \,.
\label{m_arg_sigma_leading}
\end{equation}

Let us look more closely at the photon propagator \eqref{photon_propagator_full} in the small $\wt{\mu}$ limit. We have  
\begin{equation}
	A \approx \frac{\lambda}{\Lambda}  \ln \frac{M}{\Lambda},
\end{equation}
and for the massless part of the photon propagator:
\begin{equation}
	\widehat{G}_{\gamma, \text{massless}} = 
		-i \,  \frac{g^{\mu\nu} - \frac{k^\mu k^\nu}{k^2}}{k^2} 
		\frac{4\pi}{N} \lambda\Lambda  \ln \frac{M}{\Lambda}
\end{equation}
From this Green function we calculate the electric field produced by a kink with electric charge $+1$ and  find for the vacuum energy splitting
\begin{equation}
	E_1-E_0 = \frac{1}{2 e_\gamma^2} F_{01}^2 = \frac{2\pi}{N} \left(\lambda \ln \frac{M}{\Lambda} \right)^2
\label{Evac_electical-1}	
\end{equation}

\subsubsection{Coulomb potential and vacuum energies}

In this section we study the formation  of a constant electric field in a quasivacuum generalizing a method 
developed by Witten in \cite{W79} for \ntwot supersymmetric \CP model.

 Let us start with the effective action \eqref{effaction} taking into account the presence of the trial matter charges,
\begin{equation}
	S_{\rm eff}=
	 \int d^2 x \left\{
	- \frac1{4e_{\gamma}^2}F^2_{\mu\nu} - \sqrt{2} \, b_{\gamma,\Im\sigma} \, \Im\sigma \,F^{*}
	+ j_\mu A^\mu
	  \right\},
\end{equation}
Consider a stationary point-like kink at $x=x_0$ with electric charge $+1$ described by the current $j_{\mu} = (\delta(x-x_0),\, 0)$
and $F^{*} = -\frac12\varepsilon^{\mu\nu}F_{\mu\nu} = \p_0 A_1 - \p_1 A_0$. 

 We have the equation of motion for the photon:
\begin{equation}
	 - \frac{1}{e_\gamma^2} \p_x \mathcal{E} - \sqrt{2} \, b_{\gamma,\Im\sigma} \, \p_x \Im\sigma = - j_0,
\end{equation}
where
\begin{equation}
	\mathcal{E} = F_{01}  
\end{equation}
 is the electric field strength.
(See Appendix~\ref{sec:useful_2d} for assistance.)
Integrating over the spatial coordinate we obtain
\begin{equation}
	  \frac{1}{e_\gamma^2} ( \mathcal{E}(\infty) - \mathcal{E}(-\infty)) + \sqrt{2} \, b_{\gamma,\Im\sigma} \, ( \Im\sigma(\infty) - \Im\sigma(-\infty)  )
	 = 1
\label{kink_topological_eq}	 
\end{equation}
In the supersymmetric case $\wt{\mu} = 0$ the photon is massive, so there is no constant electric field,  $\mathcal{E}(\infty) = \mathcal{E}(-\infty) = 0$. Therefore we have
\begin{equation}
	   \sqrt{2} \, b_{\gamma,\Im\sigma} \, ( \Im\sigma(\infty) - \Im\sigma(-\infty)  ) = 1
\end{equation}
Since 
\begin{equation}
	b_{\gamma,\Im\sigma} = \frac{1}{2\pi} \frac{N}{\Lambda},
\label{b_susy}	
\end{equation}
see Eq.~\eqref{eff_normalizations} for $\wt{\mu}=0$
we get
\begin{equation}
	\sqrt{2}  ( \Im\sigma(\infty) - \Im\sigma(-\infty)  ) = 2 \pi \frac{\Lambda}{N}
\end{equation}
which, if we set $ \tau (-\infty) = - \Lambda$ for the true vacuum, is an approximation of
\begin{equation}
	\tau(\infty) = - \Lambda e^{\frac{2\pi i}{N}}
\end{equation}
for the value of $\sigma$ VEV in the first quasivacuum, see \eqref{first_quasivac_strong}.
This result for the \ntwot case has been derived long ago by Witten \cite{W79} showing the presence of $N$ vacua and 
kinks interpolating between them. 

Now, consider small deformations in the Eq.~\eqref{kink_topological_eq} for  a kink interpolating between the ground state \eqref{ground_state_correction} at $x=- \infty$ and the first quasivacuum
\eqref{first_quasivacuum_correction} at $x=+ \infty$. Setting $\mathcal{E}(-\infty) = 0$  we get from 
\eqref{kink_topological_eq}	
\begin{equation}
	  \frac{1}{e_\gamma^2}  \mathcal{E}(\infty) + \sqrt{2} \, b_{\gamma,\Im\sigma} \, ( \Im\sigma(\infty) - \pi  ) = 1 
\end{equation}
Using \eqref{first_quasivacuum_correction} and \eqref{b_susy} we obtain for the electric field strength
\begin{equation}
	  \mathcal{E}(\infty) = e_\gamma^2 \, \frac{\lambda}{\Lambda}   \ln\frac{m_G}{\Lambda}. 
\end{equation}
We see that the kink  produces a constant electric field now.
This gives the contribution to the   energy density splitting between  the first quasivacuum and the true vacuum
\begin{equation}
	(E_{1} -E_0)|_{\mathcal{E}}  = \frac{1}{2 e_\gamma^2} \mathcal{E}^2 = \frac{2\pi}{N} \left(\lambda \ln \frac{m_G}{\Lambda} \right)^2
\label{Evac_electical-2}	
\end{equation}
This coincides with the result \eqref{Evac_electical-1} obtained from the photon-$\sigma$ diagonalization. This contribution is small compared to the $\sigma$-splitting   given by \eqref{splitting_1-0} at small $\wt{\mu}$.

\subsection{Second order phase transition \label{sec:strong1_strong2}}

\begin{figure}[h!]
	\centering
	\includegraphics[width=0.95\linewidth]{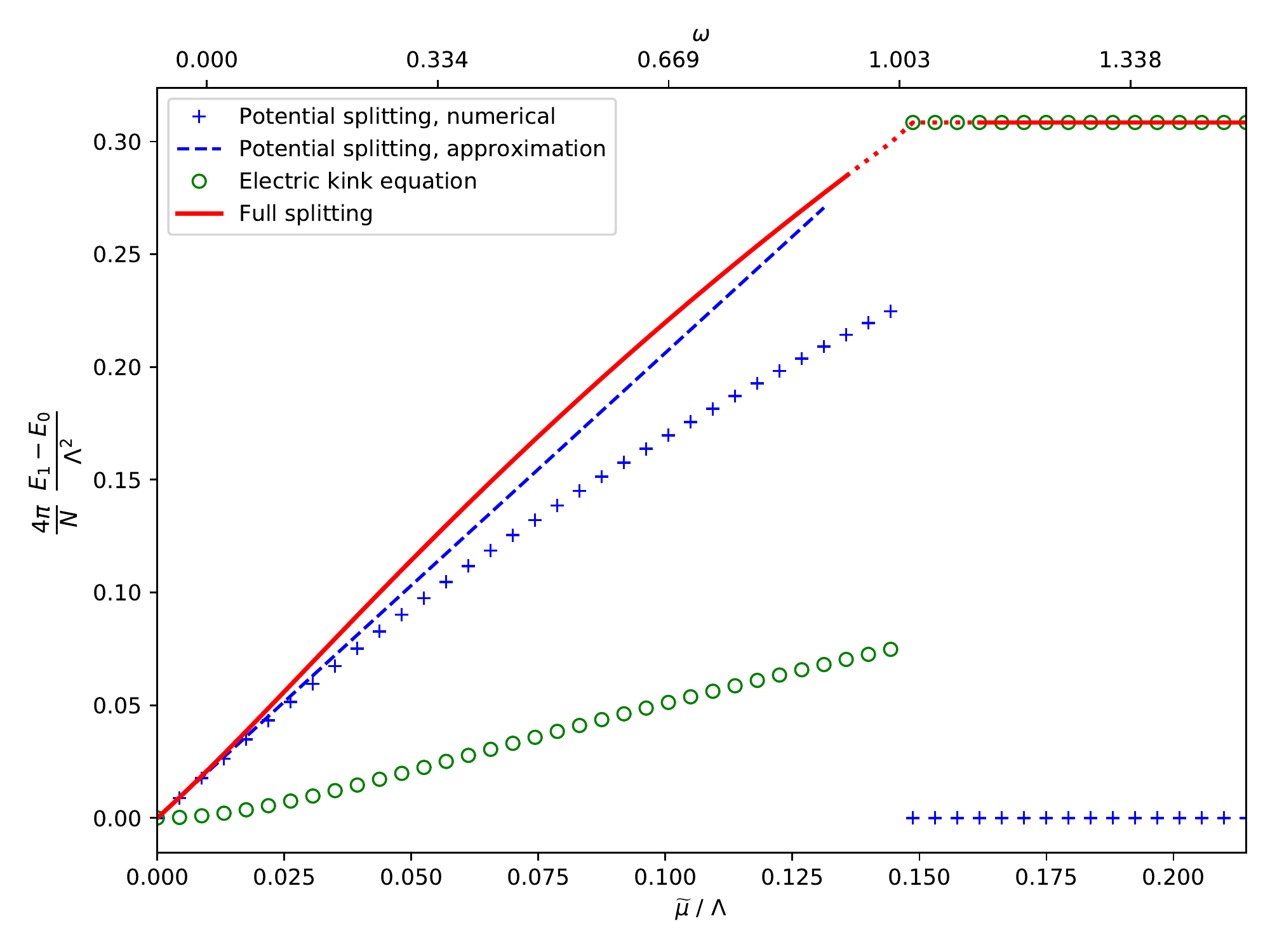}
	\caption{Different contributions to the vacuum energy. Vertical axis is labeled by the rescaled energy splitting $E_1 - E_0$. Values of the deformation parameter $\wt{\mu}$ are on the lower horizontal axis (in the units of $\Lambda$), while the upper horizontal axis represents the parameter $\omega$ \eqref{omega_splitting_parameter}. 
	Green circles denote the contribution from the electric field (solution of \eqref{kink_topological_eq}, given by \eqref{Evac_electical-2} below the phase transition point), \textquote{+} signs represent the splitting from the potential \eqref{V_eff_strong_dm=0} (the blue dashed line is the approximation \eqref{splitting_1-0}). The solid red line is the sum of these two contributions.	
	Phase transition occurs at $\omega \approx 1$ where the full energy displays a discontinuity of the first derivative. Our model does not allow us to obtain exact results in the vicinity of the phase transition point, and we have to extrapolate from the left and from the right (red dotted line continuing the solid red curve).}
\label{fig:vac_splittings} 
\end{figure}

As we learned so far, the vacuum energy (or, rather, energy splitting between the ground state and the first quasivacuum) has two contributions, which depend on the parameter
\begin{equation}
	\omega = \frac{\lambda (\wt{\mu})}{\Lambda} \, \ln\frac{m_G}{\Lambda}
\label{omega_splitting_parameter}	
\end{equation}

The first contribution is the splitting of different quasiminima  $\sigma_i$ of the effective potential \eqref{splitting_1-0}. When we turn on $\omega$ (i.e. supersymmetry breaking parameter $\wt{\mu}$), this contribution at first grows linearly with $\omega$, and then drops to zero when the $\sigma$-quasiminima  disappear.

The second contribution comes from the electric field of charged kinks interpolating between the quasivacua, see \eqref{Evac_electical-1} and \eqref{Evac_electical-2}. This contribution at first grows as $\omega^2$, and at the point when the first $\sigma$ quasivacuum disappears, electric field jumps up\footnotemark  to saturate \eqref{kink_topological_eq}. 
\footnotetext{This jumping is not seen from the propagator considerations \eqref{Evac_electical-1} since it holds only perturbatively near the true vacuum and does not take into account the presence of $\sigma$-quasivacua.
}

The jumping point is the same for these two contributions and it is where a phase transition occurs. Corresponding critical value is $\omega_c \sim 1$, i.e. (cf. \eqref{strong1_strong2_trans_arg-tau} and\eqref{tau_0=tau_1})
\begin{equation}
	\lambda_\text{crit} = \lambda (\wt{\mu}_\text{crit}) \sim \frac{\Lambda}{\ln \frac{m_G^2}{\Lambda^2}} \,.
\label{strong1_strong2_trans_electric}	
\end{equation}

Full vacuum energy is the sum of these two contributions, and on general grounds we expect that it does not jump. Rather, its first derivative is discontinuous, and the phase transition must be of the second order. Numerical calculations confirm this, see Fig.~\ref{fig:vac_splittings}. At the point where the quasivacuum disappears, the two contributions to the vacuum energy jump, and the magnitudes are just right for the total sum to stay continuous.
However, we must point out that we do not have enough accuracy for the detailed study of the vicinity of the transition point. The point is that we can trust our formula for the $\arg\tau$ potential \eqref{DeltaV}  only in the vicinities of the minima $\eqref{vac_tau_susy_dm=0}$, and we do not know the exact form of this potential in regions between any of two adjacent minima.

At small deformations, the main contribution to the vacuum energy is $\sigma$-quasivacua splitting \eqref{splitting_1-0}. After the transition point, vacuum energy is determined solely by the kink electric field. As we reviewed in Sec.~\ref{sec:no-susy} it is the electric field that is responsible for the quasivacuum energy splittings in the non-supersymmetric \CP model. This is consistent with our results, since at large $\wt{\mu}$ above  the phase transition point our model flows to the non-supersymmetric \CP model.

To conclude this section we note that  parameter $\omega$ relevant for the quasivacua splitting  is enhanced by the large logarithm $\ln{m_G/\Lambda \gg 1}$. Hence the phase transition point occurs   at  $\wt{\mu}_c \sim \lambda _c$ given by
\eqref{strong1_strong2_trans_electric}, much smaller than $\wt{\mu} \sim \Lambda$. These are even smaller values of 
$\wt{\mu}$ as compared to $m_G$ since we assume $m_G\gg \Lambda$ in order to keep the bulk theory at weak coupling.
At these small values of $\wt{\mu}$ we are way below the scale of adjoint matter decoupling in the bulk theory 
which occurs at $\wt{\mu} \gg m_G$. In particular, the scale $\Lambda $ of the world sheet theory is close to $\Lambda_{4d}$ rather than
to its large-$\wt{\mu}$ asymptotic values \eqref{Lam_2d}.

\subsection{Large deformations \label{sec:strong_dm=0_largelam}}

\begin{figure}[ht]
    \centering
    \begin{subfigure}[t]{0.5\textwidth}
        \centering
        \includegraphics[width=\textwidth]{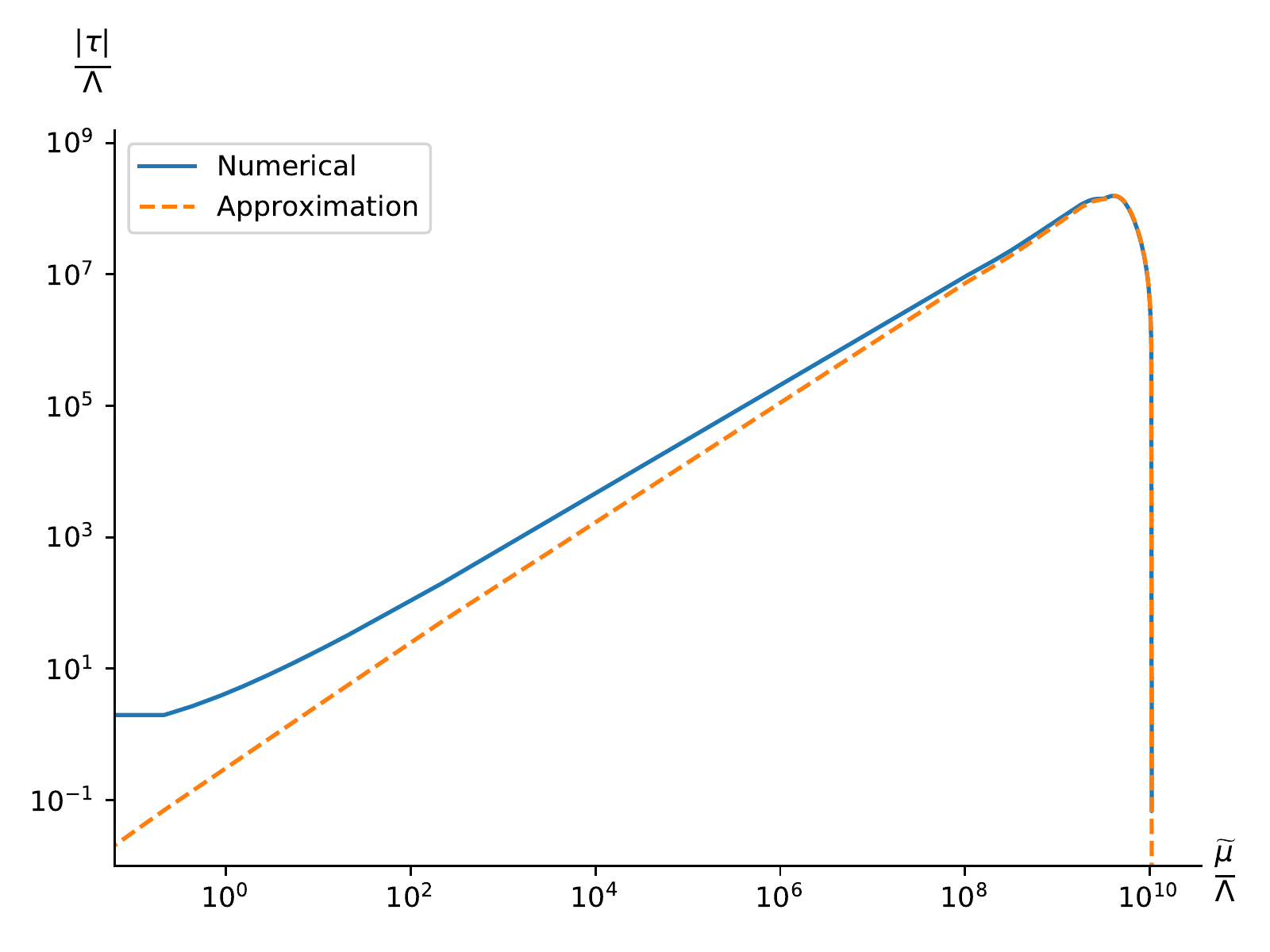}
        \caption{VEV of $\tau$ as a function of $\wt{\mu}$, double log scale. 
        	}
        \label{dm=0_large-lam_tau}  
    \end{subfigure}%
    ~ 
    \begin{subfigure}[t]{0.5\textwidth}
        \centering
        \includegraphics[width=\textwidth]{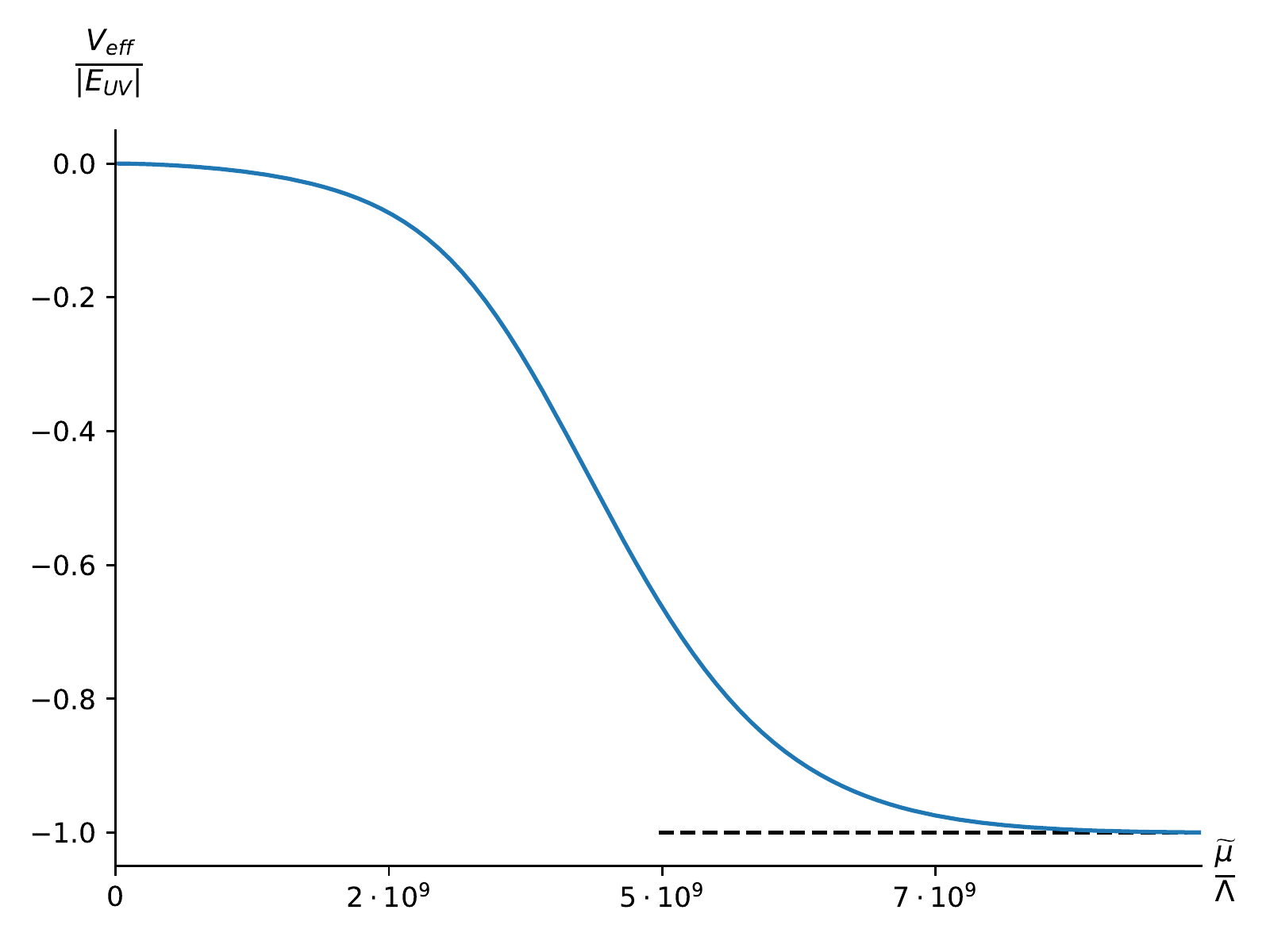}
        \caption{Vacuum energy as a function of $\wt{\mu}$, log scale}
	\label{dm=0_large-lam_Veff}          
    \end{subfigure}
    \caption{Numerical results for the VEV of $\tau$ and vacuum energy at large deformations $\lambda \gg \Lambda_{2d}$. 
    	On the figure \subref{dm=0_large-lam_tau} we have VEV of $\tau$. Dashed line shows the approximate solution \eqref{tau_approx}, while the solid line is the result of numerics. One can see that $\tau$ indeed vanishes at $\lambda(\wt{\mu}) = m_G$. 
    	On \subref{dm=0_large-lam_Veff} we have $E_\text{vac}$.  Dashed line shows its asymptotic value $E_\text{UV}$ given by \eqref{Evac_strong_largemu}.
    	In the numerical procedure we had set $m_G / \Lambda = 10^{10}$}
\label{dm=0_large-lam}    
\end{figure}

As we increase the  deformation parameter $\wt{\mu}$ , the fermion mass $\lambda$  approaches the UV cutoff scale $ m_G$ and  we can expect that the fermions become very heavy and decouple, effectively taking no part in the dynamics. Therefore, our theory should become the non-supersymmetric \CP model \eqref{cpn_lagr_simplest}. VEV of $\tau$ field should become zero.

  We can check this directly using our effective potential \eqref{V_eff_strong_dm=0}. Indeed, assume that $\tau \ll \lambda \sim m_G$. Then we can expand  \eqref{V_eff_strong_dm=0} to obtain
\begin{equation}
\begin{aligned}
	{\mathcal  V}_\text{eff} 
			&= \frac{N}{4\pi} iD\left[ 1 - \ln\frac{iD + \bigl| \tau \bigr|^2}{\Lambda^2} \right] 
			+ \frac{N}{4\pi} \bigl| \tau \bigr|^2 \left[ 1 -  \ln\frac{iD  + \bigl| \tau \bigr|^2}{m_G^2} \right] \\
			&-   \frac{N}{4\pi} \cdot 2  \Re\tau \cdot \lambda  \, \ln\frac{\lambda^2}{m_G^2}  	
			-  \frac{N}{4\pi} \lambda^2 \left( 1 - \ln\frac{\lambda^2}{m_G^2} \right) 
\end{aligned}
\label{V_eff_strong_dm=0_large-lam}	
\end{equation}
%
%
%
%
Minimizing this potential we obtain
\begin{equation}
	\tau \approx -\lambda \frac{\ln(m_G/\lambda)}{\ln (m_G/\Lambda) } \,.
\label{tau_approx}	
\end{equation}
This formula turns out to be pretty good compared to the exact numerical solution, see Fig.~\ref{dm=0_large-lam_tau}.
As $\lambda$ approaches the UV cutoff scale $m_G$, the VEV of $\tau$ vanishes. The  first term in \eqref{V_eff_strong_dm=0_large-lam} reduces to the effective potential for the non-supersymmetric \CP model, while the last term gives a vacuum energy shift.  At $\lambda = m_G$, the vacuum energy is
\begin{equation}
	E_\text{vac, UV} = \frac{N}{4\pi} \left(\Lambda^2 - m_G^2 \right) \,.
\label{Evac_strong_largemu}
\end{equation}
This is in agreement with the Appelquist-Carazzone decoupling theorem \cite{Appelquist:1974tg}, which states that the effect of heavy fields is limited to the renormalization of physical quantities. Note that since the supersymmetry is explicitly 
broken in the world sheet theory by fermion masses the vacuum energy is not positively defined.

The vacuum energy  above is a quantum correction to the classical expression for the non-Abelian string tension in the bulk theory. The latter was derived in Chapter~\ref{sec:none}, see \eqref{ten}. 
Together with \eqref{Evac_strong_largemu} it can be written as 
\begin{equation}
	T = \frac{2\pi}{\ln\frac{m_G^2}{m^2}} \frac{m_G^2}{g^2} + \frac{N}{4\pi} \left(\Lambda^2 - m_G^2\right),
\label{tension_strong}	
\end{equation}
 We see that the second term here is just an $O(g^2)$ correction to  the classical formula.

At intermediate values of $\lambda$ we were able to study this model only numerically. The results are presented on Fig.~\eqref{dm=0_large-lam}. They show the dependence of $\langle \sigma \rangle$ and $E_\text{vac}$ on the heavy fermion mass $\lambda$. 
One can see that indeed the VEV of $\tau$ vanishes at very large $\lambda$.
Note that we will have $\langle iD \rangle < 0$ in a wide range of $\lambda$, but this does not lead to an instability because, according to \eqref{master1_dm=0}, the mass of the $n$ field is always positive.

%
%

\subsection{Split mass case \label{sec:strong_split_mass}}

\begin{figure}[h!]
	\centering
	\includegraphics[width=0.6\linewidth]{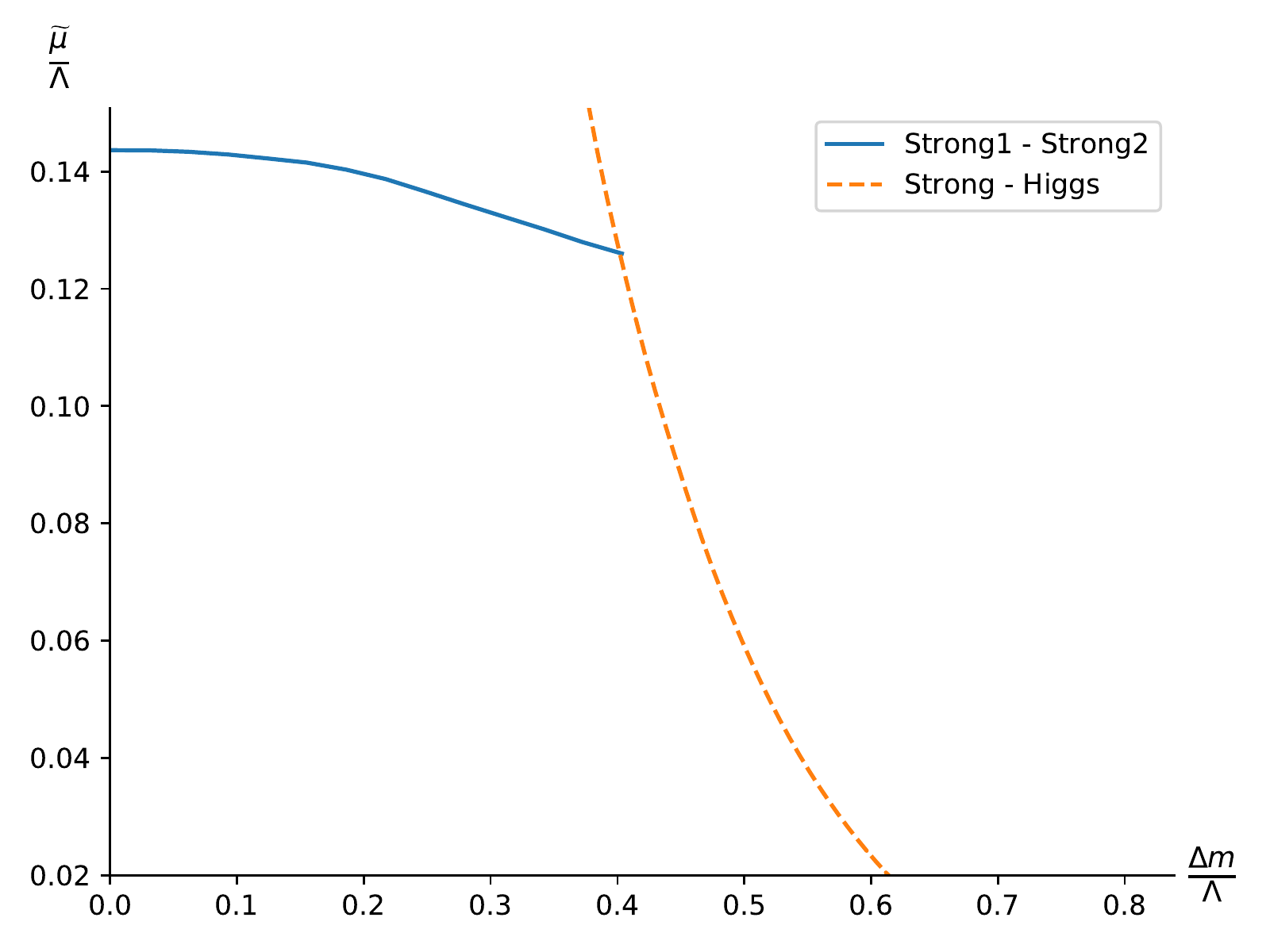}
	\caption{
		Phase transition line between two strong coupling regimes (shown in solid blue). The dashed line is the phase transition line between the Strong coupling and Higgs regimes, see Sec.~\ref{sec:trans_strong-higgs}. This plot is a result of numerical calculations for $N = 16$.
	}
\label{fig:strong1_strong2} 
\end{figure}
The results obtained in the previous section can be generalized to the case $\Delta m_{i0} \neq 0$. Consider the masses on a circle \eqref{masses_ZN}, with the radius $\Delta m$ as the mass scale of our model.

If we fix some $\Delta m$ and start increasing $\wt{\mu}$ (and, therefore, $\lambda (\wt{\mu})$), our model exhibits similar behavior as in the case $\Delta m = 0$. At $\wt{\mu} = 0$ the supersymmetry is unbroken, and there are $N$ degenerate vacua. When we switch on the deformation, the degeneracy is lifted, and eventually all lifted quasivacua decay, which signifies a phase transition.
The set of the phase transition points represents a curve on the $(\mu, \Delta m)$ plane, see Fig.~\ref{fig:strong1_strong2} 

Qualitatively, we see nothing new. However, when $\Delta m$ is large enough, the theory goes through the phase transition from the strong coupling phase into a weak coupling phase, so-called \textquote{Higgs} phase. This will be the subject of the next section.

%
%

\section{Higgs regime \label{sec:higgs}}


When the mass difference $\Delta m$ exceeds   some critical value, the theory appears in the Higgs phase. This phase is characterized by a nonzero VEV of $n$. At very weak coupling, we can use the classical Lagrangian \eqref{lagrangian_init} to find the vacuum solution,
\begin{equation}
	n_0^2 = 2 \beta \,, \quad
	\sqrt{2} \sigma = m_0 \,, \quad
	iD = 0 \,.
\label{higgs_classical}	
\end{equation}
The vacuum energy is classically zero.

In the supersymmetric case $\wt{\mu} = 0$ the solution for $\sigma$  
is exact at large $N$. Moreover, at very large $\Delta m$ the coupling constant $1/\beta$ is small (frozen at the scale $\Delta m$)
and quantum corrections to the classical vacuum solution \eqref{higgs_classical} are small.

However, at nonzero $\wt{\mu}$ and for $\Delta m \gtrsim \Lambda$, things become more complicated, as we can no longer rely on the classical equations. Generally speaking, solution \eqref{higgs_classical} receives $\Lambda /\Delta m $ and $\wt{\mu} / \Lambda$ corrections. We have to work with the quantum equations \eqref{master1} - \eqref{master3}, and most of the results presented in this section were obtained from numerical calculations%
\footnote{In order to do the computations required the author of this thesis had to develop some numerical routines which are more robust than the standard ones. Some of them can now be found at \url{https://github.com/ievlev9292/num-robust}.}.

\begin{figure}[h]
	\centering
	\includegraphics[width=0.6\linewidth]{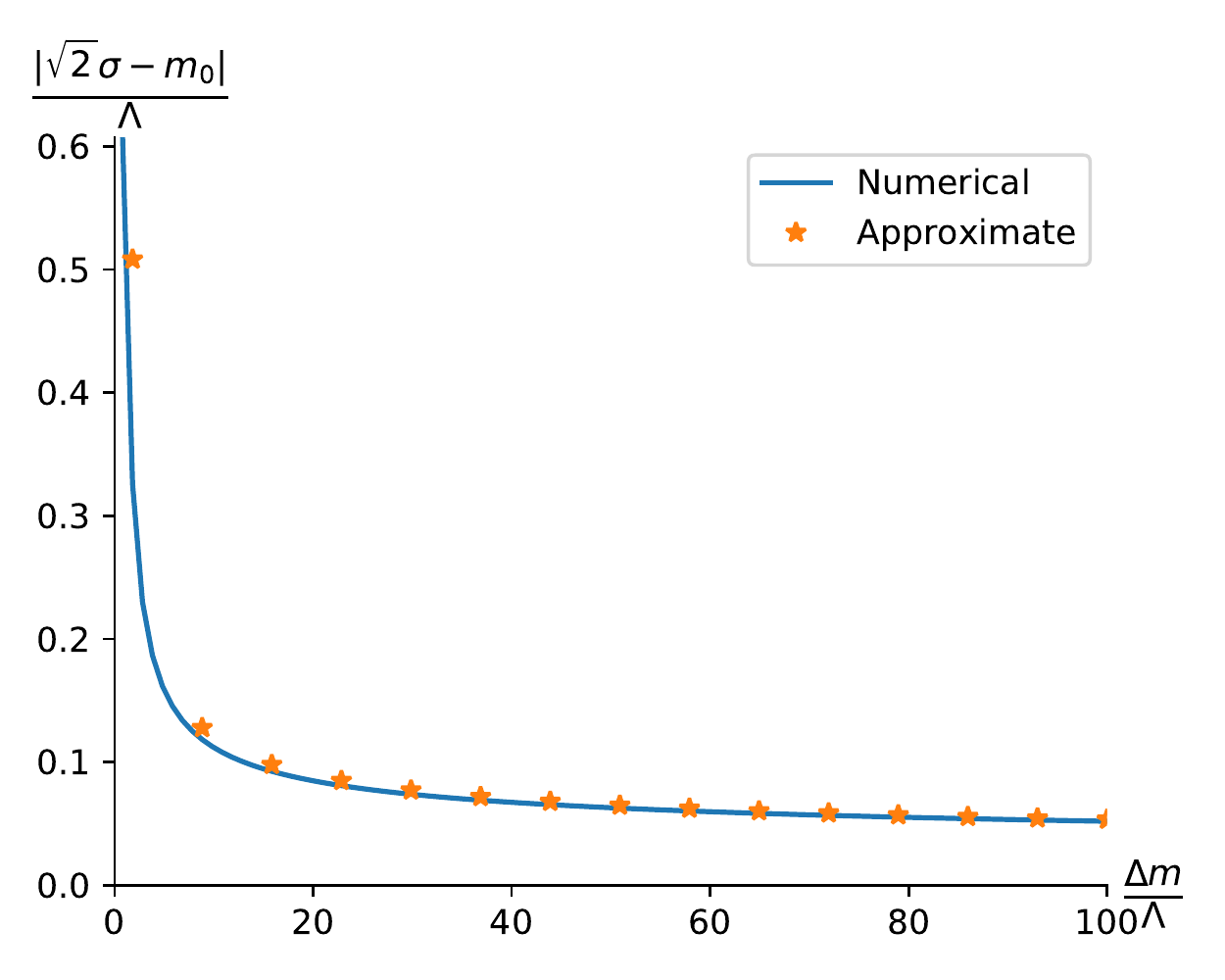}
	\caption{
		VEV of $\tau = \sqrt{2}\sigma - m_0$ as a function of $\Delta m$. 
		Solid line is the exact result of numerical calculation, while stars represent the approximate formula \eqref{higgs_tau_largedM}.
		Here $\wt{\mu} = \Lambda$. In numerical calculations we used $N = 16$.
		One can see that indeed, as $\Delta m$ grows, the VEV of $\sqrt{2}\sigma$ goes to its classical value $m_0$.
	}
\label{fig:quasiclassical_tauvev} 
\end{figure}
First of all, we wish to check that the one loop potential that we derived \eqref{v_eff} is compatible with the classical limit. Consider the limit of large $\Delta m \gg \Lambda$ with some $\wt{\mu}$ fixed. We can expand the vacuum equations \eqref{master1} - \eqref{master3} in powers of $\Lambda /\Delta m$ and easily derive an approximate solution for the ground state VEV
\begin{equation}
	\sqrt{2}\sigma - m_0 \approx - \lambda(\wt{\mu}) \, \frac{\ln \frac{m_G}{\Delta m}}{\ln\frac{\Delta m}{\Lambda}} \,.
\label{higgs_tau_largedM}	
\end{equation}

Fig.~\ref{fig:quasiclassical_tauvev} presents our results for the VEV of $\sigma$. One can see that the formula \eqref{higgs_tau_largedM} gives very good approximation (see also Fig.~\ref{fig:nokinks_trans_single_tau}). At large $\Delta m$ we indeed have 
$\sqrt{2} \sigma  \approx m_0$.

\subsection{Quasivacua \label{sec:quasivacua_higgs}}

%
\begin{figure}[h]
    \centering
    \begin{subfigure}[t]{0.5\textwidth}
		\centering
		\includegraphics[width=\linewidth]{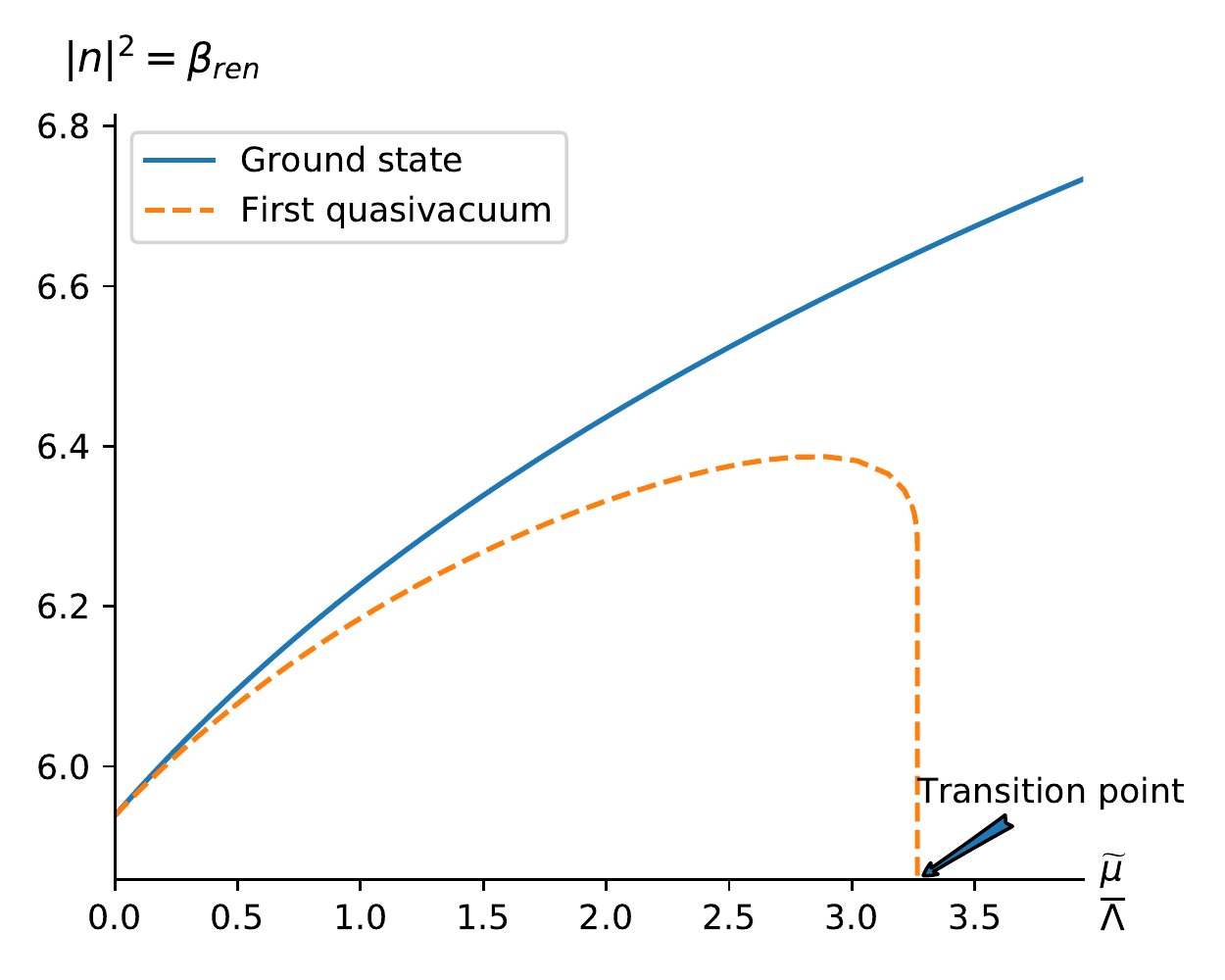}
		\caption{$|n|^2$}
	\label{fig:nokinks_trans_single_nsq} 
    \end{subfigure}%
    ~ 
    \begin{subfigure}[t]{0.5\textwidth}
		\centering
		\includegraphics[width=\linewidth]{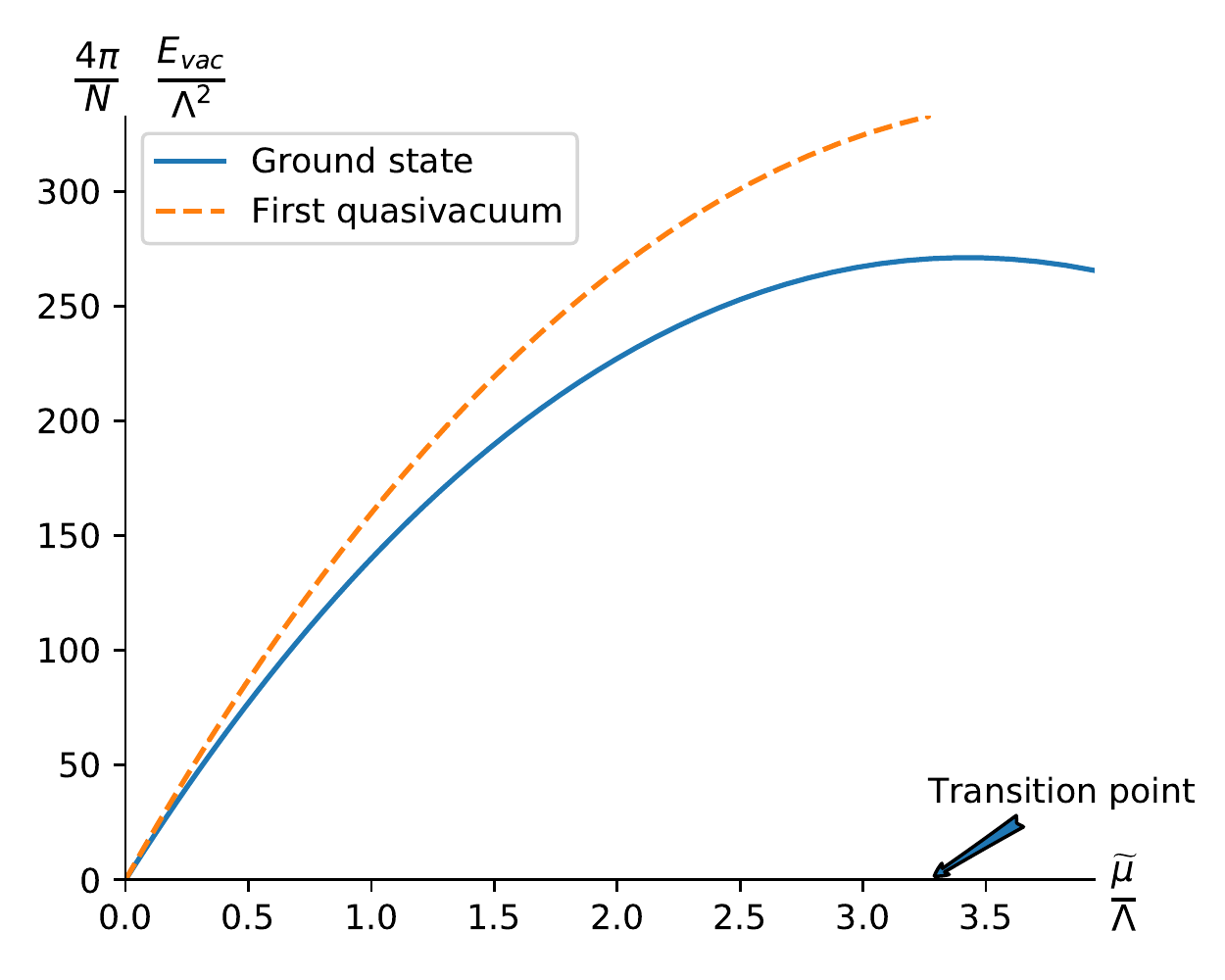}
		\caption{(Quasi)vacuum energy}
	\label{fig:nokinks_trans_single_Veff} 
    \end{subfigure}%
\caption{
	Example of Kinks-NoKinks phase transition for $\Delta m / \Lambda = 10$. Blue solid line refers to the true ground state $i_0 = 0$, orange dashed line represents the first quasivacuum $i_0 = 1$. Value of the deformation parameter $\wt{\mu}$ is on the horizontal axis (in the units of $\Lambda$), the phase transition point is indicated by an arrow.
	Both figures are the result of numerical calculations at $\Delta m / \Lambda = 10$, $N = 16$
}
\label{fig:nokinks_trans_single} 
\end{figure}
%
%
\begin{figure}[h]
    \centering
    \begin{subfigure}[t]{0.5\textwidth}
        \centering
        \includegraphics[width=\textwidth]{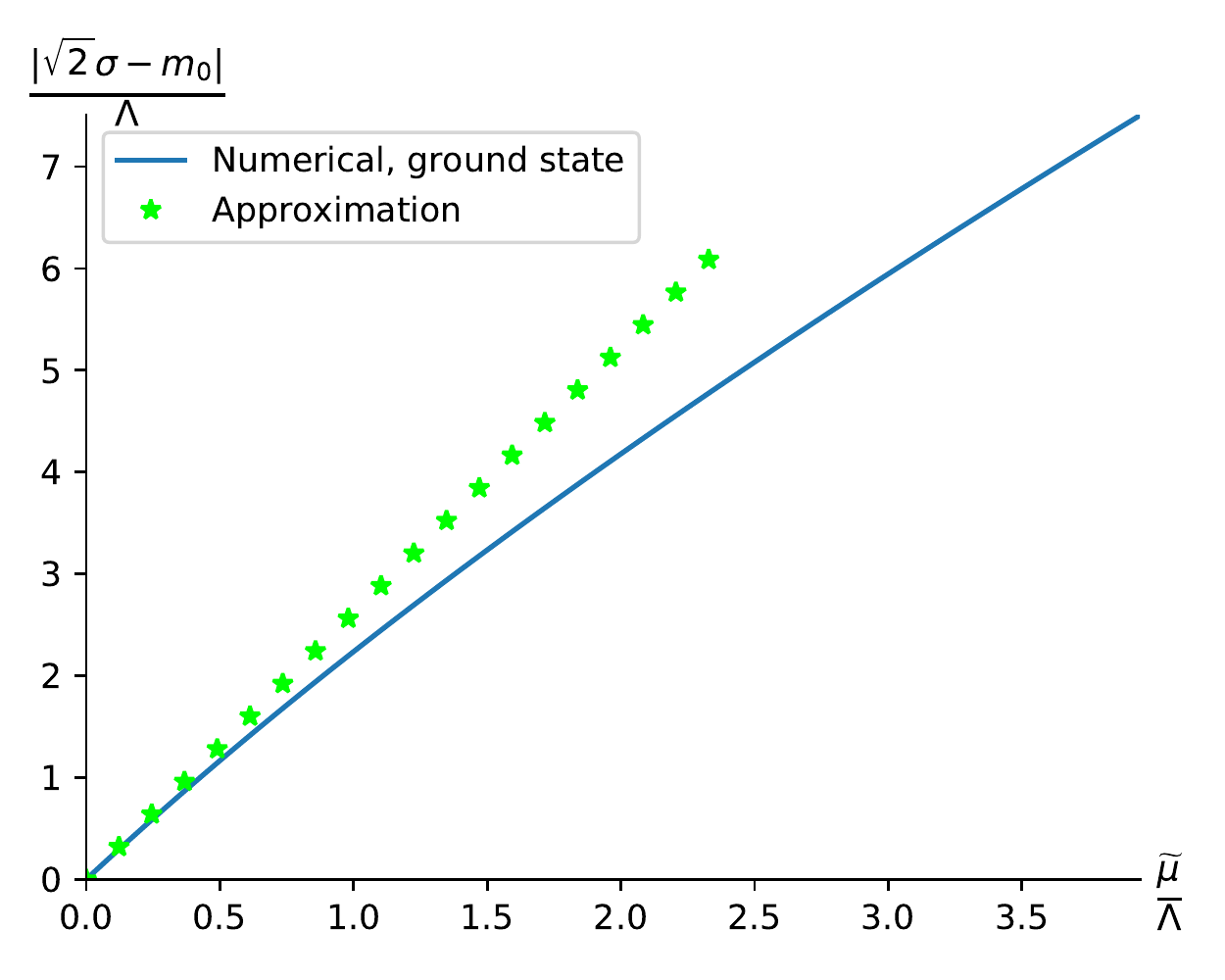}
        \caption{
        	$|\sqrt{2} \sigma - m_0|$ at small $\wt{\mu}$.
        }
    \label{fig:nokinks_trans_single_tau} 
    \end{subfigure}%
    ~ 
    \begin{subfigure}[t]{0.5\textwidth}
        \centering
        \includegraphics[width=\textwidth]{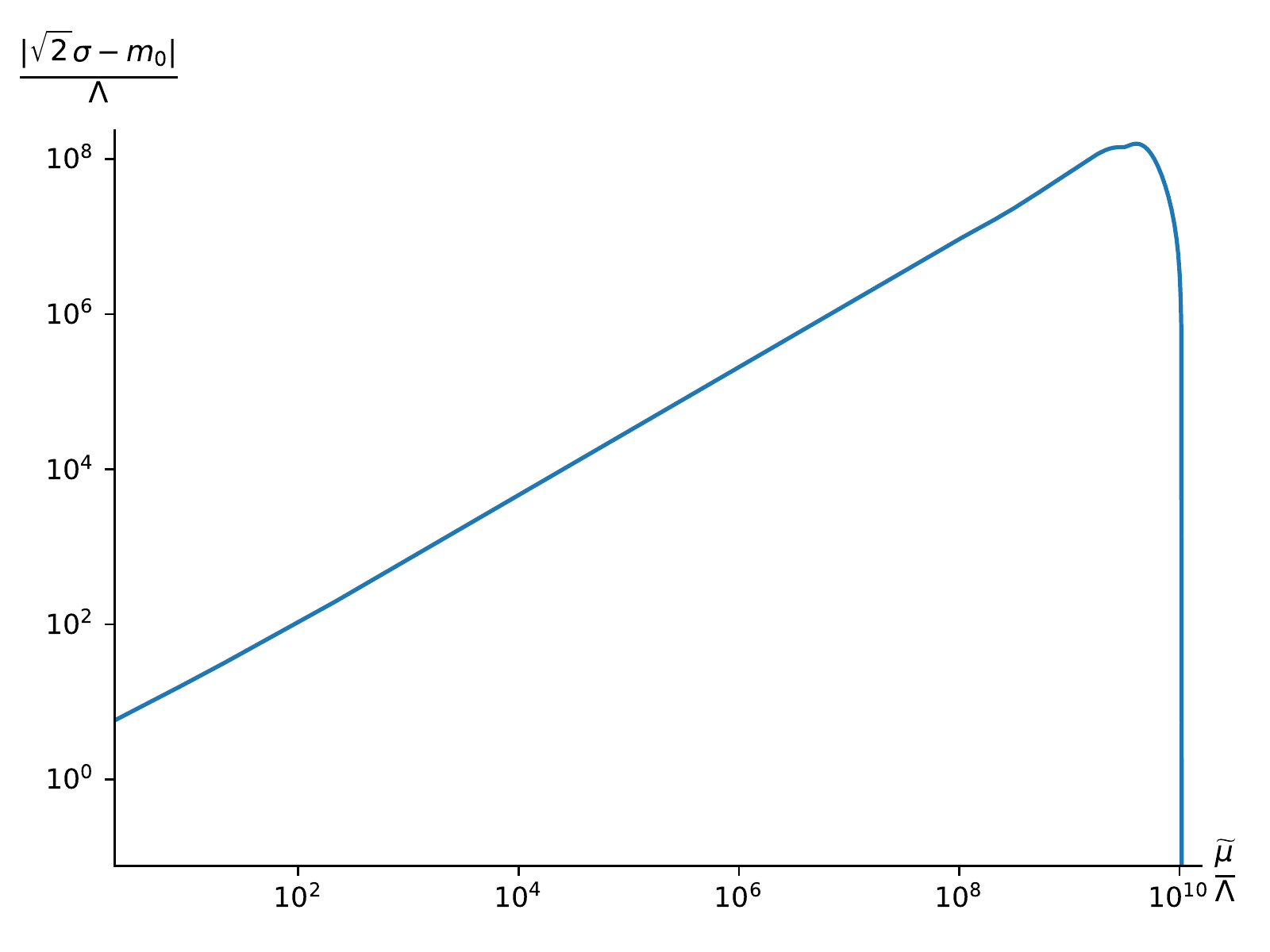}
        \caption{
        	$|\sqrt{2} \sigma - m_0|$ at large $\wt{\mu}$.
        }
    \label{fig:higgs_large_mu} 
    \end{subfigure}%
\caption{
	VEV of $\sqrt{2} \sigma - m_0$ at different scales.
	Figure \subref{fig:nokinks_trans_single_tau} shows small $\wt{\mu}$. Solid blue line is the result of numerical calculations, green stars show the approximate formula \eqref{higgs_tau_largedM}.
	Figure \subref{fig:higgs_large_mu} shows large-$\wt{\mu}$ behavior (in double log scale). One can see that as $\wt{\mu} \to m_G$ we indeed have $\sqrt{2} \langle \sigma \rangle \to m_0$.	
	The plots were made for fixed $\Delta m / \Lambda = 10$, $m_G / \Lambda = 10^{10}$, $N = 16$
}
\label{fig:higgs_tau_VEV} 
\end{figure}
\begin{figure}[h]
    \centering
    \begin{subfigure}[t]{0.5\textwidth}
        \centering
        \includegraphics[width=\textwidth]{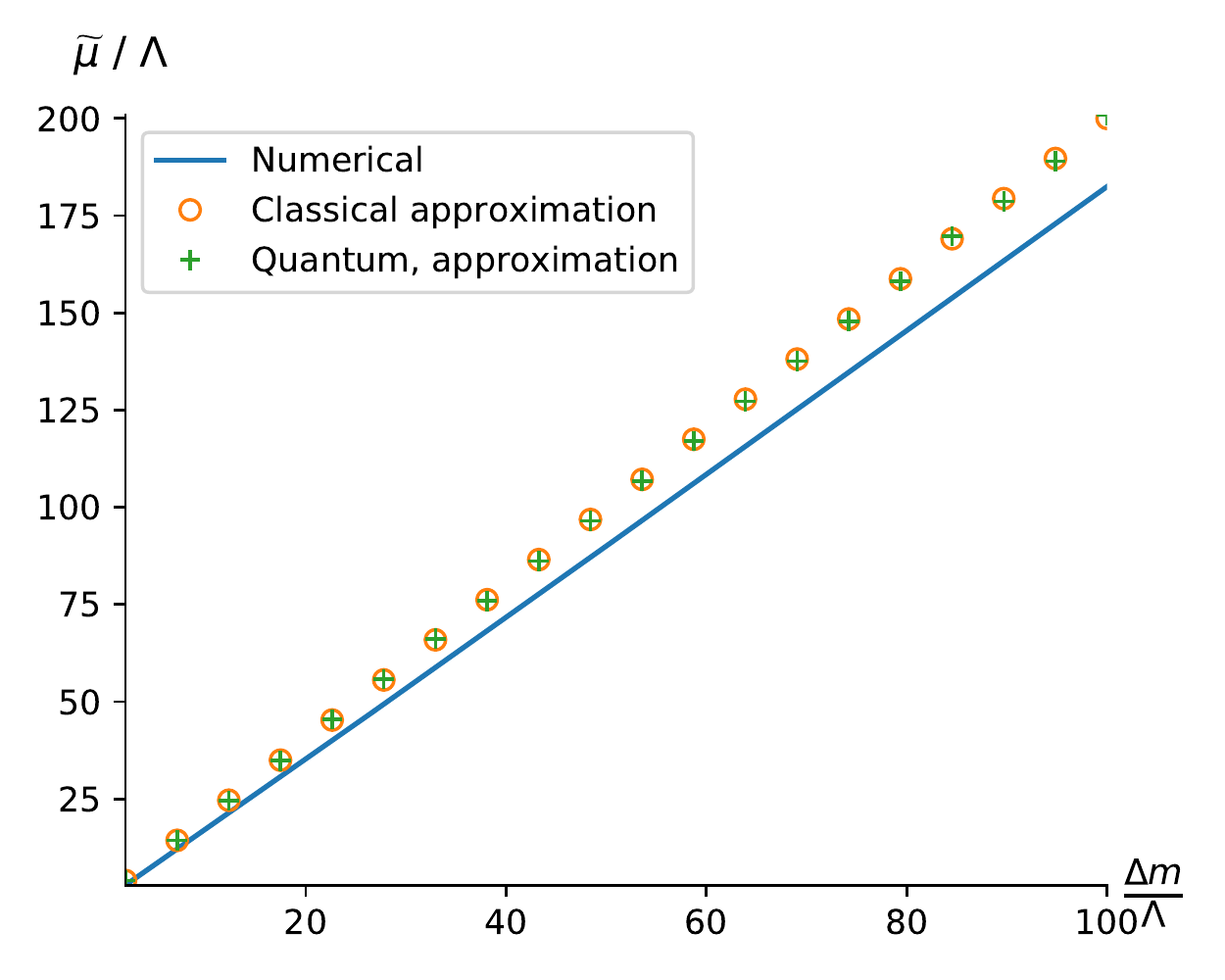}
        \caption{Agreement with the classical formula \eqref{nokink_trans_classic} if we set $\lambda = 0$}
        \label{fig:nokinks_trans_nolam}
    \end{subfigure}%
    ~ 
    \begin{subfigure}[t]{0.5\textwidth}
        \centering
        \includegraphics[width=\textwidth]{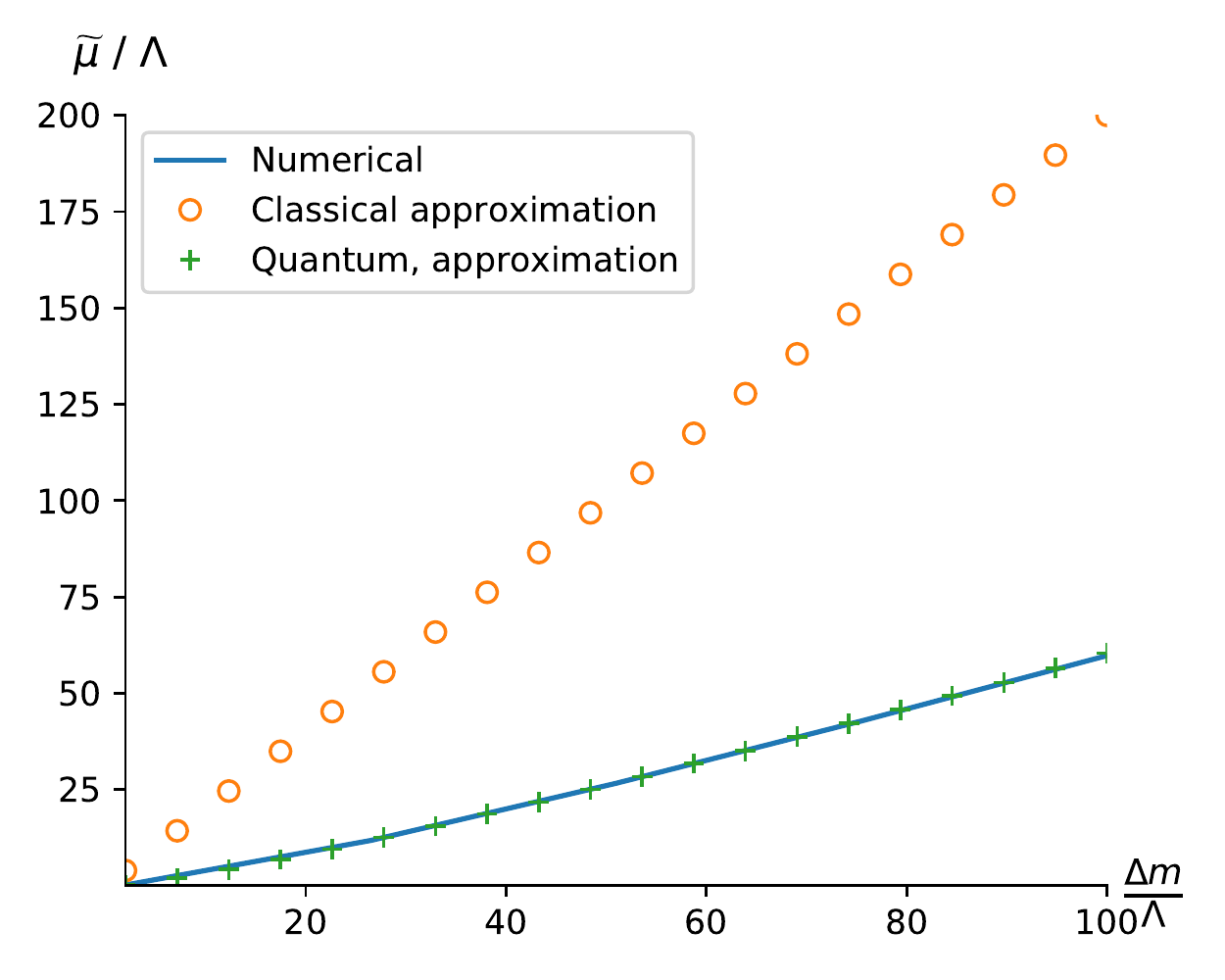}
        \caption{Presence of $\lambda$ magnifies the effect}
        \label{fig:nokinks_trans_lam}
    \end{subfigure}%
\caption{
	Kinks-NoKinks phase transition line. $\Delta m$ on the horizontal axis, $\wt{\mu}$ on the vertical axis. Solid blue line is the result of numerical calculation of the curve where all quasivacua have decayed leaving single true ground state.
	Orange circles represent the classical formula \eqref{nokink_trans_classic}, green \textquote{+} are the quantum approximation \eqref{nokink_trans_quant}.
	Figure \subref{fig:nokinks_trans_nolam} shows that if we set $\wt{\lambda}_0 = 0$ we indeed get good agreement with the classical formula \eqref{nokink_trans_classic}. However, the real scenario (figure \subref{fig:nokinks_trans_lam}) is better described by formula \eqref{nokink_trans_quant}
}
\label{fig:nokinks_trans} 
\end{figure}

Solution \eqref{higgs_classical} is just one of the possible vacuum states in the Higgs phase. In the supersymmetric case $\wt{\mu} = 0$ there are  $N$ degenerate vacua as dictated by Witten index. In \cite{Bolokhov:2010hv} it was shown that the theory at large $\Delta m$ is in the Higgs phase where different components of the $N$-plet $n_i$ develop a VEV. These vacua are characterized by
\begin{equation}
	\langle \sqrt{2} \sigma \rangle = m_{i_0}, \qquad \langle|n_{i_0}|^2\rangle = 2 \beta \,,\quad i_0 = 0 \,,\, \ldots \,, N-1
\end{equation}
Moreover, there are kinks interpolating between these vacua.

As we switch on the deformation parameter $\wt{\mu}$, these vacua split, and at small $\wt{\mu}$ we have one true ground state 
\eqref{higgs_classical} and $N-1$ quasivacua. Let us first consider this picture from the classical Lagrangian \eqref{lagrangian_init}. The classical  potential is
\begin{equation}
	{\mathcal V_\text{cl}}(n, \sigma, D) =  i\,D\left(\bar{n}_i n^i -2\beta \right)
		+  \sum_i\left|\sqrt 2\sigma-m_i\right|^2\, |n^i|^2	+ \upsilon (\mu) \sum_i \Re\Delta m_{i0} |n^i|^2
\label{classic_potential}
\end{equation}
Let us derive the mass spectrum in the vicinity of a vacuum $\sqrt{2}\sigma = m_{i_0}$ for some $i_0$. Then $n^i,\ i \neq i_0 $ are small, while
\begin{equation}
	n_{i_0} = \sqrt{2\beta} + \delta n_{i_0}
\end{equation}
From the $D$-term condition
\begin{equation}
	\delta n_{i_0} \approx - \frac{1}{2 \cdot 2\beta} \sum_{i \neq i_0} |n^i|^2
\end{equation}
and the potential \eqref{classic_potential} becomes
\begin{equation}
\begin{aligned}
	{\mathcal V_\text{cl}} &\approx 
		\sum_{i \neq i_0} |m_i - m_{i_0}|^2 \, |n^i|^2 
		+ \upsilon (\mu) \sum_{i \neq i_0} \Re(m_i - m_0) |n^i|^2 \\
		&\phantom{\sum_{i \neq i_0} |m_i - m_{i_0}|^2 \, |n^i|^2 
				+ \upsilon (\mu)} - \upsilon (\mu) \Re(m_{i_0} - m_0) \sum_{i \neq i_0} |n^i|^2 
	\\
	&= \sum_{i \neq i_0} |n^i|^2 \left[ |m_i - m_{i_0}|^2 + \upsilon (\mu) \Re(m_i - m_{i_0}) \right]
\end{aligned}	
\label{classic_potential_masses}	
\end{equation}
so that the mass of the $n^i$ particle is
\begin{equation}
	M_i^2 = |m_i - m_{i_0}|^2 + \upsilon (\mu) \Re(m_i - m_{i_0})
\end{equation}
If $M_i^2$ were to turn negative for some $i$, this would signify that the vacuum under consideration is unstable. This  happens for all $i_0\neq 0$ if the deformation is large enough because there are always some  $i$ with $\Re(m_i - m_{i_0}) < 0$. 

To be more concrete, consider our choice of the masses \eqref{masses_ZN}. Then for the vacuum $i_0 = 0$ we have $\Re(m_i - m_{0}) > 0$ for all $i \neq 0$, and this vacuum is stable. However, the vacua $0 < i_0 < N/2$ can be shown to become unstable when the deformation parameter hits the critical value
\begin{equation}
	\upsilon (\mu_{\text{crit,} i_0}) 
		= 2 \Delta m\, \frac{1 - \cos(\frac{2 \pi}{N})}{\cos(\frac{2 \pi (i_0 - 1)}{N}) - \cos(\frac{2 \pi i_0}{N})}
		\approx \frac{4 \pi}{N} \, \frac{\Delta m}{\sin(\frac{2 \pi i_0}{N})} \, 
\label{nokink_trans_classic}		
\end{equation}
The last step is the large $N$ approximation. Similar statement holds for the quasivacua $N/2 < i_0 < N$, while the quasivacuum number $i_0 = N/2$ (for even $N$) 
decays when $\upsilon (\mu_{\text{crit,} N/2}) = 2 \, \Delta m$.
When $\wt{\mu}$ is above this critical value, the theory has unique vacuum, and there are no kinks left.

These quasivacua are seen from the one loop potential as well. Following \cite{Bolokhov:2010hv}, we can study these quasivacua as follows. Recall that deriving the effective potential \eqref{v_eff} we assumed that $n\equiv n_0$ can develop a VEV.
Now to study quasivacua we assume that $n_{i_0}$ is non-zero  and integrate out the other components of $n^i$.
Numerical calculation show that the resulting effective potential has a minimum for small deformations, but this minimum fades away at large $\wt{\mu}$, see Fig.~\ref{fig:nokinks_trans_single}. On the plot \ref{fig:nokinks_trans_single_nsq}, this corresponds to the fact that $|n|^2$ rapidly drops near the phase transition point. Fig.~\ref{fig:nokinks_trans_single_Veff} shows that the quasivacua are degenerate when supersymmetry is unbroken, and that the quasivacuum energy is indeed higher than that of the true ground state.

Figure \ref{fig:nokinks_trans} shows the phase transition curve. One can see that the classical formula \eqref{nokink_trans_classic} is valid only in if we set $\lambda = 0$ in \eqref{lagrangian_init}, but it is completely inadequate when the fermions gain extra mass. As we see from Fig. \ref{fig:nokinks_trans_lam} massive fermions magnify the effect. 

Let us derive better theoretical formula for the phase transition curve.
Consider, for example, first quasivacuum $i_0 = 1$. Then in the expression for $\beta_\text{ren}$ \eqref{master1} we will have $\Delta m_{i1} = m_i - m_1$ instead of $\Delta m_{i0}$. Then $\Re\Delta m_{01} < 0$, and for the phase transition point we can take roughly the point when $\beta_\text{ren} \to -\infty$, i.e.
\begin{equation}
	iD + \upsilon(\wt{\mu})\, \Re\Delta m_{01} + |\sqrt{2}\sigma - m_0|^2 = 0.
\end{equation}
Using \eqref{upsilon}, \eqref{lambda}, \eqref{higgs_tau_largedM} and an analog of \eqref{master2} one can show that the phase transition occurs at the point
\begin{equation}
	\wt{\mu}_\text{crit} \approx \frac{2 \, \Delta m}{1 + \wt{\lambda}_0 \frac{\ln {m_G / \Delta m}}{\ln{\Delta m / \Lambda}}} \,.
\label{nokink_trans_quant}
\end{equation}

At very large values of $\wt{\mu}$ all but one vacua have decayed, and the world sheet theory flows to the non-supersymmetric model. In this limit the VEV of $\sqrt{2} \sigma$ is again tends to $m_0$. Indeed, at large $\wt{\mu}$ we can solve the vacuum equations \eqref{master1} - \eqref{master3} approximately, and using the expression for $\Lambda$ \eqref{Lam_2d}, we find that
\begin{equation}
	\sqrt{2} \sigma - m_0 \sim \frac{\Delta m\, m_G^2}{\wt{\mu}^2} \, \ln\frac{\wt{\mu}}{\Delta m}\ln\frac{\wt{\mu}}{ m_G}
\end{equation}
which vanishes at large values of $\wt{\mu}$. This is supported by numerical calculations, see Fig. \ref{fig:higgs_large_mu}.

\subsection{Strong - Higgs phase transition \label{sec:trans_strong-higgs}}

%
\begin{figure}[h]
    \centering
    \begin{subfigure}[t]{0.5\textwidth}
		\centering
		\includegraphics[width=\linewidth]{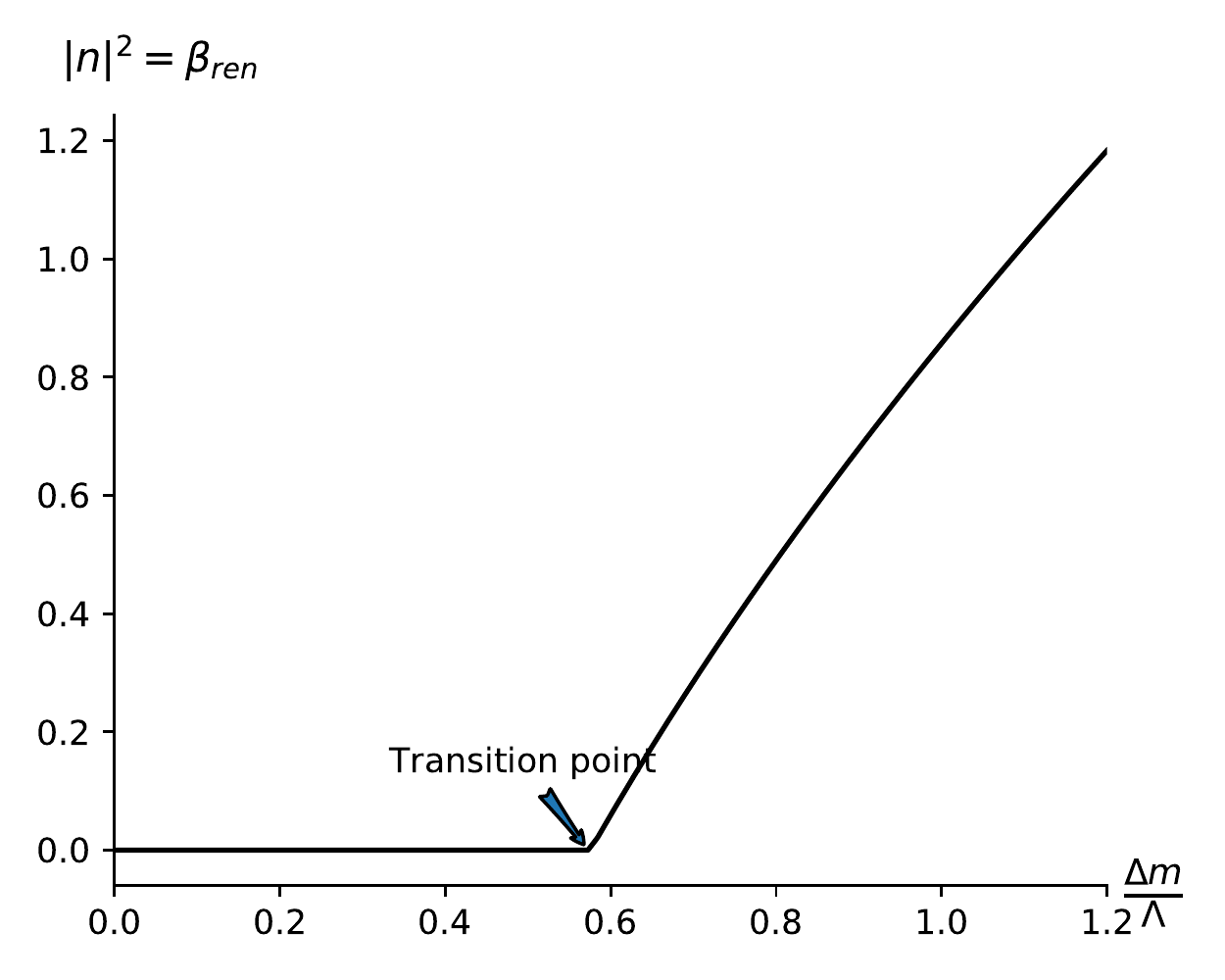}
		\caption{$|n|^2$}
	\label{fig:strong-higgs_single_nsq} 
    \end{subfigure}%
    ~ 
    \begin{subfigure}[t]{0.5\textwidth}
		\centering
		\includegraphics[width=\linewidth]{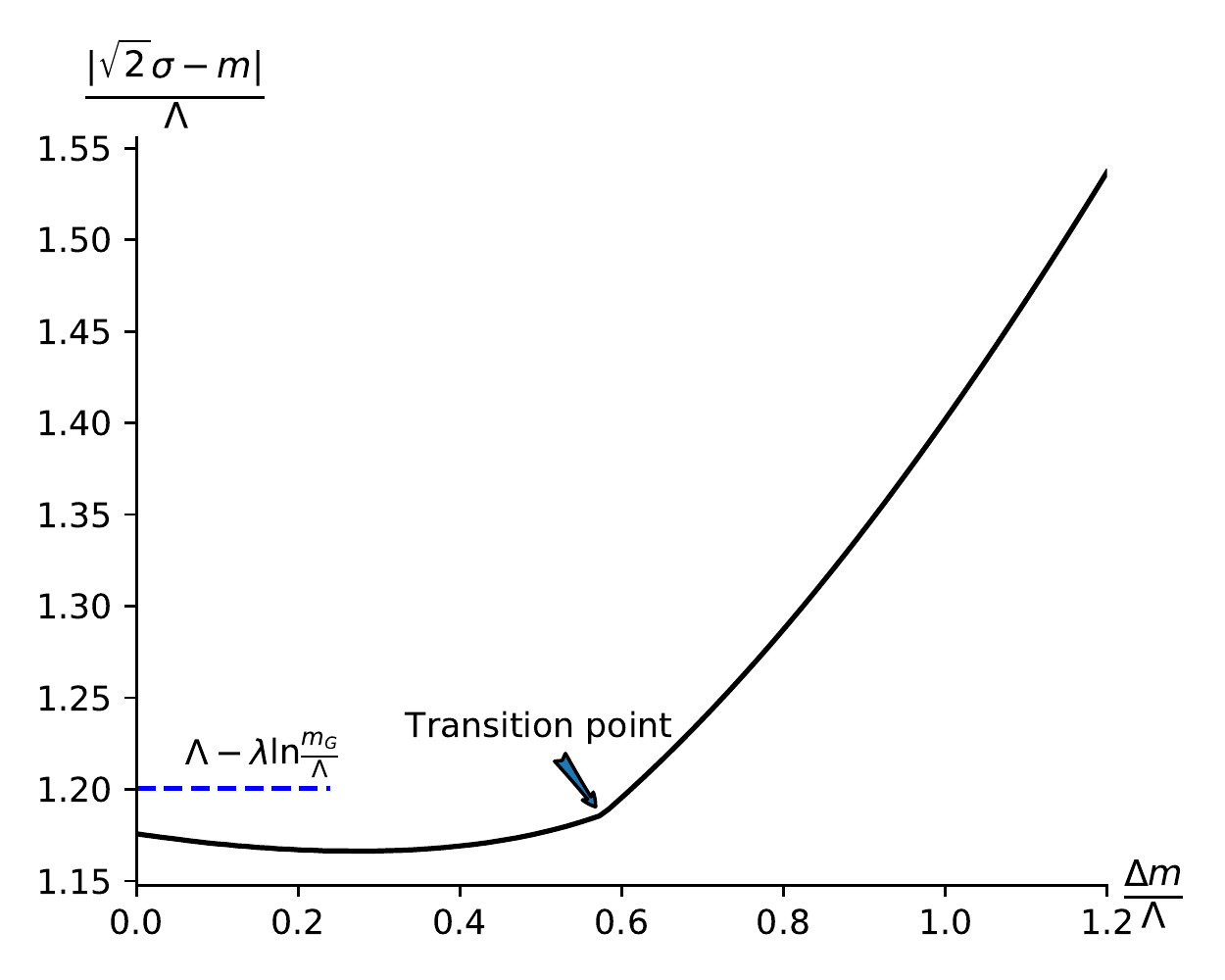}
		\caption{$\sqrt{2} \sigma - m$}
	\label{fig:strong-higgs_single_tau} 
    \end{subfigure}%
\caption{
	Strong - Higgs phase transition: VEVs. The curves show an example of the phase transition for fixed $\wt{\mu} / \Lambda = 0.03$, $N=16$. Mass scale $\Delta m$ is on the horizontal axis. Location of the phase transition point is indicated with an arrow. On the figure \subref{fig:strong-higgs_single_tau}, the position approximate strong coupling VEV \eqref{ground_state_correction}	is signified on the vertical axis by a blue dashed line.
	One can see that the character of the phase transition is qualitatively the same as in the pure non-supersymmetric \eqref{cpn_lagr_simplest} and supersymmetric \eqref{lagrangian_N=2} models, see \cite{Gorsky:2005ac,Bolokhov:2010hv}.
}
\label{fig:strong-higgs_single}
\end{figure}
\begin{figure}[h]
	\centering
	\includegraphics[width=0.7\linewidth]{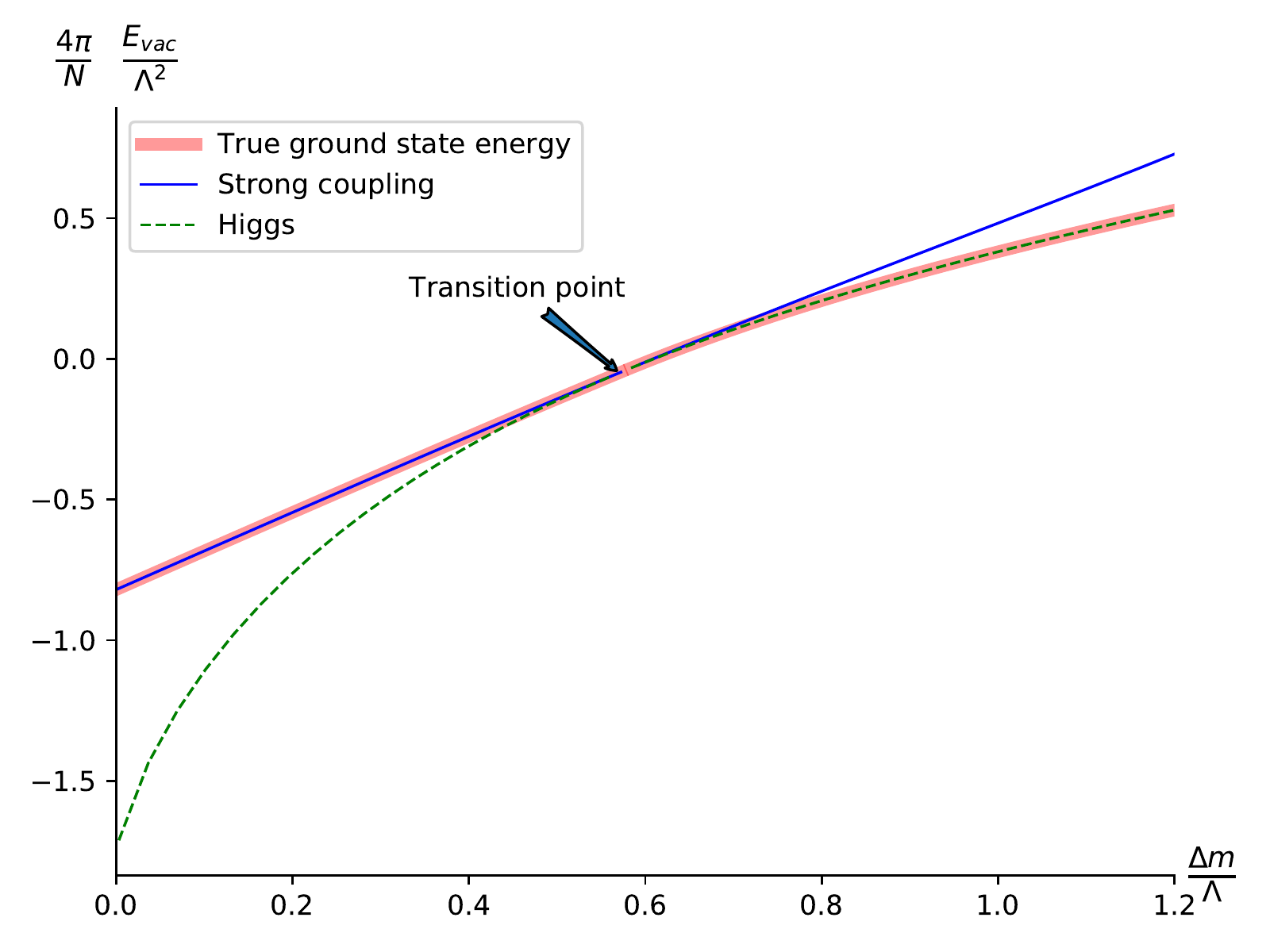}
	\caption{
		Strong-Higgs phase transition: energy. Red thick line is a numerical result for the ground state vacuum energy. Solid blue line to the right of the phase transition point is a numerical continuation of the strong coupling vacuum energy into the Higgs regime. Vice versa, dashed green line below the phase transition is a numerical continuation of the Higgs regime vacuum energy into the strong coupling (corresponds to the unphysical \textquote{state} with formally $|n|^2 < 0$. 
		At the phase transition point these two curves touch, and $|n|^2 = 0$.
		This plot is qualitatively the same as in the pure non-supersymmetric \eqref{cpn_lagr_simplest} and supersymmetric \eqref{lagrangian_N=2} models, see \cite{Gorsky:2005ac,Bolokhov:2010hv}
		In the numerical procedure we have set $\wt{\mu} / \Lambda = 0.03$, $N=16$
	}
\label{fig:strong-higgs_single_Veff} 
\end{figure}
\begin{figure}[h]
	\centering
	\includegraphics[width=0.7\linewidth]{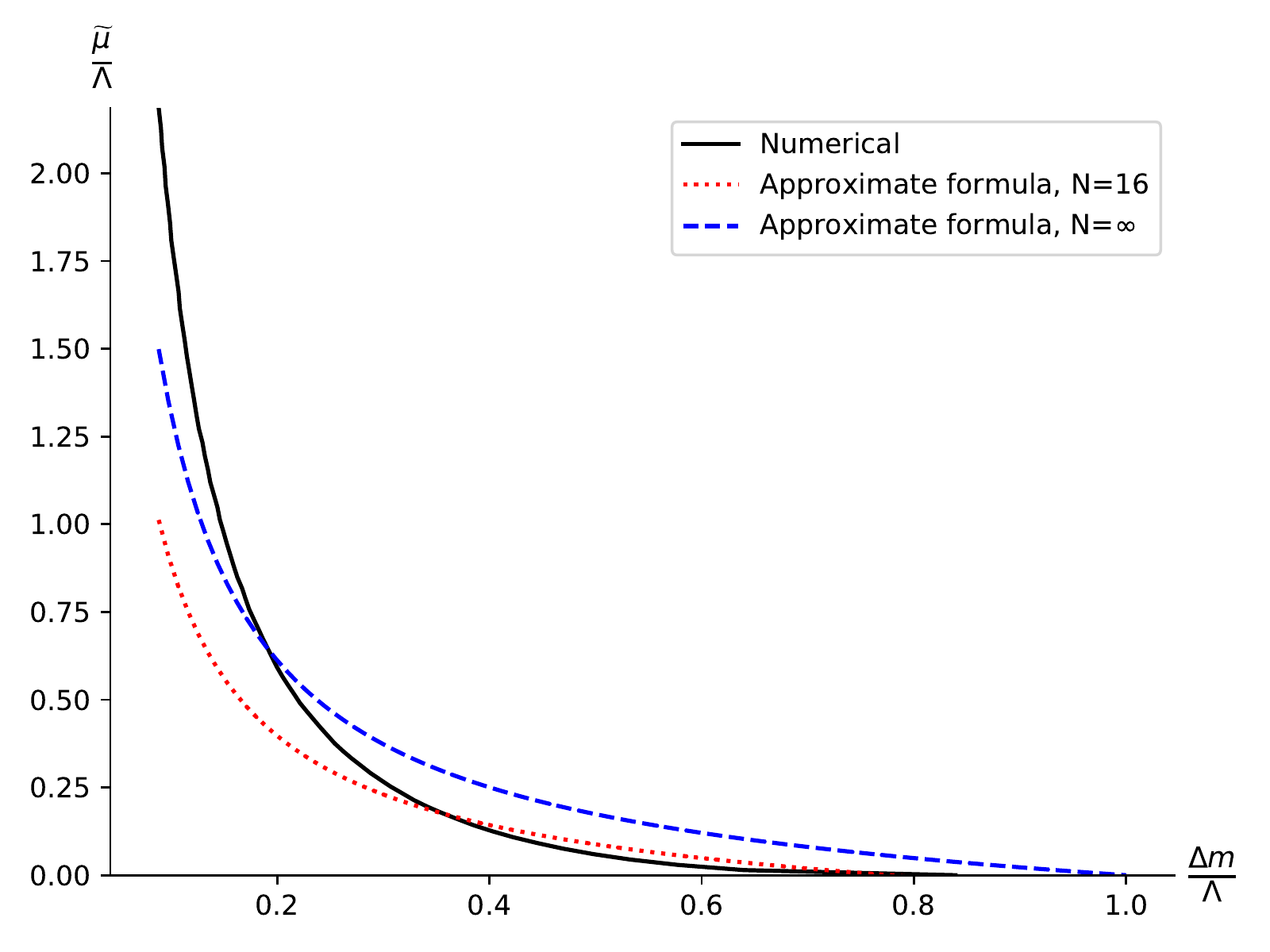}
	\caption{
		Strong-Higgs phase transition line. $\Delta m$ on the horizontal axis, $\wt{\mu}$ on the vertical axis.
		Solid black line is the numerical result for $N = 16$.
		Dotted red line is the $N=16$ approximate formula \eqref{strong_higgs_Nfin}.
		Dashed blue line is the $N\to\infty$ approximate formula \eqref{strong_higgs_Ninf}
	}
\label{fig:strong-higgs_trans} 
\end{figure}

It was found in \cite{Gorsky:2005ac,Bolokhov:2010hv} that 
for the non-supersymmetric \CP model \eqref{cpn_lagr_simplest}  and for the supersymmetric \CP model \eqref{lagrangian_N=2} a phase transition between strong coupling and Higgs phases occurs at the point $\Delta m = \Lambda$. At large $\Delta m$ the theory is weakly coupled and in the Higgs phase, while at small $\Delta m$ we have a strong coupling phase. 
We  expect similar behavior in our deformed model \eqref{lagrangian_init}.


Following  \cite{Gorsky:2005ac,Bolokhov:2010hv} we identify the Higgs-strong coupling  phase transition with a curve where $|n_0^2| =2\beta_\text{ren}$ turn negative. Thus we are looking for the solutions of the equation
\beq
\beta_\text{ren}=0,
\label{higgsphasetransit}
\eeq
where $\beta_\text{ren}$ is given by \eqref{master1}. 

In \ntwot supersymmetric model at $\wt{\mu} = 0$  a phase transition is at  $\Delta m = \Lambda$ \cite{Bolokhov:2010hv}.
The case $\wt{\mu} \neq 0$ is more complicated. We were not able to solve the vacuum equations \eqref{master1} - \eqref{master3} exactly, but an approximate calculation can be done in regions of small and very large $\wt{\mu}$.


First consider the region $\wt{\mu} \lesssim \Lambda$, and assume that the VEV of $\sigma$ is real-valued (this assumption is correct for the true ground state anyway). Then, using \eqref{master2} and the identity
\begin{equation}
	\prod_{k=1}^{N-1} \sin(\frac{\pi k}{N})  = \frac{N}{2^{N - 1}} \,,
\label{sinprod}	
\end{equation}
we can rewrite \eqref{master1} as
\begin{equation}
	2 \beta_\text{ren} = \frac{2 (N-1)}{4 \pi} \left( \ln\frac{\Delta m}{\Lambda} 
		+ \frac{1}{N-1} \ln N 
		+ \frac{1}{2} \ln\left(1 + \frac{\upsilon(\wt{\mu}) - 2(\sqrt{2}\sigma - m_0)}{2 \Delta m}\right) \right)
\end{equation}
Equating this to zero yields
\begin{equation}
	\upsilon(\wt{\mu}) - 2(\sqrt{2}\sigma - m_0) = 2 \Delta m \left( \left( \frac{\Lambda}{\Delta m} \right)^2 \, N^{- \frac{2}{N-1}} - 1 \right)
\label{trans_eq_Nfin}	
\end{equation}
At small deformations we can use the approximation $\upsilon(\wt{\mu}) \approx \wt{\mu}$, see \eqref{upsilon}. Moreover, in the strong coupling phase at fixed $\wt{\mu}$, the VEV of $\sigma$ does not depend on $\Delta m$ (this is exactly true in the supersymmetric and pure non-supersymmetric \CP models), and we can use $\Delta m = 0$ approximation \eqref{ground_state_correction} right up until the phase transition point. Then, from \eqref{trans_eq_Nfin} we can actually derive the equation for the phase transition curve:
\begin{equation}
	\wt{\mu}_\text{crit} = \frac{2 \displaystyle\frac{\Lambda^2}{\Delta m} \,  N^{- \frac{2}{N-1}} - \Lambda - \Delta m}{1 + 2 \, \wt{\lambda}_0 \ln\displaystyle\frac{m_G}{\Lambda}} \,,
\label{strong_higgs_Nfin}	
\end{equation}
or, sending $N \to \infty$,
\begin{equation}
	\wt{\mu}_\text{crit} = \frac{(2\Lambda + \Delta m) \, (\Lambda - \Delta m)}{\Delta m \left( 1 + 2 \, \wt{\lambda}_0 \ln\displaystyle\frac{m_G}{\Lambda} \right)}
\label{strong_higgs_Ninf}	
\end{equation}

These formulas give a good approximation for the phase transition curve, see Fig.~\ref{fig:strong-higgs_trans}. We see that with $\wt{\mu}_\text{crit}$ growing, $\Delta m_\text{crit}$ monotonically decreases. Moreover, comparing \eqref{strong_higgs_Nfin} and \eqref{strong_higgs_Ninf}, we can test the validity of our numerical calculations compared to the large-$N$ limit, as the numerics is done, of course, for a finite $N$\footnotemark.
\footnotetext{In the numerical calculations for this Chapter, we took $N$ = 16. Rough estimate of the accuracy from \eqref{strong_higgs_Nfin} and \eqref{strong_higgs_Ninf} is $1 - N^{- 1 / (N-1)} \approx 0.17$, i.e. qualitatively we can trust our results.}

In the region of large deformations, $\wt{\mu} \gg  m_G$.  We have
\begin{equation}
	2 \bren \sim \frac{N}{4\pi} \ln\frac{\upsilon (\mu_\text{crit})  \,\Delta m_\text{crit}  + \Delta m_\text{crit}^2 
	}{\Lambda_{2d}^2} =0,
\end{equation}
where $\Lambda_{2d}$ is exponentially small given by \eqref{beta_classical_largemu}.
From \eqref{upsilon} and \eqref{Lam_2d} we derive up to logarithmic factors
\beq
\Delta m_\text{crit} \sim \frac{(\Lambda_{4d}^{{\mathcal N}=1})^2 \ \wt{\mu}_\text{crit}}{m_G^2} \,
 \exp(- \text{const}\,\frac{\wt{\mu}_\text{crit}^2}{m_G^2}),
\label{strong-Higgs_critical}
\eeq
where  where we assumed that $\Delta m \ll m$, 
%
%
Here we see again that $\Delta m_\text{crit}$ monotonically decreases as $\wt{\mu}_\text{crit}$ becomes larger. 

%
%

\section{Phase diagram of the worldsheet theory  \label{Conclusions}}

%
\begin{figure}[h]
	\centering
	\includegraphics[width=1.0\linewidth]{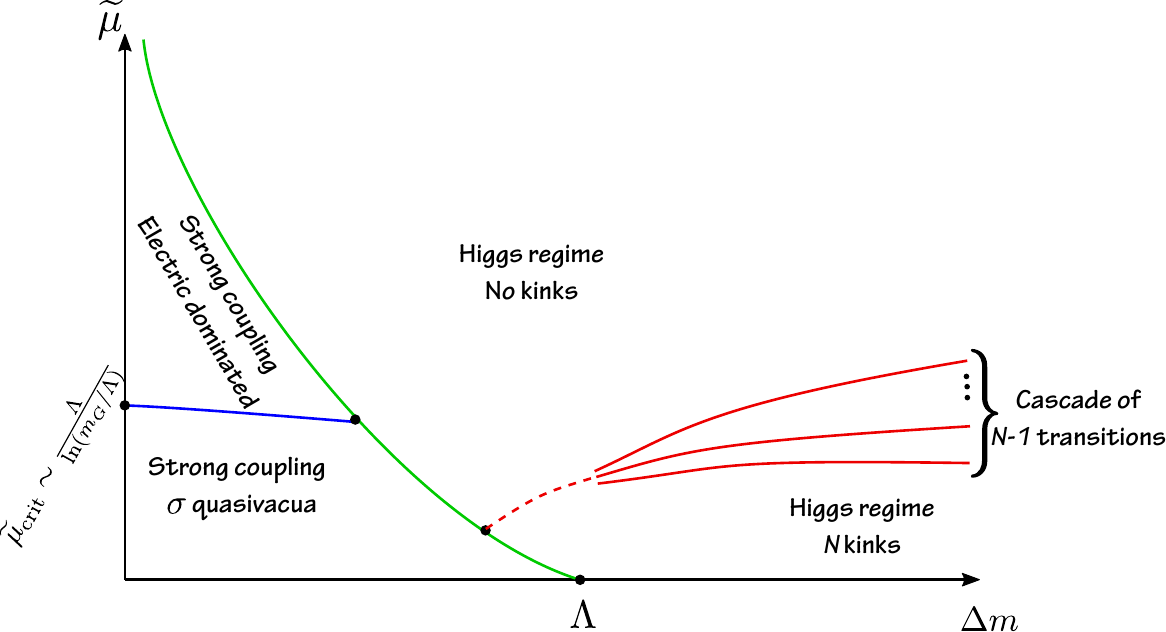}
	\caption{Whole phase diagram (schematically). $\Delta m$ on the horizontal axis, $\wt{\mu}$ on the vertical axis. Cascade of $N-1$ curves corresponds to the disappearance of kinks between the ground state and quasivacua. Dashed lines are drawn based on a general argument, since the $1/N$ expansion gives poor approximation in this region.}
\label{fig:allphases} 
\end{figure} 

In this Chapter we have studied dynamics of the $\mu$-deformed \CP model \eqref{lagrangian_init}. It arises as a world sheet theory of the non-Abelian string in \ntwo supersymmetric QCD, deformed by a mass term $\mu$ for the adjoint matter. When $\wt{\mu}$ is small, the two-dimensional theory is the \ntwot supersymmetric \CP model. As we increase the deformation parameter, the bulk theory flows to \none SQCD, while the world sheet theory becomes a non-supersymmetric $\mu$-deformed \CP model. This happens because fermion zero modes present in the bulk of the \ntwo theory are lifted when we switch on $\wt{\mu}$. As a consequence, at large $\wt{\mu}$ world sheet fermions become heavy and decouple, leaving us with the pure bosonic \CP model. In this Chapter we studied this transition in detail using the large $N$ approximation.

$\mu$-Deformed \CP model has two $N$-independent parameters, the deformation $\wt{\mu}$ (see \eqref{tildem}) and the mass scale $\Delta m$ which is the scale of the quark mass differences in the bulk theory. We obtained  a non-trivial phase diagram in the 
$(\Delta m, \,\wt{\mu} )$ plane, with two strong coupling phases  and  two Higgs phases separated by  three critical curves with two tricritical points. This phase diagram is shown on Fig. \ref{fig:allphases}. 

When $\wt{\mu}$ goes to zero, the supersymmetry is unbroken, and the theory is either in the strong coupling phase ( at small $\Delta m$ ) or in the Higgs phase (large $\Delta m$, weak coupling). In both phases there are $N$ degenerate vacua, and kinks interpolating between neighboring vacua are not confined. In the strong coupling phase at  small $\Delta m$ the photon becomes dynamical and 
acquire mass due to the chiral anomaly.

As we switch  on the deformation parameter   degenerate vacua split. 
At strong coupling we get a unique ground state and $N-1$ quasivacua, while the photon develops a small massless component. Kinks are now confined. When the deformation $\wt{\mu}$ is  small, the confinement is due to the splitting of the $\sigma$-quasivacua energies. As $\wt{\mu}$ gets larger eventually we cross the critical line where  original $\sigma$-quasivacua decay.  Now the quasivacua splitting and confinement of kinks is only due to the constant electric field.

In the Higgs phase at large $\Delta m$ the theory is at weak coupling. The $n$ field develop a VEV, photon is unphysical and heavy due to the Higgs mechanism.   When $\wt{\mu}$ is small enough
 energies of  $N$ degenerate vacua split, and  kinks interpolating between the neighboring quasivacua are confined. However as we increase $\wt{\mu}$ it crosses   critical  lines where (see e.g. \eqref{nokink_trans_classic})  quasivacua decay one by one leaving the theory with a single ground state, and thus without kinks. 

In this Chapter we have shown that results obtained in Chapter~\ref{sec:none} for the  $\mu$-deformed  bulk theory  agree with the world sheet considerations. We can either go to the world sheet in the \ntwo theory and then take the large $\wt{\mu}$ limit, or first apply the large deformation in the bulk and then go to the world sheet theory.
In other words, the following diagram is commutative:
\begin{equation}
  \begin{tikzcd}[column sep=large]
  	\text{4d \ntwo SQCD} \arrow[swap]{r}{\text{worldsheet} } \arrow[swap]{d}{\text{large } \wt{\mu} } & \text{2d \ntwot \CP} \arrow[swap]{d}{\text{large } \wt{\mu} } \\
    \text{4d \none SQCD} \arrow[swap]{r}{\text{worldsheet} } & \text{2d } {\mathcal N}=0 \text{ \CP}      
  \end{tikzcd}	
\end{equation}
We note however that  a derivation of the world sheet theory at intermediate values of $\wt{\mu}$ is still absent.

As we already discussed we interpret kinks of the world sheet theory  as confined monopoles of the four-dimensional SQCD.
Our results show, in particular, that at large $\wt{\mu}$, when the bulk theory  basically becomes  \none SQCD,  monopoles survive  only in the strong coupling phase at very small mass differences below the critical line \eqref{strong-Higgs_critical}. In the Higgs phase quasivacua decay at large $\wt{\mu}$
which means that confined monopole and antimonopole forming a ''meson'' on the string ( see Fig.~\ref{fig:conf}) 
annihilate each other and disappear.
This confirms a similar  conclusion of Chapter~\ref{sec:none}.

The results of this Chapter are published in the paper \cite{Gorsky:2019shz} and in the graduation work \cite{VKR}.

%
%

\chapter{String \textquote{baryon} of the \ntwo supersymmetric QCD} \label{sec:b_meson}

%


In the previous Chapters, we were mostly concerned with the \none limit of the supersymmetric QCD.
In this Chapter we are going to take a step in a \textquote{perpendicular} direction and consider the \ntwo theory in another special limit, when the $\beta$-function in the 4d vanishes, and the bulk theory becomes superconformal\footnote{In fact, it is not exactly conformal; this will be explained below.}. This highly symmetric setting allows one to make far going advances and even derive the hadron spectrum (at least in part).

%
%

\section{Overview}  \label{sec:b_meson:intro}

In 2015 a non-Abelian semilocal vortex string was discovered possessing a world-sheet theory which is both superconformal and critical \cite{SYcstring}. 
This string is supported in four-dimensional  ${\cal N} = 2$ super-QCD with the U$(N=2)$ gauge group, $N_f=4$ flavors 
of quarks and a Fayet-Iliopoulos term \cite{FI}.  
Due to the extended supersymmetry,  the gauge coupling  in the 4D bulk could be renormalized only at one loop.
With our judicial choice of the matter sector ($N_f=2N$) the one-loop renormalization cancels. No dynamical scale parameter 
$\Lambda$ is generated in the bulk \footnote{However, conformal invariance of 4D SQCD is broken by the Fayet-Iliopoulos term.}.

This is also the case in the world-sheet theory described by the weighted $\mathbb{CP}$  model ($\mathbb{WCP}(2,2)$), see Sec.~\ref{sec:wcp} and here below. Its $\beta$ function vanishes,  and the overall Virasoro central charge is critical \cite{SYcstring}.  
This happens because in
addition to four translational moduli, non-Abelian string has six orientational and size
moduli described by $\mathbb{WCP}(2,2)$ model. Together, they form a ten-dimensional target space required for a superstring to be critical. The target
space of the string sigma model is $\mathbb{R}_4\times Y_6$, a product of the flat four-dimensional space and a Calabi-Yau non-compact threefold $Y_6$, namely, the conifold.

This allows one to apply string theory for consideration of the closed string spectrum and its interpretation  as a spectrum of hadrons in 4D \ntwo SQCD. The vortex string at hand was identified as the string theory of Type IIA \cite{KSYconifold}.

The study of the above vortex string from the standpoint of string theory, with the focus on massless states in four dimensions has been started   in \cite{KSYconifold,KSYcstring}. Later the low lying massive string states were found by virtue of little string theory \cite{SYlittles}.
Generically,  most of massless modes have  non-normalizable wave functions over the conifold $Y_6$, i.e. they are not localized in 4D 
and, hence, cannot be interpreted as dynamical states in 4D SQCD.  In particular, no massless 4D gravitons or vector fields were found  in the physical spectrum in \cite{KSYconifold}. However, a single massless BPS hypermultiplet in the 4D bulk was detected 
at a self-dual point (at strong coupling). It is associated with deformations of a complex structure of the conifold and was  interpreted  as a composite 4D \textquote{baryon.}\footnote{If the gauge group is U(2), as is our case, there are no {\em bona fide} baryons. We still use the term baryon because of a particular value of its  charge $Q_B$(baryon) = 2 with respect to the global unbroken U(1)$_B$, see Sec.~\ref{conifold}.}

Our general strategy is as follows. We explore the BPS protected sector of the world-sheet model, two-dimensional \wcpt, starting from weak coupling $\beta \gg 1$, where $\beta$ is the inverse coupling\footnote{Note that in Chapters~\ref{sec:none} and \ref{sec:large_N} the 2D coupling was denoted as $2\beta$, while here it's just $\beta$.}. 
This procedure requires an infra-red  regularization.  To this end we introduce  masses of quarks in 4D SQCD. They translates into four twisted masses in the world-sheet \wcpt (two for $n^P$  and two for $\rho^K$) which we arrange in a certain hierarchical order. We find both vacua of the theory, and study distinct kinks (in the mirror representation).  
Thus the  vacuum structure   and kink spectrum of this theory are  known exactly, and so are all curves (walls) of the marginal stability (CMS). Then we move towards strong coupling $\beta \sim 0$ 
carefully identifying CMS in the complex $\beta$ plane.
At each step we determine which kinks decay on CMS and which are stable upon crossing and establish their relation to four-dimensional monopoles using the so called 2D-4D correspondence, the coincidence of BPS spectra in 4D \ntwo SQCD and in the string world-sheet theory, see Sec.~\ref{2-4} and \cite{SYmon,HT2,Dorey}.

At strong coupling we use 2D-4D correspondence to confirm that our 4D SQCD enters the so called \textquote{instead-of-confinement} phase found earlier in  asymptotically free versions of SQCD \cite{SYdual}, see \cite{SYdualrev} for a review.
This phase  is qualitatively similar  to
the conventional QCD confinement: the quarks and gauge bosons screened at
weak coupling, at strong coupling evolve into monopole-antimonopole pairs
confined by non-Abelian strings. They form monopole mesons and baryons shown in Fig.~\ref{monmb}  (on page \pageref{monmb}). The role of the constituent quark in this phase is played by the confined monopole.

Needless to say, the quark  masses break the global symmetry
\begin{equation}
	 {\rm SU}(N)_{C+F}\times {\rm SU}(\tN)\times {\rm U}(1)_B\,,
\label{b:globgroup_d=4}
\end{equation}
cf. (\ref{globgroup_d=4}). At the very end we tend them to zero, restoring the global symmetries, as well as conformal invariance of the world-sheet theory on the string. 
Moreover, If we introduce non-zero quark masses the Higgs branch \eqref{dimH}
\begin{equation}
	{\rm dim}\,{\cal H}= 4N \tN.
\label{b:dimH}
\end{equation}
is lifted and 
bifundamental quarks acquire masses $(m_P-m_K)$, $P=1,2$, $K=3,4$. Note that bifundamental quarks form  short BPS multiplets
and their masses do not receive quantum corrections, see \cite{SYrev} for details.

Our main result is the emergence (at $\beta =0$) of a short BPS massless \textquote{baryon} supermultiplet with the U(1)$_{B}$ charge $Q_B=2$.  In this way we demonstrate that the massless \textquote{baryon} state which had been previously observed using string theory
arguments \cite{KSYconifold} is seen in the field-theoretical approach too. We believe this is the first example of this type.

To obtain this result we use the following strategy. It is known that \wcpt 
model at $\beta=0$ has a marginal deformation
associated with the deformation of the complex structure of the conifold. Since \wcpt model is a world-sheet theory on the non-Abelian string the natural question to address is what is  the origin of this deformation in 4D SQCD. On general grounds one expects that this could be some parameter of the 4D theory such as a coupling constant. Another option is that it could be a modulus, a vacuum expectation value of a certain dynamical field. We show using 2D-4D correspondence (see Sec.~\ref{2-4}) that the latter option is realized in the case at hand. A new non-perturbative Higgs branch opens up at $\beta=0$ in  4D SQCD. The modulus parameter on this Higgs branch is  the VEV of the massless BPS baryon
constructed from four monopoles connected by confining strings as shown in Fig.~\ref{monmb}b.

So, the theory that we will be working with in this Chapter is the \wcpt sigma model, see Sec.~\ref{sec:wcp}. 
It can be  defined  as a low energy limit of the  U(1) gauge theory \cite{W93}.  The bosonic part of the action reads
 \footnote{Equation 
(\ref{wcp22}) and similar expressions below are given in Euclidean notation.}
\begin{equation}
\begin{aligned}
	&S = \int d^2 x \left\{
	\left|\nabla_{\alpha} n^{P}\right|^2 
	+\left|\tilde{\nabla}_{\alpha} \rho^K\right|^2
	+\frac1{4e^2}F^2_{\alpha\beta} + \frac1{e^2}\,
	\left|\pt_{\alpha}\sigma\right|^2
	\right.
	\\[3mm]
	&+\left.
	2\left|\sigma+\frac{m_P}{\sqrt{2}}\right|^2 \left|n^{P}\right|^2 
	+ 2\left|\sigma+\frac{m_{K}}{\sqrt{2}}\right|^2\left|\rho^K\right|^2
	+ \frac{e^2}{2} \left(|n^{P}|^2-|\rho^K|^2 - r \right)^2
	\right\},
	\\[4mm]
	&
	P=1,2\,,\qquad K=3,4\,.
\end{aligned}
\label{wcp22}
\end{equation}
Here, $m_A$ ($A=1,..,4$) are the so-called twisted masses (they come from 4D quark masses),
while $r$ is the inverse coupling constant (2D FI term). Note that $r$ is the real part of the complexified coupling constant 
introduced in Eq. (\ref{beta_complexified}), $$r = {\rm Re}\,\beta\,.$$

From action \eqref{wcp22} for $\mathbb{WCP}(2,2)\;$ it is obvious that this model is self-dual. The duality transformation
\begin{equation}
\begin{aligned}
	\beta &\to \wt{\beta} = - \beta \, \\
	m_{1,2} &\to \wt{m}_{1,2} = - m_{3,4} \, \\
	m_{3,4} &\to \wt{m}_{3,4} = - m_{1,2} \, \\
	\sigma &\to \wt{\sigma} = - \sigma
\end{aligned}
\label{2d_S-duality}
\end{equation}
exchanges the roles of the orientation moduli $n^P$ and size moduli $\rho^K$. The point $\beta = 0$ is the {\em self-dual} point.

%
%

\section {Massless 4D baryon from  string theory}
\label{conifold}

 The world-sheet \wcpt model \eqref{wcp22} is conformal and due to \ntwot supersymmetry the metric of its target space is 
K\"ahler. The conformal invariance of the model also ensures that this metric is Ricci flat. Thus the target space of model \eqref{wcp22} is a Calabi-Yau manifold. 

Moreover, as we explained in the previous Section the world-sheet \wcpt 
model has six real bosonic degrees of freedom.
Its target space defined by the  $D$-term condition
\beq
|n^P|^2-|\rho^K|^2 = \beta\,
\label{Fterm}
\eeq 
plus gauge invariance,
is a six dimensional non-compact Calabi-Yau space $Y_6$ known as  conifold, see \cite{NVafa} for a review. Together with
four translational moduli of the non-Abelian vortex it forms a ten dimensional target space $\mathbb{R}^4\times Y_6$ required for 
a superstring to be critical \cite{SYcstring}. 

In this section we briefly review the only 4D massless state found from the string theory of the critical non-Abelian vortex \cite{KSYconifold}. It is associated 
with the deformation of the conifold complex structure. 
 As was already mentioned, all other massless string modes  have non-normalizable wave functions over the conifold. In particular, 4D graviton associated with a constant wave
function over the conifold $Y_6$ is
absent \cite{KSYconifold}. This result matches our expectations since we started with
\ntwo SQCD in the flat four-dimensional space without gravity.

We can construct the U(1) gauge-invariant \textquote{mesonic} variables
\beq
w^{PK}= n^P \rho^K.
\label{w}
\eeq
These variables are subject to the constraint
\beq
{\rm det}\, w^{PK} =0. 
\label{coni}
\eeq

Equation (\ref{coni}) defines the conifold $Y_6$.  
It has the K\"ahler Ricci-flat metric and represents a non-compact
 Calabi-Yau manifold \cite{W93,NVafa,Candel}. It is a cone which can be parametrized 
by the non-compact radial coordinate 
\beq
\widetilde{r}^{\, 2} = {\rm Tr}\, \bar{w}w\,
\label{tilder}
\eeq
and five angles, see \cite{Candel}. Its section at fixed $\widetilde{r}$ is $S_2\times S_3$.

At $\beta =0$ the conifold develops a conical singularity, so both $S_2$ and $S_3$  
can shrink to zero.
The conifold singularity can be smoothed out
in two distinct ways: by deforming the K\"ahler form or by  deforming the 
complex structure. The first option is called the resolved conifold and amounts to keeping
a non-zero value of $\beta$ in (\ref{Fterm}). This resolution preserves 
the K\"ahler property and Ricci-flatness of the metric. 
If we put $\rho^K=0$ in (\ref{wcp22}) we get the $\mathbb{CP}(1)$ model with the $S_2$ target space
(with the radius $\sqrt{\beta}$).  
The resolved conifold has no normalizable zero modes. 
In particular, 
the modulus $\beta$  which becomes a scalar field in four dimensions
 has non-normalizable wave function over the 
$Y_6$ and therefore is not dynamical \cite{KSYconifold}.  

If $\beta=0$ another option exists, namely a deformation 
of the complex structure \cite{NVafa}. 
It   preserves the
K\"ahler  structure and Ricci-flatness  of the conifold and is 
usually referred to as the {\em deformed conifold}. 
It  is defined by deformation of Eq.~(\ref{coni}), namely,   
\beq
 {\rm det}\, w^{PK} = b\,,
\label{deformedconi}
\eeq
where $b$ is a complex number.
Now  the $S_3$ can not shrink to zero, its minimal size is 
determined by
$b$. 

The modulus $b$ becomes a 4D complex scalar field. The  effective action for  this field was calculated in \cite{KSYconifold}
using the explicit metric on the deformed conifold  \cite{Candel,Ohta,KlebStrass},
\beq
S(b) = T\int d^4x |\pt_{\mu} b|^2 \,
\log{\frac{T^2 L^4}{|b|}}\,,
\label{Sb}
\eeq
where $L$ is the  size of $\mathbb{R}^4$ introduced as an infrared regularization of 
logarithmically divergent $b$ field 
norm.\footnote{The infrared regularization
on the conifold $\widetilde{r}_{\rm max}$ translates into the size $L$ of the 4D space 
 because the variables  $\rho$ in \eqref{tilder} have an interpretation of the vortex string sizes,
$\widetilde{r}_{\rm max}\sim TL^2$ .}

We see that the norm of
the $b$ modulus turns out to be  logarithmically divergent in the infrared.
The modes with the logarithmically divergent norm are at the borderline between normalizable 
and non-normalizable modes. Usually
such states are considered as \textquote{localized} ones. We follow this rule. 
This scalar mode is localized near the conifold singularity  in the same sense as the orientational 
and size zero modes are localized on the vortex-string solution.
   
 The field $b$  being massless can develop a VEV. Thus, 
we have a new Higgs branch in 4D \ntwo SQCD which is developed only for the critical value of 
the 4D coupling constant \footnote{The complexified 4D coupling constant $\tau=1$ at this point, see Sec.~\ref{sec:2D_4D}.} associated with $\beta=0$.

 In \cite{KSYconifold} the massless state $b$ was interpreted as a baryon of 4D \ntwo QCD.
Let us explain this.
 From Eq.~(\ref{deformedconi}) we see that the complex 
parameter $b$ (which is promoted to a 4D scalar field) is a singlet with respect to both SU(2) factors in 
the global world-sheet group\footnote{Which is isomorphic to the 4D
global group \eqref{b:globgroup_d=4}.}
\begin{equation}
	 {\rm SU}(2)\times {\rm SU}(2)\times {\rm U}(1)_B \,,
\label{b:globgroup}
\end{equation}
cf. (\ref{globgroup}). Recall from \eqref{repsnrho} that the fields $n$ and $\rho$ transform in the following representations: 
\begin{equation}
	n:\quad \left(\textbf{2},\,\textbf{1},\, 0\right), \qquad \rho:\quad \left(\textbf{1},\, \textbf{2},\, 1\right)\,.
\label{b:repsnrho}
\end{equation} 
 What about the baryonic charge of the field $b$? From \eqref{b:repsnrho} and \eqref{deformedconi}
we see that the $b$ state transforms as 
\beq
({\bf 1},\,{\bf 1},\,2).
\label{brep}
\eeq
 In particular it has the baryon charge $Q_B(b)=2$.

To conclude this section let us note that in type IIA superstring the complex scalar 
associated with deformations of the complex structure of the Calabi-Yau
space enters as a 4D \ntwo BPS hypermultiplet. Other components of this hypermultiplet can be restored 
by \ntwo supersymmetry. In particular, 4D \ntwo hypermultiplet should contain another complex scalar $\tilde{b}$
with baryon charge  $Q_B(\tilde{b})=-2$. In the stringy description this scalar comes from ten-dimensional
three-form, see \cite{Louis} for a review.

Below in this Chapter we study the BPS kink spectrum of the world-sheet model \eqref{wcp22} using purely field theory
methods. Besides other results we use the  2D-4D correspondence to confirm  the emergence of 4D baryon with quantum numbers
\eqref{brep} and the presence of the associated non-perturbative Higgs branch at $\beta=0$.

%
%

\section{Kink mass from the exact superpotential \label{sec:kink_mass}}

As was mentioned above, the \wcpt model \eqref{wcp22} supports BPS saturated kinks interpolating between different vacua. 
In this section we will obtain the kink central charges and, consequently, their masses.

\subsection{Exact central charge}

For the model at hand we can obtain an exact formula for the BPS kink central charge. 
This is possible because for this model an exact twisted superpotential obtained by integrating out $n$ and $\rho$ supermultiplets  is known. It is a generalization \cite{HaHo,DoHoTo}
of the \CP model superpotential \cite{Dorey,W93,AdDVecSal,ChVa} of the  Veneziano-Yankielowicz  type \cite{VYan}.
In the present case $N_f = 2N = 4$ it reads:
\begin{multline}
	 {\cal W}_{\rm WCP}(\sigma)= \frac{1}{4\pi}\Bigg\{ 
	 	\sum_{P=1,2} \left( \sqrt{2} \, \sigma + m_P \right) \ln\left( \sqrt{2} \, \sigma + m_P \right)
	 	\\
	 	- \sum_{K=3,4} \left( \sqrt{2} \, \sigma + m_K \right) \ln\left( \sqrt{2} \, \sigma + m_K \right)
	 	+ 2 \pi \,  \sqrt{2} \, \sigma  \, \beta
	 	+ \text{const}
	 \Bigg\}\,,
\label{WCPsup}
\end{multline}
where we use one and the same notation $\sigma$ for the  twisted superfield \cite{W93} and its lowest scalar
component. 
To study the vacuum structure of the theory we minimize this superpotential with respect to $\sigma$ to obtain the 2D vacuum equation
\begin{equation}
	\prod_{P=1,2}\left(\sqrt{2} \, \sigma + m_P \right) 
		= e^{- 2 \pi \beta} \cdot \prod_{K = 3,4} \left(\sqrt{2} \, \sigma + m_K \right) \,.
\label{2d_equation}	
\end{equation}
The invariance of equation \eqref{2d_equation}	under the duality transformation \eqref{2d_S-duality} is evident. 

The vacuum equation \eqref{2d_equation} has two solutions (VEVs) $\sigma_{1,2}$, which means that generically there are two degenerate vacua in our theory. 
Therefore, there are BPS kinks interpolating between these two vacua.
Their masses are given by the absolute value of the central charge, 
\beq
M_{\rm BPS} = |Z|. 
\label{MZ}
\eeq
The central charge can be found by taking the appropriate difference of the superpotential \eqref{WCPsup} calculated at distinct roots \cite{Dorey,HaHo,DoHoTo}.
Say, for the kink interpolating between the vacua $\sigma_2$ and $\sigma_1$, the central charge is given by
\begin{equation}
\begin{aligned}
	Z_{\rm BPS} 
		&= 2\left[{\cal W}_{\rm WCP}(\sigma_{1})-{\cal W}_{\rm WCP}(\sigma_{2})\right] 	\\
		&= \frac{1}{2\pi}\Bigg[ 
			\sum_{P=1}^{N_c} m_P \ln\frac{\sqrt{2}\sigma_1 + m_P}{\sqrt{2}\sigma_2 + m_P}
			- \sum_{K=N_c+1}^{N_f} m_K \ln\frac{\sqrt{2}\sigma_1 + m_K}{\sqrt{2}\sigma_2 + m_K}
		\Bigg]
		\,.
\end{aligned}
\label{BPSmass}
\end{equation}
Note that in order for this equation to transform well under the $S$ duality transformation, we must assume that the masses are transformed as $m_P \to - m_K$, $m_K \to - m_P$.

The central charge formula \eqref{BPSmass} contains logarithms, which are multivalued. Distinct choices differs by contributions $im_A \times {\rm integer}$.  In addition to the
topological charge, the kinks can carry Noether charges with respect to the global group \eqref{b:globgroup} broken down to
U(1)$^3$ by the mass differences. This produces a whole family of dyonic kinks. We stress that all these kinks  interpolate
between the same pair of vacua $\sigma_1$ and $\sigma_2$. In Eq.~\eqref{BPSmass} we do not specify these dyonic contributions. 
Below in this Chapter we present a detail study of the BPS kink spectrum in different regions of the coupling constant $\beta$.

The $\sigma$ vacua are found by solving equation \eqref{2d_equation},
\begin{equation}
	\sqrt{2} \sigma_{\pm} = - \frac{\Delta m}{2} \, \frac{1 + e^{- 2 \pi \beta}}{1 - e^{- 2 \pi \beta}} 
		~\pm~ \sqrt{ \frac{(\delta m_{12})^2 - e^{- 2 \pi \beta} \, (\delta m_{34})^2}{4 (1 - e^{- 2 \pi \beta}) } + \Delta m^2 \, \frac{e^{- 2 \pi \beta}}{(1 - e^{- 2 \pi \beta})^2}}
		\,.
\label{roots_symmetric}	
\end{equation}
In writing down this formula we have used the following parametrization of the masses:
\begin{equation}
\begin{aligned}
	\Delta m      &~=~  \ov{m} - \wt{m} \,,\quad { \Delta m^2      ~=~  \big(\ov{m} - \wt{m}\big)^2\,,}\\
	\delta m_{12} &~=~ m_1 - m_2 \,, \\
	\delta m_{34} &~=~ m_3 - m_4 \,,
\end{aligned}
\label{mass_parametrization}
\end{equation}
where $\ov{m}$ and $\wt{m}$ are the averages of bare masses of the $n^{P}$ and $\rho^{K}$ fields, respectively, 
\begin{equation}
	\ov{m} = \frac{m_1 + m_2}{2} 
	\,, \quad
	\wt{m} = \frac{m_3 + m_4}{2} \,.
\label{mbar_mtilde}
\end{equation}
From \eqref{roots_symmetric} we immediately observe that generically one of the roots grows indefinitely near the self-dual point $\beta=0$, while the other remains finite. This will turn out to be important for consideration of kinks at strong coupling.

The Argyres-Douglas (AD) points \cite{AD} correspond to
fusing the two vacua. In these points certain kinks become massless.  Given the solution \eqref{roots_symmetric}, the AD points arise when  the expression under the square root 
vanishes. 
The formula for the positions of the AD points in the $\beta$ plane can be expressed as
\begin{equation}
	e^{- 2 \pi \beta_{AD}} =  \big( 2 P - 1 \pm 2 \sqrt{P (P - 1)} \big) \cdot \frac{m_1 - m_2}{m_3 - m_4} \,,
\label{AD_general}
\end{equation}
where $P$ is a conformal cross-ratio,
\begin{equation}
	P[m_1, m_4, m_3, m_2] 
		= \frac{(m_1 - m_4) (m_3 - m_2)}{(m_1 - m_2) (m_3 - m_4)} \,.
\end{equation}
Formula \eqref{AD_general} may have singularities. Values $P=0,1$ correspond to a Higgs brunch opening up, while at $P\to\pm\infty$ one of AD points runs away to $\beta\to\pm\infty$. There is not much interesting going on at these singularities, and at generic masses formula \eqref{AD_general} is perfectly fine. Therefore, we will not consider  these points here.

\subsection{$\mathbb{CP}(1)$ limit}

To make contact with the well understood kink spectrum of $\mathbb{CP}(1)\;$ model 
 we consider the following limit\footnote{In this section and below in writing similar inequalities involving $\gg$ or $\ll$ we actually assume that on the l.h.s. we take the absolute value of masses and real part of $\beta$, e.g. \eqref{CP1_limit} actually means $| \Delta m | \gg | \delta m_{12}|  \,, \ | \delta m_{34} | $; $\Re\beta \gg 1$.}:
\begin{equation}
	\Delta m \gg \delta m_{12} \,, \ \delta m_{34}
	\,;	\quad 
	\beta \gg 1 \,.
\label{CP1_limit}
\end{equation}
Most of the general features (with the exception of the weak coupling bound states, see Sec.~\ref{sec:weak}) of the $\mathbb{WCP}(2,2)\;$ model are still preserved in this limit, but calculations simplify greatly.
Moreover, results of this section easily generalize to the case $\beta \ll -1$.

By an appropriate redefinition of the $\sigma$ field we can shift the masses to
\begin{equation}
\begin{aligned}
	m_1 &= \delta m_{12} / 2 \,, \quad
	m_2 = - \delta m_{12} / 2 \,, \\
	m_3 &= - \Delta m + \delta m_{34} / 2 \,, \quad
	m_4 = - \Delta m - \delta m_{34} / 2 \,.
\end{aligned}
\label{CP1_masses}
\end{equation}
In this representation it is evident that in the limit \eqref{CP1_limit} the $\rho^K$ fields are heavy and decouple at energies below $\Delta m$, and the theory at low energies reduces to the ordinary $\mathbb{CP}(1)\;$ model with mass scale $\delta m_{12}$. 
The effective coupling constant is no longer constant. It runs  below $\Delta m$ and freezes at the scale $\delta m_{12}$,
\beq
2\pi\beta_{CP(1)} = 2\pi \beta - 2 \ln \frac{\Delta m }{\delta m_{12}}= 2 \log{\left\{\frac{\delta m_{12}}{\Lambda_{CP(1)}}\right\}},
\label{CP1_beta}
\eeq
where the factor 2 in the r.h.s. is the first coefficient of the $\beta$ function (for  $\mathbb{CP}(N-1)\;$ this coefficient is 
$N$),
while $\Lambda_{CP(1)}$ is the dynamical scale of the low-energy $\mathbb{CP}(1)\;$ model,
\begin{equation}
	\Lambda_{CP(1)} = \Delta m \, e^{- \pi \beta} \,.
\label{CP1_scale}	
\end{equation}

The vacuum equation \eqref{2d_equation} becomes 
\begin{equation}
	(\sqrt{2} \sigma - \delta m_{12} / 2) \, (\sqrt{2} \sigma + \delta m_{12} / 2) \approx 
	e^{- 2 \pi \beta} (- \Delta m)^2 \, = \Lambda_{CP(1)}^2.
	\label{k414}
\end{equation}
In the limit \eqref{CP1_limit},  Eq. (\ref{k414})  fits the $\mathbb{CP}(N-1)\;$ vacuum equation
\begin{equation}
	\prod_{P=1}^{N}\left(\sqrt{2}\sigma + m_P \right) = \left( \Lambda_{CP(N-1)} \right)^{N}
\end{equation}
(for $N=2$). 
In the limit \eqref{CP1_limit} the AD points \eqref{AD_general} are given by $\pm \beta_{AD}$ with
\begin{equation}
	\beta_{AD} \approx \frac{1}{\pi} \ln\frac{2 \Delta m}{\delta m_{12}} \pm \frac{i}{2}  \,.
\label{AD_roots_simplecase_beta1_approx}
\end{equation}
We see that the $\mathbb{CP}(1)\;$ weak coupling condition $\Lambda_{CP(1)} \ll \delta m_{12}$ directly translates to $\beta \gg \beta_{AD}$,
see \eqref{CP1_scale}. Let us stress that this  is more restrictive condition then just $\beta \gg 1$. If $\beta \to \beta_{AD}$
the effective coupling \eqref{CP1_beta} hits the infrared pole.

Now, since we are in the $\mathbb{CP}(1)\;$ limit, the BPS kink central charge must be given by the well known formula \cite{Dorey}. Indeed, the 2D vacua are approximately given by
\begin{equation}
	\sqrt{2}\sigma_\pm  \approx \pm \frac{1}{2} \sqrt{\delta m_{12}^2 + 4 \, \Lambda_{CP(1)}^2} \,.
\end{equation}
Substituting this and \eqref{CP1_masses} into the \wcpt central charge formula \eqref{BPSmass} and neglecting terms $\delta m_{12} / \Delta m$ and $\Lambda_{CP(1)} / \Delta m$
we obtain for the central charge
\begin{equation}
	Z_\text{kink} = \frac{1}{2\pi}\left[ 
		2 \sqrt{\delta m_{12}^2 + 4 \, \Lambda_{CP(1)}^2}
		- \delta m_{12} \, \ln\frac{\delta m_{12} + \sqrt{\delta m_{12}^2 + 4 \, \Lambda_{CP(1)}^2}}{\delta m_{12} - \sqrt{\delta m_{12}^2 + 4 \, \Lambda_{CP(1)}^2}}
		\right]
		\,.
\label{CP1_Z}
\end{equation}
This is exactly Dorey's  formula \cite{Dorey} for $\mathbb{CP}(1)$.

The above central charge \eqref{CP1_Z} tends to zero at the AD point \eqref{AD_roots_simplecase_beta1_approx}. This ensures that 
the BPS kink becomes massless at this point. We will see later that at two AD points $\beta = \beta_{AD}$ with ${\rm Re}\, \beta >0$ two kinks with distinct
dyonic charges become massless. 

Moreover, the central charge \eqref{CP1_Z} has a singularity at the AD point. Indeed, near this point we have $\delta m_{12}^2 + 4 \Lambda_{CP(1)}^2 \approx 0$. Expanding \eqref{CP1_Z} we get
\begin{equation}
\begin{aligned}
	Z_\text{kink} 
		&\approx - \frac{1}{3\pi \, \delta m_{12}^2} \Big( \delta m_{12}^2 + 4 \,\Lambda_{CP(1)}^2 \Big)^{3/2} \\
		&\approx - \frac{2 \sqrt{2\pi}}{3} \, \delta m_{12} \cdot (\beta - \beta_{AD1})^{3/2} \,.
\label{CP1_Z_AD_2}
\end{aligned}
\end{equation}
This shows that locally the central charge has a root-like singularity near the AD point.

In the quasiclassical limit $\Lambda_{CP(1)} \ll \delta m_{12}$ (or, equivalently, $\beta \gg \beta_{AD}$) the central charge \eqref{CP1_Z} is
\begin{equation}
\begin{aligned}
	Z_\text{kink} 
		&\approx - \frac{\delta m_{12}}{\pi} \, \ln\frac{\delta m_{12}}{\Lambda_{CP(1)}} + i \, \frac{\delta m_{12}}{2} + \frac{\delta m_{12}}{\pi} \\
		&\approx -  \beta_{CP(1)} \cdot (m_1 - m_2) + i \, (m_1 - \ov{m})  + \frac{m_1 - m_2}{\pi} 
\label{CP1_Z_quasiclassical}
\end{aligned}
\end{equation}
where $\ov{m}$ is the average of the first two masses, see \eqref{mbar_mtilde}.
The second term represents the fractional U(1) charge of the soliton \cite{ShVanZwi}. 
Indeed \eqref{CP1_Z_quasiclassical} can be compared to the Dorey quasiclassical formula \cite{Dorey} for the central charge
\begin{equation}
	Z_\text{kink} = -( \beta_{CP(1)} \, T - i  \, q ) \, \delta m_{12}
\label{dorey_quasiclassical}
\end{equation}
where $T=+1$ is the topological charge, while $q$ is the kink global (or \textquote{dyonic}) charge. Comparing \eqref{CP1_Z_quasiclassical} and \eqref{dorey_quasiclassical} we see that the kink dyonic charge is $q = 1/2$. The last term in \eqref{CP1_Z_quasiclassical} is the central charge anomaly \cite{ShVanZwi}. For details see e.g. \cite{SYrev,ShifmanLectureTop}.

In the limit when we are far from any AD point, $\beta \gg \beta_{AD}$, the expression in the second line in \eqref{CP1_Z_quasiclassical} is actually valid for any mass parameters, not just in the $\mathbb{CP}(1)\;$ limit \eqref{CP1_limit}.

%
%

\section{Weak coupling spectrum}  
\label{sec:weak}

Now let us discuss the weak coupling spectrum. 
In the $\mathbb{CP}(1)\;$ limit \eqref{CP1_limit} at weak coupling, $\Lambda_{CP(1)} \ll \delta m_{12}$, a part of the spectrum coincides with the ordinary $\mathbb{CP}(1)\;$ model spectrum coming from the $n^P$ fields. 

The $\mathbb{CP}(1)\;$ spectrum \cite{Dorey} consists of elementary perturbative excitations and a tower of BPS dyonic kinks.
The perturbative states have a mass $|i (m_1 - m_2) |$. This can be understood on the classical level from the action 
\eqref{wcp22}. Suppose that field $n^1$ classically develops VEV equal to $\sqrt{\beta}$. Then  the first term with $P=1$ in the second line in \eqref{wcp22} forces $\sigma$ to acquire the classical value $\sqrt{2}\sigma =-m_1$, while the term with $P=2$ gives  the mass $|m_1-m_2|$ to $n^2$. Note, that this result obtained in the quasiclassical limit is in fact exact because of the 
BPS nature of this perturbative state.

The mass of a kink interpolating between the two vacua is $M_\text{kink} = |Z_\text{kink}|$ where the central charge is given by 
 \eqref{CP1_Z_quasiclassical}.
This kink is in fact a part of a dyonic tower with central charges
\begin{equation}
	D^{(n)} = Z_\text{kink} + n \cdot i (m_1 - m_2)
	\,, \quad
	n \in \mathbb{Z} \,,
\label{M2_tower}
\end{equation}
which can be interpreted as a bound state of the kink and $n$ quanta of perturbative states with the central charge 
 $i (m_1 - m_2)$.  
The number $n$ in \eqref{M2_tower} is a manifestation of the multiple logarithm brunches in \eqref{BPSmass}. 
It gives a contribution to the kink dyonic charge $q$, see the quasiclassical expression \eqref{dorey_quasiclassical}. 
The total dyonic  charge $q = n + 1/2$ also has a contribution coming from $Z_\text{kink}$ which makes  it non-integer. 
The presence of the tower \eqref{M2_tower} in the weak coupling region of $\mathbb{CP}(1)\;$ model was found in \cite{Dorey} using quasiclassical methods.

In our \wcpt model extra states are present too, coming from the $\rho^K$ fields. 
They include perturbative BPS states with masses $|i (m_P - m_K) |,\ P=1,2,\ K=3,4$. These states are seen at the classical level from the action \eqref{wcp22}. Say, in the classical vacuum $n^1=\sqrt{\beta}$, $\sqrt{2}\sigma =-m_1$, the  fields $\rho^K$
acquire masses $|m_1-m_3|$ and $|m_1-m_4|$ given by the second term in the second line in \eqref{wcp22}.
We will call these states \textquote{bifundamentals.} They are 2D \textquote{images} of bifundamental quarks of 4D SQCD upon 2D-4D
correspondence, see Sec.~\ref{sec:ntwo}.

If we relax  the $\mathbb{CP}(1)\;$ conditions  \eqref{CP1_limit}, the spectrum described above stays intact.
However we  get some extra states.
States from the dyonic tower \eqref{M2_tower} might form bound states with
\textquote{bifundamental} fermions $\wt{\psi}^{P}_K$. 
The central charge of the resulting state is given by \cite{DoHoTo}
\begin{equation}
	Z^{(n)}_\text{bound} = Z_\text{kink} + n \cdot i (m_1 - m_2) + i (m_1 - m_K) \,.
\label{Z_bound_state}
\end{equation}
These states are formed if the condition
\begin{equation}
	0 < \Re{\frac{m_1 - m_K}{m_1 - m_2}} \equiv 1 - \Re{\frac{m_2 - m_K}{m_2 - m_1}} < 1
\label{bound_stability_condition}	
\end{equation}
is satisfied for some $K \in \{3,4\}$ (and any $P$) \cite{DoHoTo}.
From the stability condition \eqref{bound_stability_condition} it is evident that there are no such bound states 
for our choice of quark masses, $\Delta m \gg \delta m_{12},\,\delta m_{34}$, see the first condition in  \eqref{CP1_limit}.
We will not consider these bound states here.

Here we have just described the spectrum at weak coupling  $\beta \gg \beta_{AD}$. It literally translates into the spectrum in the dual weak coupling region at $\beta \ll - \beta_{AD}$ with substitution of indices $P=1,2 \leftrightarrow K=3,4$.

%
%

\section{Mirror description and the strong coupling spec\-trum} 
\label{sec:mirror}

\begin{figure}
	\centering
	\includegraphics[width=0.7\linewidth]{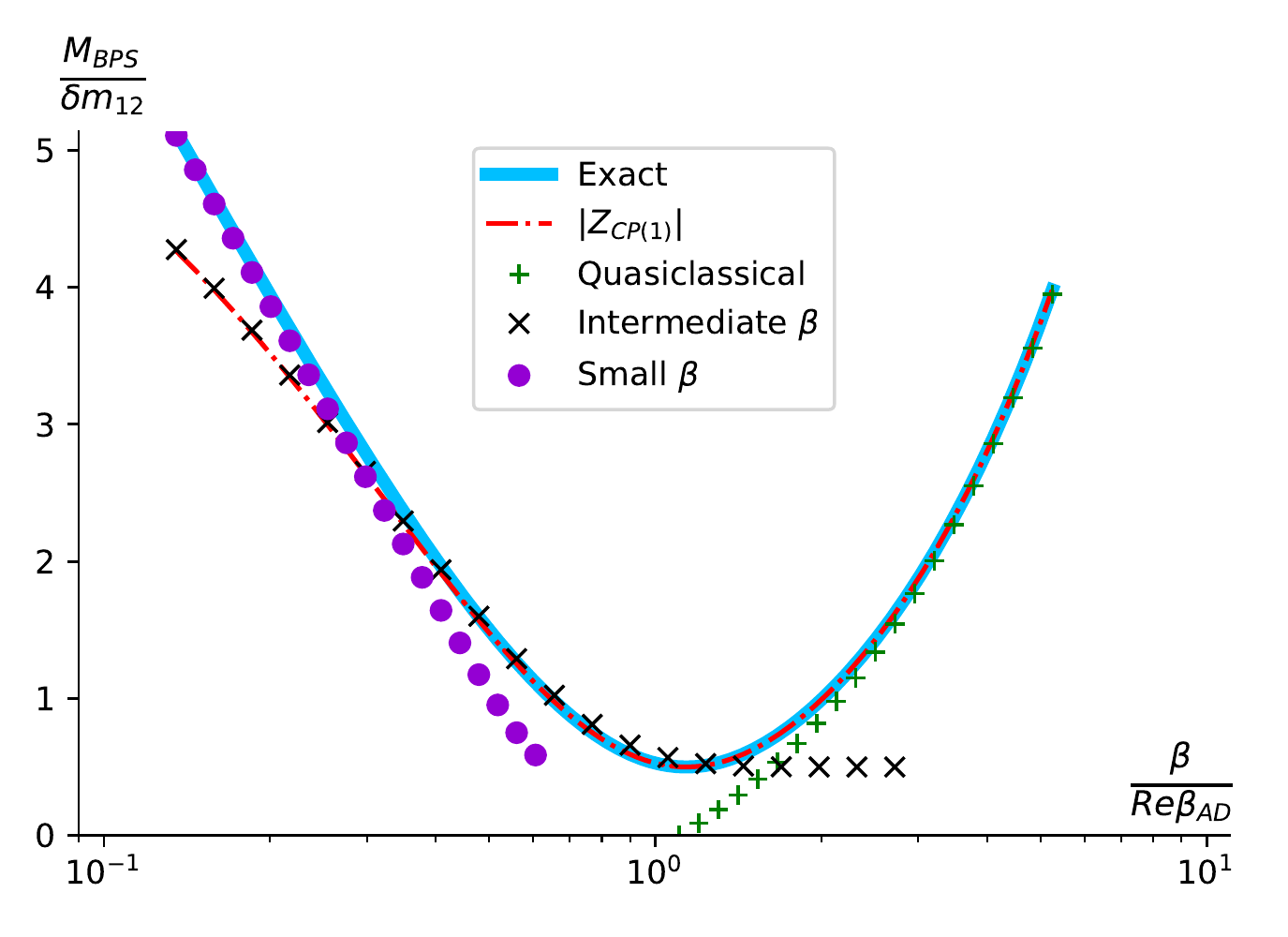}
	\caption{
		Various approximations of the kink mass $M_1$ (absolute value of the central charge): 
		$\mathbb{CP}(1)\;$ limit \eqref{CP1_Z}, 
		quasiclassical \eqref{CP1_Z_quasiclassical}, 
		intermediate $\beta$ \eqref{kink_mass_P_CP1_1},
		small $\beta$ \eqref{kink_mass_P_beta0}.
		Fixed $\Delta m / \delta m_{12} = 10$.
	}
\label{fig:M_approximations}
\end{figure}

In this section we will investigate the BPS kink spectrum in the strong coupling domain where $\beta$ is small, $\beta \ll \beta_{AD}$. 
(For comparison of various limits of the kink mass, see Fig.~\ref{fig:M_approximations}.)
We will generalize the analysis of  \cite{Shifman:2010id} carried out for asymptotically free \wcp models to the present case of 
conformal \wcpt model.

\subsection{Mirror superpotential}

To this end we will implement the mirror description of kinks \cite{FFS,HoVa} of the \wcpt model \eqref{wcp22}. 
The formula for the mirror superpotential is
\begin{equation}
	{\cal W}_{\rm mirror}(X_P, Y_K)
		= - \frac{\Lambda}{4 \pi} \left[ \sum_P X_P - \sum_K Y_K - \sum_P \frac{m_P}{\Lambda} \ln X_P + \sum_K \frac{m_K}{\Lambda} \ln Y_K \right] \,.
\label{W_mirror}		
\end{equation}
Here, the indices run as $P = 1, 2$, $K = 3, 4$. Parameter $\Lambda$ is an auxiliary parameter of dimension of mass which will cancel in the very end. 

The fields $X_P, Y_K$ are subject to the  constraint
\begin{equation}
	\prod_P X_P = e^{- 2 \pi \beta} \prod_K Y_K \,.
\label{mirror_constraint}	
\end{equation}
The VEVs  of $X_P, Y_K$ can be obtained by minimizing the superpotential \eqref{W_mirror} and using the above constraint
\cite{Shifman:2010id,HoVa}. Below we use a simplified approach which utilizes the relation of $X_P, Y_K$ to the $\sigma$ solutions of the vacuum equation \eqref{2d_equation} \cite{Shifman:2010id,HoVa},
\begin{equation}
	X_P = \frac{\sqrt{2}\sigma + m_P}{\Lambda}
	\,, \quad 
	Y_K = \frac{\sqrt{2}\sigma + m_K}{\Lambda} \,.
\label{mirror-x_map}	
\end{equation}

For a kink interpolating between two vacua $Vac_1$ and $Vac_2$,  the central charge is given by an exact formula
\begin{equation}
	Z_\text{kink} = 2 \left[ {\cal W}_{\rm mirror}(Vac_2) - {\cal W}_{\rm mirror}(Vac_1) \right] \,,
\label{mirror_kink_mass}	
\end{equation}
while its mass $M_\text{kink} =|Z_\text{kink}|$, see \eqref{MZ}.

\subsection{Kinks at intermediate $\beta$}

As a warm-up exercise we are going to consider the $\mathbb{CP}(1)\;$ limit \eqref{CP1_limit}. 
In the intermediate domain  $\delta m_{12} \ll \Lambda_{CP(1)} \ll \Delta m$ (or, equivalently, $1 \ll \beta \ll \beta_{AD}$), the effective \cpone model is at strong coupling, but at the same time we can use the large-$\beta$ expansion. 
The solutions of the vacuum equation \eqref{2d_equation} are given by $\sqrt{2}\sigma_\pm \approx - \Delta m / 2 \pm \Lambda_{CP(1)}$,
which yields two mirror vacua:
\begin{center}
\begin{tabular}{ c | c }
  $Vac_1$ at $\sigma = \sigma_+$ & $Vac_2$ at $\sigma = \sigma_-$ \\[2mm]
  \hline
  $X_1 \approx X_2 \approx \frac{\Lambda_{CP(1)}}{\Lambda}$ & $X_1 \approx X_2 \approx - \frac{\Lambda_{CP(1)}}{\Lambda}$ \\[2mm]
  $Y_3 \approx Y_4 \approx - \frac{\Delta m}{\Lambda}$ & $Y_3 \approx Y_4 \approx - \frac{\Delta m}{\Lambda}$
\end{tabular}
\end{center}
For both vacua the constraint \eqref{mirror_constraint} is satisfied:
\begin{equation}
	\prod_P X_P = e^{- 2 \pi \beta} \prod_K Y_K \approx \frac{\Lambda_{CP(1)}^2}{\Lambda^2} \,.
\end{equation}
There are different types of kinks interpolating between these vacua.


\begin{figure}[h!]
    \centering
    \begin{subfigure}[t]{0.4\textwidth}
        \centering
        \includegraphics[width=\textwidth]{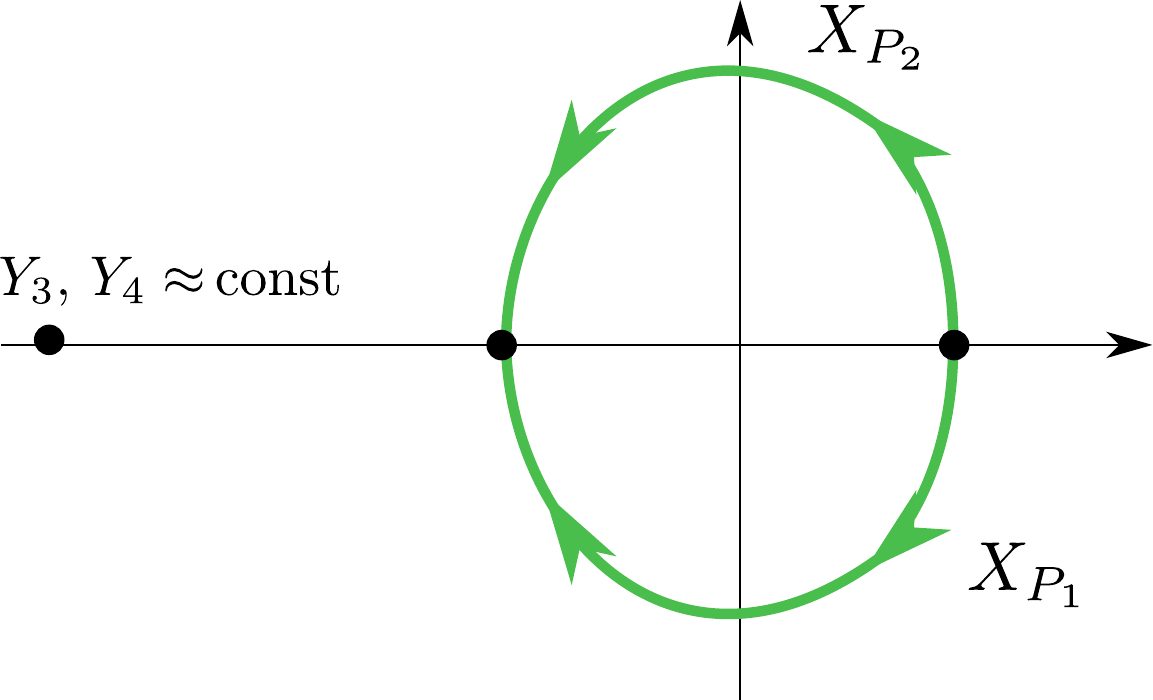}
        \caption{$P$-kinks}
        \label{fig:mirror_kinks_intermediate:P}
    \end{subfigure}%
    ~
    \begin{subfigure}[t]{0.4\textwidth}
        \centering
        \includegraphics[width=\textwidth]{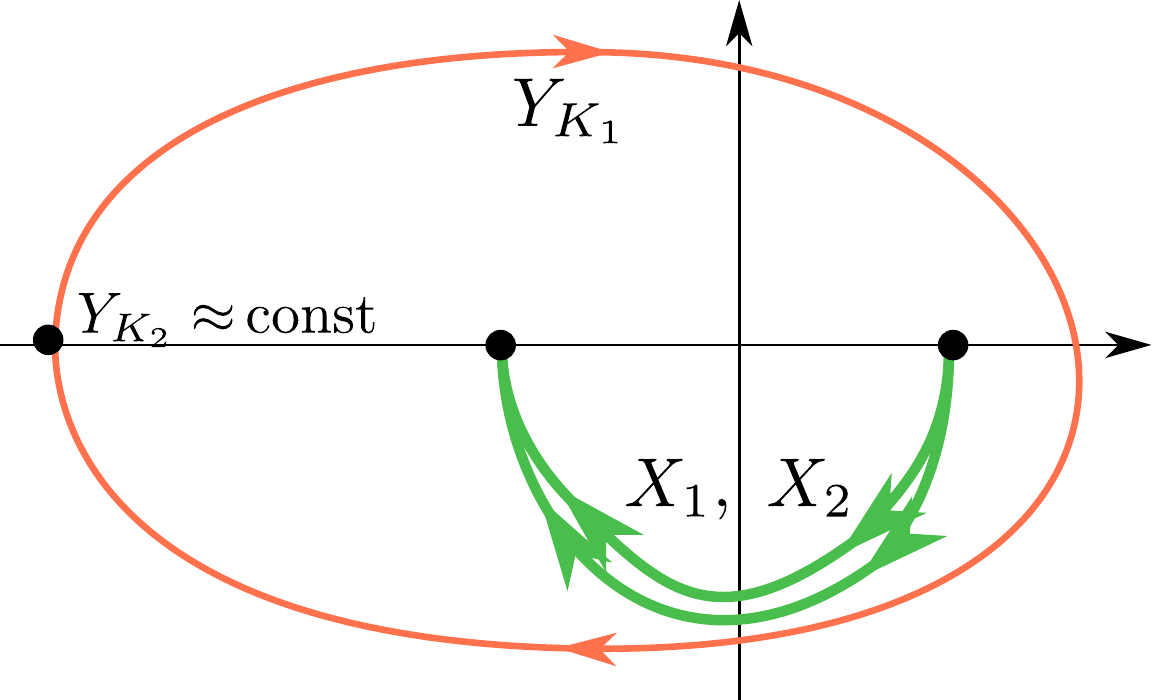}
        \caption{$K$-kinks}
        \label{fig:mirror_kinks_intermediate:K}
    \end{subfigure}%
\caption{
	Trajectories of $X_P$ and $Y_K$ in the mirror representation of a kink at intermediate $1 \ll \beta \ll \beta_{AD}$
}
\label{fig:mirror_kinks_intermediate}
\end{figure}

\paragraph{$n$-kinks} For these kinks, the two $X_P$ wind in the opposite directions, while the $Y_K$ stay intact to preserve the constraint \eqref{mirror_constraint}; see Fig.~\ref{fig:mirror_kinks_intermediate:P}. Since the arguments of $X_P$ change, the logarithms $\ln X_K$ in 
\eqref{W_mirror} acquire imaginary parts. There are two kinks of this type, depending on which flavor winds clockwise and which counter clockwise. 
From the central charge formula \eqref{mirror_kink_mass} we obtain for  these kinks
\begin{equation}
	Z_P =  \frac{2 \Lambda_{CP(1)}}{\pi} + i(m_P - \ov{m})  \,,
	\quad
	P = 1,2 \,.
\label{kink_mass_P_CP1_1}	
\end{equation}
The average mass $\ov{m}$ is defined in \eqref{mbar_mtilde}, and we used that $m_1 - m_2 = 2 (m_1 - \ov{m}) = - 2 (m_2 - \ov{m})$. 
The corresponding kink masses are  given by  the absolute values of central charges in \eqref{kink_mass_P_CP1_1}.

This formula is known in strongly coupled $\mathbb{CP}(1)\;$ and can be derived by expanding the central charge \eqref{CP1_Z} in powers of the small parameter $\delta m_{12} / \Lambda_{CP(1)}$. Namely, the  central charge \eqref{CP1_Z} reduces to the central charge of $P=1$ kink in \eqref{kink_mass_P_CP1_1} at small $\delta m_{12}$.

In the limit of equal $m_1$ and $m_2$  ($\delta m_{12}=0$) two kinks in \eqref{kink_mass_P_CP1_1} degenerate and form a doublet of the first SU(2) in the global group \eqref{b:globgroup}, namely
\beq
n{\rm -kinks}: \quad \left(\textbf{2},\,\textbf{1},\, 0 \right)
\label{n_kink_rep}
\eeq
The fact that kinks of the $\mathbb{CP}(N-1)\,$ model  at strong coupling form a fundamental representation of SU$(N)$ and transform as $n^P$ fields was discovered by Witten long ago \cite{W79}. Later it was confirmed by Hori and Vafa  \cite{HoVa} using the mirror representation. This is reflected in our notation of kinks in \eqref{kink_mass_P_CP1_1} as $n$-kinks. 

\paragraph{$\rho$-kinks} For these kinks, the two $X_P$ wind in one directions, while exactly one of the $Y_K$ winds double in the same direction according to \eqref{mirror_constraint}; see Fig.~\ref{fig:mirror_kinks_intermediate:K}. Then the corresponding logarithms in \eqref{mirror_kink_mass} acquire imaginary parts. There are again two kinks of this type, depending on which flavor $Y_K$ winds.
The kink central charges  are given by
\begin{equation}
	Z_K =  \frac{2 \Lambda_{CP(1)}}{\pi} + i(m_K - \ov{m})  \,,
	\quad
	K = 3,4 \,.
\label{kink_mass_K_CP1_1}
\end{equation}
These are new states, not present in $\mathbb{CP}(1)$. 
At $\beta \gg 1$ these states are much heavier than the $n$-kinks.

In the limit of equal $m_3$ and $m_4$  ($\delta m_{34}=0$) the two kinks in \eqref{kink_mass_K_CP1_1} degenerate and form a doublet of the second SU(2) in \eqref{b:globgroup}, namely
\beq
\rho{\rm -kinks}: \quad \left(\textbf{1},\,\textbf{2},\, 1\right).
\label{rho_kink_rep}
\eeq
These kinks behave as $\rho$ fields, see \eqref{b:repsnrho}. In what follows we will heavily use the fact that $n$-kinks and $\rho$-kinks transforms as
$n^P$ and $\rho^K$  fields.

Note that the BPS spectrum of \wcpt model at strong coupling is very different from that at weak coupling. First, there are no
perturbative states at strong coupling. Second, instead of the infinite tower of dyonic kinks \eqref{M2_tower} present at weak coupling
at strong coupling we have just four kinks which belong to representations  \eqref{n_kink_rep} and \eqref{rho_kink_rep}
of the global group \eqref{b:globgroup}. Note also that  global charges of kinks in the perturbative tower \eqref{M2_tower}
associated with the single mass difference $(m_1 -m_2)$. In contrast the kink global charges at  strong coupling are associated with all masses $m_A$ present in the model.
We study CMS where the transformations of the BPS spectra occurs in Sec.~\ref{sec:CMS}.

The above  results can be directly generalized to the dual domain of negative $\Re\beta$.  
When $\beta$ is in the intermediate domain between $- |\beta_{AD}|$ and $-1$, the  $n^P$  fields are heavy and decouple, and we are again left with a $\mathbb{CP}(1)\;$ model, only this time comprised of the  $\rho^K$ fields and a new strong coupling scale
\begin{equation}
	\Lambda_{\wt{CP}(1)} = \Delta m \, e^{+ \pi \beta} \,.
\end{equation}
Roles of $n$-kinks and $\rho$-kinks are reversed. Their central charges  are given by
\begin{equation}
	Z_P =  \frac{2 \Lambda_{\wt{CP}(1)}}{\pi} + i(m_P - \wt{m}) 
	\ , \quad
	Z_K =  \frac{2 \Lambda_{\wt{CP}(1)}}{\pi} + i(m_K - \wt{m}) 
	\,.
\label{kink_mass_wtCP1_1}	
\end{equation}
with $\wt{m}$ defined in \eqref{mbar_mtilde}. As we can see, now the $Z_K$-kinks are light. 
Note that these results match with the $S$-duality transformation \eqref{2d_S-duality}.

Finally, we note that apart from the kinks just described, there can be kinks described by $X_P,\ Y_K$ fields winding in the opposite direction. Say, for $\rho$-kinks on Fig.~\ref{fig:mirror_kinks_intermediate:K} the $X_P$ may wind in the upper half plane, with $Y_K$ winding counter clockwise. These kinks turn out to be $n=+1$ states from  the strong coupling tower of higher winding  states discussed in  the next subsection.

\subsection{Kinks near the origin $\beta = 0$}

\begin{figure}[h]
    \centering
    \begin{subfigure}[t]{0.4\textwidth}
        \centering
        \includegraphics[width=\textwidth]{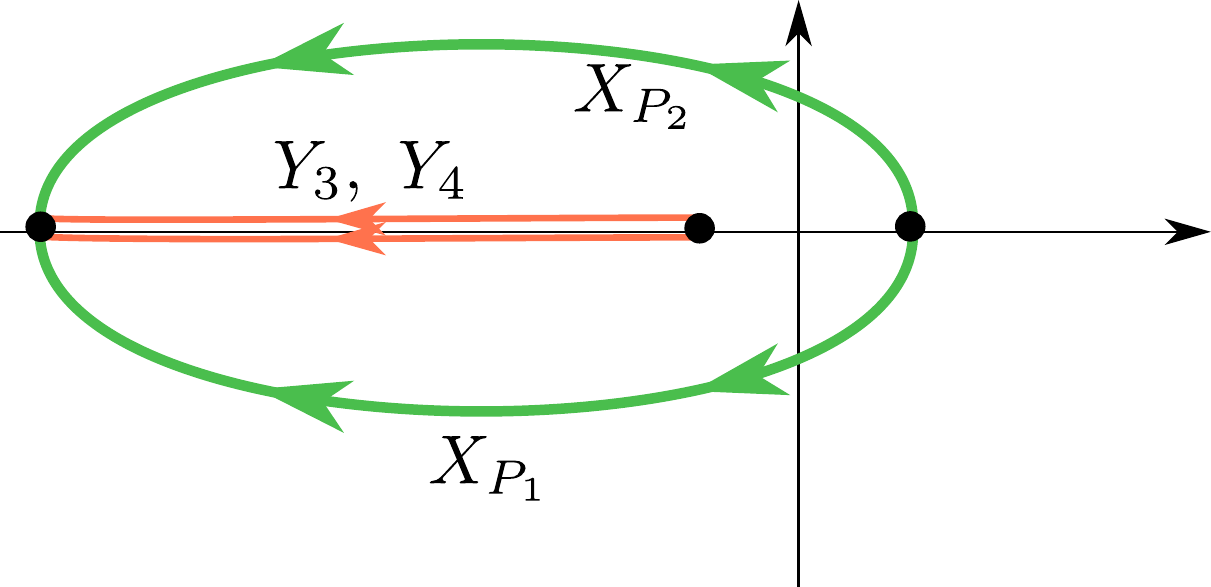}
        \caption{$P$-kinks}
        \label{fig:mirror_kinks_origin:P}
    \end{subfigure}%
    ~
    \begin{subfigure}[t]{0.4\textwidth}
        \centering
        \includegraphics[width=\textwidth]{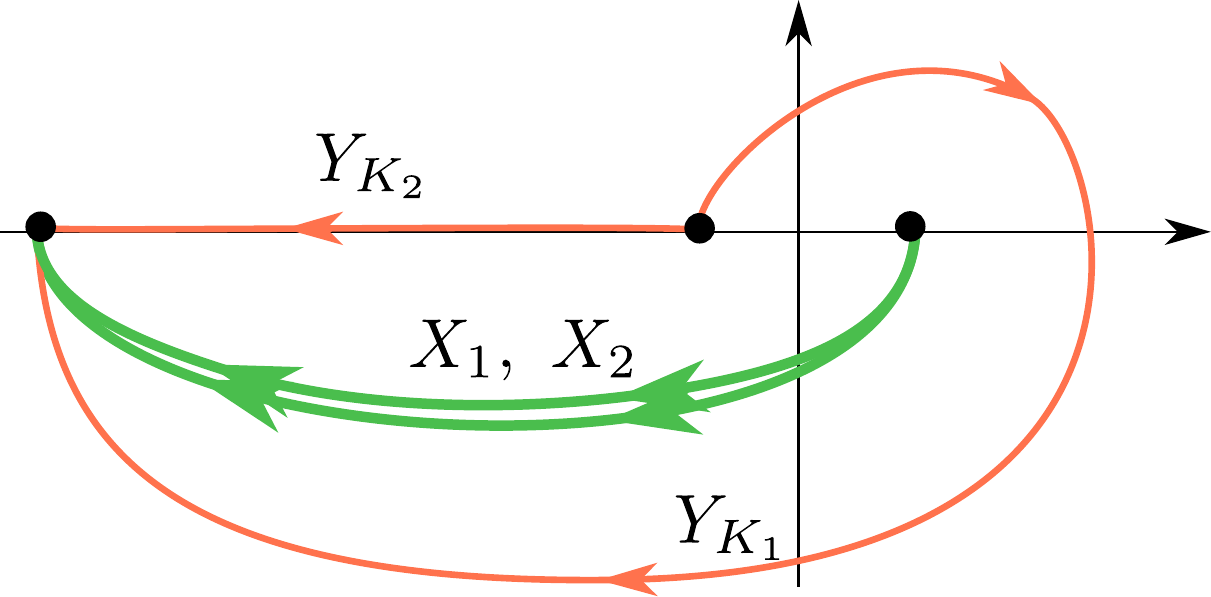}
        \caption{$K$-kinks}
        \label{fig:mirror_kinks_origin:K}
    \end{subfigure}%
\caption{
	Trajectories of $X_P$ and $Y_K$ in the mirror representation of a kink at $\beta \to 0$.
}
\label{fig:mirror_kinks_origin}
\end{figure} 

Now consider the limit $\beta \to 0$. 
In the vicinity of the origin the last condition in \eqref{CP1_limit} is badly broken, and all of the \wcpt fields $n^P$, 
$\rho^K$ \eqref{wcp22} play an important role.

In this limit we can use the small-$\beta$ expansion. We have $e^{- 2 \pi \beta} \approx 1 - 2 \pi \beta$, and the $\sigma$-vacua \eqref{roots_symmetric} are approximately
\begin{equation}
	\sqrt{2} \sigma_+ \approx \frac{\delta m_{12}^2 - \delta m_{34}^2}{8 \Delta m ^2} 
	\,, \quad 
	\sqrt{2} \sigma_- \approx - \frac{\Delta m}{\pi \beta} \,.
\label{mirror_roots_approx_beta=0}
\end{equation}
Without loss of generality we can consider the limit when $\sigma_+ \approx 0$. Then, the two mirror vacua are given by
\begin{center}
\begin{tabular}{ c | c }
  $Vac_1$ at $\sigma = \sigma_+$ & $Vac_2$ at $\sigma = \sigma_-$ \\
  \hline
  $X_P \approx m_P  / \Lambda$ & $X_1 \approx X_2 \approx - \frac{\Delta m}{\pi \beta \Lambda}$ \\
  $Y_K\approx m_K  / \Lambda$ & $Y_3 \approx Y_4 \approx - \frac{\Delta m}{\pi \beta \Lambda}$
\end{tabular}
\end{center}

Again, there are two types of kinks.
$n$-kinks are obtained when, say, $X_1$ picks up the phase $+i\pi$, $X_2$ picks up $-i\pi$, while the phases of $Y_K$ remain intact; see Fig.~\ref{fig:mirror_kinks_origin:P}. There is also a kink for which the roles of $X_1$ and $X_2$ are reversed. We have the total of two kinks with central charges
\begin{equation}
	Z_P =  \frac{m_1 + m_2 - m_3 - m_4}{2 \pi} \ln \frac{2}{\pi \beta} + i (m_P - \ov{m})
\label{kink_mass_P_beta0}
\end{equation}
where $\ov{m}$ is defined in \eqref{mbar_mtilde}.

Similarly, the $\rho$-kinks are obtained when two $X_P$ wind with the same phase, while exactly one of the $Y_K$ winds twice as much in accord with \eqref{mirror_constraint}; see Fig.~\ref{fig:mirror_kinks_origin:K}. The central charges  of these kinks are given by
\begin{equation}
	Z_K =   \frac{m_1 + m_2 - m_3 - m_4}{2 \pi} \ln \frac{2}{\pi \beta} + i (m_K - \ov{m})  \,.
\label{kink_mass_K_beta0}
\end{equation}
We immediately observe that the kink masses are singular at the self-dual point $\beta=0$. They are very heavy in the vicinity of this point.

To get the kink spectrum at ${\rm Re}\beta <0$ we can analytically continue  \eqref{kink_mass_P_beta0} and
\eqref{kink_mass_K_beta0} to $\beta \to \tilde{\beta}=-\beta$. The log terms in \eqref{kink_mass_P_beta0}, 
\eqref{kink_mass_K_beta0} give $i(m_1 + m_2 - m_3 - m_4)/2$ which converts $\bar{m}$ into $\tilde{m}$.
Note, that this matches with the $S$-duality transformation \eqref{2d_S-duality}.

Now observe that the central charges of $n$ and $\rho$-kinks \eqref{kink_mass_P_beta0} and \eqref{kink_mass_K_beta0} have a branching point at $\beta=0$. This is a new feature absent in asymptotically free versions of \wcp models. What is the meaning of this branching point? Below in this section we will argue that the self-consistency
of the BPS spectrum requires the presence of a new tower of  higher winding states in our conformal \wcpt model.
This tower is present  only at strong coupling and decays as we move to large $\beta$. This can be seen as follows.

Consider changing the coupling constant  $\beta$ along some trajectory in the complex plane. This trajectory may stretch from the weak coupling region $\beta \gg \beta_{AD}$ through the strong coupling domain into the dual weak coupling region $\beta \ll -\beta_{AD}$. This trajectory may also encircle an AD point and go through a cut on a different sheet. The charges of various BPS states change, but there are CMS starting at the AD points, and the BPS spectrum as a whole stays intact. The would-be 
\textquote{extra} states decay on CMS \cite{DoHoTo}.

However, this trajectory may also go full circle around the singularity $\beta=0$. It can also encircle this point several times. There are no CMS starting at $\beta=0$ and extending outwards. What we end up with is another set of BPS states. From the expressions for the kink central charges \eqref{kink_mass_P_beta0}, \eqref{kink_mass_K_beta0} we see that if we go around the origin $n$ times, then the central charge of the BPS kinks becomes
\begin{equation}
\begin{aligned}
	Z_A^{[n]} &=  \frac{m_1 + m_2 - m_3 - m_4}{2 \pi} \ln \frac{2}{\pi \beta} + i (m_A - \ov{m}) + i \, n \cdot (m_1 + m_2 - m_3 - m_4) \,, \\
	&\frac{\pi}{2} \leqslant \arg\beta < - \frac{3 \pi}{2} \,.
\end{aligned}
\label{Z_higher_windings}
\end{equation}
Here the argument of $\beta$ is constrained so as to account for the cut, see Fig.~\ref{fig:CMS_right_13}.
Does it mean that the full BPS spectrum changes as we go to other sheets?

The way to resolve this issue is to assume that that in fact {\em all} of the states \eqref{Z_higher_windings} are already present at strong coupling on the first sheet. When we wind circles around the origin, this tower of states $Z_A^{[n]}$ simply shifts in the index $n$. Since this index runs over all integers and the number of states in the tower is infinite, the whole BPS spectrum is in fact $2\pi$-periodic with respect to $\arg\beta$.

The new tower \eqref{Z_higher_windings} is present only at strong coupling. At weak coupling it decays. We study associated CMS
and decay processes in Appendix~\ref{sec:higher_winding}.

%
%

\section{Curves of marginal stability}
\label{sec:CMS}

In this section we will present the curves of marginal stability (CMS) for various decays of BPS states.

As was stated above, in \wcpt theory under consideration  the coupling $\beta$ does not run.
We want to understand transformations of the BPS spectrum at different values of $\beta$, particularly weak vs. strong coupling regions as well as at  $\Re\beta < 0$. In order to better capture the relevant effects, we are going to investigate more closely how the particle spectrum depends on $\beta$ while holding the masses\footnote{Or, rather, their ratios since the CMS positions on the $\beta$ plane can depend only on dimensionless parameters, and there is no dynamical strong coupling scale $\Lambda$ in \wcpt.} 
$m_A$ fixed. 
To this end we will study curves of marginal stability (CMS) on the complex $\beta$ plane.
Since the $\theta_{2d}$ angle is $2\pi$ periodic, the whole picture of spectra will be periodic as well.

In Sections \ref{sec:weak} and \ref{sec:mirror} we saw that the strong and weak coupling spectra are different. At weak coupling $\beta \gg \beta_{AD}$, we observed  the dyonic tower \eqref{M2_tower} as well as \textquote{perturbative} states with the central charge $i(m_1 - m_2)$. They are not present at strong coupling and must decay on a CMS separating the strong and  weak coupling regions. We will refer to these CMS as the {\em primary curves}.

Moreover, at strong coupling we have $\rho$-kinks not present at weak coupling.
Correspondingly, CMS must exist on which these states will decay. We will call these the {\em secondary curves}.

Finally, we saw that at weak coupling there are the so-called bifundamentals -- the perturbative  states with masses $|i(m_P - m_K)|,\ P=1,2,\ K=3,4$. These states do not decay even at strong coupling. They are present everywhere on the $\beta$ plane. To see that  this is the case
suffice it to note that in the massless limit $m_A \to 0$ 4D SQCD has a Higgs branch formed by bifundamental quarks. This Higgs branch is protected by supersymmetry and present at all couplings. Through the 2D-4D correspondence we conclude that 2D bifundamentals are also present at all $\beta$.

Below in this section we study the primary CMS while the are secondary CMS discussed in Appendix~\ref{sec:secondary_CMS}.

\begin{figure}[h!]
	\centering
	\includegraphics[width=0.7\linewidth]{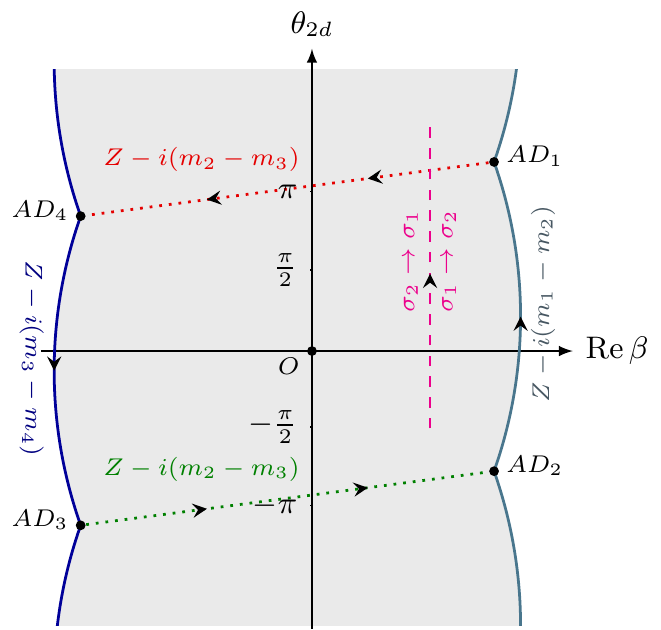}
	\caption{
		Primary CMS (solid lines on the left and on the right, schematically) and central charge shifts (see Appendix~\ref{sec:Z_windings}).
		$AD_A$ are the Argyres-Douglas points where the corresponding central charge $Z_A$ vanishes.
		The  masses $m_A$ are generic.
		Grey region is the strong-coupling domain.
		 When going from one AD point to the next shown in this  figure one observes phase shifts.
		There is also a $\mathbb{Z}_2$ 
			transformation (exchanging the $\sigma$ roots) when shifting $\theta_{2d} \to \theta_{2d} + 2\pi$.
		Apart from that, the picture is $2 \pi$ periodic with respect to $\theta_{2d}$.
		$AD_2 = AD_1 - 2 \pi i$, $AD_3 = AD_4 - 2 \pi i$.  
	}
\label{fig:CMS_type1}
\end{figure}

\subsection{Primary curves in the \boldmath{$\beta$} plane}

As was discussed above, when we pass from large $\beta \gg \beta_{AD}$ to strong coupling $\beta \sim 1$, the perturbative states with the central charge $i(m_1 - m_2)$ decay on CMS producing (dyonic) kink-antikink pairs. We can write the decay processes schematically as%
\footnote{Here and further on we use the notation with square brackets $[Z_A]$, $[i (m_1 - m_2)]$ to represent particles with the corresponding central charges. The central charge of an antiparticle equals negative of that of the particle.}
\begin{equation}
	\underbrace{[ i (m_1 - m_2) ]}_{\text{elementary quantum}} \to 
		\underbrace{[Z_1]}_{\text{dyon}}
		+ \underbrace{[- Z_2]}_{\text{antidyon}} \,.
\label{right_decay_W}		
\end{equation}
On the CMS, the central charges of the decaying particles must have the same argument, i.e. they must be collinear vectors in the complex plane. From this we can derive the equation for the CMS,
\begin{equation}
	\Im \left( \frac{ Z_P }{i (m_1 - m_2)} \right) = 0 
	\Leftrightarrow
	\Re \left( \frac{ Z_P }{m_1 - m_2} \right) = 0
	\,, \quad
	P = 1, \, 2 \,.
\label{CMS_right_curve}		
\end{equation}
The same decay curve describes the decay of the dyonic tower  \eqref{M2_tower} into the strong coupling states.

This curve separates the weak coupling region $\beta \gg \beta_{AD}$ from the strong coupling region. It passes through the AD points \eqref{AD_roots_simplecase_beta1_approx} with ${\rm Re}\,\beta >0$ where the mass of one of the $n$-kinks $[Z_P]$   vanishes. We denote these AD points AD$_P$, $${\rm AD}_2 ={\rm AD}_1 - 2\pi i\,,$$ see Appendix~\ref{sec:Z_windings} for a detailed discussion. Of course, the corresponding CMS is $2\pi$ periodic in $\theta_{2d}$. 

We solve \eqref{CMS_right_curve} numerically. The result is presented by 
 the r.h.s. curve on Fig.~\ref{fig:CMS_type1}. Note that $[Z_1]$ and  $[Z_2]$ kinks 
\eqref{kink_mass_P_CP1_1} present at strong coupling survive at the weak coupling region at positive $\beta\gg \beta_{AD}$.
In this region they belong to the tower  \eqref{M2_tower} with $n=0$ and $n=-1$ respectively. This is a well-known behavior: the states that can become massless at some points on CMS, are present both at weak and strong coupling \cite{SW1,Ferrari:1996sv}.  

Note that the CMS curve for \cpone model in the complex plane $\delta m_{12}$ is well-known \cite{ShVanZwi}. 
It is a closed curve around the origin which passes through the AD points. Our  curve in Fig.~\ref{fig:CMS_type1} (more exactly, its right branch at  ${\rm Re}\,\beta >0$) is a translation of the  curve in  \cite{ShVanZwi} into the 
$\beta$ plane. In the $\beta$ plane the curve is not closed. It is periodic in $\theta_{2d}$.

Analogously, when $\Re\beta$ is large but negative (l.h.s. on Fig.~\ref{fig:CMS_type1}), there are perturbative states whose central charge is $i(m_3-m_4)$ and a corresponding dyonic tower. Their decay curves satisfy
\begin{equation}
	\Re \left( \frac{ Z_K }{m_3 - m_4} \right) = 0
	\,, \quad
	K = 3, \, 4 \,.
\label{CMS_left_curve}		
\end{equation}
This CMS separates the dual weak coupling region $\beta \ll - \beta_{AD}$ from the strong coupling region. 
On Fig.~\ref{fig:CMS_type1} it is drawn on the left side.

We see two weak coupling regions in the complex plane of $\beta$ . They are separated by a strong coupling region which resembles a band stretched along the $\theta_{2d}$ direction. This is illustrated on Fig.~\ref{fig:CMS_type1}.

%
%

\section{Instead-of-confinement phase }
\label{sec:instead_of_conf}

In this section we use 2D-4D correspondence to confirm the instead-of-confinement phase in the bulk 4D SQCD at strong coupling. This phase  was discovered earlier in asymptotically free versions of SQCD \cite{SYdual}, see \cite{SYdualrev} for a review.

To this end we first consider our world-sheet \wcpt model on the non-Abelian string. In the previous sections  we have learned that 
the BPS spectrum of states is very different at weak and strong coupling. In particular, the perturbative states with
mass $|m_1-m_2|$ decay into say, $[Z_1]$ kink and $[-Z_2]$ antikink on the CMS on the the r.h.s. in Fig.~\ref{fig:CMS_type1} when we 
pass from the weak coupling region into the strong coupling one. At strong coupling these perturbative states do not exist.

The 2D-4D correspondence tells us that a similar process occurs on the Coulomb branch (at $\xi =0$) in 4D SQCD when we pass from the weak coupling region to the strong coupling one. The 2D perturbative states with mass $|m_1-m_2|$ correspond in the bulk theory (4D SQCD) to
BPS off-diagonal quarks $q^{kP}$, $P=1,2$, and  gluons. They do not exist at strong coupling. They decay into monopole and 
anti-monopole pair\,\footnote{We call all 4D states with non-zero magnetic charge monopoles although they can be dyons  carrying
also electric and global charges \cite{SW2}.}. 

Moreover, since the $n$-kinks of the 2D theory form doublets with respect to
the first SU(2) factor of the global group \eqref{b:globgroup}, see \eqref{n_kink_rep} and \eqref{rho_kink_rep}, the  monopoles and anti-monopoles formed as a result of the quark/gluon decay also transform as doublets and anti-doublets of the first SU(2) factor of the global group.

As we turn on $\xi$ at  weak coupling the 4D theory goes into the Higgs phase. Quarks $q^{kP}$, $P=1,2$, get screened by the condensate \eqref{Nf_qvev}. They combine with massive gluons to form a long non-BPS \ntwo multiplets with mass $g\sqrt{\xi}$,
see the review \cite{SYrev} for details. Moreover, at non-vanishing values of  $\xi$
the monopoles become confined by non-Abelian strings.

 Now, if we move to the strong coupling domain, the monopole and anti-monopole
created as a result of the quark/gluon decay cannot move apart. They are attached to two confining strings and form 
a monopole-antimonopole stringy meson shown in Fig.~\ref{monmb}a. Of course this meson is a non-BPS state. Its mass is of the
order of $\sqrt{\xi}$. Note, that this meson is formed also in the massless limit $m_A\to 0$. The mass scale in 4D SQCD is set by the FI 
parameter $\xi$. 

Thus we see that the screened quarks and gluons present in 4D SQCD in the Higgs phase at weak coupling do not survive when we move to strong coupling. They evolve into monopole-antimonopole stringy mesons. The phase which emerges at the strong coupling is called  {\em instead-of-confinement} phase \cite{SYdualrev}.

This phase is  an alternative to the ordinary confinement phase in QCD. The role of the constituent quarks in this phase is played by confined monopoles. Moreover, since monopoles and antimonopoles  transform as doublets and antidoublets of the first SU(2) factor of the global group \eqref{b:globgroup_d=4} the stringy mesons  appear in the singlet  or adjoint representations 
of the first SU(2) subgroup. This is similar to what happens in QCD: quark-antiquark mesons form the singlet or adjoint representation of the flavor group.

The same instead-of-confinement mechanism  works  if we start at large negative $\beta $ and pass through l.h.s CMS into strong coupling.
The monopole-antimonopole stringy mesons formed on this CMS appear in the singlet  or adjoint representations 
of the second SU(2) subgroup of the global group.
The strong coupling region between the r.h.s and l.h.s. CMS in Fig.~\ref{fig:CMS_type1} in 2D theory corresponds to 
the strong coupling domain around the large semicircle in Fig.~\ref{fig:coupling_traj} in terms of the complexified 4D
coupling $\tau$, see \eqref{tau_def} in Sec.~\ref{sec:2D_4D}. This is the region of instead-of-confinement phase in 4D
SQCD.

%
%

\section{Stringy baryon from field theory} \label{sec:baryon_2d}


In this section we show that the presence of the baryonic state \eqref{brep} found as a  massless string state of the critical string theory on the non-Abelian vortex in our 4D \ntwo SQCD can be confirmed  using purely field-theoretical methods.
Let us start with the world-sheet \wcpt model on the string at strong coupling near the origin in the $\beta$ plane.
The baryonic charge $Q_B=2$ and the absence of the Cartan charges with respect to both SU(2) factors of the global group 
\eqref{b:globgroup} suggests  that this state can be formed as a BPS bound state of two different $n$-kinks and two different $\rho$-kinks arranged  on the infinite straight  string  in the following order
\beq
[Z_P]|_{1\to 2} + [Z_K]|_{2\to 1} + [Z_{P'}]|_{1\to 2} + [Z_{K'}]|_{2\to 1}, \qquad P\neq P', \quad K\neq K'\,,
\label{4_kinks}
\eeq
where the subscript $|_{1\to 2}$ ($|_{2\to 1}$) denotes the   kink  interpolating from vacuum 1  to vacuum 2 (vacuum 2  to vacuum 1).
The central charges of the second and last kinks come with the minus sign, see \eqref{mirror_kink_mass}, and the net central charge 
of the bound state \eqref{4_kinks} is 
\beq
Z_b = i(m_1 + m_2 -m_3 -m_4),
\eeq
see \eqref{kink_mass_P_beta0} and \eqref{kink_mass_K_beta0}.
Note that this state cannot have a net topological charge. 
The 2D  topological charge translates into  4D  magnetic charge of a monopole.
Clearly, the baryon (or any other hadron) cannot have 
color-magnetic charge because  magnetic charges are confined in 4D SQCD.

The 4-kink composite state \eqref{4_kinks} transforms under the global group \eqref{b:globgroup} as 
\begin{equation}
	n^P \rho^K n^{P'} \rho^{K'} = w^{PK} w^{P'K'},
\label{baryon_candidate}		
\end{equation}
where we use the gauge invariant mesonic variables \eqref{w}.
  It is clear that \eqref{baryon_candidate} is symmetric with respect to indices $P,P'$ and $K,K'$. Thus, this state is in 
	the triplet representation  ({\bf 3}, {\bf 3}, 2) of the global group. This is not what we need.
	
The singlet representation ({\bf 1}, {\bf 1}, 2) \eqref{brep} we are looking for would correspond to $\det(w)$.
But it is zero, see \eqref{coni}! 

However, recall that it is zero only in \wcpt model formulated in terms
of  $n$'s and $\rho$'s. 
Let us take the massless limit $m_A\to 0$ and go to the point $\beta=0$.
Our world-sheet \wcpt  theory on the conifold allows a marginal deformation of the conifold complex structure
 at $\beta =0$ \cite{NVafa,Candel}, namely 
\beq
\det (w) = b
\label{deformedconi1},
\eeq
where $b$ is a complex parameter, see \eqref{deformedconi} in Sec.~\ref{conifold}.
  This deformation preserves Ricci-flatness which ensures that 2D world-sheet  theory is still conformal and has no
dynamical	$\Lambda$ scale, so the   baryonic state $\det (w)$ which emerges in the deformed theory is massless. 
	
Next we use the 2D-4D correspondence that ensures that at $\beta=0$ and non-zero $b$ there is a similar massless baryonic BPS state in
 4D SQCD formed by  four monopoles.  At non-zero values of $\xi$, the monopoles are confined and  this baryon is represented by  a
necklace configuration formed by four monopoles connected by confining strings, see Fig.~\ref{monmb}b. At non-zero $\xi$ this state becomes a well-defined localized state in 4D SQCD. Its size is determined by $1/\sqrt{\xi}$. Note, that this 
baryon is still a short massless BPS hypermultiplet at nonvanishing $\xi$ because there is no other massless BPS state with the same quantum numbers to combine with to form a long multiplet \footnote{This is similar to what happens with the bifundamental quarks which
remain massless BPS states as we switch on a non-zero $\xi$.}.

Now we can address the question: what is the origin of the marginal deformation parameter $b$ in 4D SQCD? As was already mentioned in  
Sec.~\ref{sec:b_meson:intro}, it can be a marginal coupling constant which respects \ntwo supersymmetry or a VEV of a dynamical state. The coupling constant
$\beta$ is associated with the deformation of the K\"ahler class of the conifold rather then its complex structure. Moreover, note
that the deformation parameter $b$ cannot be a coupling associated with gauging of any symmetry of the global group 
\eqref{b:globgroup_d=4}
because it has non-zero $Q_B$.
This leads us to the conclusion that $b$ is a VEV of a dynamical state, namely the VEV of the massless stringy four-monopole baryon discussed above.

The baryon $b$ exists only at the origin $\beta = 0$. As we move away from $\beta=0$, it must decay on a point-like degenerate CMS which tightly wraps the origin. It decays into two massless bifundamental quarks which belong to the representation ({\bf 2}, {\bf 2}, 1) of the global group.

Thus we confirm that a new non-perturbative Higgs branch of real dimension ${\rm dim}\,{\cal H}= 4$ opens up in our 4D SQCD at the point $\beta=0$ (up to $2\pi$ periodicity of the $\theta_{2d}$ angle) in the massless limit.
Most likely the perturbative Higgs branch \eqref{b:dimH} formed by bifundamental quarks is  lifted  at $b\neq 0$.
The point $\beta=b=0$ is a  phase transition point, a singularity where two Higgs branches meet. This issue needs future clarification.

%
%

\section{Detailing the 2D-4D correspondence} \label{sec:2D_4D}

As was stated above, the sigma model \eqref{wcp22} is an effective world-sheet theory on the semilocal non-Abelian string in four-dimensional \ntwo SQCD. Generally speaking, if we consider the bulk theory with the gauge group U($N$) and $N < N_f \leqslant 2N$ flavors of quarks, then the world-sheet theory is the weighted sigma model \wcp. In this Chapter we focus on the case $N_f = 2N = 4$.

The mass parameters $m_A$ of the world-sheet theory \eqref{wcp22} are the same as quark masses in the bulk 4D SQCD. The two-dimensional coupling $\beta$ \eqref{beta_complexified} is also related to the four-dimensional complexified coupling constant $\tau_\text{SW}$ which is defined as
\begin{equation}
	\tau_\text{SW} = i \, \frac{8\pi}{g^2}  + \frac{\theta_{4d}}{\pi} \,.
\label{tau_def}	
\end{equation}
Here $\theta_{4d}$ is the four-dimensional $\theta$ angle. 
We will start this section from derivation of the corresponding relation.

\subsection{Relation between the couplings}
\label{rbc}

In the weak coupling limit, the known classical-level relation between the couplings of the bulk and world-sheet theories
is \cite{ABEKY,SYmon}
\begin{equation}
	\Re\beta \approx \frac{4 \pi}{g^2} \,.
\end{equation}
But what is the exact formula?

To establish a relation applicable at the quantum level, we are going to use the 2D-4D correspondence -- the coincidence 
of the BPS spectra of monopoles in 4D SQCD and  kinks in 2D world-sheet $\mathbb{WCP}(N,\tN)$ model, see Sec.~\ref{2-4}. As was already noted  the key technical reason behind  this
coincidence is that the VEVs of $\sigma$ given by the exact twisted superpotential  coincide
with the double roots of the Seiberg-Witten curve \cite{SW2} in the quark vacuum of
the 4D SQCD \cite{Dorey,DoHoTo}. Below we use this coincidence  to derive the exact relation between 4D coupling $\tau_{SW}$ and 2D coupling $\beta$ in the theory at hand, $N_f = 2N = 4$, where both couplings do not run.

 Mathematically, this can be formulated as follows. Consider the Seiberg-Witten curve of the bulk SQCD.
The Seiberg-Witten (SW) curve for the SU($N$)  gauge theory with $N_f = 2 N$ flavors was derived in \cite{ArgPlessShapiro,APS}. 
It has the form
\begin{equation}
	y^2 = \prod_{a=1}^{N}(x-\phi_a)^2 + h(h+2) \prod_{i=1}^{N_f} (x + h m_S + m_i)
	\,, \quad
	N_f = 2 N \,.
\label{4d_curve_sun}	
\end{equation}
Here, $$h \equiv h(\tau_\text{SW})$$  is a modular function \eqref{h_def_APS},
see Appendix~\ref{sec:modular_101}.
Moreover, $\tau_\text{SW}$ is defined in \eqref{tau_def}. The parameter $m_S$ in (\ref{4d_curve_sun}) is the average mass, 
\begin{equation}
	m_S = \frac{1}{N_f} \sum_{i=1}^{N_f} m_i \,.
\label{m_avg}	
\end{equation}
The combination $h(h+2)$ is invariant under $S$ and $T$ duality transformations.

In fact, we are interested in the case when the gauge group is actually 
\beq
\mbox{U$(N)=$SU($N$) $\times$ U(1)}\,.
\eeq
 Therefore we can make a shift
$x\to (x + hm_S)$, $\phi_a\to (\phi_a+hm_S)$ and get rid of $m_S$. Note, that  in the U($N$) theory -- in contrast to the SU($N$) case --
the $\sum _a\phi_a$ does not have to vanish.
The SW curve \eqref{4d_curve_sun} then becomes
\begin{equation}
	y^2 = \prod_{a=1}^{N}(x-\phi_a)^2 + h(h+2) \prod_{i=1}^{N_f} (x + m_i)
	\,, \quad
	N_f = 2 N \,.
\label{4d_curve}	
\end{equation}
Our quark vacuum is  a singular point on the Coulomb branch where all the Seiberg-Witten roots are double roots, so
the diagonal quarks $q^{kP}$ with $k=P$ are massless. Upon switching on a nonvanishing $\xi$  this singularity transforms into an isolated vacuum  where the diagonal quarks develop VEVs \eqref{Nf_qvev}.
 
To guarantee the coincidence of the BPS spectra,
we require that the double roots of the four-dimensional Seiberg-Witten curve \eqref{4d_curve} coincide with the solutions of the two-dimensional vacuum equation \eqref{2d_equation}. In asymptotically free versions of the theory the SW curve is simply  the square of the vacuum equation of the two dimensional theory \cite{Dorey}. This ensures the coincidence of roots.
We use the same idea for the conformal case at hand.

Consider the square of \eqref{2d_equation} in the following form:
\begin{equation}
	\wt{y}^2 = \left[ 
			\prod_{P=1}^{N}\left(\sqrt{2} \sigma + m_P \right) 
			- e^{- 2 \pi \beta} \prod_{K = N + 1}^{2 N} \left(\sqrt{2} \sigma + m_K \right)
		\right]^2		\,.
\label{2d_to_4d_1}		
\end{equation}
We want to make a connection with the SW curve \eqref{4d_curve}. Equation \eqref{2d_to_4d_1} can be rewritten as
\begin{equation}
	\wt{y}^2 = \left[ 
			\prod_{P=1}^{N}\left(\sqrt{2} \sigma + m_P \right) 
			+ e^{- 2 \pi \beta} \prod_{K = N + 1}^{2 N} \left(\sqrt{2} \sigma + m_K \right)
		\right]^2	
		- 4 \, e^{- 2 \pi \beta} \, \prod_{A = 1}^{2 N} \left(\sqrt{2} \sigma + m_A \right)
		\,.
\label{2d_to_4d_2}		
\end{equation}
Let us compare this to the four-dimensional curve \eqref{4d_curve}. We immediately identify
\begin{equation}
\begin{aligned}
	x &= \sqrt{2} \sigma \,, \\[2mm]
	h (h + 2) &= - \frac{4 e^{- 2 \pi \beta}}{(1 + e^{- 2 \pi \beta})^2}  \,, \\[2mm]
	y^2 &=  \frac{ \wt{y}^2 }{ (1 + e^{- 2 \pi \beta})^2 }  \,.
\end{aligned}
\label{2d_4d_map_1}	
\end{equation}
We can also find the Coulomb branch parameters,
\begin{equation}
	\phi_{1,2} = - \frac{\Delta m}{2} \, \frac{1 - e^{- 2 \pi \beta}}{1 + e^{- 2 \pi \beta}} 
		~\pm~ \sqrt{ \frac{(\delta m_{12})^2 + e^{- 2 \pi \beta} \, (\delta m_{34})^2}{4 (1 + e^{- 2 \pi \beta}) } - \Delta m^2 \, \frac{e^{- 2 \pi \beta}}{(1 + e^{- 2 \pi \beta})^2}}\,,
\end{equation}
where the mass notation is according to \eqref{mass_parametrization}.
Note that one of these Coulomb parameters diverges at $\beta = i k / 2$, $k \in \mathbb{Z}$ (cf. our discussion in Appendix~\ref{sec:dual_couplings}).

The second  relation in \eqref{2d_4d_map_1} can be viewed as a quadratic equation with respect to $e^{- 2 \pi \beta}$.
Solving it, we obtain two solutions
\begin{equation}
\begin{aligned}
	e^{- 2 \pi \beta_1} &= \lambda(\tau_\text{SW} + 1)  \,,  \\
	e^{- 2 \pi \beta_2} &= \frac{1}{\lambda(\tau_\text{SW} + 1)}  \,,
\end{aligned}
\label{2d_4d_coupling_two_solutions}	
\end{equation}
where we used \eqref{h_combination} and \eqref{lambda_trans_halfshift}. See Appendix~\ref{klambdaf} for the definition of the $\lambda$ functions.
These two solutions are interchanged by the $S$ duality transformation, see \eqref{lambda_p1_S}. 

In the weak coupling limit $\Im \tau_\text{SW} \gg 1$ the $\lambda$ functions in \eqref{2d_4d_coupling_two_solutions} can be expanded according to  \eqref{lambda_function}. For the first option in \eqref{2d_4d_coupling_two_solutions} we have
\begin{equation}
	e^{- 2 \pi \beta} \approx 16 e^{\pi i (\tau_\text{SW} + 1)} \,.
\end{equation}
Recalling the definitions of the complexified couplings \eqref{beta_complexified} and \eqref{tau_def}, we can write down the weak coupling relation as
follows:
\begin{equation}
\begin{aligned}
	r &\approx \frac{4 \pi}{g^2} - \frac{2\ln(2)}{\pi}	\,, \\[2mm]
	\theta_{2d} &\approx - \theta_{4d} - \pi   \,,
\end{aligned}
\label{beta_of_tau_weak_coupling}
\end{equation}
cf. Eq. (\ref{beta_complexified}).
This is compatible with the known quasiclassical results.
From this analysis we see that out of two options \eqref{2d_4d_coupling_two_solutions}, the first one gives a correct weak coupling limit.
Thus we can write down our final formula for the relation between the world-sheet and bulk couplings,
\begin{equation}
	e^{- 2 \pi \beta} = \lambda(\tau_\text{SW} + 1) \,.
\label{2d_4d_coupling_my}	
\end{equation}
To visualize this relation between 4D and 2D couplings see Fig.~\ref{fig:coupling_traj} and Fig.~\ref{fig:beta_traj}.

The result \eqref{2d_4d_coupling_my} corrects the relation claimed previously in \cite{SYlittles} without derivation.   It should be compared with the result 
\cite{Karasik} obtained in 2017 by Gerchkovitz and Karasik. The latter is not quite identical to (\ref{2d_4d_coupling_my}). See the explanation below Eq. (\ref{2d_4d_coupling_my_noshift}).

Note, that different possible forms of the SW curve can lead to different relations between 4D and 2D couplings. For example, 
in \cite{ArgPlessShapiro} the authors claimed that the shift $\tau_\text{SW} \to \tau_\text{SW} + 1$ is basically a change of the origin of the $\theta$ angle by $\pi$, so supposedly it does not change physics, but only changes the form of the SW curve.
The curve \eqref{4d_curve} corresponds to the choice
\begin{equation*}
	g = \frac{\theta_2^4 + \theta_1^4}{\theta_2^4 - \theta_1^4} ,
	\quad
	h (h + 2) = - (1 - g^2) \,.
\end{equation*}
for the function $g$ from \cite{ArgPlessShapiro}. One could have also chosen this function differently, 
\begin{equation*}
	g = \frac{\theta_3^4 - \theta_1^4}{\theta_3^4 + \theta_1^4} \,,
\end{equation*}
which would lead to 
\begin{equation}
	e^{- 2 \pi \beta} = \lambda(\tau_\text{SW})
\label{2d_4d_coupling_my_noshift}	
\end{equation}
instead of \eqref{2d_4d_coupling_my}. Relation \eqref{2d_4d_coupling_my_noshift} between 4D and 2D couplings has been  obtained in \cite{Karasik} using localization. However, this formula uses an unconventional definition of the origin of the bulk $\theta_{4d}$ angle ($\theta_{4d}$ is shifted by $\pi$).
In this Chapter we  use the relation \eqref{2d_4d_coupling_my}.

Finally we note that the formula relating the world sheet and bulk coupling constants may actually depend on renormalization scheme; see e.g.
\cite[Sec. 3.4 and Sec. 3.5]{Jaewon:2012wsa} and \cite[Sec. 9.2.1]{Tachikawa:2013kta}.
Here we have obtained the formula \eqref{2d_4d_coupling_my}. Consider an alternative formula
\begin{equation}
	e^{- 2 \pi \beta} \big|_\text{alt} = - h(\tau_\text{SW})[h(\tau_\text{SW}) +2] \,.
\label{2d_4d_coupling_alternative}	
\end{equation}
Similar (but not quite the same) expressions have appeared previously in e.g. \cite{SYlittles}. Using the identity \eqref{h_tau_relation}, we can see that the $\beta$ couplings in \eqref{2d_4d_coupling_my} and \eqref{2d_4d_coupling_alternative} are related,
\begin{equation}
	e^{- 2 \pi \beta} \big|_\text{alt} 
		= \frac{4 \, e^{- 2 \pi \beta} \big|_\text{\eqref{2d_4d_coupling_my}}}%
			{ \left(1 + e^{- 2 \pi \beta} \big|_\text{\eqref{2d_4d_coupling_my}}\right)^2}
\end{equation}
These two couplings describe equivalent theories, see \cite[eq. (3.104)]{Jaewon:2012wsa}. A nontrivial relation between them seems to merely reflect different choices of a renormalization scheme.

\subsection{Dualities}
\label{dualities}

\begin{figure}[h]
	\centering
	\includegraphics[width=0.5\textwidth]{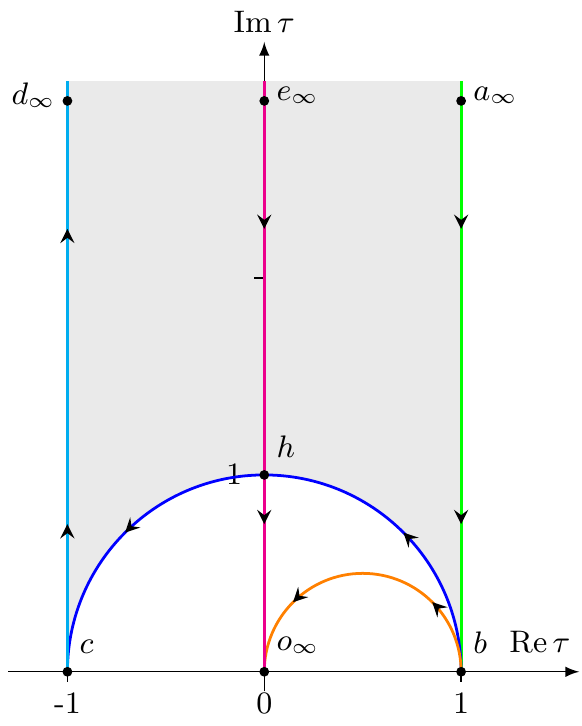}
	\caption{
		Fundamental domain of the duality group on the $\tau$ plane (shaded region).
		Shown are some particular trajectories in the space of the $\tau$ coupling. For the corresponding paths in the space of $\beta$ see Fig.~\ref{fig:beta_traj}. 
		The path $b \to o_\infty$ is an $ST^{-1}$ image of $b \to a_\infty$.
		The path $h \to o_\infty$ is an $S$-image of $h \to e_\infty$.
	}
\label{fig:coupling_traj}
\end{figure}

\begin{figure}[h]
	\centering
	\includegraphics[width=0.6\textwidth]{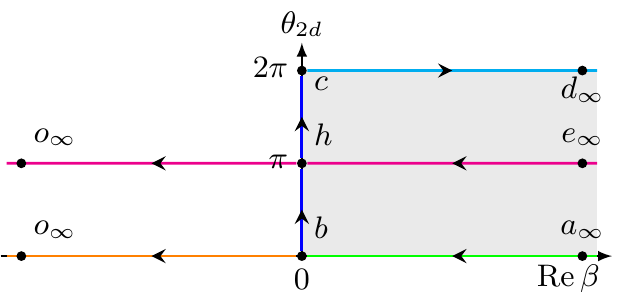}
	\caption{
		Fundamental domain of the duality group on the $\beta$ plane (shaded region).
		Shown are some particular trajectories in the space of the $\beta$ coupling. For the corresponding paths in the space of $\tau$ see Fig.~\ref{fig:coupling_traj}. 
		The path $b \to o_\infty$ is an $ST^{-1}$ image of $b \to a_\infty$.
		The path $h \to o_\infty$ is an $S$ image of $h \to e_\infty$.
		The trajectories are drawn modulo the relation $\theta_{2d} \sim \theta_{2d} + 2 \pi$.
	}
\label{fig:beta_traj}
\end{figure}

We have already seen that the world-sheet \wcpt model \eqref{wcp22} respects the duality transformation \eqref{2d_S-duality}. 
In this subsection we will see how this transformation is connected to the $S$ duality of the 4D SQCD, and discuss  other dualities as well.
We will define the $S$ duality and $T^\frac{1}{2}$ transformations as
\begin{equation}
\begin{aligned}
		S: \ \tau_\text{SW} &\to \frac{-1}{\tau_\text{SW}} \,,\\[2mm]
		T^\frac{1}{2}: \ \tau_\text{SW} &\to \tau_\text{SW} + 1 \,.
\end{aligned}
\label{ST_transformation}
\end{equation}
The conventional $T$ duality transformation $\tau_\text{SW} \to \tau_\text{SW} + 2$ is just a square of $T^\frac{1}{2}$.

The above transformations in Eq. \eqref{ST_transformation} generate the modular group SL($2,{\mathbb Z}$).
For the theory with the SU$(2)$ gauge group the  duality group is not a full SL($2,{\mathbb Z}$), but rather a subgroup generated by $S$ and $T$, the so-called $\Gamma^0(2)$ congruence subgroup of SL($2,{\mathbb Z}$).
It is not difficult to find the fundamental domain, see Fig.~\ref{fig:coupling_traj}.

In \cite{Karasik} it was shown  that 4D \ntwo SQCD with the U$(N)$ gauge group is not invariant under $S$ duality.
Say, our theory with the equal U(1) charges of four  quarks is mapped onto a SQCD with different U(1) quark charges. However, 
SQCD with the U(2) gauge group and equal U(1) charges  is invariant under the $ST^\frac{1}{2}S$ transformation \cite{Karasik}.  This transformation   in the world-sheet theory language  means that the theory is invariant under the sign change $\beta \to - \beta$. 

In our convention for the $\theta_{4d}$ angle, see Eq. \eqref{2d_4d_coupling_my}, the corresponding duality transformation is in fact 
an $S$ transformation.
Indeed, the $\theta_{4d}$ angle of \cite{Karasik} differs from ours by a shift by $\pi$ (cf. \textquote{+1} in \eqref{2d_4d_coupling_my}), which is a $T^\frac{1}{2}$ transformation. Since without this shift, the duality transformation would be $ST^\frac{1}{2}S$, our duality transformation is in fact
\begin{equation}
	T^\frac{1}{2} \cdot ST^\frac{1}{2}S \cdot T^\frac{1}{2} = S  \,.
\end{equation}
This identity can be checked explicitly.

Let us have a closer look at the $S$ duality.
Under the $S$ transformation the 4D coupling is transformed as
\begin{equation}
	\tau_\text{SW} \xrightarrow{S} \frac{-1}{\tau_\text{SW}} \,,
\end{equation}
and the $\lambda$ function in Eq. \eqref{2d_4d_coupling_my} becomes (see Eq. \eqref{lambda_p1_S}):
\begin{equation}
	\lambda(\tau_\text{SW} + 1) \xrightarrow{S}  \frac{1}{\lambda(\tau_\text{SW} + 1)} \,,
\end{equation} 
so that under the $S$ duality (cf. \eqref{2d_S-duality})
\begin{equation}
	\beta \xrightarrow{S} - \beta \,.
\label{selfdual_beta_eq}	
\end{equation}
Thus we have shown that the world-sheet duality \eqref{2d_S-duality} exactly corresponds to the $S$ duality of the bulk theory.
For an illustration, see Fig.~\ref{fig:coupling_traj} and~\ref{fig:beta_traj}.

The \wcpt self-dual point $\beta=0$ corresponds to $\tau_\text{SW} = 1$. Under the four-dimensional $S$-duality transformation, this maps to $\tau_\text{SW} = -1$, which differs from the initial value $\tau_\text{SW} = 1$ by a $2\pi$ shift of the $\theta_{4d}$ angle. The four-dimensional self-dual point $\tau=i$ corresponds to $\beta = i/2$ in two dimensions, see also Appendix~\ref{sec:dual_couplings}.

%
%

\section{Discussion}  \label{sec:b_meson:conclusions}

It has been known for a while now that the non-Abelian vortex string in four-dimensional \ntwo SQCD can become critical \cite{SYcstring}. 
This happens because, in addition to four translational zero modes of a usual ANO vortex, this string exhibits six orientational and  size zero modes. 
The target space of the effective world-sheet theory becomes $\mathbb{R}^4\times Y_6$, where $Y_6$ is a non-compact six-dimensional Calabi-Yau manifold, the so-called resolved conifold.

This has opened a way to quantize the solitonic string and to study the underlying gauge theory in terms of an \textquote{effective} string theory -- a kind of a \textquote{reverse holography} picture. 
It made possible quantitative description of the hadron spectrum \cite{KSYconifold,KSYcstring,SYlittles,SYlittmult}.
In particular, in \cite{SYlittles,SYlittmult}, the \textquote{Little String Theory} approach was used, namely a duality between the critical string  on the conifold 
and the non-critical $c=1$ string with the Liouville field and a compact scalar at the self-dual radius.
At the self-dual point $\beta=0$ of the world-sheet theory, the presence of the  massless 4D baryonic hypermultiplet $b$ was confirmed   and low-lying massive string states were also found.

In view of these spectacular results, the question arises: can we see these states directly from the field theory? 
In the present Chapter we managed to do just that.
To this end we employ the so-called 2D-4D correspondence. In the present case it means coincidence of the BPS spectra in the two-dimensional weighted sigma model, \wcpt \eqref{wcp22}, with the BPS spectrum in four-dimensional \ntwo SQCD with the U(2) gauge group  and four quark flavors  in the quarks vacuum. 
This coincidence was observed in \cite{Dorey,DoHoTo} and later explained 
in \cite{SYmon,HT2} using the picture of confined bulk monopoles which are seen as kinks in the 
world-sheet theory.
Then, we can reduce the problem to study of the BPS spectrum of the two-dimensional model {\em per se}.

Starting from weak coupling, we progressed into the strong coupling domain and further into the dual weak coupling domain. We managed to build a consistent picture of the BPS spectra in these regions and curves of marginal stability separating these domains. 
 
Consideration of the world-sheet kinks near the self-dual point $\beta=0$ led us to a rediscovery of a non-perturbative Higgs branch emerging at that point. The multiplet that lives on this branch turns out to be exactly the baryon multiplet $b$ found from 
string theory. 
Thus we have confirmed the consistency of the string theory picture describing the underlying gauge theory.

Moreover, in this model it was possible to observe the \textquote{instead-of-confinement} mechanism in action (see \cite{Shifman:2009mb,Shifman:2012yi} and a review \cite{SYdualrev}). At weak coupling $\beta \gg 1$ ($\beta$ being the sigma model coupling) there are perturbative states which look like $\mathbb{CP}(1)$ model excitations. At strong coupling $\beta \sim 1$ they decay into kink-antikink pairs. As we move further, we enter the dual weak coupling domain $\beta \ll -1$, with its own kinks and perturbative excitations. 
This evolution was described in the course of the present Chapter.

This world-sheet picture directly translates to the bulk theory. 
At weak coupling $g^2 \ll 1$ the perturbative spectrum of the four-dimensional \ntwo SQCD contains screened quarks and Higgsed gauge bosons. There are also solitonic states -- monopoles connected with non-Abelian flux tubes, forming mesons; but they are very heavy. As we progress into the strong coupling domain $g^2 \sim 1$, the screened quarks and Higgsed gauge bosons decay into confined monopole-antimonopole pairs. The \textquote{instead-of-confinement} phase is an alternative to the conventional confinement phase in QCD.

Similar instead-of-confinement phase appears if we move from large negative $\beta$ towards the strong coupling at $\beta \sim -1$.
In 4D SQCD this corresponds to moving  from the origin in the $\tau$-plane towards the upper semicircle shown in 
Fig.~\ref{fig:coupling_traj}. 
It is important  that $S$ dualities in the world-sheet and bulk theories are directly related, see Sec.~\ref{sec:2D_4D}.

 The results of this Chapter are published in the paper \cite{Ievlev:2020qch}.

%
%

\chapter*{Conclusion} \label{sec:conclusion}
\addcontentsline{toc}{likechapter}{Conclusion}

This thesis is devoted to strong coupling phenomena and confinement in supersymmetric gauge theories.
The central object of our investigations was the non-Abelian string responsible for confinement of monopoles in supersymmetric cousins of QCD.

We started by recalling basic facts about the non-Abelian strings in \ntwo supersymmetric QCD with the gauge group U$(N)$ and $N_f \geqslant N$ quark flavors. These strings are similar to the Abrikosov-Nielsen-Olesen string, but possess additional \textquote{orientational} internal degrees of freedom (bosonic $n^l$ as well as fermionic $\xi^l$).

We then tried to understand what happens to the non-Abelian strings and confined monopoles when we go to the \none SQCD. 
This was done by deforming the original \ntwo theory by a mass 
term $\mu$ for adjoint matter. In the limit of large $\mu$ this theory flows to \none SQCD. 

We started with the case $N_f = N$.
We found a
solution for the non-Abelian string and derived a two dimensional effective theory on the string world sheet
which describes the dynamic of its orientational zero modes. This theory turns out to be 
bosonic $\mathbb{CP}(N-1)$ model with a shallow potential generated by small quark
mass differences. The fermionic superpartners $\xi^l$ of the bosonic orientational moduli $n^l$ present in 
the \ntwo limit become heavy at large $\mu$ and decouple.

We addressed the question of what happen to confined 't Hooft-Polyakov monopoles
at large $\mu$. We showed that, if the quark mass differences are larger than (exponentially small) $\Lambda_{CP}$,
the confined monopoles become unstable at large $\mu$. However, if 
the quarks have equal masses, the confined monopoles survive in the \none QCD limit. 
 This result is quite remarkable since \none QCD is in the non-Abelian regime and quasiclassically we 
do not expect monopoles in this theory. It also supports the picture of \textquote{instead-of-confinement}
phase for \none QCD at strong coupling \cite{SYdualrev}.

We then went on to generalize this construction to the case $N_f > N$. In the \ntwo theory the non-Abelian string is semilocal in this case, i.e. it has additional size moduli corresponding to massless fields living on the string. It was found that after the $\mu$-deformation the size moduli develop a potential and decouple, leaving behind a \textquote{local} non-Abelian string.

Next, we considered the non-Abelian string from the point of view of the world sheet model per se. 
We found that the $\mu$-deformation induced from the bulk leads to the same consequences. Namely, the confined monopoles, seen in the world sheet theory as kinks interpolating between different vacua, indeed survive in the large $\mu$ limit provided that the quark mass differences $\Delta m$ are zero (the latter play the role of a mass scale in the effective theory). Along the way we found the whole phase diagram on the $(\mu, \Delta m)$ plane, which shows a rich phase structure of the theory.

After that we took a U-turn at engaged into more symmetric developments. Namely, we studied non-Abelian string in the \ntwo theory with the U$(N=2)$ gauge group, $N_f=4$ flavors 
of quarks and a Fayet-Iliopoulos term \cite{FI}.
In this setting the world sheet theory is superconformal and critical. 
We have studied the BPS protected spectrum of the world sheet theory and then, using the 2D-4D correspondence, translated the results into the 4d SQCD terms. 

At weak coupling the observed states are the perturbative gauge bosons and heavy solitonic objects (towers of dyons), together with their superpartners. 
When we enter into the strong coupling, the perturbative states and high-lying dyons decay into monopole-antimonopole pairs, each of which then stays tied by non-Abelian strings. This is a clear illustration of the \textquote{instead-of-confinement} mechanism.

Moreover, we have confirmed the existence of a massless $b$-baryon found earlier by treating the solitonic vortex as a critical superstring.
This is an important result providing evidence for the consistency of such string theory approach.

The field theory approach developed here may also give us a way how to generalize these constructions to arbitrary $N$ and $N_f$. This is certainly desirable. 
There is also an intriguing question of a connection with the AdS/CFT program. There are hints that may allow us to understand the reasons for or even prove some variants of the AdS/CFT hypothesis.

In fairness, it should be noted that here we did not consider the question of the spontaneously broken chiral symmetry. 
The thing is that in the \ntwo gauge theory chiral symmetry is broken already by the Yukawa couplings. Although this theory is called \textquote{supersymmetric QCD}, it is of course a bit different from the \textquote{original} QCD.
It makes sense to investigate chiral symmetry in the \none SQCD, but this question is quite difficult and deserves a separate discussion.

This thesis broadens the understanding of non-Abelian strings in supersymmetric gauge theories, and of strong coupling phenomena in general.
While it is not certain that this road will lead us to solving the confinement problem in the \textquote{real-world} QCD, it will certainly take us far.
On the journey of understanding confinement in strongly coupled gauge theories, there are three major steps. The first one is the understanding of the basic nature of confinement. We can now say that we reached this stage in \ntwo and \none theories.
The second step is calculation of the hadron spectrum. 
In the light of the latest results (including some presented here), this step is underway in \ntwo theories.
The third step is derivation of the low energy theory of pion-nucleon interaction directly from the first principles.
We are not there just yet, but this third step may be just within a hand's reach.

\vspace{60pt}

\begin{figure*}[h]
	\centering
	\includegraphics[width=0.6\textwidth]{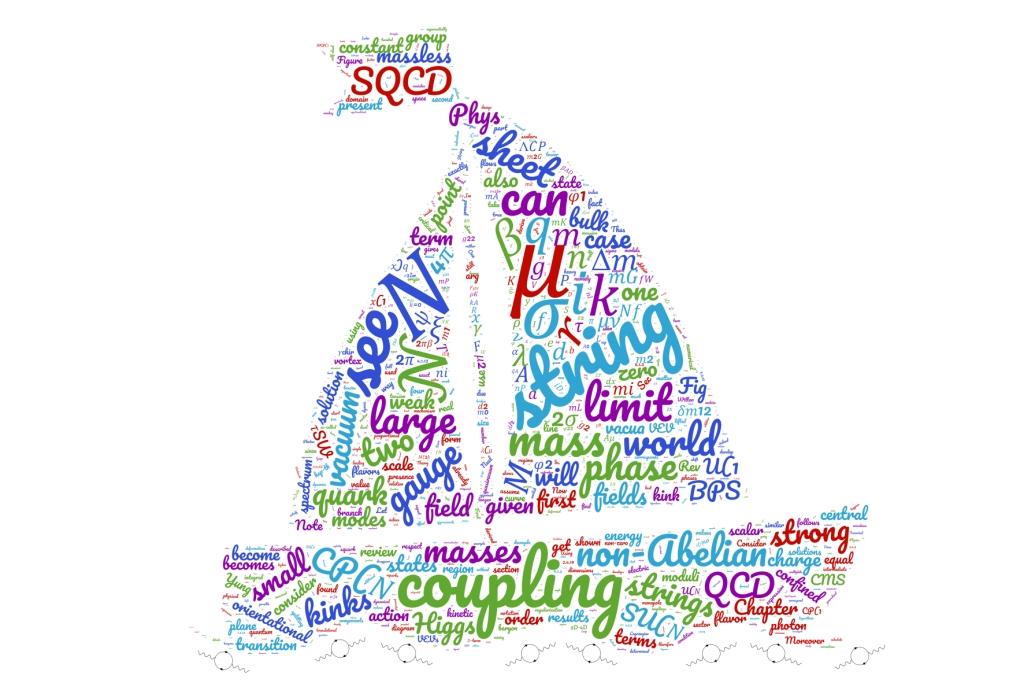}

	{ 
	\scriptsize
	Strong coupling voyage. Created with the help of \href{https://www.wordclouds.com}{wordclouds.com}.}
\end{figure*}

%
%

\appendix 

\setupname{Appendix}

%
%


\chapter{Useful formulas in two spacetime dimensions}
\label{sec:useful_2d}
\setcounter{section}{1}

This Appendix contains some equations and conventions frequently used in two-dimensional physics.

Tensors in 2d:
\begin{equation}
	g^{\mu\nu} = \text{diag} (+,-) \,,
\end{equation}
\begin{equation}
	\epsilon_{01} = +1, \quad \epsilon^{01} = -1 \,,
\end{equation}
\begin{equation}
	\epsilon^{\mu\nu} \epsilon^{\alpha\beta} = - g^{\mu\alpha} g^{\nu\beta} + g^{\mu\beta} g^{\nu\alpha} \,.
\end{equation}
Dual gauge field strength:
\begin{equation}
	F^{*}=\frac12\epsilon_{\mu\nu}F^{\mu\nu} = \epsilon_{\mu\nu} \p^\mu A^\nu = F^{01} = - F_{01} \,,
\end{equation}
\begin{equation}
	F_{01} = \mathcal{E} \,,
\end{equation}
where $\mathcal{E}$ is the electric field strength, and $\Box = \p_\mu \p^\mu$.

Pure photon action:
\begin{equation}
	\int d^2 x \left\{- \frac{1}{4} F_{\mu\nu}F^{\mu\nu} \right\} = \int d^2 x \frac{1}{2} A^\mu \left( \Box g_{\mu\nu} - \p_\mu \p_\nu \right) A^\nu \,,
\end{equation}

Useful identity:
\begin{equation}
\begin{aligned}
	F^* \frac{1}{\Box} F^* &= \epsilon^{\mu\nu} \p_\mu A_\nu \frac{1}{\Box} \epsilon^{\alpha\beta} \p_\alpha A_\beta \\
		&= \p_\mu A_\nu \frac{1}{\Box} (- g^{\mu\alpha} g^{\nu\beta} + g^{\mu\beta} g^{\nu\alpha}) \p_\alpha A_\beta  \\
		&\simeq - A_\nu \frac{1}{\Box} ( - \Box g^{\nu\beta} + \p^\nu \p^\beta )  A_\beta	\\
		&= A_\mu A^\mu - A_\mu \frac{\p^\nu \p^\beta}{\Box} A_\beta
		\,,
\end{aligned}	
\label{F-dual-identity}	
\end{equation}
where $\simeq$ involves integration by parts, i.e. it is an equality up to surface terms.

Gamma matrices:
\begin{equation}
	\gamma^0 =
		\begin{pmatrix}
			0 & 1 \\
			1 & 0
		\end{pmatrix}
		\ , \quad
	\gamma^1 =
		\begin{pmatrix}
			0 & -1 \\
			1 & 0
		\end{pmatrix}
		\ , \quad
	\gamma_\text{chir} = \gamma^0 \gamma^1 = 
		\begin{pmatrix}
			-1 & 0 \\
			0 & 1
		\end{pmatrix}
		\,,
\end{equation}
\begin{equation}
	\{ \gamma^\mu, \, \gamma_\text{chir} \} = 0 \,.
\end{equation}
Derivatives:
\begin{equation}
	\pt_L = \pt_{t}+\pt_z\,,\qquad \pt_R = \pt_{t}- \pt_z\,,
\end{equation}
\begin{equation}
	\nabla_\mu = \p_\mu - i \, A_\mu \,,
\end{equation}
\begin{equation}
	\slashed{\nabla} = \gamma^\mu \nabla_\mu  
	\,.
\end{equation}
Dirac fermion:
\begin{equation}
	\Xi = \begin{pmatrix}
		\xi_L \\
		\xi_R
	\end{pmatrix}
	\ , \quad
	\ov{\Xi} = \Xi^\dagger \gamma_0 = \left(\ov{\xi}_R, \ \ov{\xi}_L\right)
	\,,
\end{equation}
\begin{equation}
	\xi_L = \frac{1}{2} (1 - \gamma_\text{chir}) \Xi \,,
	\quad
	\xi_R = \frac{1}{2} (1 + \gamma_\text{chir}) \Xi \,.
\end{equation}

\paragraph{Relationship between Euclidean and Minkowski formulation.}

$ $

Coordinates:
\begin{equation}
\begin{aligned}
	x^\mu_M &=\{t_M,\,z\}\,, \\
	x^\mu_E &=\{t_E,\,z\}\,.
\end{aligned}	
\end{equation}
\begin{equation}
	g^{\mu\nu} = \text{diag} (+,-) \,.
\end{equation}
The path integral in Minkowski formulation:
\begin{equation}
	\mathcal{A}_M = \int D\varphi \ e^{i S_M} = \int D\varphi \ e^{i \int \mathcal{L}_M d t_M} \,.
\end{equation}
In passing to Euclidean, we substitute
\begin{equation}
\begin{aligned}
	(x^0)_M = t_M &\longrightarrow - i t_E = -i (x^0)_E \,,\\
	(k_0)_M &\longrightarrow i (k_0)_E  \,,\\
	(k_1)_M &\longrightarrow   (k_1)_E  \,,\\
	\Box &\longrightarrow - \Delta = - (\p_0^2 + \p_1^2)	 \,,\\
	\mathcal{L}_M &\longrightarrow - \mathcal{L}_E		 \,,\\
	i S_M &\longrightarrow - S_E	 \,,\\
	\text{Oscillating path integrand} &\longrightarrow \text{Exponentialy decaying path integrand} \,.
\end{aligned}	
\end{equation}
For example, for the free scalar
\begin{equation}
	\mathcal{L}_M = - \frac{1}{2} \varphi (\Box + m^2) \varphi  \,,
\end{equation}
\begin{equation}
	\mathcal{L}_E =  \frac{1}{2} \varphi (- \Delta + m^2) \varphi  \,.
\end{equation}
The situation with the Dirac fermion is a bit trickier:
\begin{equation}
\begin{aligned}
	\gamma^0_M &\longrightarrow \gamma^0_E	 \,,\\
	\gamma^1_M &\longrightarrow i\gamma^1_E	 \,,\\
\end{aligned}	
\end{equation}
so that
\begin{equation}
\begin{aligned}
	\{\gamma^\mu_M, \gamma^\nu_M\} = 2g^{\mu\nu} &\longrightarrow \{\gamma^\mu_E, \gamma^\nu_E \} = 2\delta^{\mu\nu}	 \,,\\
	(\gamma^\mu k_\mu)_M &\longrightarrow i (\gamma^\mu k_\mu)_E \,.
\end{aligned}	
\end{equation}
Furthermore,
\begin{equation}
\begin{aligned}
	\pt_L^M &= \pt_{t_M}+\pt_z\,,\qquad \pt_R^M &= \pt_{t_M}- \pt_z\,,	\\
	\pt_L^E &= \pt_{t_E} - i \pt_z\,,\qquad \pt_R^E &= \pt_{t_E} + i \pt_z\,.	
\end{aligned}	
\end{equation}
\begin{equation}
\begin{aligned}
	\Psi^M &\longrightarrow \Psi^E  \,,\\
	\ov{\Psi}^M &\longrightarrow i \, \ov{\Psi}^E	 \,.
\end{aligned}	
\end{equation}
The fields $\Psi^E$ and $\ov{\Psi}^E$ are no longer related by  conjugation (in fact, there is no notion of  conjugation in Euclidean spacetime).

For the photon,
\begin{equation}
\begin{aligned}
	(A_0)_M &\longrightarrow i (A_0)_E  \,,\\
	(A_1)_M &\longrightarrow   (A_1)_E  \,,\\
	F^M_{01} &\longrightarrow iF^E_{01}	 \,,\\
	(F^*)_M &\longrightarrow i(F^*)_E	 \,.
\end{aligned}	
\end{equation}
\begin{equation}
	{\cal L}_M~=~-\frac{1}{4}F^M_{\mu\nu}F_M^{\mu\nu}~=~\frac{1}{2}F^M_{0j}F^M_{0j}-\frac{1}{4}F_{jk}F_{jk},
\end{equation}
The Lagrangians:
\begin{equation}
	{\cal L}_E~=~\frac{1}{4}F^E_{\mu\nu}F_E^{\mu\nu}~=~\frac{1}{2}F^E_{0j}F^E_{0j}+\frac{1}{4}F_{jk}F_{jk}.
\end{equation}

%
%


\chapter{Solution of the Dirac equation for superorientational modes}
\label{app:dirac}
\setcounter{section}{1}

In this Appendix we solve Dirac equations \eqref{fermeqs_ovpsi1_lplus_small-1}.
After a substitution
\begin{align}
\notag
\ov{\psi}_{\dot{1}-}(r) &~~=~~ \frac{1}{r\phi_2(r)}\Psi(r),\\
\notag
\lambda_{(1)}(r) &~~=~~ i g_2^2 \Lambda(r)
\end{align}
equations \eqref{fermeqs_ovpsi1_lplus_small-1} reduce to
\begin{equation}
\begin{aligned}
\frac{1}{r g_2^2 \phi_1\phi_2} \p_r \Psi &~~=~~ \Lambda \,,\\[2mm]
r\p_r\Lambda ~+~ f_N\Lambda
~-~ \frac{\phi_1}{\phi_2}\Psi &~~=~~  ~-~  \frac{\mu_2 f_N}{2} \frac{\phi_1}{\phi_2} \, ,
\end{aligned}
\label{newvariables}
\end{equation}
which in turn gives an equation of second order for $\Psi$:
\begin{equation}
\p_r^2 \Psi ~-~ \frac{1}{r}\left( 1 ~+~ \frac{2}{N}(f-f_N) \right)\, \p_r \Psi ~-~ g_2^2\phi_1^2\Psi ~~=~~ -\,\frac{\mu_2 f_N}{2} g_2^2\phi_1^2 \,.
\label{fermeqs_Psi_small}
\end{equation}
First let us solve the homogeneous version of \eqref{fermeqs_ovpsi1_lplus_small-1}, i. e. 
put $\mu_2 ~=~ 0$. The solutions are
\begin{align*}
\ov{\psi}_{\dot{1}-} ~~=~~ c\frac{f_N}{r\phi_2}, \\[2mm]
\lambda_{(1)} ~~=~~ c\frac{i g_2^2}{2} \left( \frac{\phi_1}{\phi_2} ~-~ \frac{\phi_2}{\phi_1} \right).
\end{align*}
with some constant $c$. %
They correspond to $\Psi ~=~ f_N$; indeed, this is a solution for homogeneous version of \eqref{fermeqs_Psi_small}. With the help of it we can reduce the order of this equation. Let us take
\[
\Psi(r) ~~=~~ \mu_2\, f_N(r)\left( \int\limits_0^r {\mathrm d}x\, \chi(x) + c_1 \right),
\]
with some constant $c_1$, then from \eqref{fermeqs_Psi_small} it follows that
\begin{equation}
\p_r \chi ~+~ \frac{1}{r}\left( \frac{1}{f_N}r^2g_2^2(\phi_1^2 ~-~ \phi_2^2) ~-~ 1 ~-~ \frac{2}{N}(f-f_N) \right)\, \chi ~~=~ -\,\frac{1}{2} g_2^2  \phi_1^2 \,.
\label{fermeqs_theta_small}
\end{equation}
This is just an equation of the first order; its solution can be found very easily as
\begin{equation}
\chi ~~=~~ -\, \frac{g_2^2 r \phi_2^2}{2 f_N^2} \left(  \int\limits_0^r \frac{{\mathrm d}y}{y}\, \frac{\phi_1^2}{\phi_2^2} f_N^2 ~+~ c_2 \right).
\end{equation}
Putting all this together, we obtain:
\begin{equation}
\ov{\psi}_{\dot{1}-}(r) ~~=~~ -\,\mu_2\, g_2^2\, \frac{f_N(r)}{r\phi_2(r)} \left( \int\limits_0^r {\mathrm d}x\,
\frac{ x \phi_2^2(x)}{2 f_N^2(x)} \left(  \int\limits_0^x \frac{{\mathrm d}y}{y}\, \frac{\phi_1^2(y)}{\phi_2^2(y)}f_N^2(y) ~+~ c_2 \right) + c_1 \right)   .
\label{fermsol_ovpsi1_lplus_small-const}
\end{equation}
with some new constant $c_1$.

For this solution to behave well at the origin we have to put $c_1 ~=~ 0$. Considering the infinity, we should also require that
\[
c_2 ~~=~~ -\, \int\limits_0^\infty \frac{{\mathrm d}y}{y}\, \frac{\phi_1^2(y)}{\phi_2^2(y)}f_N^2(y).
\]
This gives
\begin{equation}
\ov{\psi}_{\dot{1}-}(r) ~~=~~ \frac{\mu_2\, g_2^2}{2}\, \frac{f_W(r)}{r\phi_2(r)}  \int\limits_0^r {\mathrm d}x\,
\frac{ x \phi_2^2(x)}{f_W^2(x)}   \int\limits_x^\infty \frac{{\mathrm d}y}{y}\,
 \frac{\phi_1^2(y)}{\phi_2^2(y)}f_W^2(y)   .
\label{fermsol_ovpsi1_lplus_small}
\end{equation}
for $\ov{\psi}_{\dot{1}-}$ and
\begin{equation}
\begin{multlined}
\lambda_{(1)}(r) ~~\equiv~~ \lambda^{22}_- ~+~ \lambda^{21}_- ~~=~~ \\
\phantom{\lambda^{22}_- ~+~ \lambda^{21}_-} =~~ \frac{i\, \mu_2\, g_2^2}{2}\, \Bigg( \frac{g_2^2}{2}
\left( \frac{\phi_1}{\phi_2} ~-~ \frac{\phi_2}{\phi_1} \right)
\int\limits_0^r {\mathrm d}x\, \frac{ x \phi_2^2(x)}{f_W^2(x)}   \int\limits_x^\infty \frac{{\mathrm d}y}{y}\, \frac{\phi_1^2(y)}{\phi_2^2(y)}f_W^2(y) \\
~+~ \frac{\phi_2}{\phi_1 f_W} \int\limits_r^\infty \frac{{\mathrm d}y}{y}\, \frac{\phi_1^2(y)}{\phi_2^2(y)}f_W^2(y)
\Bigg) .
\end{multlined}
\label{fermsol_lamplus_lplus_small}
\end{equation}
 for  $\lambda_{(1)}$.
By direct substitution we  verified that these modes indeed satisfy the Dirac equations.

Now let us consider Dirac equations \eqref{fermeqs_ovpsi1_lplus_small-2} with the non-zero eigenvalue 
$m_{or}$. Applying the method developed above we find the solutions
\begin{equation}
\ov{\psi}_{\dot{1}-}(r) ~~=~~ -\left(m_{or } - \mu_2\frac{g_2^2}{2} \right)\,\frac{f_N(r)}{r\phi_2(r)}\int\limits_0^r {\mathrm d}x \frac{x\phi_2^2(x)}{f_N^2(x)}  \int\limits_x^\infty {\mathrm d}y \frac{f_N^2(y)\phi_1^2(y)}{y\phi_2^2(y)}
\label{solution_orient_ov_psi_1-}
\end{equation}
for $ \ov{\psi}_{\dot{1}-}$ and  
\begin{equation}
\begin{multlined}
\lambda_{(1)}(r) ~~=~~ 
 -i\frac{m_{or }}{2}\left( \frac{\phi_1}{\phi_2} ~-~ \frac{\phi_2}{\phi_1}  \right)
~-~
ig_2^2\left(m_{or } - \mu_2\frac{g_2^2}{2} \right)\,\frac{1}{2}\left( \frac{\phi_1}{\phi_2} ~-~ 
\frac{\phi_2}{\phi_1}  \right)\, \times \\
\int\limits_0^r {\mathrm d}x \frac{x\phi_2^2(x)}{f_N^2(x)}  
\int\limits_x^\infty {\mathrm d}y \frac{f_N^2(y)\phi_1^2(y)}{y\phi_2^2(y)}
~-~
i\left(m_{or } - \mu_2\frac{g_2^2}{2} \right)\,\frac{\phi_2}{f_N\phi_1}
\int\limits_r^\infty {\mathrm d}y \frac{f_N^2(y)\phi_1^2(y)}{y\phi_2^2(y)}
\end{multlined}
\label{solution_orient_lambda(1)}
\end{equation}
 for $\lambda_{(1)}$.

One can see, that the first and the last terms in the last expression behave at the origin as $1/r$. 
We can choose eigenvalue $m_{or }$ to insure that $1/r$ terms cancel out. This gives the expression
\eqref{solution_orient_alpha} for the mass $m_{or }$.

%
%



\chapter{Coefficients of the effective action of the \CP ~model}
\label{sec:loops}

In this Appendix we derive the effective action \eqref{effaction}.

%
%

\section{Brief overview}

We consider the masses on a circle \eqref{masses_ZN}. The effective action is derived in the vicinity of the vacuum with $\Im\sigma = 0$.

Consider bosonic loops. In the Lagrangian \eqref{lagrangian_init} we can expand the $\sigma - n$ interaction term as
\begin{equation}
\begin{aligned}
	\left|\sqrt 2\sigma-m_i\right|^2 |n^i|^2 \approx 
		&\phantom{+} \left|\sqrt 2 \langle\sigma\rangle - m_i\right|^2  |n^i|^2	\\
		&+ 2 \Re(\sqrt{2}\delta\sigma) \cdot \left(\sqrt 2 \langle\sigma\rangle - \Re m_i\right)  |n^i|^2 \\
		&- 2 \Im(\sqrt{2}\delta\sigma) \cdot \Im m_i \,	|n^i|^2	
\end{aligned}		
\end{equation}
where $\delta\sigma$ are the vacuum fluctuations around the vacuum with $\Im\sigma = 0$. The diagram for the $\Re\sigma$ kinetic term is then proportional to 
$\left(\sqrt 2 \langle\sigma\rangle - \Re m_i\right)^2$, while the kinetic term for $\Im\sigma$ is proportional to $\left(\Im m_i\right)^2$. Calculation of the diagrams itself is straightforward, see below.

Calculation of the fermion loops is a bit trickier. The fermion mass matrix can be read off from \eqref{lagrangian_init}. Say, for the flavor number $i$, 
\begin{equation}
	M_i = \left(\sqrt{2}\langle\sigma\rangle - \Re m_i \right) \cdot \text{Id} + i \left(\Im m_i\right) \cdot \gamma_\text{chir}
\end{equation}
where $\text{Id}$ is the $2 \times 2$ identity matrix, and $\gamma_\text{chir}$ is the two-dimensional analogue of the $\gamma_5$. This $\gamma_\text{chir}$ interferes with the traces over the spinorial indices. Say, the fermionic contribution to the $\Re\sigma$ kinetic term coming from the diagram on Fig.~\ref{fig:loops:scalar} is
\begin{equation}
\begin{aligned}
\begin{gathered}
	\includegraphics[width=0.15\textwidth]{sigma_sigma.pdf}
\end{gathered}	
	 &= - (i \sqrt{2})^2 \sum_i \int \frac{d^2 k}{(2\pi)^2} \Tr \left[ 
		\frac{i}{\slashed{k} - M_i} \frac{i}{\slashed{k} + \slashed{q} - M_i}
	\right] \\
	&= - 2 \sum_i  \int \frac{d^2 k}{(2\pi)^2} \Tr \left[ 
			\frac{\slashed{k} + M_i^\dagger}{k^2 - |M_i|^2} \frac{\slashed{k} + \slashed{q} + M_i^\dagger}{(k+q)^2 - |M_i|^2}
		\right]	\\
	&= - 4 \sum_i  \int \frac{d^2 k}{(2\pi)^2} \Tr \left[ 
			\frac{\left( k \cdot (k + q) \right) + \left(\sqrt{2}\langle\sigma\rangle - \Re m_i \right)^2 - \left(\Im m_i\right)^2}{(k^2 - |M_i|^2)((k+q)^2 - |M_i|^2)}
		\right]		
\end{aligned}	
\end{equation}
where $|M_i|^2 = \left(\sqrt{2}\langle\sigma\rangle - \Re m_i \right)^2 + \left(\Im m_i\right)^2$. Calculation of the integral itself is straightforward, see below. The rest of the diagrams with fermionic loops are treated the same way. In the end we arrive at \eqref{eff_normalizations}.

Note that in the limit $\wt{\mu} \to 0$ supersymmetry is restored. In this case, in the vacuum $D = 0$, $\Im\sigma = 0$ we have
\begin{equation}
	M^2_{\xi_k} = m_{n_k}^2 = |\sqrt{2}\langle\sigma\rangle - m_k|^2 \,,
\end{equation}
and the coefficients \eqref{eff_normalizations} reduce to
\begin{equation}
	\frac{1}{e^2_{\Re\sigma}} = \frac{1}{e^2_{\Im\sigma}} = \frac{1}{e^2_{\gamma}} 
		= \frac{1}{4\pi}\,\sum_{k=0}^{N-1} \frac{1}{|\sqrt{2}\langle\sigma\rangle - m_k|^2}
\end{equation}
\begin{equation}
	b_{\gamma,\Im\sigma} = \frac{1}{2\pi}\,\sum_{k=0}^{N-1} \frac{1}{\sqrt{2}\langle\sigma\rangle - m_k} \,.
\end{equation}

The rest of this Appendix contains details of this derivation.

%
%

\section{Fermionic loops}

Lagrangian for one fermionic flavor interacting with the photon:
\begin{equation}
	\mathcal L_\text{ferm-A} = i \, \ov{\Xi} \, \slashed{\nabla} \, \Xi 
		\ - \  \ov{\Xi} \,  M \, \Xi
\end{equation}
Here, the fermion mass 
\begin{equation}
\begin{aligned}
	M &= \begin{pmatrix}
			\sqrt{2}\langle\sigma\rangle - m - \lambda(\mu) & 0 \\
			0 & \sqrt{2}\langle\ov{\sigma}\rangle - \ov{m} - \ov{\lambda(\mu)}
		\end{pmatrix}
		\\
		&= \Re\left(\sqrt{2}\langle\sigma\rangle - m\right) \cdot \text{Id} - i \Im\left(\sqrt{2}\langle\sigma\rangle - m\right) \cdot \gamma_\text{chir} \,,
\end{aligned}
\end{equation}
where $\text{Id}$ is the $2 \times 2$ identity matrix. We will use the following shorthand notation:
\begin{equation}
	M = R - I \cdot i \gamma_\text{chir}
	\,, \quad 
	M^\dagger = R + I \cdot i \gamma_\text{chir}
\label{M_shorthand}	
\end{equation}
\begin{equation}
	|M|^2 = R^2 + I^2
\end{equation}
Furthermore, we have to include the interaction with vacuum fluctuations of the $\sigma$ field:
\begin{equation}
	\mathcal L_\text{ferm} = i \, \ov{\Xi} \, \slashed{\nabla} \, \Xi 
		\ - \  \ov{\Xi} \,  M \, \Xi
		\ - \  \sqrt{2}\Re(\sigma) \ov{\Xi} \, \Xi
		\ + \  i\sqrt{2}\Im(\sigma) \ov{\Xi} \, \gamma_\text{chir} \, \Xi
\end{equation}

For a particular flavor number $k$, the mass is given by the formula
\begin{equation}
\begin{aligned}
	|M_{k^{th}\text{ flavor}}|^2 &= |\sqrt{2}\langle\sigma\rangle - m_k - \lambda(\mu)|^2	\\
	R_{k^{th}\text{ flavor}} &= \Re \left(\sqrt{2}\langle\sigma\rangle - m_k \right) - \lambda(\mu)	\\
	I_{k^{th}\text{ flavor}} &= \Im \left(\sqrt{2}\langle\sigma\rangle - m_k \right)
\end{aligned}	
\label{xi_masses}
\end{equation}
Recall that we use the mass parameters $m_k$ on the circle \eqref{masses_ZN}.

\subsection{Photon kinetic term}

\begin{figure}[h]
	\centering
     \includegraphics[width=0.3\textwidth]{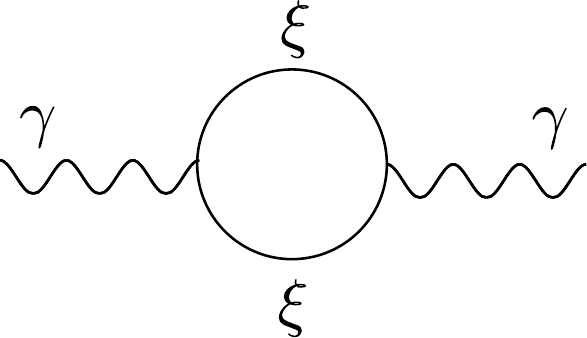}
     \caption{Photon wave function renormalization}
     \label{fig-app:loops:photon}
\end{figure}

Consider the diagram from Fig. \ref{fig-app:loops:photon}. Corresponding contribution is
\begin{equation}
	i \Pi^{\mu\nu} = - (+i )^2 \int \frac{d^2 k}{(2\pi)^2} \Tr \left[ 
		\gamma^\mu \frac{i}{\slashed{k} - M} \gamma^\nu \frac{i}{\slashed{k} + \slashed{q} - M}
	\right]
\label{polarization_ferm_init}	
\end{equation}
We will calculate this integral using the dimensional regularization.
Now, using the identity
\begin{equation}
	\left(\gamma^\mu k_\mu - M\right) \left(\gamma^\mu k_\mu + M^\dagger\right) 
		= k^2 - |M|^2 + k_\mu \underbrace{\left(\gamma^\mu M^\dagger - M \gamma^\mu\right)}_{=0}
\end{equation}
we can rewrite the fermion propagators as
\begin{equation}
	\frac{i}{\slashed{k} - M} = i \frac{\slashed{k} + M^\dagger}{k^2 - |M|^2}
\end{equation}
Let's calculate the trace in \eqref{polarization_ferm_init}. We will use the following identities in $d=2$ dimensions:
\begin{equation}
\begin{aligned}
	\Tr \left[ \gamma^\mu \gamma^\alpha \gamma^\nu \gamma^\beta \right] 
		&= d \left( g^{\mu\alpha}g^{\nu\beta} - g^{\mu\nu}g^{\alpha\beta} + g^{\mu\beta}g^{\nu\alpha} \right)	\\
	\Tr \left[ \gamma^\mu \gamma^\nu \right] 
		= - \Tr \left[ \gamma^\mu \gamma_\text{chir} \gamma^\nu \gamma_\text{chir} \right] 
		&= d \, g^{\mu\nu}	\\
	\Tr \left[ \gamma^\mu \gamma^\nu \gamma_\text{chir} \right] 
		&= - d \, \epsilon^{\mu\nu}	\\
	\Tr \left[ \text{odd \#  } \gamma^\mu  \right] 
		= \Tr \left[ \gamma_\text{chir} \cdot \text{odd \#  } \gamma^\mu \right] 
		&= 0
\end{aligned}
\label{trace_id}	
\end{equation}
Therefore,
\begin{equation}
\begin{aligned}
	\Tr \left[ \gamma^\mu \slashed{k} \gamma^\nu \slashed{q} \right] 
		&= d \left( k^\mu q^\nu + k^\nu q^\mu  - (k \cdot q) g^{\mu\nu}\right)	\\
	\Tr \left[ \gamma^\mu  M^\dagger \gamma^\nu M^\dagger \right]
		&= d \, |M|^2 \, g^{\mu\nu}
\end{aligned}	
\end{equation}
So, the trace in the numerator of \eqref{polarization_ferm_init} becomes
\begin{equation}
\begin{aligned}
	\Tr & \left[  \gamma^\mu (\slashed{k} + M^\dagger) \gamma^\nu (\slashed{k} + \slashed{q} + M^\dagger) \right] \\
		&= \Tr \left[ \gamma^\mu \slashed{k}  \gamma^\nu (\slashed{k} + \slashed{q}) \right]
			+ \Tr \left[ \gamma^\mu  M^\dagger \gamma^\nu M^\dagger \right]
			+ \Tr \left[ \gamma^\mu \slashed{k} \gamma^\nu M^\dagger \right]
			+ \Tr \left[ \gamma^\mu M^\dagger \gamma^\nu (\slashed{k} + \slashed{q}) \right]
			\\
		&= d \, \left[ k^\mu (k+q)^\nu + k^\nu (k+q)^\mu - g^{\mu\nu} (k \cdot (k + q) - |M|^2)
			\right]
\end{aligned}	
\end{equation}

The derivation to come follows closely \cite[Sec. 7.5]{PeskSchr}.
We will calculate the integral in \eqref{polarization_ferm_init} using the Feynman prescription:
\begin{equation}
\begin{aligned}
	\frac{1}{(k^2 - |M|^2) \, ((k + q)^2 - |M|^2)} 
		&= \int\limits_0^1 \, dx \, \frac{1}{\left(k^2 + 2 x k \cdot q + x q^2 - |M|^2\right)^2} \\
		&= \int\limits_0^1 \, dx \, \frac{1}{\left(l^2 + x (1-x) q^2  - |M|^2\right)^2}
\end{aligned}	
\label{feynman_prescription}
\end{equation}
where $l = k + x q$. In terms of $l$, the numerator of \eqref{polarization_ferm_init} is $d$ times
\begin{multline}
	2 l^\mu l^\nu - g^{\mu\nu} l^2 - 2 x (1-x) q^\mu q^\nu + g^{\mu\nu} (|M|^2 + x (1-x) q^2) 
		\\
		+ \text{terms linear in  }	l
\end{multline}
Let us perform a Wick rotation:
\begin{equation}
	l^0 = i l^0_E
	\,, \quad
	l_\mu l^\mu = - (l_E)^2
\label{Wick_simple}	
\end{equation}
In terms of $l_E$ the integral \eqref{polarization_ferm_init} becomes
\begin{multline}
		i \Pi^{\mu\nu} = - i \, d \, \int\limits_0^1 \, dx \, \int \frac{d^d l_E}{(2\pi)^d}
		\\
		\times
		\frac{- g^{\mu\nu} \frac{2}{d} l_E^2 + g^{\mu\nu} l_E^2 - 2 x (1-x) q^\mu q^\nu + g^{\mu\nu} (|M|^2 + x (1-x) q^2)}{\left(l_E^2 + \Delta\right)^2}
\label{polarization_ferm_2}		
\end{multline}
where we introduced $\Delta = |M|^2 - x (1-x) q^2$, and used the fact that in $d$ dimensions (see \cite[p. 251 eq. (7.87)]{PeskSchr})
\begin{equation}
	\int d^d l \ l^\mu l^\nu = \int d^d l \ \frac{1}{d} g^{\mu\nu} l^2 
\end{equation}
Now, we can evaluate the integrals over the momentum $l_E$ (see \cite[p. 251 eq. (7.85)]{PeskSchr}):
\begin{equation}
	\int \frac{d^d l_E}{(2\pi)^d} \frac{1}{\left(l_E^2 + \Delta\right)^2}
		= \frac{1}{(4\pi)^{d/2}} \Gamma \left(2 - \frac{d}{2} \right) \frac{1}{\Delta^{2 - d/2}}
\label{pleasant_int_1}		
\end{equation}
\begin{equation}
	\int \frac{d^d l_E}{(2\pi)^d} \frac{l_E^2}{\left(l_E^2 + \Delta\right)^2}
		= \frac{1}{(4\pi)^{d/2}} \frac{d}{2} \Gamma \left(1 - \frac{d}{2} \right) \frac{1}{\Delta^{1 - d/2}}
\label{unpleasant_int_1}		
\end{equation}
The integral \eqref{unpleasant_int_1} is divergent and would give a pole at $d=2$, if it weren't for the factor $1 - d/2$ in \eqref{polarization_ferm_2}:
\begin{equation}
\begin{aligned}
	\int \frac{d^d l_E}{(2\pi)^d} \frac{\left(- \frac{2}{d} + 1 \right) l_E^2 g^{\mu\nu}}{\left(l_E^2 + \Delta\right)^2}
		&= - g^{\mu\nu} \frac{1}{(4\pi)^{d/2}} \left(1 - \frac{d}{2} \right) \Gamma \left(1 - \frac{d}{2} \right) \frac{1}{\Delta^{1 - d/2}} \\
		&= g^{\mu\nu} \frac{1}{(4\pi)^{d/2}} (- \Delta) \Gamma \left(2 - \frac{d}{2} \right) \frac{1}{\Delta^{2 - d/2}} \\
		&= g^{\mu\nu} \frac{1}{(4\pi)^{d/2}} (- |M|^2 + x (1-x) q^2) \Gamma \left(2 - \frac{d}{2} \right) \frac{1}{\Delta^{2 - d/2}}
\end{aligned}	 
\label{pleasant_int_2}
\end{equation}
Combining this with \eqref{pleasant_int_1} we get for the polarization operator \eqref{polarization_ferm_2}:
\begin{equation}
\begin{aligned}
	\Pi^{\mu\nu}(q) &= - d \int\limits_0^1 \, dx \, \frac{1}{(4\pi)^{d/2}} \frac{\Gamma \left(2 - \frac{d}{2} \right)}{\Delta^{2 - d/2}} \\
		&\phantom{- d \int\limits_0^1 \, dx } \times \big[ g^{\mu\nu} \, (- |M|^2 + x (1-x) q^2) - 2 x (1-x) q^\mu q^\nu 
		\\
		&\phantom{- d \int\limits_0^1 \, dx }  + g^{\mu\nu} \, (|M|^2 + x (1-x) q^2) \big]
		\\
		&= (- q^2 g^{\mu\nu} + q^\mu q^\nu) \cdot \Pi(q^2)
\end{aligned}	
\end{equation}
where
\begin{equation}
\begin{aligned}
	\Pi(q^2) 
		&=  \frac{2 \, d \, \Gamma \left(2 - \frac{d}{2} \right)}{(4\pi)^{d/2}} \int\limits_0^1 \, dx \, \frac{x (1 - x)}{\Delta^{2 - d/2}} \\
		&\xrightarrow[d=2]{}  \frac{4}{4\pi} \int\limits_0^1 \, dx \, \frac{x (1 - x)}{|M|^2 - x (1-x) q^2} \\
		&\underset{q \to 0}{\approx}  \frac{1}{4\pi} \frac{2}{3} \frac{1}{|M|^2}
\end{aligned}	
\label{polarization_ferm_ans}
\end{equation}

To get the full fermionic contribution to the photon kinetic term, we have to sum over all fermionic flavors. Using \eqref{masses_ZN} and \eqref{xi_masses} we get from \eqref{polarization_ferm_ans} the photon normalization
\begin{equation}
	\left( \frac{1}{e^2_{\gamma}} \right)_\text{ferm} =  \frac{1}{4\pi}  \sum_{k=0}^{N-1} \left[\frac{2}{3} \frac{1}{M^2_{\xi_k}} \right]
\label{polarization_ferm_factor}	
\end{equation}
where we used notation
\begin{equation}
	M^2_{\xi_k} \equiv |M_{k^{th}\text{ flavor}}|^2 \,.
\end{equation}

\subsection{$\Re\sigma$ kinetic term}

\begin{figure}[h]
	\centering
     \includegraphics[width=0.3\textwidth]{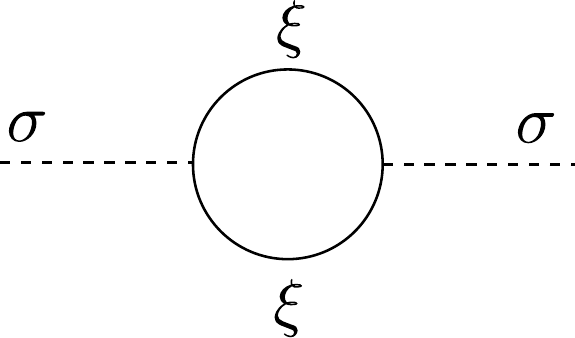}
     \caption{Scalar wave function renormalization}
     \label{fig-app:loops:scalar}
\end{figure}

Consider the diagram from Fig. \ref{fig-app:loops:scalar} with $\Re\sigma$ external legs. Corresponding contribution is
\begin{equation}
\begin{aligned}
	i D_{\Re\sigma} &= - (i \sqrt{2})^2 \int \frac{d^2 k}{(2\pi)^2} \Tr \left[ 
		\frac{i}{\slashed{k} - M} \frac{i}{\slashed{k} + \slashed{q} - M}
	\right] \\
	&= - 2 \int \frac{d^2 k}{(2\pi)^2} \Tr \left[ 
			\frac{\slashed{k} + M^\dagger}{k^2 - |M|^2} \frac{\slashed{k} + \slashed{q} + M^\dagger}{k^2 - |M|^2}
		\right]
\end{aligned}	
\label{Re_sigma_ferm_init}	
\end{equation}
Using the trace identities \eqref{trace_id}	we can evaluate the traces:
\begin{equation}
\begin{aligned}
	\Tr \left[ \slashed{k} \slashed{q} \right] 
		&= d \left( k \cdot q \right)	\\
	\Tr \left[  M^\dagger  M^\dagger \right]
		&= d \, (R^2 - I^2)
\end{aligned}	
\end{equation}
where the real and imaginary masses $R$ and $I$ are defined in \eqref{M_shorthand}.

Again, we will evaluate the integral in dimensional regularization using the Feynman prescription \eqref{feynman_prescription}. Introducing $l = k + x q$, we can rewrite the numerator of \eqref{Re_sigma_ferm_init} as 
\begin{equation}
\begin{aligned}
	\text{Num} &= d \left(  \left( k \cdot (k + q) \right) + (R^2 - I^2) \right) \\
		&= d \left( l^2 - x (1-x) q^2 + (R^2 - I^2) \right) + \text{terms linear in } l
\end{aligned}		
\end{equation}
Introducing again $\Delta = |M|^2 - x (1-x) q^2$, performing a Wick rotation \eqref{Wick_simple} and using \eqref{pleasant_int_1} and \eqref{unpleasant_int_1} we arrive at 
\begin{equation}
\begin{aligned}
	 D_{\Re\sigma} &=  - 2 d \, \int\limits_0^1 dx \, \int \frac{d^d l_E}{(2\pi)^d} 
			\frac{- l_E^2 - x (1-x) q^2 + (R^2 - I^2)}{(l_E^2 + \Delta)^2} \\
		&= - 2 d \, \int\limits_0^1 dx \, 
			\frac{1}{(4\pi)^{d/2}} \Gamma \left(2 - \frac{d}{2} \right)
			\\
			&\phantom{AAAA} \times
			\left( - \frac{1}{\Delta^{1 - d/2}} \frac{d}{2} \frac{1}{1 - \frac{d}{2}} + \frac{- x (1-x) q^2 + (R^2 - I^2)}{\Delta^{2 - d/2}} \right)
\end{aligned}	
\label{Re_sigma_ferm_1}	
\end{equation}
Denote
\begin{equation}
	1 - \frac{d}{2} = \varepsilon \,, 
	\quad
	d = 2 - 2 \varepsilon
\end{equation}
Using the gamma function decomposition
\begin{equation}
	\Gamma(\varepsilon) \approx \frac{1}{\varepsilon} - \gamma
\end{equation}
\begin{equation}
	\Gamma(1 + \varepsilon) = \varepsilon \Gamma(\varepsilon) \approx 1 - \varepsilon\gamma
\end{equation}
the first (singular) term of \eqref{Re_sigma_ferm_1} can be rewritten as
\begin{equation}
\begin{aligned}
	\text{singular} &=  \frac{d}{(4\pi)^{d/2}} \Gamma \left(2 - \frac{d}{2} \right)\frac{1}{\Delta^{1 - d/2}} \frac{d}{2} \frac{1}{1 - \frac{d}{2}} \\
		&=  \frac{2 - 2 \varepsilon}{(4\pi)^{1 - \varepsilon}} \Gamma \left(1 + \varepsilon \right)\frac{1}{\Delta^{\varepsilon}} (1 - \varepsilon) \frac{1}{\varepsilon} \\
		&=  \frac{2}{4\pi} (1 - \varepsilon)^2 \Gamma \left(1 + \varepsilon \right) \left(\frac{4\pi}{\Delta}\right)^\varepsilon \frac{1}{\varepsilon} \\
		&\approx   \frac{2}{4\pi} \frac{1}{\varepsilon} (1 - 2\varepsilon) (1 - \varepsilon\gamma) \left(1 + \varepsilon \ln\frac{4\pi}{\Delta} \right) \\
		&\approx  \frac{2}{4\pi} \left( \frac{1}{\varepsilon} - \gamma - 2 + \ln 4\pi - \ln{\Delta} \right)
\end{aligned}		
\label{Re_sigma_ferm_singular}
\end{equation}
We see that this expression is divergent. If we were to use Pauli-Villars regularization, then instead of \eqref{Re_sigma_ferm_singular} we would get
\begin{equation}
	\text{singular} =  \frac{2}{4\pi} \left( -1 + \ln M_\text{uv}^2 - \ln{\Delta} \right)
\end{equation}
Either way, we will be interested only in $q^2 \to 0$ behavior, particularly in $O(q^2)$ terms, so we can throw the divergent terms out. The remaining integral is
\begin{equation}
\begin{aligned}
	D_{\Re\sigma}^\text{fin} &= - 4 \, \int\limits_0^1 dx \, \frac{1}{4\pi}
			\left( \ln \frac{\Delta}{|M|^2} + \frac{- x (1-x) q^2 + (R^2 - I^2)}{\Delta} \right) \\
		&= - 4 \, \int\limits_0^1 dx \, \frac{1}{4\pi}
			\left(\frac{- x^2 q^2 + (R^2 - I^2)}{|M|^2 - x (1-x) q^2} \right) \\
		&\underset{q \to 0}{\approx} - 4 \frac{1}{4\pi} \ q^2 \, \left(- \frac{1}{3}\frac{1}{|M|^2} + \frac{1}{6} \frac{R^2 - I^2}{|M|^4}  \right) \\
		&= 2 \frac{1}{4\pi} \, q^2 \, \frac{1}{3} \, \frac{|M|^2 + 2 I^2}{|M|^4}
\end{aligned}	
\label{Re_sigma_ferm_2}	
\end{equation}

To get the full fermionic contribution to the kinetic term, we have to sum over all fermionic flavors. Using \eqref{masses_ZN} and \eqref{xi_masses} we get from \eqref{Re_sigma_ferm_2} the normalization factor
\begin{equation}
	\left( \frac{2}{e^2_{\Re\sigma}} \right)_\text{ferm} = \sum_{k=0}^{N-1} \left[ 2 \frac{1}{4\pi}  \, \frac{1}{3} \, 
		\frac{|M_{k^{th}\text{ flavor}}|^2 + 2 I_{k^{th}\text{ flavor}}^2}{|M_{k^{th}\text{ flavor}}|^4} \right]
\end{equation}
or, equivalently,
\begin{equation}
	\left( \frac{1}{e^2_{\Re\sigma}} \right)_\text{ferm} = \sum_{k=0}^{N-1} \left[  \frac{1}{4\pi}  \, \frac{1}{3} \, 
		\frac{M^2_{\xi_k} + 2 \left(\Im \left(\sqrt{2}\langle\sigma\rangle - m_k \right) \right)^2}{M^4_{\xi_k}} \right]
\end{equation}
where we used notation
\begin{equation}
	M^2_{\xi_k} \equiv |M_{k^{th}\text{ flavor}}|^2
\end{equation}
In the vacuum where $\Im\sigma = 0$ we have
\begin{equation}
	\left( \frac{1}{e^2_{\Re\sigma}} \right)_\text{ferm} = \frac{1}{4\pi}  \, \sum_{k=0}^{N-1} \left[  
		 \frac{1}{3} \,\frac{M^2_{\xi_k} + 2 \left(\Im  m_k  \right)^2}{M^4_{\xi_k}} \right]
\label{Re_sigma_ferm_factor}	
\end{equation}

\subsection{$\Im\sigma$ kinetic term}

The derivation follows closely that of the previous section.

Consider the diagram from Fig. \ref{fig-app:loops:scalar}, only with $\Im \sigma$ as external lines. Corresponding contribution is
\begin{equation}
\begin{aligned}
	i D_{\Im\sigma} &= - (- \sqrt{2})^2 \int \frac{d^2 k}{(2\pi)^2} \Tr \left[ 
		\gamma_\text{chir} \frac{i}{\slashed{k} - M} \gamma_\text{chir} \frac{i}{\slashed{k} + \slashed{q} - M}
	\right] \\
	&=  2 \int \frac{d^2 k}{(2\pi)^2} \Tr \left[ 
			\gamma_\text{chir} \frac{\slashed{k} + M^\dagger}{k^2 - |M|^2} \gamma_\text{chir} \frac{\slashed{k} + \slashed{q} + M^\dagger}{k^2 - |M|^2}
		\right]
\end{aligned}	
\label{Im_sigma_ferm_init}	
\end{equation}
Using the trace identities \eqref{trace_id}	we can evaluate the traces:
\begin{equation}
\begin{aligned}
	\Tr \left[\gamma_\text{chir} \slashed{k} \gamma_\text{chir} \slashed{q} \right] 
		&= - d \left( k \cdot q \right)	\\
	\Tr \left[\gamma_\text{chir}  M^\dagger \gamma_\text{chir} M^\dagger \right]
		&= d \, (R^2 - I^2)
\end{aligned}	
\end{equation}
where the real and imaginary masses $R$ and $I$ are defined in \eqref{M_shorthand}.

Again, we will evaluate the integral in dimensional regularization using the Feynman prescription \eqref{feynman_prescription}. Introducing $l = k + x q$, we can rewrite the numerator of \eqref{Re_sigma_ferm_init} as 
\begin{equation}
\begin{aligned}
	\text{Num} &= d \left( - \left( k \cdot (k + q) \right) + (R^2 - I^2) \right) \\
		&= d \left( - l^2 + x (1-x) q^2 + (R^2 - I^2) \right) + \text{terms linear in } l
\end{aligned}		
\end{equation}
Introducing again $\Delta = |M|^2 - x (1-x) q^2$, performing a Wick rotation \eqref{Wick_simple} and using \eqref{pleasant_int_1} and \eqref{unpleasant_int_1} we arrive at 
\begin{equation}
\begin{aligned}
	i D_{\Im\sigma} &=  2 d \, \int\limits_0^1 dx \, \int \frac{d^d l_E}{(2\pi)^d} 
			\frac{ l_E^2 + x (1-x) q^2 + (R^2 - I^2)}{(l_E^2 + \Delta)^2} \\
		&=  2 d \, \int\limits_0^1 dx \, 
			\frac{1}{(4\pi)^{d/2}} \Gamma \left(2 - \frac{d}{2} \right)
			\left(  \frac{1}{\Delta^{1 - d/2}} \frac{d}{2} \frac{1}{1 - \frac{d}{2}} + \frac{ x (1-x) q^2 + (R^2 - I^2)}{\Delta^{2 - d/2}} \right)
\end{aligned}	
\label{Im_sigma_ferm_1}	
\end{equation}
The story with singularities repeats again. We throw out the $q^2$-independent terms and obtain
\begin{equation}
\begin{aligned}
	i D_{\Im\sigma}^\text{fin} &= 4 \, \int\limits_0^1 dx \, \frac{1}{4\pi}
			\left( - \ln \frac{\Delta}{|M|^2} + \frac{ x (1-x) q^2 + (R^2 - I^2)}{\Delta} \right) \\
		&=  4 \, \int\limits_0^1 dx \, \frac{1}{4\pi}
			\left(\frac{+ x^2 q^2 + (R^2 - I^2)}{|M|^2 - x (1-x) q^2} \right) \\
		&\underset{q \to 0}{\approx}  4 \frac{1}{4\pi} \ q^2 \, \left(+ \frac{1}{3}\frac{1}{|M|^2} + \frac{1}{6} \frac{R^2 - I^2}{|M|^4}  \right) \\
		&= 2 \frac{1}{4\pi} \, q^2 \, \frac{1}{3} \, \frac{3 |M|^2 - 2 I^2}{|M|^4}
\end{aligned}	
\label{Im_sigma_ferm_2}	
\end{equation}

To get the full fermionic contribution to the kinetic term, we have to sum over all fermionic flavors. Using \eqref{masses_ZN} and \eqref{xi_masses} we get from \eqref{Im_sigma_ferm_2} the normalization factor
\begin{equation}
	\left( \frac{2}{e^2_{\Im\sigma}} \right)_\text{ferm} = \sum_{k=0}^{N-1} \left[ 2 \frac{1}{4\pi}  \, \frac{1}{3} \, 
		\frac{3 |M_{k^{th}\text{ flavor}}|^2 - 2 I_{k^{th}\text{ flavor}}^2}{|M_{k^{th}\text{ flavor}}|^4} \right]
\end{equation}
or, equivalently,
\begin{equation}
	\left( \frac{1}{e^2_{\Im\sigma}} \right)_\text{ferm} = \sum_{k=0}^{N-1} \left[  \frac{1}{4\pi}  \, \frac{1}{3} \, 
		\frac{3 M^2_{\xi_k} - 2 \left(\Im \left(\sqrt{2}\langle\sigma\rangle - m_k \right) \right)^2}{M^4_{\xi_k}} \right]
\end{equation}
where we used notation
\begin{equation}
	M^2_{\xi_k} \equiv |M_{k^{th}\text{ flavor}}|^2
\end{equation}
In the vacuum where $\Im\sigma = 0$ we have
\begin{equation}
	\left( \frac{1}{e^2_{\Im\sigma}} \right)_\text{ferm} = \frac{1}{4\pi}  \, \sum_{k=0}^{N-1} \left[  
		\frac{1}{3} \, \frac{3 M^2_{\xi_k} - 2 \left(\Im  m_k  \right)^2}{M^4_{\xi_k}} \right]
\label{Im_sigma_ferm_factor}	
\end{equation}

\subsection{$A_\mu  -  \Im\sigma$ mixing}

\begin{figure}[h]
	\centering
     \includegraphics[width=0.3\textwidth]{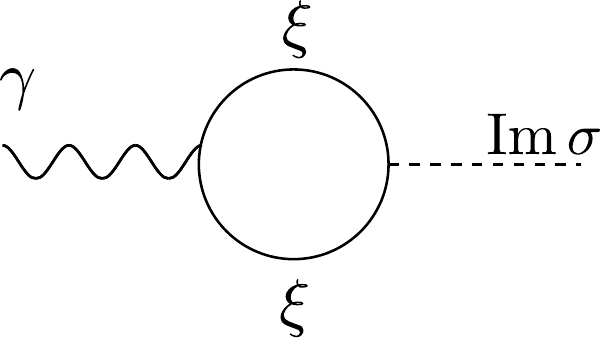}
     \caption{Photon-scalar mixing}
     \label{fig-app:loops:mixing}
\end{figure}

Consider the diagram from Fig. \ref{fig-app:loops:mixing}. Corresponding contribution is
\begin{equation}
\begin{aligned}
	i V_\text{mix} &= - \,(+ i) \, (- \sqrt{2}) \int \frac{d^2 k}{(2\pi)^2} \Tr \left[ 
		\frac{i}{\slashed{k} - M} \gamma^\mu \frac{i}{\slashed{k} + \slashed{q} - M} \gamma_\text{chir}
	\right] \\
	&=  \sqrt{2} \, i \, \int \frac{d^2 k}{(2\pi)^2} \Tr \left[ 
			\frac{\slashed{k} + M^\dagger}{k^2 - |M|^2} \gamma^\mu \frac{\slashed{k} + \slashed{q} + M^\dagger}{k^2 - |M|^2} \gamma_\text{chir}
		\right]
\end{aligned}	
\label{mix_ferm_init}	
\end{equation}
Using the trace identities \eqref{trace_id}	we can evaluate the traces:
\begin{equation}
\begin{aligned}
	\Tr \left[ \slashed{k} \gamma^\mu (\slashed{k} + \slashed{q}) \gamma_\text{chir}\right] 
		&= 0	\\
	\Tr \left[ \slashed{k} \gamma^\mu M^\dagger \gamma_\text{chir}\right] 
		&= d \, k_\nu \left( - R \epsilon^{\nu\mu}  + i I g^{\nu\mu} \right)	\\
	\Tr \left[ M^\dagger \gamma^\mu (\slashed{k} + \slashed{q}) \gamma_\text{chir}\right] 
		&= d \, \left( - R \epsilon^{\mu\nu}  + i I g^{\mu\nu} \right) (k_\nu + q_\nu)	\\
	\Tr \left[  M^\dagger \gamma^\mu  M^\dagger \gamma_\text{chir} \right]
		&= 0
\end{aligned}	
\end{equation}
where the real and imaginary masses $R$ and $I$ are defined in \eqref{M_shorthand}.

Again, we will evaluate the integral in dimensional regularization using the Feynman prescription \eqref{feynman_prescription}. Introducing $l = k + x q$, we can rewrite the numerator of \eqref{Re_sigma_ferm_init} as 
\begin{equation}
\begin{aligned}
	\text{Num} 
		&= d \left(  k_\nu \left( - R \epsilon^{\nu\mu}  + i I g^{\nu\mu} \right) 
				+ \left( - R \epsilon^{\mu\nu}  + i I g^{\mu\nu} \right) (k_\nu + q_\nu) \right) \\
		&= d \left(  k_\nu \left(  R \epsilon^{\mu\nu}  + i I g^{\mu\nu} \right) 
				+ \left( - R \epsilon^{\mu\nu}  + i I g^{\mu\nu} \right) (k_\nu + q_\nu) \right) \\
		&= d \left( -  R q_\nu \, \epsilon^{\mu\nu}  + i I (2 k_\nu + q_\nu) g^{\mu\nu} \right) \\
		&= d \left( -  R q_\nu \, \epsilon^{\mu\nu}  + i I  (1 - 2x) \, q_\nu \, g^{\mu\nu} \right)
			+  \text{terms linear in } l
\end{aligned}		
\end{equation}
Introducing again $\Delta = |M|^2 - x (1-x) q^2$, performing a Wick rotation \eqref{Wick_simple} and using \eqref{pleasant_int_1} and \eqref{unpleasant_int_1} we arrive at 
\begin{equation}
\begin{aligned}
	V_\text{mix}  &=   \sqrt{2} d \, \int\limits_0^1 dx \, \int \frac{d^d l_E}{(2\pi)^d} 
			\frac{-  R q_\nu \, \epsilon^{\mu\nu}  + i I  (1 - 2x) \, q_\nu \, g^{\mu\nu}}{(l_E^2 + \Delta)^2} \\
		&= - \sqrt{2} d \, \int\limits_0^1 dx \, 
			\frac{1}{(4\pi)^{d/2}} \Gamma \left(2 - \frac{d}{2} \right)
			\left(  \frac{-  R q_\nu \, \epsilon^{\mu\nu}  + i I  (1 - 2x) \, q_\nu \, g^{\mu\nu}}{\Delta^{2 - d/2}} \right) \\
		&\underset{d=2}{=} - 2 \sqrt{2} \, \frac{1}{4\pi} \, \int\limits_0^1 dx  \,
					\frac{-  R q_\nu \, \epsilon^{\mu\nu}  + i I  (1 - 2x) \, q_\nu \, g^{\mu\nu}}{|M|^2 - x (1-x) q^2} \\
		&= - 2 \sqrt{2} \, \frac{1}{4\pi} \, \int\limits_0^1 dx   \,
					\frac{-  R q_\nu \, \epsilon^{\mu\nu}  }{|M|^2 - x (1-x) q^2} \\
		&\underset{q \to 0}{\approx}   \sqrt{2} \, \frac{1}{2\pi} \, \frac{R}{|M|^2} \, q_\nu \, \epsilon^{\mu\nu}
\end{aligned}	
\label{mix_ferm_1}	
\end{equation}

To get the full fermionic contribution to the kinetic term, we have to sum over all fermionic flavors. Using \eqref{masses_ZN} and \eqref{xi_masses} we get from \eqref{mix_ferm_1} the mixing coupling constant
\begin{equation}
	\sqrt{2} b_{\gamma,\Im\sigma}  = \sum_{k=0}^{N-1} \left[
		\sqrt{2} \, \frac{1}{2\pi} \, \frac{R}{M^2_{\xi_k}}   \right]
\end{equation}
or, equivalently,
\begin{equation}
	b_{\gamma,\Im\sigma}  = \frac{1}{2\pi} \sum_{k=0}^{N-1} \left[
		 \frac{\Re \left(\sqrt{2}\langle\sigma\rangle - m_k \right) - \lambda(\mu)}{M^2_{\xi_k}}  \right]
\end{equation}
where we used notation
\begin{equation}
	M^2_{\xi_k} \equiv |M_{k^{th}\text{ flavor}}|^2
\end{equation}
In the vacuum where $\Im\sigma = 0$ this becomes
\begin{equation}
	b_{\gamma,\Im\sigma}  = \frac{1}{2\pi} \sum_{k=0}^{N-1} \left[
		 \frac{\sqrt{2}\langle\sigma\rangle - m_k  - \lambda(\mu)}{M^2_{\xi_k}}  \right]
\label{mix_ferm_constant}
\end{equation}

This corresponds to the term in the effective Lagrangian
\begin{equation}
	{\mathcal L}_\text{eff}^\text{mix} 
		= \sqrt{2} \, b \, \Im\sigma \epsilon^{\mu\nu} \p_\mu A_\nu 
		= - \sqrt{2} \, b \, \Im\sigma F^* 
\end{equation}

\subsection{Would-be $A_\mu  -  \Re\sigma$ mixing}

Consider the diagram like on Fig. \ref{fig-app:loops:mixing}, only with $\Re\sigma$ instead of $\Im\sigma$. Corresponding contribution is
\begin{equation}
\begin{aligned}
	i V_\text{mix Re} &= - \,(+ i) \, (- i \sqrt{2}) \int \frac{d^2 k}{(2\pi)^2} \Tr \left[ 
		\frac{i}{\slashed{k} - M} \gamma^\mu \frac{i}{\slashed{k} + \slashed{q} - M} 
	\right] \\
	&=  - \sqrt{2}  \, \int \frac{d^2 k}{(2\pi)^2} \Tr \left[ 
			\frac{\slashed{k} + M^\dagger}{k^2 - |M|^2} \gamma^\mu \frac{\slashed{k} + \slashed{q} + M^\dagger}{k^2 - |M|^2}
		\right]
\end{aligned}	
\label{Remix_ferm_init}	
\end{equation}
Using the trace identities \eqref{trace_id}	we can evaluate the traces:
\begin{equation}
\begin{aligned}
	\Tr \left[ \slashed{k} \gamma^\mu (\slashed{k} + \slashed{q}) \right] 
		&= 0	\\
	\Tr \left[ \slashed{k} \gamma^\mu M^\dagger \right] 
		&= d \, k_\nu \left( R g^{\nu\mu}  - i I \epsilon^{\nu\mu} \right)	\\
	\Tr \left[ M^\dagger \gamma^\mu (\slashed{k} + \slashed{q}) \right] 
		&= d \, \left(  R g^{\mu\nu}  - i I \epsilon^{\mu\nu} \right) (k_\nu + q_\nu)	\\
	\Tr \left[  M^\dagger \gamma^\mu  M^\dagger  \right]
		&= 0
\end{aligned}	
\end{equation}
where the real and imaginary masses $R$ and $I$ are defined in \eqref{M_shorthand}.

Again, we will evaluate the integral in dimensional regularization using the Feynman prescription \eqref{feynman_prescription}. Introducing $l = k + x q$, we can rewrite the numerator of \eqref{Re_sigma_ferm_init} as 
\begin{equation}
\begin{aligned}
	\text{Num} 
		&= d \left(  k_\nu \left( R g^{\nu\mu}  - i I \epsilon^{\nu\mu} \right)
				+ \left(  R g^{\mu\nu}  - i I \epsilon^{\mu\nu} \right) (k_\nu + q_\nu) \right) \\
		&= d \left(  k_\nu \left( R g^{\mu\nu}  + i I \epsilon^{\mu\nu} \right)
				+ \left(  R g^{\mu\nu}  - i I \epsilon^{\mu\nu} \right) (k_\nu + q_\nu) \right) \\
		&= d \left(  R g^{\mu\nu} (2 k_\nu + q_\nu)  - i I \epsilon^{\mu\nu} q_\nu \right)		\\
		&= d \left(  R g^{\mu\nu} q_\nu (1 - 2x)  - i I \epsilon^{\mu\nu} q_\nu \right)				
			+  \text{terms linear in } l
\end{aligned}		
\end{equation}
Introducing again $\Delta = |M|^2 - x (1-x) q^2$, performing a Wick rotation \eqref{Wick_simple} and using \eqref{pleasant_int_1} and \eqref{unpleasant_int_1} we arrive at 
\begin{equation}
\begin{aligned}
	 V_\text{mix Re}  &=  i \sqrt{2} d \, \int\limits_0^1 dx \, \int \frac{d^d l_E}{(2\pi)^d} 
			\frac{R g^{\mu\nu} q_\nu (1 - 2x)  - i I \epsilon^{\mu\nu} q_\nu}{(l_E^2 + \Delta)^2} \\
		&= i \sqrt{2} d \, \int\limits_0^1 dx \, 
			\frac{1}{(4\pi)^{d/2}} \Gamma \left(2 - \frac{d}{2} \right)
			\left(  \frac{R g^{\mu\nu} q_\nu (1 - 2x)  - i I \epsilon^{\mu\nu} q_\nu}{\Delta^{2 - d/2}} \right) \\
		&\underset{d=2}{=} i 2 \sqrt{2} \, \frac{1}{4\pi} \, \int\limits_0^1 dx  \,
					\frac{R g^{\mu\nu} q_\nu (1 - 2x)  - i I \epsilon^{\mu\nu} q_\nu}{|M|^2 - x (1-x) q^2} \\
		&= i 2 \sqrt{2} \, \frac{1}{4\pi} \, \int\limits_0^1 dx   \,
					\frac{- i I \epsilon^{\mu\nu} q_\nu  }{|M|^2 - x (1-x) q^2} \\
		&\underset{q \to 0}{\approx}   \sqrt{2} \, \frac{1}{2\pi} \, \frac{I}{|M|^2} \, q_\nu \, \epsilon^{\mu\nu}
\end{aligned}	
\label{Remix_ferm_1}	
\end{equation}

To get the full fermionic contribution to the kinetic term, we have to sum over all fermionic flavors. Using \eqref{masses_ZN} and \eqref{xi_masses} we get from \eqref{Remix_ferm_1} the corresponding coupling constant
\begin{equation}
	b_\text{mix Re} = \sum_{k=0}^{N-1} \sqrt{2} \, \frac{1}{2\pi} \, \frac{I_{k^{th}\text{ flavor}}}{M^2_{\xi_k}} 
\end{equation}
or, equivalently,
\begin{equation}
	b_\text{mix Re} = \sum_{k=0}^{N-1} \sqrt{2} \, \frac{1}{2\pi} \, \frac{\Im \left(\sqrt{2}\langle\sigma\rangle - m_k \right)}{M^2_{\xi_k}} 
\end{equation}
This vanishes exactly in the vacuum where $\Im\sigma = 0$.

For the same reason, the diagram Fig. \ref{fig-app:loops:scalar} with one external leg $\Re\sigma$ and another $\Im\sigma$ (i.e. a would-be $\Re - \Im$ mixing) also vanishes.

\section{Bosonic loops}

\subsection{Preliminaries}

The bosonic Lagrangian is (Euclidean version)
\begin{multline}
\mathcal L_\text{bos} 
	= \left|\nabla_\mu n^i\right|^2 + i\,D\left(\bar{n}_i n^i -2\beta \right)
	\\
	+ \sum_i\left|\sqrt 2\sigma-m_i\right|^2\, |n^i|^2
	+ \upsilon (\mu) \sum_i \Re\Delta m_{i0} |n^i|^2
\end{multline}
We can expand the $\sigma$ term as
\begin{multline}
	\left|\sqrt 2\sigma-m_i\right|^2 \approx \left|\sqrt 2 \langle\sigma\rangle - m_i\right|^2 
		+ 2 \Re(\sqrt{2}\delta\sigma) \cdot \Re\left(\sqrt 2 \langle\sigma\rangle - m_i\right)
		\\
		+ 2 \Im(\sqrt{2}\delta\sigma) \cdot \Im\left(\sqrt 2 \langle\sigma\rangle - m_i\right)		
\end{multline}
where $\delta\sigma$ is the vacuum fluctuations.

Introducing the masses of the $n^k$ fields
\begin{equation}
	m_{n_k}^2 = i\langle D \rangle + \upsilon (\mu)\Delta m_k  + \bigl| \sqrt{2}\langle\sigma\rangle - m_k \bigr|^2
\label{n_mass}	
\end{equation}
we can write down the relevant Lagrangian (Minkowski version) 
\begin{equation}
\begin{aligned}
\mathcal L_\text{bos} 
	= \left|\nabla_\mu n^i\right|^2 &-  \sum_i m^2_{n_k} |n^i|^2	\\
		&+ \sum_i 2 \Re(\sqrt{2}\delta\sigma) \cdot \Re\left(\sqrt 2 \langle\sigma\rangle - m_i\right) |n^i|^2	\\
		&+ \sum_i 2 \Im(\sqrt{2}\delta\sigma) \cdot \Im\left(\sqrt 2 \langle\sigma\rangle - m_i\right) |n^i|^2
\end{aligned}		
\label{lagrangian_bos}
\end{equation}
For simplicity, in loop calculations we will be calculating Feynman diagrams for one single flavor, and only then summing up over all flavors. So, we will work with the Lagrangian
\begin{equation}
\begin{aligned}
\mathcal L_\text{single flavor} 
	= \left|\nabla_\mu n\right|^2 &-  M^2 |n|^2	\\
		&+  2 \Re(\sqrt{2}\delta\sigma) \cdot R\ |n|^2	\\
		&+  2 \Im(\sqrt{2}\delta\sigma) \cdot I\ |n|^2
\end{aligned}		
\label{lagrangian_bos_single}
\end{equation}
where we have denoted
\begin{equation}
\begin{aligned}
	R &= \Re\left(\sqrt 2 \langle\sigma\rangle - m_i\right) \\
	I &= \Im\left(\sqrt 2 \langle\sigma\rangle - m_i\right)
\end{aligned}	
\end{equation}

\subsection{Photon kinetic term}

Consider the diagram like on Fig. \ref{fig-app:loops:photon} only with the bosonic loop. Corresponding contribution is
\begin{equation}
\begin{aligned}
	i \Pi^{\mu\nu} 
		&=   \int \frac{d^2 k}{(2\pi)^2}  
			i (- (k^\mu + q^\mu) - k^\mu) \frac{i}{k^2 - M^2} i (- (k^\nu + q^\nu) - k^\nu) \frac{i}{(k + q)^2 - M^2}	\\
		&=   \int \frac{d^2 k}{(2\pi)^2}  
			 (2k^\mu + q^\mu) (2 k^\nu + q^\nu)  \frac{1}{k^2 - M^2}  \frac{1}{(k + q)^2 - M^2}	\\
\end{aligned}		
\label{polarization_bos_init}	
\end{equation}
We will calculate this integral using the dimensional regularization.
The derivation to come follows closely \cite[Sec. 7.5]{PeskSchr}.
We will calculate the integral in \eqref{polarization_bos_init} using the Feynman prescription \eqref{feynman_prescription}:
\begin{equation}
\begin{aligned}
	\frac{1}{(k^2 - |M|^2) \, ((k + q)^2 - |M|^2)} 
		&= \int\limits_0^1 \, dx \, \frac{1}{\left(k^2 + 2 x k \cdot q + x q^2 - |M|^2\right)^2} \\
		&= \int\limits_0^1 \, dx \, \frac{1}{\left(l^2 + x (1-x) q^2  - |M|^2\right)^2}
\end{aligned}	
\end{equation}
where $l = k + x q$. In terms of $l$, the numerator of \eqref{polarization_bos_init} is 
\begin{equation}
\begin{aligned}
	\text{Num} 
		&= (2k^\mu + q^\mu) (2 k^\nu + q^\nu) \\
		&= (2l^\mu + (1-2x) q^\mu) (2 l^\nu + (1-2x) q^\nu) \\
		&= 4 l^\mu l^\nu + (1-2x)^2 q^\mu q^\nu 
			+ \text{terms linear in  }	l
\end{aligned}	
\end{equation}
Let us perform a Wick rotation:
\begin{equation}
	l^0 = i l^0_E
	\,, \quad
	l_\mu l^\mu = - (l_E)^2
\end{equation}
In terms of $l_E$ the integral \eqref{polarization_bos_init} becomes
\begin{equation}
		i \Pi^{\mu\nu} =  i  \, \int\limits_0^1 \, dx \, \int \frac{d^d l_E}{(2\pi)^d}
		\frac{ - l_E^2 \frac{4}{d} g^{\mu\nu} + (1-2x)^2 q^\mu q^\nu}{\left(l_E^2 + \Delta\right)^2}
\label{polarization_bos_2}		
\end{equation}
where we introduced $\Delta = |M|^2 - x (1-x) q^2$, and used the fact that in $d$ dimensions (see \cite[p. 251 eq. (7.87)]{PeskSchr})
\begin{equation}
	\int d^d l \ l^\mu l^\nu = \int d^d l \ \frac{1}{d} g^{\mu\nu} l^2 
\end{equation}
Now, we can evaluate the integrals over the momentum $l_E$ (see \cite[p. 251 eq. (7.85)]{PeskSchr}):
\begin{equation}
	\int \frac{d^d l_E}{(2\pi)^d} \frac{1}{\left(l_E^2 + \Delta\right)^2}
		= \frac{1}{(4\pi)^{d/2}} \Gamma \left(2 - \frac{d}{2} \right) \frac{1}{\Delta^{2 - d/2}}
\end{equation}
\begin{equation}
	\int \frac{d^d l_E}{(2\pi)^d} \frac{l_E^2}{\left(l_E^2 + \Delta\right)^2}
		= \frac{1}{(4\pi)^{d/2}} \frac{d}{2} \Gamma \left(1 - \frac{d}{2} \right) \frac{1}{\Delta^{1 - d/2}}
\end{equation}
We get for the polarization operator \eqref{polarization_bos_2}:
\begin{equation}
\begin{aligned}
	\Pi^{\mu\nu}(q) 
		&=   \int\limits_0^1 \, dx \, \frac{1}{(4\pi)^{d/2}} \frac{1}{\Delta^{1 - d/2}} \\
		&\phantom{- d \int\limits_0^1 \, dx } 
			\times \left[ - 2  g^{\mu\nu}  \Gamma \left(2 - \frac{d}{2} \right) 
				+  \frac{(1-2x)^2 q^\mu q^\nu \Gamma \left(2 - \frac{d}{2} \right)  }{ |M|^2 - x (1-x) q^2 }
			\right]
\end{aligned}	
\end{equation}
This expression has a pole at $d=2$, but we can ignore it since we are interested only in the quadratic in $q^\mu$ part, and the pole doesn't contain such terms\footnotemark.
\footnotetext{This pole cancels anyway with the corresponding pole from the 1-loop diagram with two external legs and 4-point coupling.}
So, the finite part at $d=2$ is:
\begin{equation}
\begin{aligned}
	\Pi^{\mu\nu}_\text{fin}(q) 
		&=   \int\limits_0^1 \, dx \, \frac{1}{4\pi}  
			 	\left[  2 g^{\mu\nu} \ln\frac{|M|^2 - x (1-x) q^2}{|M|^2}  +  \frac{(1-2x)^2 q^\mu q^\nu }{ |M|^2 - x (1-x) q^2 } \right]	\\ 
 		&=   \int\limits_0^1 \, dx \, \frac{1}{4\pi}  
	 			\left[ - 2 g^{\mu\nu} \frac{- x (1- 2x) q^2}{|M|^2 - x (1-x) q^2}  +  \frac{(1-2x)^2 q^\mu q^\nu }{ |M|^2 - x (1-x) q^2 } \right]	\\
 		&=   (- q^2 g^{\mu\nu} + q^\mu q^\nu) \cdot \int\limits_0^1 \, dx \, \frac{1}{4\pi}  
	 			  \frac{(1-2x)^2 }{ |M|^2 - x (1-x) q^2 } 	\\
	 	&\underset{q \to 0}{\approx}  (- q^2 g^{\mu\nu} + q^\mu q^\nu) \cdot \frac{1}{4\pi} \frac{1}{3} \frac{1}{|M|^2}
\end{aligned}	
\label{polarization_bos_ans}
\end{equation}

To get the full bosonic contribution to the photon kinetic term, we have to sum over all flavors. Using \eqref{masses_ZN} and \eqref{n_mass} we get from \eqref{polarization_bos_ans} the photon normalization
\begin{equation}
	\left( \frac{1}{e^2_{\gamma}} \right)_\text{bos} =  \frac{1}{4\pi}  \sum_{k=0}^{N-1} \left[\frac{1}{3} \frac{1}{m_{n_k}^2} \right]
\label{polarization_bos_factor}	
\end{equation}
where we used notation
\begin{equation}
	m_{n_k}^2 \equiv M_{k^{th}\text{ flavor}}^2
\end{equation}

\subsection{$\Re\sigma$ kinetic term}

Consider the diagram like on Fig. \ref{fig-app:loops:scalar} only with the scalar $n$ loop and $\Re\sigma$ external legs. Corresponding contribution is
\begin{equation}
\begin{aligned}
	i D_{\Re\sigma} 
		&= (i R)^2 \int \frac{d^2 k}{(2\pi)^2} \Tr \left[ \frac{i}{k^2 - M^2} \frac{i}{(k+q)^2 - M^2}	\right] \\
		&= R^2 \int \frac{d^2 k}{(2\pi)^2} \Tr \left[ \frac{1}{k^2 - M^2} \frac{1}{(k+q)^2 - M^2}	\right] \\
\end{aligned}	
\label{Re_sigma_bos_init}	
\end{equation}
Introducing again $\Delta = M^2 - x (1-x) q^2$, performing a Wick rotation \eqref{Wick_simple} and using \eqref{pleasant_int_1} we arrive at 
\begin{equation}
\begin{aligned}
	 D_{\Re\sigma} &=  R^2 \, \int\limits_0^1 dx \, \int \frac{d^d l_E}{(2\pi)^d} 
			\frac{1}{(l_E^2 + \Delta)^2} \\
		&= R^2 \, \int\limits_0^1 dx \, 
			\frac{1}{(4\pi)^{d/2}} \Gamma \left(2 - \frac{d}{2} \right)
			\frac{1}{\Delta^{2 - d/2}} \\
		&\underset{d=2}{=}  R^2 \, \int\limits_0^1 dx \, \frac{1}{4\pi} \frac{1}{M^2 - x (1-x) q^2}	\\
		&\underset{q \to 0}{\approx} \frac{1}{4\pi} \, \frac{R^2}{M^4} \, \frac{1}{6} \, q^2
\end{aligned}	
\label{Re_sigma_bos_1}	
\end{equation}

To get the full bosonic contribution to the kinetic term, we have to sum over all flavors. Using \eqref{masses_ZN} and \eqref{n_mass} we get from \eqref{Re_sigma_bos_1} the normalization factor
\begin{equation}
	\left( \frac{2}{e^2_{\Re\sigma}} \right)_\text{bos} = \sum_{k=0}^{N-1} \left[ 2 \frac{1}{4\pi}  \, \frac{4}{6} \, 
		\frac{\left(\Re\left(\sqrt 2 \langle\sigma\rangle - m_i\right)\right)}{m_{n_k}^4} \right]
\end{equation}
or, equivalently,
\begin{equation}
	\left( \frac{1}{e^2_{\Re\sigma}} \right)_\text{bos} = \frac{1}{4\pi} \, \sum_{k=0}^{N-1} \left[ \frac{2}{3} \, 
		\frac{\left(\Re\left(\sqrt 2 \langle\sigma\rangle - m_i\right)\right)^2}{m_{n_k}^4} \right]
\end{equation}

In the vacuum where $\Im\sigma = 0$ we have
\begin{equation}
	\left( \frac{1}{e^2_{\Re\sigma}} \right)_\text{bos} = \frac{1}{4\pi} \, \sum_{k=0}^{N-1} \left[ \frac{2}{3} \, 
		\frac{\left(\sqrt 2 \langle\sigma\rangle - \Re m_i\right)^2}{m_{n_k}^4} \right]
\label{Re_sigma_bos_factor}	
\end{equation}

\subsection{$\Im\sigma$ kinetic term}

Calculation of the $\Im\sigma$ kinetic term is almost the same as for the $\Re\sigma$, except for the vertex coefficient $I$ instead of $R$. So, for a single flavor we get (cf. \eqref{Re_sigma_bos_1})
\begin{equation}
\begin{aligned}
	 D_{\Im\sigma} &\underset{q \to 0}{\approx}  \frac{1}{4\pi} \, \frac{I^2}{M^4} \, \frac{1}{6} \, q^2
\end{aligned}	
\label{Im_sigma_bos_1}	
\end{equation}
while for the full bosonic contribution in the vacuum where $\Im\sigma = 0$ we have
\begin{equation}
	\left( \frac{1}{e^2_{\Im\sigma}} \right)_\text{bos} = \frac{1}{4\pi} \, \sum_{k=0}^{N-1} \left[ \frac{2}{3} \, 
		\frac{\left(\Im m_i\right)^2}{m_{n_k}^4} \right]
\label{Im_sigma_bos_factor}	
\end{equation}

\subsection{Would-be mixings}

There are a few more diagrams with bosonic loops which, however, do not contribute.

First, there is the $\Im\sigma - \Re\sigma$ mixing diagram. However, it is proportional to
\begin{equation}
	\left[\Im\sigma - \Re\sigma\right]_\text{mixing} \sim \sum_{k=0}^{N-1} \left\{
		\left(\Re\left(\sqrt 2 \langle\sigma\rangle - m_i\right)\right) \cdot \left(\Im\left(\sqrt 2 \langle\sigma\rangle - m_i\right)\right)
	\right\}
\end{equation}
which vanishes in the vacuum with $\Im\sigma = 0$.

Second, the are $A_\mu - \Re\sigma$ mixing, and the $A_\mu - \Im\sigma$ mixing diagrams. However, these are proportional to $q^\mu$ without any $\epsilon^{\mu\nu}$, and therefore they must vanish due to gauge invariance. (Actually we checked explicitly that they do indeed vanish.)

\section{Final result}

Collecting all the above results,
we arrive at the effective action \eqref{effaction} (Minkowski notation)   
\begin{multline}
	S_{\rm eff}=
	 \int d^2 x \Big\{
	- \frac1{4e_{\gamma}^2}F^2_{\mu\nu} + \frac1{e_{\Im\sigma}^2}	|\pt_{\mu}\Im\sigma|^2 + \frac1{e_{\Re\sigma}^2}	|\pt_{\mu}\Re\sigma|^2
	\\
	- V(\sigma) - \sqrt{2} \, b_{\gamma,\Im\sigma} \, \Im\sigma \,F^{*}
	  \Big\}.
\end{multline}
The constants of these effective action are given by
	\eqref{polarization_ferm_factor} and \eqref{polarization_bos_factor},    
	\eqref{Re_sigma_ferm_factor} and     \eqref{Re_sigma_bos_factor},
	\eqref{Im_sigma_ferm_factor} and   \eqref{Im_sigma_bos_factor},    
	\eqref{mix_ferm_constant},
and we get \eqref{eff_normalizations}.

%
%


\chapter{Photon propagator in two spacetime dimensions}
\label{app:2d_photon}

In this Appendix we discuss an issue of the photon mass compatible with gauge invariance.

\section{Photon propagator in the generalized gauge}

Let's start from a \textquote{bare} photon propagator in a generalized Feynman gauge:
\begin{equation}
\begin{aligned}
	G_{\mu\nu}^0 &= -i\, e^2_\gamma  \frac{1}{k^2} \,  \left(g^{\mu\nu} - (1 - \nu^{-1}) \frac{k^\mu k^\nu}{k^2}\right) \\
		 &= -i\, e^2_\gamma \left[  \frac{1}{k^2} \,  \left(g^{\mu\nu} - \frac{k^\mu k^\nu}{k^2}\right) + \nu^{-1} \frac{k^\mu k^\nu}{k^4}  \right]
\end{aligned}	
\label{photon_bare}
\end{equation}
As one can see, it generally contains the transverse and the longitudinal parts.

Now consider the photon-scalar mixing of Sec.~\eqref{sec:effact} in the supersymmetric case. 
(The question considered here is most clear in this setting.)
To derive the full photon propagator, we must consider a sum of diagrams iterating the photon and scalar $\Im\sigma$ lines (see Fig. \ref{fig:iter_gamma}).
Such an iteration gives for the full propagator
\begin{equation}
	\widehat{G}_{\mu\nu} = G_{\mu\nu}^0 + G_{\mu\nu'}^0  \omega^{\nu' \mu'}  G_{\mu'\nu}^0 + \ldots
\end{equation}
where the quantity $\omega^{\nu' \mu'}$ is given by the $\Im\sigma$ line and two vertices (an \textquote{amputated} photon-scalar-photon diagram),
\begin{equation}
\begin{aligned}
	\omega^{\alpha \beta} &= i\sqrt{2} b_{\gamma,\Im\sigma} \varepsilon^{\alpha \beta'} k_{\beta'} 
		\cdot \, \left(-\frac{i}{2}\right)\, e^2_{\Im\sigma} \frac{1}{k^2} \, \cdot
		i\sqrt{2} b_{\gamma,\Im\sigma}  k_{\alpha'} \varepsilon^{\alpha' \beta}
		\\
		&= - i e^2_{\Im\sigma} b_{\gamma,\Im\sigma}^2 \left(g_{\alpha \beta} - \frac{k^\alpha k^\beta}{k^2} \right)
\end{aligned}		
\end{equation} 
This is a purely transverse quantity, and therefore only transverse part of \eqref{photon_bare} gets renormalized. We find for the full photon propagator:
\begin{equation}
	\widehat{G}_{\mu\nu} = -i\, e^2_\gamma \left[  \frac{1}{k^2 - m_\gamma^2} \,  \left(g^{\mu\nu} - \frac{k^\mu k^\nu}{k^2}\right) + \nu^{-1} \frac{k^\mu k^\nu}{k^4}  \right]	
\label{photon_ren_full}
\end{equation}
where the photon mass equals that of $\Im\sigma$,
\begin{equation}
	m_\gamma^2 = m_{\Im\sigma}^2 = e^2_\gamma\, e^2_{\Im\sigma}\, b^2_{\gamma,\Im\sigma} = 4 \Lambda^2 \,.
\label{m_gamma}	
\end{equation}
We obtained the result \eqref{photon_ren_full} at first by direct calculation, and then by applying a general formula from \cite[page 325 equation 7-17]{ItzZu}\footnotemark, the result is of course the same. The bare propagator \eqref{photon_bare} is renormalized in such a way that the pole in front of the transversal part is shifted, while the longitudinal part stays unchanged.
\footnotetext{Note a different photon field normalization in \cite{ItzZu}.}

Formula \eqref{photon_ren_full} can be compared with a massive photon propagator in a general gauge (see, for example, \cite[page 619 equation 12-226]{ItzZu})
\begin{equation}
	G_{\mu\nu}^\text{mass} = -i\, e^2_\gamma \, \frac{1}{k^2 - m_\gamma^2} \left(g^{\mu\nu} - (1 - \nu^{-1}) \frac{k^\mu k^\nu}{k^2 - m_\gamma^2 / \nu}\right)  \,.
\label{photon_massive}	
\end{equation}
One can see that \eqref{photon_ren_full} and \eqref{photon_massive} do not generally coincide.
Why so?
Turns out that there is more than one way to give the photon a mass in two spacetime dimensions. 
The propagator \eqref{photon_massive} corresponds to just one of them, namely the Higgs mechanism.

\section{Photon masses}

Consider usual action of a massive vector field in Minkowski spacetime
\begin{equation}
	S = \int d^2 x \left\{ - \frac1{4}F_{\mu\nu}F^{\mu\nu} + \frac{m_\gamma^2}{2} A_\mu A^\mu \right\} \,.
\end{equation}
This action by itself is not gauge invariant. 
(It could have been turned into a gauge invariant action by means of Higgs mechanism.)
The generalized photon propagator is given by \eqref{photon_massive}.

However, in two spacetime dimensions there is another way to introduce the photon mass. This other option does not destroy gauge invariance. Consider the effective action for the Schwinger model (with fermions integrated out):
\begin{equation}
\begin{aligned}
	S_\text{schw} &= \int d^2 x \left\{ - \frac1{4}F_{\mu\nu}F^{\mu\nu} + \frac{m_\gamma^2}{2} F^* \frac{1}{\Box} F^* \right\} \\
		&= \int d^2 x \left\{ - \frac1{4}F_{\mu\nu}F^{\mu\nu} + \frac{m_\gamma^2}{2} A_\mu A^\mu 
			- \frac{m_\gamma^2}{2} A_\mu \frac{\p^\nu \p^\beta}{\Box} A_\beta \right\} \,.
\end{aligned}
\label{schw_vect}	
\end{equation}
For the transition here see Eq. \eqref{F-dual-identity}.
Adding to the Lagrangian in \eqref{schw_vect} a gauge fixing term 
\begin{equation}
	- \frac{\nu}{2} (p_\mu A^\mu)^2
\end{equation} 
we can write generalized photon propagator
\begin{equation}
	\widehat{G}_{\mu\nu} = -i\,  \left[  \frac{1}{k^2 - m_\gamma^2} \,  \left(g^{\mu\nu} - \frac{k^\mu k^\nu}{k^2}\right) + \nu^{-1} \frac{k^\mu k^\nu}{k^4}  \right]	
\label{photon_ren_1}
\end{equation}

The action \eqref{schw_vect} appears to be non-local, but actually it is not.
In two dimensions, the vector field can be parametrized as
\begin{equation}
	A_\mu = \p_\mu \alpha + \epsilon_{\mu\nu} \p^\nu \phi \,.
\end{equation}
In this parametrization \eqref{schw_vect} becomes
\begin{equation}
\begin{aligned}
	S_\text{schw} &= \int d^2 x \left\{ \frac1{2} (\Box \phi) (\Box \phi) + \frac{m_\gamma^2}{2} \phi \Box \phi \right\} \\
		&= \int d^2 x \, \frac1{2} \,  \phi \left( \Box + m_\gamma^2 \right) \Box \, \phi 
\end{aligned}		
\label{schw_scal}	
\end{equation}
Equation of motion for the $\phi$ field is
\begin{equation}
	\Box (\Box + m_\gamma^2) \phi = 0 \,.
\end{equation}
So, it would appear that the field $\phi$ has two modes, one massless and the other massive. However, the massless mode is unphysical, since it corresponds to a trivial vector field strength:
\begin{equation}
	F^* = - F_{01} = - (\p_0 A_1 - \p_1 A_0) = \Box \phi \,,
\end{equation}
\begin{equation}
	\Box \phi = 0 \Longleftrightarrow F_{\mu\nu} = 0 \,.
\end{equation}

\section{Our model}

Let's start from our effective action \eqref{effaction} in Minkowski spacetime.
As we derived previously, if we start from the photon propagator in a generalized gauge \eqref{photon_bare}
and then diagonalize 
our effective action, we will get exactly \eqref{photon_ren_1} multiplied by $e^2_\gamma$. We do not get \eqref{photon_massive}, {\em and this is a good thing}, since it is \eqref{photon_ren_1} that corresponds to the gauge invariant effective action in our case, and not \eqref{photon_massive}.

Let's look at this situation more closely. First, drop from \eqref{effaction} the terms with $\Re\sigma$ and $V(\sigma)$, which are irrelevant for our purposes:
\begin{equation}
	S_{\rm eff}=
	 \int d^2 x \left\{
	- \frac1{4e_{\gamma}^2} F_{\mu\nu}F^{\mu\nu} + \frac1{e_{\Im\sigma}^2}	|\pt_{\mu}\Im\sigma|^2  + \sqrt{2} \, b_{\gamma,\Im\sigma} \, \Im\sigma \,F^{*}
	  \right\},
\label{effaction_drop}	  
\end{equation}
%
%
The scalar field enters \eqref{effaction_drop} at most quadratically, so it can be integrated out using its equation of motion:
\begin{equation}
	- \frac{2}{e_{\Im\sigma}^2} \Box \Im\sigma + \sqrt{2} \, b_{\gamma,\Im\sigma} \,F^{*} = 0
		\Longrightarrow \Im\sigma = \frac{1}{\Box} \frac{1}{\sqrt{2}} \, e_{\Im\sigma}^2 \, b_{\gamma,\Im\sigma} \,F^{*}
\label{Im_sigma_eq}		
\end{equation}
Substituting this back into \eqref{effaction_drop} gives
\begin{equation}
\begin{aligned}
	S_{\rm eff} &=
	 \int d^2 x \left\{
	- \frac1{4e_{\gamma}^2} F_{\mu\nu}F^{\mu\nu} + \frac{1}{2}  \, e_{\Im\sigma}^2 \, b_{\gamma,\Im\sigma}^2 F^{*} \frac{1}{\Box} F^{*}
	  \right\}
	  \\
	  &= 
 	 \frac1{e_{\gamma}^2} \int d^2 x \left\{
  	- \frac1{4} F_{\mu\nu}F^{\mu\nu} + \frac{m_\gamma^2}{2}   F^{*} \frac{1}{\Box} F^{*}
	  \right\}
\end{aligned}	  
\label{effaction_drop_photon}	  
\end{equation}
where $m_\gamma^2 = e_{\gamma}^2 \,e_{\Im\sigma}^2 \, b_{\gamma,\Im\sigma}^2$.
This is obviously equivalent to \eqref{schw_vect}.

%
%


\chapter{Modular functions}
\label{sec:modular_101}

In this Appendix we discuss some properties of the modular functions used in Chapter~\ref{sec:b_meson}.

\section{$\theta$ functions}

Let us introduce the nome
\begin{equation}
	q = e^{i \pi \tau_\text{SW}} = e^{2 i \pi \tau}
\label{nome}	
\end{equation}
where $\tau_\text{SW}$ is the gauge coupling defined in \eqref{tau_def}.
We define $\theta$-functions as in \cite{SW2}. In terms of the nome \eqref{nome} they are
\begin{equation}
\begin{aligned}
	\theta_1 (q) &= \sum_{n \in \mathbb{Z}} q^{(n + 1/2)^2}		= 2 q^{1/4} (1 + q^2 + \ldots)  \,, \\
	\theta_2 (q) &= \sum_{n \in \mathbb{Z}} (-1)^n \, q^{n^2}	= 1 - 2 q + \ldots  \,, \\
	\theta_3 (q) &= \sum_{n \in \mathbb{Z}} q^{n^2}				= 1 + 2 q + \ldots  \,. \\
\end{aligned}
\label{theta_def}
\end{equation}
There are many relations among these, e.g. \cite{Chandrasekharan}
\begin{equation}
	\theta_3^4 = \theta_2^4 + \theta_1^4 \,.
\end{equation}

The $\theta$ functions \eqref{theta_def} are obviously invariant under $T$ transformation $\tau_\text{SW} \to \tau_\text{SW} + 2$ \eqref{ST_transformation}. Moreover, the following identities \cite[eq. (8.10)]{Chandrasekharan} hold for the $T^\frac{1}{2}$ \eqref{ST_transformation} transformation:
\begin{equation}
\begin{aligned}
	\theta_1 \left( \tau_\text{SW} + 1 \right) &= e^\frac{i \pi}{4} \ \theta_1(\tau_\text{SW})  \,,  \\[2mm]
	\theta_2 \left( \tau_\text{SW} + 1 \right) &= \theta_3(\tau_\text{SW})  \,, \\[2mm]
	\theta_3 \left( \tau_\text{SW} + 1 \right) &= \theta_2(\tau_\text{SW})  \,. \\
\end{aligned}
\label{half_shift}
\end{equation} 
Under $S$ \eqref{ST_transformation}, we have \cite[eq. (8.9)]{Chandrasekharan}
\begin{equation}
\begin{aligned}\\
	\theta_1 \left(- \frac{1}{\tau_\text{SW}} \right) &= \sqrt{- i \tau_\text{SW}} \ \theta_2(\tau_\text{SW})  \,, \\[2mm]
	\theta_2 \left(- \frac{1}{\tau_\text{SW}} \right) &= \sqrt{- i \tau_\text{SW}} \ \theta_1(\tau_\text{SW})  \,, \\[2mm]
	\theta_3 \left(- \frac{1}{\tau_\text{SW}} \right) &= \sqrt{- i \tau_\text{SW}} \ \theta_3(\tau_\text{SW})  \,, \\
\end{aligned}
\label{theta_S_APS}
\end{equation} 
where $\sqrt{-i \tau_\text{SW}} = +1$ for $\tau_\text{SW} = i$.
Here we slightly abused notation by using the same letter $\theta$ as in \eqref{theta_def}.

\section{The \boldmath{$h$} function}

From the $\theta$ functions we can build modular functions. In the SW curve \eqref{4d_curve_sun} the $h$ function was used, which is defined as \cite{APS}
\begin{equation}
	h(\tau_\text{SW}) = \frac{2 \theta_1^4(\tau_\text{SW})}{\theta_2^4(\tau_\text{SW}) - \theta_1^4(\tau_\text{SW})}
\label{h_def_APS}	
\end{equation}
or, in terms of the nome \eqref{nome},
\begin{equation}
	h(q) = 32 \, q + O(q^2) \,.
\end{equation}
The $S$-transformation acts on \eqref{h_def_APS} as
\begin{equation}
	h\left(- \frac{1}{\tau_\text{SW}} \right) = - 2 - h(\tau_\text{SW}) \,,
\end{equation}
and the combination
\begin{equation}
	h \cdot (h + 2) = \frac{4 \theta_1^4 \theta_2^4}{(\theta_2^4 - \theta_1^4)^2}
\label{h_combination}	
\end{equation}
is invariant with respect to $S$ and $T$ transformations. Under the half shift $T^\frac{1}{2}$ it becomes
\begin{equation}
	h(\tau_\text{SW} + 1) \cdot (h(\tau_\text{SW} + 1) + 2)  
		= - \frac{4 \theta_3^4(\tau_\text{SW}) \cdot \theta_2^4(\tau_\text{SW})}{(\theta_3^4(\tau_\text{SW}) + \theta_1^4(\tau_\text{SW}))^2}
		\,.
\end{equation}

\section{The \boldmath{$\lambda$} function}
\label{klambdaf}

In Chapter~\ref{sec:b_meson} we have also used the modular $\lambda$ function (see e.g. \eqref{2d_4d_coupling_my}) which can be expressed as 
\begin{equation}
	\lambda(\tau_\text{SW}) 
		= \frac{\theta_1^4(\tau_\text{SW}) }{\theta_3^4(\tau_\text{SW}) }
		= 16 q - 128 q^2 + O(q^3)
\label{lambda_function}		
\end{equation}
where $q$ is the nome \eqref{nome}.
This function is again invariant under the $T$ transformation, while under $S$ it transforms as
\begin{equation}
	\lambda\left(- \frac{1}{\tau_\text{SW}} \right) = 1 - \lambda(\tau_\text{SW}) \,.
\label{lambda_trans_S}	
\end{equation}
Under the half shift \eqref{half_shift} $\lambda$ becomes
\begin{equation}
	\lambda(\tau_\text{SW} + 1) = \frac{\lambda(\tau_\text{SW})}{\lambda(\tau_\text{SW}) - 1} = - \frac{\theta_1^4(\tau_\text{SW}) }{\theta_2^4(\tau_\text{SW}) } \,.
\label{lambda_trans_halfshift}	
\end{equation} 
From \eqref{lambda_trans_S} and \eqref{lambda_trans_halfshift} we see that under the $S$ transformation
\begin{equation}
	\lambda(\tau_\text{SW} + 1) \xrightarrow{S} \lambda\left(- \frac{1}{\tau_\text{SW}} + 1 \right) = \frac{1}{\lambda(\tau_\text{SW} + 1)} \,.
\label{lambda_p1_S}	
\end{equation} 
Using \eqref{h_combination} and \eqref{lambda_trans_halfshift}	we can write down a relation between $\lambda$ and $h$ function,
\begin{equation}
	- h(\tau_\text{SW})[h(\tau_\text{SW}) +2] = \frac{4 \, \lambda(\tau_\text{SW} + 1)}{(1 + \lambda(\tau_\text{SW} + 1))^2} \,.
\label{h_tau_relation}	
\end{equation}

The inverse of $\lambda(\tau)$ is given in terms of the hypergeometric functions
\begin{equation}
	  \tau = i~\frac{{}_2F_1\left(1/2,1/2;1;1-\lambda\right)}{ {}_2F_1\left(1/2,1/2;1;\lambda\right)} \,.
\label{lambda_inv_2F1}	  
\end{equation}
In terms of the complete elliptic integral of the first kind $K(k)$, 
\begin{equation}
	  \tau = i~\frac{K(\sqrt{1 - \lambda})}{K(\sqrt{\lambda})} \,.
\label{lambda_inv_K}	  
\end{equation}

%
%


\chapter{More on the central charge of the $\mathbb{WCP}(2,2)$ model}

In this Appendix we discuss some additional properties of the central charge and secondary curves of marginal stability in the \wcpt model.

\section{Secondary curves}
\label{sec:secondary_CMS}

Now we will investigate decays of particles, not covered in Sec.~\ref{sec:CMS}, and draw the corresponding CMS. Do this end, one has to keep in mind that the BPS kink central charge \eqref{BPSmass} is, generally speaking, a multi-branched function. 
On the $\beta$ plane, it can have branch cuts originating at the points where the kink mass develops some kind of a singularity. 
(This could be ignored while considering the primary CMS \eqref{CMS_right_curve} and \eqref{CMS_left_curve}, but not for the present task.)
From the explicit expressions for the kink mass we see that it is singular at the origin (see Eq. \eqref{kink_mass_P_beta0} and \eqref{kink_mass_K_beta0}) 
and at the AD points (see Eq. \eqref{CP1_Z_AD_2}). Therefore there are cuts originating from these points (modulo the $2\pi$ periodicity in the $\theta_{2d}$ direction).

\subsection{\textquote{Extra} kink decays}

When we go from strong coupling into the weak coupling domain $\beta \gg 0$, the kinks $[Z_P],\ P=1,2$ do not decay (they become massless at the AD points on the right curve \eqref{CMS_right_curve}, and they can be \textquote{dragged} through these points, where these kinks are the only massless particles and therefore absolutely stable \cite{Ferrari:1996sv}). On the other hand, the kinks  $[Z_K],\ K=3,4$ have masses of the order $~ |m_K - \ov{m}|$, and therefore they could decay into, say, a pair $[Z_P]$ + bifundamental. They can decay via the process
\begin{equation}
	[Z_K] \to [Z_P] + [- i (m_P - m_K)] \,.
\label{MK-MP_reaction}	
\end{equation}
In the dual weak coupling domain $\beta \ll 0$, the $P$-kinks decay via
\begin{equation}
	[Z_P] \to [Z_K] + [ i (m_P - m_K)] \,.
\label{MP-MK_reaction}	
\end{equation}

\begin{figure}[h]
	\centering
	\includegraphics[width=0.7\linewidth]{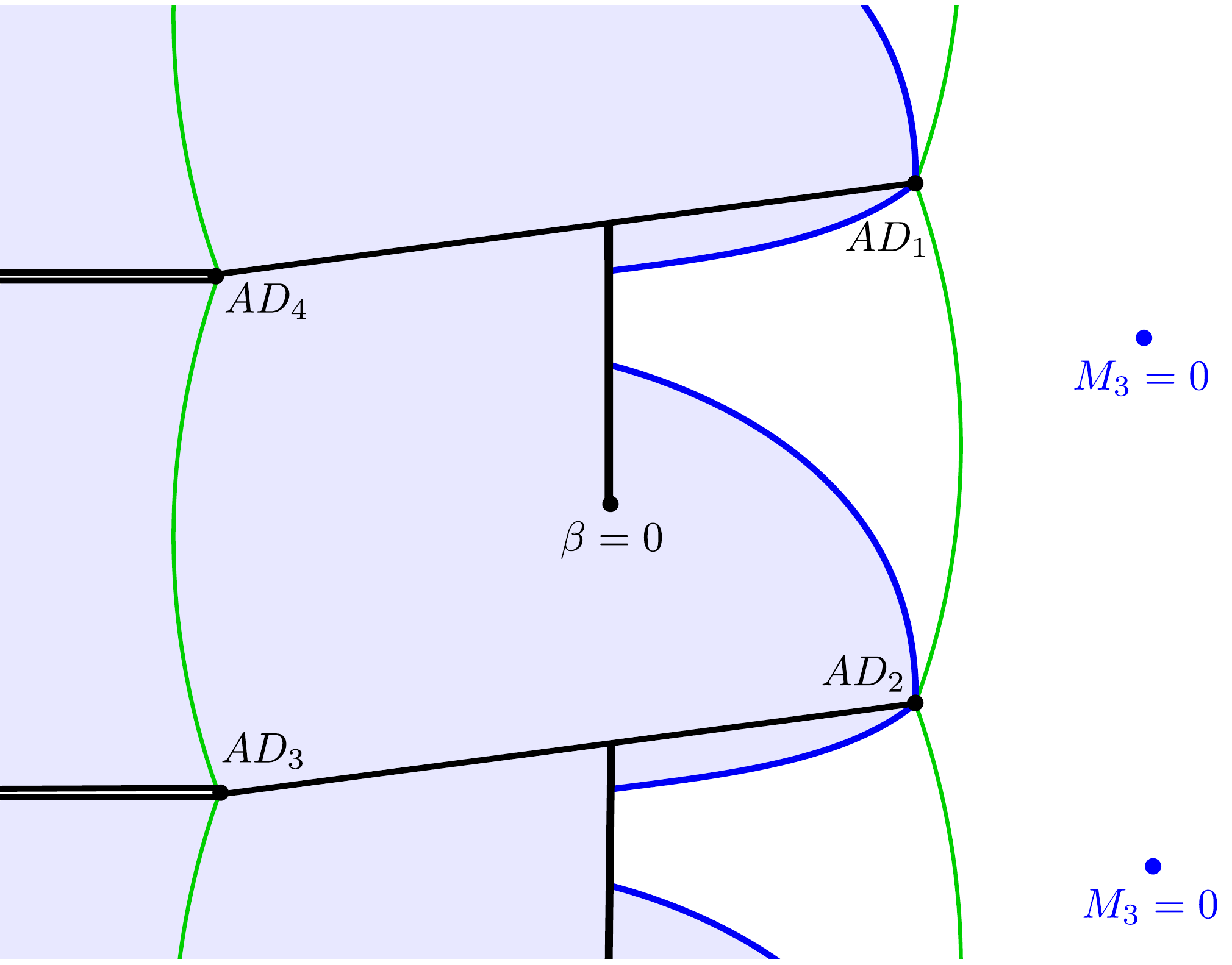}
	\caption{
		Complex plane of $\beta$.
		Schematic representation of the CMS structure for the $M_3$ kink decay. (Picture for $M_4$ is qualitatively the same.)
		Thin green lines are the primary curves.
		Thick black lines are the cuts.
		Thick blue lines are the CMS for the decays of $M_3$.
		Blue-shaded region is the domain of existence of $M_3$.
		The blue points on the right are where the mass of would-be $M_3$ state would vanish, see \eqref{M_K_large_beta}.
		}
\label{fig:CMS_right_13}
\end{figure}

Moreover, under some conditions these kinks must decay, otherwise they could become massless at some points at weak coupling. 
To see this, consider a $K$-kink central charge at weak coupling in the $\mathbb{CP}(1)$ limit \eqref{CP1_limit}. Formula \eqref{CP1_Z_quasiclassical} can be straightforwardly generalized for $K$-kinks as
\begin{equation}
	Z_K \approx  - \beta_{CP(1)} \cdot \delta m_{12} + i \, (m_K - \ov{m})  + \frac{\delta m_{12}}{\pi} \,,
\end{equation}
so that the mass of the corresponding state is the limit $\Re\beta \equiv r \gg 1$ is given by
\begin{equation}
	M_K \approx |\delta m_{12}| \cdot \left( r - \frac{1}{\pi} \ln \abs{\frac{\Delta m }{\delta m_{12}}} - \frac{1}{\pi} - \Im\frac{m_K - \ov{m}}{\delta m_{12}} \right) \,.
\label{M_K_large_beta}
\end{equation}

We see that for certain mass choices there are points at weak coupling (r.h.s. domain $\beta \gg \beta_{AD}$) where $M_K$ vanish . Therefore, these states must decay.
Analogously, $M_P$  kinks ($P=1,2$) may become massless in the dual weak coupling domain $\beta \ll - \beta_{AD}$. Their mass in this limit is given by
\begin{equation}
	M_P \approx |\delta m_{34}| \cdot \left( r - \frac{1}{\pi} \ln \abs{\frac{\Delta m }{\delta m_{34}}} - \frac{1}{\pi} + \Im\frac{m_P - \wt{m}}{\delta m_{34}} \right) \,.
\end{equation}

The  CMS equation for both decays \eqref{MK-MP_reaction} and \eqref{MP-MK_reaction} is
\begin{equation}
	\Re \left( \frac{ Z_P }{m_P - m_K} \right) = 0  
	\Leftrightarrow
	\Re \left( \frac{ Z_K }{m_P - m_K} \right) = 0 \,.
\label{CMS_K_decay}
\end{equation}
We must add to this equation  the condition that a particle cannot decay into heaver particles:
\begin{equation}
\begin{aligned}
	| Z_K | &= | Z_P | + | - i ( m_P - m_K ) | \quad \text{for the decay \eqref{MK-MP_reaction}} \,, \\[2mm]
	| Z_P | &= | Z_K | + |   i ( m_P - m_K ) | \quad \text{for the decay \eqref{MP-MK_reaction}} \,.
\end{aligned}
\label{abs_decay_condition}
\end{equation}
In the case when $m_1,\ m_2,\ m_K$ for some $K$ lie on a straight line in the complex plane, the CMS for the decay \eqref{MK-MP_reaction} {\em coincides with the primary curve} \eqref{CMS_right_curve}. Conversely, when for some $P$ the masses $m_P,\ m_3,\ m_4$ are aligned, the CMS for \eqref{MP-MK_reaction} coincides with the dual primary curve \eqref{CMS_left_curve}.
%

The CMS equation \eqref{CMS_K_decay} simplifies in the $\mathbb{CP}(1)$ limit \eqref{CP1_limit}. 
Using a simple generalization of the approximate central charge formula \eqref{CP1_Z_AD_2} we can rewrite this equation near an AD point as:
\begin{equation}
	\Re \left[ \frac{m_1 - m_2}{m_P - m_K} \cdot (\beta - \beta_{AD_P})^{3/2} \right] = 0
\end{equation}
for  $P=1,2$. (The indices of AD points follow Fig.~\ref{fig:CMS_type1}.) This equation is equivalent to
\begin{equation}
	\cos \left( \frac{3}{2} \arg(\beta - \beta_{AD_P}) + \phi_{PK} \right) = 0
	\,, \quad
	\phi_{PK} = \arg\left( \frac{m_1 - m_2}{m_P - m_K} \right) \,.
\end{equation}
The solution is represented by lines originating from the $AD_P$ point and going out at angles
\begin{equation}
	\arg(\beta - \beta_{AD_P}) = -  \frac{2}{3} \phi_{PK} - \frac{\pi}{3} + \frac{2}{3} \pi \, n
	\,, \quad
	n \in \mathbb{Z} \,.
\label{CMS_13_near_AD1}	
\end{equation}
From this equation we see that, generally speaking, three different CMS originate in the Argyres-Douglas point $AD_P$ (this is very similar to the $\mathbb{CP}(1)$ case). However, only some of them satisfy the additional condition \eqref{abs_decay_condition}. Namely, for \eqref{abs_decay_condition} to be true, we have to impose
\begin{equation}
	\arg{Z_P} - \arg(-i(m_P - m_K))  \in 2 \pi \mathbb{Z} \,.
\end{equation}
This condition leaves us with even $n$ in \eqref{CMS_13_near_AD1}.
These are the CMS for the decay \eqref{MK-MP_reaction} near the point $AD_P$. 

Let us look at the equation \eqref{CMS_13_near_AD1} more closely. 
Depending on $\phi_{PK}$, the qualitative picture changes. When it is zero, then the CMS is stretched between $AD_1$ and $AD_2$ and coincides with the primary curve \eqref{CMS_right_curve}.
When $\phi_{PK} \in [-\pi/2, 0)$, the CMS bends into the strong coupling domain, and the kink $M_K$ cannot penetrate into the weak coupling domain at $\beta > 0$.
If $\phi_{PK} \in (0, \pi/2]$, the CMS for the $K$-kink decay goes into the weak coupling domain, and the $K$-kink is present in some subregion of the weak coupling at $\beta > 0$. But of course it cannot reach the region where its mass would vanish, see \eqref{M_K_large_beta}. Other values of $\phi_{PK}$ are recovered by relabeling $1 \leftrightarrow 2$. On Fig.~\ref{fig:CMS_right_13} we present the CMS for the $[Z_3]$ kink decays (CMS for decays of $[Z_4]$ is qualitatively the same).
 
The same argument can be applied to the $P$-kinks near the dual weak coupling domain at $\beta < 0$. If $\arg\left( (m_4 - m_3) / (m_P - m_K) \right)$ is zero, the CMS for the decay \eqref{MP-MK_reaction} coincides with the dual primary curve \eqref{CMS_left_curve}. When this $\arg$ is positive, the $P$-kinks cannot penetrate into the dual weak coupling region. When it is negative, the $P$-kinks are present in a subregion of the dual weak coupling domain, but they never reach the regions where their masses would vanish.

\subsection{Decay of strong coupling tower of  higher winding states 
\label{sec:higher_winding}}

Now we briefly discuss decays of the $n \neq 0$ states of the tower \eqref{Z_higher_windings}.
In the limit $\Delta m \gg \delta m_{12} \,, \ \delta m_{34}$ they can decay into the states of lower winding with emission of bifundamentals. For example, if $n > 0$, some of the decays are
\begin{equation}
\begin{aligned}
	[Z_1^{[n]}] &~\to~ [Z_4^{[ n ]}] ~+~ [i (m_1 - m_4)]  \,, \\[2mm]
	[Z_4^{[n]}] &~\to~ [Z_2^{[ n-1 ]}] ~+~ [i (m_1 - m_3)]  \,.  
\end{aligned}
\end{equation}
More generally, the $n > 0$ states can decay as
\begin{equation}
\begin{aligned}
	[Z_P^{[n]}] &~\to~ [Z_K^{[ n ]}] ~+~ [i (m_P - m_K)]  \,, \\[2mm]
	[Z_K^{[n]}] &~\to~ [Z_P^{[ n-1 ]}] ~+~ [i (m_{\wt{P}} - m_{\wt{K}})]  \,,
\end{aligned}
\end{equation}
where $P, \wt{P}$ is some permutation of indices $1,\ 2$, and $K, \wt{K}$ is a permutation of $3,\ 4$. The states with $n < 0$ decay similarly.

The corresponding CMS satisfies the equation
\begin{equation}
	\Re \left( \frac{ Z_P^{[n]} }{m_P - m_K} \right) = 0  
	\,, \quad 
	\Re \left( \frac{ Z_K^{[n]} }{m_{\wt{P}} - m_{\wt{K}}} \right) = 0 \,.
\label{CMS_higherwind_decay}
\end{equation}
Far from the origin, when $\beta \gg 1$ in the $\mathbb{CP}(1)\;$ limit \eqref{CP1_limit}, the equations \eqref{CMS_higherwind_decay} differ from \eqref{CMS_K_decay} only by $O(\delta m_{12} / \Delta m ,\, \delta m_{34} / \Delta m)$ terms; therefore, the corresponding CMS should be close on each other, at least in some region.

Careful numerical studies show that there are two possibilities: either the CMS \eqref{CMS_higherwind_decay} form closed curves lying inside the strong coupling domain, or they form spirals that go to the origin. In any case, it follows that the higher winding states considered here live exclusively inside the strong coupling domain and cannot get into the weak coupling regions.

%
%

\section{Central charge windings at strong coupling \label{sec:Z_windings}}

In this Section we are going to derive different windings ot the central charge \eqref{BPSmass} indicated on Fig.~\ref{fig:CMS_type1}.

\begin{figure}[h]
    \centering
    \begin{subfigure}[t]{0.5\textwidth}
        \centering
        \includegraphics[width=\textwidth]{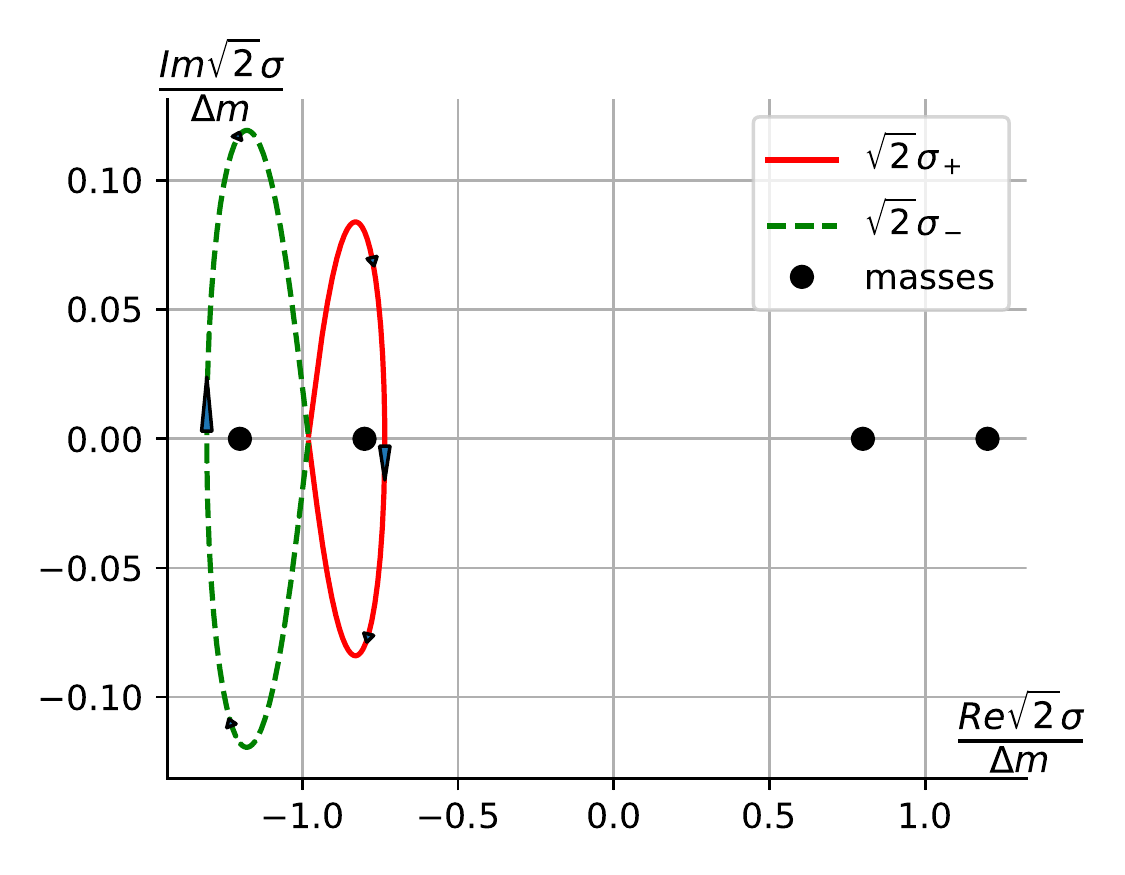}
        \caption{Trajectory of 2D roots $\sigma$ along \eqref{beta_traj_along_theta}}
        \label{fig:Z_windings_Imbeta}
    \end{subfigure}%
    ~
    \begin{subfigure}[t]{0.5\textwidth}
        \centering
        \includegraphics[width=\textwidth]{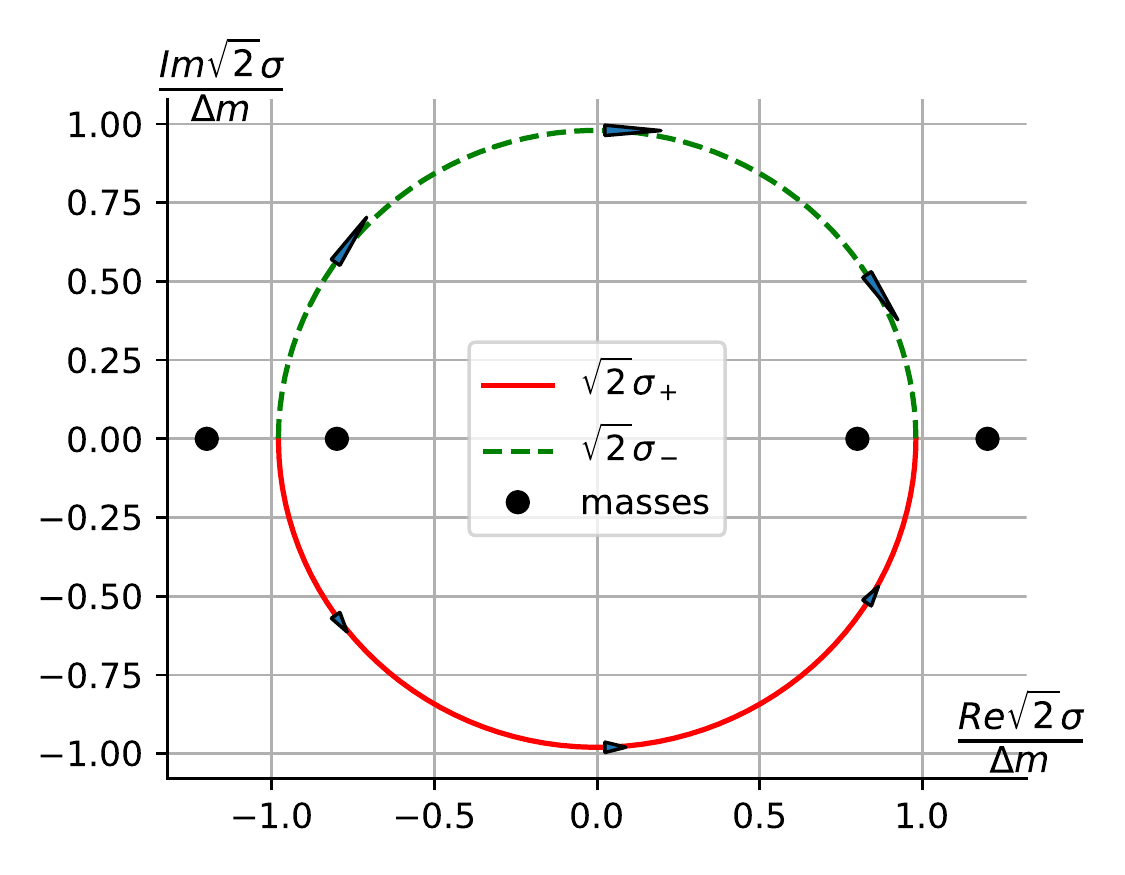}
        \caption{Trajectory of 2D roots $\sigma$ along \eqref{beta_traj_alongRe}}
        \label{fig:Z_windings_Rebeta}
    \end{subfigure}%
\caption{
	Trajectories of 2D roots $\sigma$ along different $\beta$-paths.
	Numerical results.
	Complex plane of $\sqrt{2}\sigma$.
	We see that $\sigma$-roots encircle the masses $m_A$ (represented by bullets).
}
\label{fig:Z_windings}
\end{figure}

\subsection{Winding along \boldmath{$\theta_{2d}$} \label{sec:along_theta}}

Now we are going to derive the $AD_1 \to AD_2$ phase shift from Fig.~\ref{fig:CMS_type1}. 
For simplicity we consider the $\mathbb{CP}(1)$ limit \eqref{CP1_limit}.
Positions of AD points $AD_1$ and $AD_2$ are approximately given by \eqref{AD_roots_simplecase_beta1_approx}.
Consider a trajectory in the $\beta$ plane, where the coupling flows continuously from one AD point at $\beta_{AD}$ to another at $\beta_{AD} + i$
\begin{equation}
	\beta = \frac{1}{\pi} \ln\frac{2 \, \Delta m}{\delta m_{12}}   
		+ \frac{i (t - \pi)}{2 \pi} -\varepsilon
		\, ,
		\quad
		1 \gg \varepsilon > 0 \, , \  t \in [0, 2\pi] \,.
\label{beta_traj_along_theta}		
\end{equation}
Here, $\varepsilon$ is just a regularization parameter.
Then, 
\begin{equation}
	e^{-2 \pi \beta} \approx - \left( \frac{\delta m_{12}}{2 \, \Delta m} \right)^2 \, e^{- i t} \, (1 + 2 \pi \varepsilon) \,,
\end{equation}
and for the expression under the square root in \eqref{roots_symmetric} (i.e. the discriminant) we get
\begin{equation}
	D \approx \frac{\delta m_{12}^2}{4} + \Lambda_{CP(1)}^2 =\frac{\delta m_{12}^2}{4} \, \left( 1 - (1 + \wt{\varepsilon}) e^{- i t} \right)
	\,, \quad
	1 \gg \wt{\varepsilon} > 0 \,.
\end{equation}
This expression winds around $1$ with the radius $(1 + \wt{\varepsilon})$ clockwise. Then the $\sigma$ vacua, which are approximately given by
\begin{equation}
	\sqrt{2}\sigma_\pm \approx  \pm \frac{\Delta m}{2} + \sqrt{D} \,,
\label{temp_xpm}
\end{equation}
wind, see Fig.~\ref{fig:Z_windings_Imbeta}. In the limit $\varepsilon \to 0$, the root $\sqrt{2}\sigma_+$ winds around $(- \Delta m + \delta m_{12})/2 = - m_2$ clockwise, while $\sqrt{2}\sigma_-$ winds around $(- \Delta m - \delta_{12}) / 2 = - m_1$ clockwise, both with radius $\delta m_{12} \, (1 + \wt{\varepsilon} / 2)$.

This yields nontrivial phase shifts in the mirror variables, see \eqref{mirror-x_map}. While $Y$'s stay intact, the $X_1$ winds because of $\sqrt{2}\sigma_-$ and picks up $- 2 \pi i$. $X_2$ winds because of $\sqrt{2}\sigma_+$ and picks up $- 2 \pi i$. Then, the complexified kink central charge defined by $Z = 2 ({\cal W}_{\rm mirror}(Vac_-) - {\cal W}_{\rm mirror}(Vac_+))$ is shifted by $- i (m_1 - m_2)$. Therefore if $Z_2=0$  at $AD_2$ than $Z_1=Z_2 +i(m_1-m_2)$ 
becomes zero    at  $AD_1$, see \eqref{kink_mass_P_CP1_1}. In other words $[Z_1]$ and $[Z_2]$ kinks are massless at AD points $AD_1$ and $AD_2$ respectively.

With the same reasoning we can prove the $AD_3 \to AD_4$ shift from Fig.~\ref{fig:CMS_type1}.

\subsection{From positive to negative \boldmath{$\beta$}}

Now, consider the trajectory in the $\beta$ plane going from right to left, i.e. from $AD_2$ to $AD_3$ on Fig.~\ref{fig:CMS_type1}.
For simplicity we consider the limit of real $\Delta m \gg \delta m_{12} = \delta m_{34} > 0$.
\begin{equation}
	\beta \approx t \left[ \frac{1}{\pi} \ln\frac{2 \,  \Delta m}{\delta m_{12}}   - \varepsilon \right] - \frac{i}{2}
		\, ,
		\quad
		1 \gg \varepsilon > 0 \, , \  t \in [1, -1]
		\,.
\label{beta_traj_alongRe}
\end{equation}
When $t$ changes from $1$ to $-1$, the value of $\beta$ flows from $AD_2$ to $AD_3$. Then we have
\begin{equation}
	e^{- 2 \pi \beta} \approx - \left(\frac{\delta m_{12}}{2 \, \Delta m}\right) ^ {2 t} \ (1 + 2 \pi t \varepsilon) \,,
\end{equation}
and for the expression under the square root in \eqref{roots_symmetric} (i.e. the discriminant) we get
\begin{equation}
	D \approx \delta m_{12}^2 \left( 
		1 - \frac
				{\left( \frac{2 \, \Delta m}{\delta m_{12}} \right) ^ {2 (1-t)} \ (1 + 2 \pi t \varepsilon)}
				{\left( 1 + \left( \frac{\delta m_{12}}{2 \, \Delta m} \right) ^ {2 t} \ (1 + 2 \pi t \varepsilon) \right)^2} 
	\right) \,.
\label{temp_discr}
\end{equation}
When $\delta m_{12} < \Delta m$, this expression always gives negative $D$. 
There is no nontrivial windings of the roots. However the first term in the root formula \eqref{roots_symmetric} changes smoothly from $- \Delta m / 2$ to $+ \Delta m / 2$ as $t$ is varied from $1$ to $-1$. Therefore, both SW roots evolve from the vicinity of $- \Delta m / 2$ at $\beta \sim AD_2$ to the vicinity of $+ \Delta m / 2$ at $\beta \sim AD_3$. Turns out that in terms of $\sqrt{2}\sigma_\pm$ from \eqref{temp_xpm}, the root $\sqrt{2}\sigma_+$ travels in the lower half plane, while the root $\sqrt{2}\sigma_-$ travels in the upper half plane, see Fig.~\ref{fig:Z_windings_Rebeta}.

From this and the map \eqref{mirror-x_map} it follows that, when $\beta$ flows from $AD_2$ to $AD_3$, the mirror variable $X_1$ stays in the right half plane $\Re X_1 > 0$, $Y_4$ stays in the left half plane $\Re Y_4 < 0$. $X_2$ and $Y_3$ each pick up $+i \pi$ because of the $\sigma_+$ change, while because of the $\sigma_-$ they each pick up $- i \pi$. All in all, the complexified kink central charge defined by $Z = 2 ({\cal W}_{\rm mirror}(Vac_-) - {\cal W}_{\rm mirror}(Vac_+))$ is shifted by $- i (m_2 - m_3)$, which is exactly the shift indicated on Fig.~\ref{fig:CMS_type1}. Thus $[Z_2]$ and $[Z_3]$ kinks are massless at AD points $AD_2$ and $AD_3$ respectively.

With the same reasoning we can prove the $AD_4 \to AD_1$ monodromy from Fig.~\ref{fig:CMS_type1}. 
Note that these results are consistent with the $\mathbb{Z}_2$ transformation, see Fig.~\ref{fig:CMS_type1}.

%
%

\chapter{More on self-dual couplings} \label{sec:dual_couplings}
\setcounter{section}{1}

Consider 4D self-dual points. Corresponding $\tau_\text{SW}$ should satisfy the equation
\begin{equation}
	\tau_\text{SW} = \frac{-1}{\tau_\text{SW}} \,.
\label{selfdual_eq_naive}	
\end{equation}
The solution in the upper half plane is
\begin{equation}
	\tau_0 = i \,.
\label{selfdual_tau_naive}	
\end{equation}
However, if we take into account also $T$ duality, then the equation \eqref{selfdual_eq_naive} is modified:
\begin{equation}
	\tau_\text{SW} = \frac{-1}{\tau_\text{SW}} + 2 \, k \,,
	\quad
	k \in {\mathbb Z} \,.
\label{selfdual_eq_T}	
\end{equation}
Solving this equation, we obtain a whole series of self-dual points,
\begin{equation}
	\tau_{\pm k} = k \pm \sqrt{k^2 - 1}\,,
	\quad
	k \in {\mathbb Z} \,,
\end{equation} 
or, equivalently,
\begin{equation}
	\tau_{\pm k} = \pm (k - \sqrt{k^2 - 1} ) \,,
	\quad
	k \in \{0, \, 1, \, 2, \, \ldots \} \,.
\label{selfdual_tau_T}		
\end{equation} 
For $k=0$ this gives \eqref{selfdual_tau_naive}. For $k=1$, this gives a point 
\begin{equation}
	\tau_1 = 1 \,.
\label{selfdual_tau=1}
\end{equation}

Now consider the 2D self-dual points. An obvious point \eqref{selfdual_beta_eq} is
\begin{equation}
	\beta_0 = 0 \,, 
	\quad
	e^{-2 \pi \beta_0} = +1 \,.
\label{selfdual_beta_naive}
\end{equation}
But if we take into account the 2d $T$ duality $\beta \to \beta + i$, we see that in fact there is 
a whole series of the points self-dual under $S$ \eqref{selfdual_beta_eq},
\begin{equation}
	\beta_k =  \frac{i}{2} \, k\,, 
	\quad
	e^{-2 \pi \beta_1} = (-1)^k \,,
	\quad
	k \in \mathbb{Z} \,.
\label{selfdual_beta_T}	
\end{equation}
We have seen some of them in Sec.~\ref{sec:2D_4D}:
\begin{equation}
\begin{aligned}
	\tau_0 = i &\leftrightarrow \beta_1 =  \frac{i}{2}  \,,\\[2mm]
	\tau_1 = 1 &\leftrightarrow \beta_0 = 0  \,.\\
\end{aligned}	
\end{equation}

%
%


\label{last_one_ever}

\end{document}